\documentclass[12pt]{article}
\usepackage{amsmath,amssymb,amsfonts,amsthm,graphicx,placeins, float}
\usepackage{color}
\usepackage{url}
\usepackage{longtable}
\usepackage{booktabs}
\usepackage{multirow}
\usepackage{natbib}

\title{A flexible Bayesian non-confounding spatial model for analysis of dispersed count data in clinical studies}

\newcommand{\N}{{\rm{N}}}
\newcommand{\E}{{\rm{E}}}

\newcommand{\bA}{\mbox{\boldmath$A$\unboldmath}}

\newcommand{\bB}{\mbox{\boldmath$B$\unboldmath}}

\newcommand{\bD}{\mbox{\boldmath$D$\unboldmath}}

\newcommand{\bH}{\mbox{\boldmath$H$\unboldmath}}

\newcommand{\bI}{\mbox{\boldmath$I$\unboldmath}}

\newcommand{\bP}{\mbox{\boldmath$P$\unboldmath}}

\newcommand{\bs}{\mbox{\boldmath$s$\unboldmath}}
\newcommand{\bS}{\mbox{\boldmath$S$\unboldmath}}

\newcommand{\bW}{\mbox{\boldmath$W$\unboldmath}}
\newcommand{\by}{\mbox{\boldmath$y$\unboldmath}}
\newcommand{\bY}{\mbox{\boldmath$Y$\unboldmath}}
\newcommand{\bx}{\mbox{\boldmath$x$\unboldmath}}
\newcommand{\bX}{\mbox{\boldmath$X$\unboldmath}}

\newcommand{\zero}{\mbox{\boldmath$0$\unboldmath}}

\newcommand{\bpsi}{\mbox{\boldmath$\psi$\unboldmath}}

\newcommand{\bbeta}{\mbox{\boldmath$\beta$\unboldmath}}

\newcommand{\btheta}{\mbox{\boldmath$\theta$\unboldmath}}

\newcommand{\bmu}{\mbox{\boldmath$\mu$\unboldmath}}

\newcommand{\bPhi}{\mbox{\boldmath$\Phi$\unboldmath}}

\renewcommand*\dot[1]{%
  \placeaccent{\acc@dot}{#1}%
  }
\renewcommand*\ddot[1]{%
  \placeaccent{\acc@dot\mkern1.4mu\acc@dot}{#1}%
  }
\begin{document}
\maketitle
\begin{center}
\author{Mahsa Nadifar$^{*}$, Hossein Baghishani$^*$,  Afshin Fallah$^{**}$\\ 
$^*$ Department of Statistics, Faculty of Mathematical Sciences, Shahrood University of Technology, Iran\\
 $^{**}$ Department of Statistics, Faculty of Sciences, Imam Khomeini International University, Iran \\}
\end{center}

\noindent
\textbf{Abstract}: In employing spatial regression models for counts, we usually meet two issues. First, ignoring the inherent collinearity between covariates and the spatial effect would lead to causal inferences. Second, real count data usually reveal over- or under-dispersion where the classical Poisson model is not appropriate to use. We propose a flexible Bayesian hierarchical modeling approach by joining non-confounding spatial methodology and a newly reconsidered dispersed count modeling from the renewal theory to control the issues. Specifically, we extend the methodology for analyzing spatial count data based on the gamma distribution assumption for waiting times. The model can be formulated as a latent Gaussian model, and consequently, we can carry out the fast computation by using the integrated nested Laplace approximation method. We also examine different popular approaches for handling spatial confounding and compare their performances in the presence of dispersion. We use the proposed methodology to analyze a clinical dataset related to stomach cancer incidence in Slovenia and perform a simulation study to understand the proposed approach's merits better. 

\section{Introduction}\label{sec1} 
Spatial count data models are broadly exploited in various disciplines such as disease mapping, environment, ecology, earth science, weather forecasting, biostatistics, and sociology. Such data are observations from random variables with non-negative integer values for which each observation is spatially indexed, displaying the number of times an event occurs \citep{VerHoef2018}. 
\cite{Besag1991} recommended the spatial generalized linear mixed model (SGLMM) for spatial data with discrete (areal) domain, which is a hierarchical model that includes data aggregated over spatially indexed units, such as regions, districts or countries. It introduces spatial dependency through a latent Gaussian Markov random field (GMRF) \citep{Rue2005}. 
Despite the widespread and flexible applications of SGLMM, this model fails when there is a correlation or multicollinearity between the fixed and random effects. In the spatial context, this problem is known as confounding. Indeed, confounding specifies the scheme when a likely latent variable is correlated with both the response variable and one or more covariates \citep{Thaden2018}. Since it is unobservable in most applications, it may interfere with the estimation procedure. In the spatial background, \citet{Clayton1993} and \citet{Reich2006} determined the presence of confounding between the fixed and random effects in the SGLMM. In their work, \citet{Reich2006} demonstrates that covariates taking a spatial correlation may be confounded with the spatial random effects, appearing in fixed effects estimates that are unobservable. \citet{Hughes2013} introduced an alternative model that alleviates spatial confounding and, at the same time, dimension reduction to assess both confounding and high dimensional problems. Moreover, they recognize an orthogonal projection of the spatial effects that take into account the covariates and the spatial effect. 
\citet{Thaden2018} used structural equation for negotiating with spatial confounding. As regards structural equation properties, their approach is not satisfactory for counts. Also, \cite{Azevedo2020} covered some introduced non-confounding spatial models for disease mapping, and 
developed an R software package, called \texttt{RASCO}. Here, our special interest in dealing with spatial confounding is to work with the following alternatives: 
\begin{enumerate}
\item
\citet{Reich2006} proposed a model, called the RHZ, to mitigate the confounding problem. The RHZ model projects the spatial effects into the orthogonal space spanned by the covariates.
\item
\citet{Prates2019} developed an approach called SPOCK to contract with spatial confounding. They removed spatial confounding by projecting the areas' spatial coordinates into the orthogonal space of the covariates, producing a new set of geographical coordinates.
\item
\citet{Dupont2020} introduced a novel approach, spatial+, which is based on the thin-plate spline (TPS) method. They developed TPS for spatial confounded data from a likelihood point of view.
\end{enumerate} 

A prevalent model for analyzing count data is the Poisson regression from generalized linear or generalized additive models \citep{Hastie1990}. However, in many practical data analyses, the conditional variance and mean of data are not equivalent, so the Poisson model is not satisfying. In real applications, count data can exhibit further characters, specifically under-dispersion and over-dispersion. During the years, several extended count regression models have been developed for contracting with these problems. Some concepts such as adopting a generalized linear mixed model (GLMM) \citep{Breslow1993} considered to control the over-dispersion problem in the count data, such as a negative binomial (NB) model. However, the NB model is not a suitable substitute for under-dispersed data \citep{Cameron2013}. Two alternative classes of models for accounting unobserved heterogeneity are finite mixture models \citep{Pearson1894} and hurdle models \citep{Baetschmann2014}. Likewise, we can use the hurdle model for modeling both over-dispersion and under-dispersion. Other methods include weighting the Poisson distribution \citep{Ridout2004}, the COM-Poisson distribution \citep{Lord2010}, the generalized Poisson inverse Gaussian family \citep{Zhu2009}, and the Poisson-Tweedie distribution \citep{Bonat2016} to name a few. 

In this paper, we consider the analysis of the non-confounding spatial over- or under-dispersed counts by mixing renewal theory \citep{Cox1962}, and the above-mentioned non-confounding approaches. \citet{Winkelmann1995} proposed an alternative approach for an over- or under-dispersed counts model that relates non-exponential duration (waiting times) between events. He associated the models for counts and models for the duration, relaxing the equi-dispersion assumption at the cost of an extra parameter, with renewal theory. He replaced the independently and identically exponentially distributed waiting times (which would lead to the Poisson distribution for counts) by a less restrictive non-negative distribution with a non-constant hazard function. Moreover, he noticed that if the hazard function is a decreasing (increasing) function of time, the distribution reveals a negative (positive) duration dependence. These explain that negative duration dependence causes over-dispersion and positive duration dependence would cause under-dispersion. Different analysts have offered some models dealing with this methodology. Some constructed models with this view are the gamma-count (GC) model \citep{Winkelmann1995}, the Weibull-count model \citep{McShane2008}, and the lognormal-count model \citep{Gonzales2011}. Several recent works have focused on \citeauthor{Winkelmann1995}'s point, and some R packages have been presented for it, including \texttt{countr} package \citep{Kharrat2019}. Here, we focus on the GC model, which \cite{nadifar2019} extended it for analysis of spatially correlated count data without considering confounding problem.
  
The purposes of this paper are twofold. The first is to model count responses with non-equivalent dispersion when there is collinearity between fixed and random effects, motivated by spatial confounding problems. The second purpose is to develop a Bayesian cousin of the spatial+ approach for count data. Then, we can fit the model using the INLA methodology. Furthermore, we expose how effective spatial interaction within the data influences the inferences during simulation study. Finally, we apply our proposed model to analyze a well-known clinical dataset in the spatial confounding context.

The plan for the rest of the paper is as follows. In Section \ref{Sec2}, we briefly explain the suggested confounding and non-confounding approaches. The fundamental methodology for Bayesian non-confounding spatial GC regression analysis is developed in Section \ref{Sec3}. The performance of the proposed approach is examined in simulation studies in Section \ref{Sec4} under various scenarios. Section \ref{Sec5} applies the methodology to a clinical data related to stomach cancer incidence in Slovenia \citep{Zadnik2006}. Finally, we discuss the results in Section \ref{Sec6}.
 
\section{Some existing approaches}\label{Sec2} 
We briefly describe confounding and some non-confounding spatial models for data with discrete (areal or regional) domain in the following subsections. 
\subsection{Confounding spatial model} 
We consider the hierarchical spatial modeling of areal count data that include data aggregated over spatially indexed units, such as regions, districts, or countries. This modeling approach allows the incorporation of area-specific random effects to capture unobserved spatial heterogeneity or spatial correlation that cannot be explained by the available covariates. Let $\bY=(Y_1, \ldots, Y_n)'$ be the response vector, where $Y_i$ is the  response in the $i$th area,  $i=1\ldots,n$, from the exponential families with $\E(\bY)= \bmu=(\mu_1,\ldots, \mu_n)'$. A linear predictor can be formulated as 
\begin{eqnarray} 
\label{sglmm1} 
\eta_i = g(\mu_i) = \bx'_i\bbeta + \phi_i,\qquad i=1,\ldots,n 
\end{eqnarray} 
where $\bbeta=(\beta_0, \beta_1,\ldots,\beta_{p-1})'$ is the vector of regression coefficients, $\bx_i=(1, x_{i1},\ldots,x_{ip-1})'$ is the vector of covariates for the $i$th region, $g(\cdot)$ is an appropriate link function and $\bPhi=(\phi_1,\ldots,\phi_n)'$ represents spatial effects related to regions $i=1,\ldots,n$. Conditional autoregressive (CAR) models \citep{Besag1974,Rue2005} are often used to describe the spatial heterogeneity or correlation. CAR models have numerous applications in spatial statistics \citep{Cressie1989,Pettitt2002}, and especially in disease mapping \citep{Lawson2018,Riebler2016}. These models consider spatial dependence locally across neighboring areas. The most common definition of neighborhoods is to consider two regions $i$ and $j$ as neighbors if they share a common border, denoted here as $i \sim j$. Let $\delta_i$ denote the set of neighbours of region $i$ and $|\delta_i|$ be its size. 
Intrinsic conditional autoregressive (ICAR) model \citep{Besag1991, Besag1995} is a particular case of the CAR model that has several advantages over the CAR model, both conceptually and in practice \citep[see][]{Besag1995} for the details). These models are broadly used as priors to model underlying dependency structures in Bayesian spatial hierarchical models \citep{Sorbye2013}. 
Consider a geographic region that is partitioned into sub-regions indexed by integers $1,2,\ldots,n$. The spatial interactions between regions can be modeled conditionally using the spatial effects $\bPhi$. The joint distribution for $\bPhi$ is 
\begin{eqnarray} 
\label{CAR} 
\bPhi | \tau_{\phi} \sim \mathrm{N}\left(\zero, (\tau_{\phi}\bA)^{-1}\right) 
\end{eqnarray} 
where $\tau_{\phi}$ is the spatial precision and $\bA= (\bD-\bW)$ is the precision matrix in which $\bD$ is a diagonal matrix with elements $|\delta_i|$, for $i=1,\ldots,n$, and $\bW=(w_{ij})$ is the neighborhood (proximity) matrix with entries 
\begin{eqnarray} \label{adjacency}
w_{ij}=\left\{\begin{array}{lll} 
1 & \mbox{if $i\sim j$}\\ 
0 & \mbox{if i=j} \\ 
0 & \mbox{else.} 
\end{array}\right. 
\end{eqnarray} 
The model is intrinsic in the sense that $\bA$ is singular. Moreover, the joint distribution is non-identifiable; that means the joint density is invariant to the addition of a constant. Adding a sum-to-zero constraint solves the problem. 
\subsection{Non-confounding spatial models} 
We now review the models previously listed for alleviating the spatial confounding. \\
\textbf{RHZ model.} 
\citet{Reich2006} revisited the problem of spatial confounding and proposed an alternative method to mitigate the confounding in the spatial models. The model looks at a random effect that belongs to the orthogonal space of the fixed effects predictors. In this standpoint, the spatial effect is divided into two parts. That is $\bPhi = \bPhi^{\bx} + \bPhi^{\perp}=\bH\bPhi_1 + \bB\bPhi_2$, where $\bH$ is a $n\times q$ matrix that has the same span as the design matrix $\bX$, $\bB$ is a $n\times (n-q)$ matrix whose columns lie in the orthogonal space of $\bX$ and $\bPhi_1$ and $\bPhi_2$  are $\bPhi$ divisions that are vectors with dimensions $q$ and $n - q$, respectively. 
Hence,  the linear predictor \eqref{sglmm1} could be rewritten as 
\begin{eqnarray*} 
\eta_i = g(\mu_i) = \bx’_i\bbeta + \bH_i\bPhi_1 + \bB_i\bPhi_2 
\end{eqnarray*} 
where $\bH_i$ and $\bB_i$ are the $i$th rows of $\bH$ and $\bB$, respectively.
\citet{Reich2006}  represented that $\bH$ causes confounding in the model. Accordingly, they advocated to take out the $\bH$ component letting to the RHZ model as 
\begin{eqnarray*} 
\label{rhz} 
\eta_i = \bx'_i\bbeta + \bB_i\bPhi_{2},~~~~\bPhi_2 |\tau_{\phi} \sim \N\left(\zero, \left(\tau_{\phi}\bB'\bA\bB\right)^{-1}\right). 
\end{eqnarray*} 
\textbf{SPOCK model.} 
Although the RHZ model is adequate to mitigate the confounding, there are two demerits: \textbf{1)} The  model does not hold parameters in the precision matrix $\bA$; \textbf{2)} It did not take advantage of the Markov property, and hence, sparsity in the original spatial model \eqref{sglmm1} \citep{Prates2019}. Therefore, \cite{Prates2019} proposed a new approach for alleviating confounding, called the Spatial Orthogonal Centroid Korrection (SPOCK) model. This model specifies a projected image of the original graph into the orthogonal space of the design matrix $\bX$. Indeed, the main idea of this methodology is based on misplacing the original centroids and creating the new adjacency matrix. While restoring the spatial information from the model, this current arrangement does not share it with the fixed effects. 
Let $\bS =\{\bs_i=(s_{i1},s_{i2})', i=1,\ldots,n\}$ be a set of geographical centroids with corresponding adjacency matrix, $\bW$, in \eqref{adjacency}. The first step is multiplying  $\bS$ by the projection matrix ($\bP^{\bot}= \bI - \bX(\bX^{\prime}\bX)^{-1}\bX^{\prime}$) into the orthogonal space to $\bX$ and obtain the new geographical centroids, $\bS^{*}$. In the second step, we specify the new neighborhood structure using two alternative methods: 1) Knn: fixing the number of neighbors of each area in the original graph; 2) Delaunay: defining the number of neighbors automatically using Delaunay triangulation. \cite{Prates2019} showed that the Knn method is preferable to the Delaunay approach; therefore, we consider it to achieve a new graph, $\bW^{*}$, and replace it in \eqref{CAR}. \\
\textbf{Spatial+ model.} 
\citet{Dupont2020} proposed a novel approach, called the spatial+ model. Their main idea is formulating model as a partial TPS model. Preliminary, the linear predictor in the equation \eqref{sglmm1} could be rewritten as 
\begin{eqnarray*} 
\label{spatial+1} 
\eta_i = \bx'_i\bbeta + f^+_{tps}(s_i),
\end{eqnarray*} 
where $f^+_{tps}(\cdot)$ is a TPS function as a smooth term for spatial effect and $\bs = (s_1,\ldots,s_n)$ denotes 
 the spatial locations. To obtain the spatial+ model, \citet{Dupont2020} assumed the covariates $\bx_i$ is formulated as 
\begin{eqnarray} 
\label{spatial+2} 
\bx_i = f_{tps_x} (s_i) + \varepsilon~~~~ \varepsilon\sim\N(0, \tau_x),
\end{eqnarray} 
where $f_{tps_x}(\cdot)$ means that $\bx$ is correlated with the smooth term $f^+_{tps}(\cdot)$ through the component $f_{tps_x}(\cdot)$. Finally, the linear predictor for spatial+ model can be written as 
\begin{eqnarray} 
\label{spatial+3} 
\eta_i = r(\bx_i)\bbeta + f^+_{tps}(s_i),
\end{eqnarray} 
where $r(\bX)=(r(\bx_1),\ldots , r(\bx_n))' = \bX-\hat{f}_{tps_x}(\bs)$ are the residuals in the TPS regression model \eqref{spatial+2} and $\hat{f}_{tps_x}(\bs)$ are the corresponding fitted values. 

\section{Non-Confounding Spatial Dispersed Count Model}\label{Sec3} 
Here, we develop a Bayesian non-confounding spatial dispersed count model based on the gamma-count distribution.
\subsection{Model}
We briefly review the essential properties of the GC model. As \citet{Winkelmann2013} has noticed, the count and the duration view are just two different representations of the same underlying stochastic process. From a statistical viewpoint, the distribution of cumulative waiting times uniquely determine the distribution of counts and vice versa. This relationship can be employed to derive new count data distributions \citep{Winkelmann1995, McShane2008, Gonzales2011, Ong2015}. For example, the Poisson distribution corresponds to exponential inter-arrival times between events. The GC distribution has been proposed based on gamma-distributed inter-arrival times by \cite{Winkelmann1995}. 

Let $\{u_k, k\geq 1\}$ be the waiting times between the $(k-1)$th and $k$th events. Therefore, the arrival time of the $n$th event is given by
\[\vartheta_n=\sum_{k=1}^n u_k,~~~n=1,2,\ldots .\]
Let $Y_t$ denote the total number of events that have occurred between $0$ and $t$. Hence, $\{Y_t, ~ t>0\}$ is a counting process and for a fixed $t$, $Y_t$ is a count variable. The stochastic properties of the counting process (and consequently of the count variable) are entirely determined once we know the joint distribution function of the waiting times, $\{u_k,~k\geq 1\}$. In particular, $Y_t<n$ if and only if $\vartheta_n>t$. Therefore,
$f_{Y_t}(n)=F_n(t)-F_{n+1}(t)$,
in which $F_n(T)$ is the distribution function of $\vartheta_n$. Generally, $F_n(t)$ is a complicated convolution of the underlying densities of $u_k$'s, which makes it analytically intractable. However, by using the theory of renewal processes \citep{Cox1962},
 a significant simplification arises if $u_k$'s are identically and independently distributed with a common distribution.
Here, we assume that $\{u_k,~k\geq 1\}$ is a sequence of independently and identically gamma distributed variables, $Gamma(\alpha, \gamma)$, with mean ${\rm E}(u_k)=\alpha/\gamma$ and variance ${\rm Var}(u_k)=\alpha/\gamma^2$.
It can be shown that if $Y_t$ denotes the number of events within $(0,t)$ interval, it is a GC distributed variable with parameters $\alpha$ and $\gamma$, denoted by $Y_t\sim {\rm GC}(\alpha, \gamma)$. The probability mass function of $Y_t$ is given by 
\begin{eqnarray}\label{f2}
f_{Y_t}(y)=G(y\alpha,\gamma t)-G((y+1)\alpha,\gamma t), ~~~~~ y=0,1,2,\ldots,
\end{eqnarray}
where
$G(n\alpha,\gamma t)=\frac{1}{\Gamma(n\alpha)}\int_0^{\gamma t}v^{n\alpha-1}e^{-v}dv$,
and $G(0,\gamma t)=1$. For non-integer $\alpha$, no closed form expression is available for $G(y\alpha,\gamma t)$ and thus for $f_{Y_t}(y)$.
For $\alpha=1$, the distribution of $u_k$ reduces to the exponential, and \eqref{f2} simplifies to the Poisson distribution with the parameter $\gamma t$. More importantly, for positive duration dependence, $\alpha>1$ and the GC distribution is under-dispersed; for negative duration dependence, $0<\alpha<1$ and the GC distribution is over-dispersed. We refer readers to \cite{Winkelmann1995}
 for more details about the definition and properties of the GC distribution.

For developing a GC regression model, we can relax the assumption of a homogeneous population by formulating a conditional model in which the mean of the count variable depends on a vector of covariates, $\bx=(1, x_1,\ldots,x_{p-1})^\prime$. The mean of GC distribution is
$
{\rm E}(Y_t)=\sum_{k=1}^{\infty}G(k\alpha ,\gamma t)
$,
that has no closed form. Therefore, extending a regression model based on the mean is not straightforward.
Assuming that the length of the time interval is the same for all observations, we can set $t$ to unity, without loss of generality. This results in the following regression model \citep{Zeviani2014}:
\begin{eqnarray*}\label{f4}
{\rm E}(u_{k_i}|\bx_i)=\frac{\alpha}{\gamma_i}=\exp\left(-\bx_i^{'}\bbeta\right),
\end{eqnarray*}
where $\bbeta$ is the $p\times 1$ vector of regression coefficients with the first element as the intercept.
We should notice that the regression model is defined on the waiting times $u_{k_i}$ instead of $Y_i$, where $u_{k_i}$ is the generic representation of waiting times for the $i$th observation. Its origin is for failure to establish the equality ${\rm E}(Y_i|\bx_i)=\big({\rm E}(u_{k_i}|\bx_i)\big)^{-1}$ unless for $\alpha=1$. Indeed, given the inverse relationship between gaps and the number of occurrences, the minus sign behind $\bbeta$ is due to the reverse effect of covariates on waiting times instead of counts; the longer the expectation of time interval, the fewer the number of occurrences. Therefore, the GC regression model is developed from inherent parametric assumptions that nest the Poisson regression model by a singular parametric constraint.

From \eqref{f4}, one can write 
$
\gamma_i =\alpha\exp(\bx_i^{'}\bbeta)
$.
Therefore, given a sample of independent observations $\{(y_i,\bx_i),i=1,\ldots, n\}$, the GC regression model can be written as 
$
Y_i|\bx_i ;\alpha,\bbeta\sim {\rm GC}(\alpha,\alpha\exp(\bx_i^{'}\bbeta))
$.
\citet{nadifar2019} introduced the spatial GC regression model as the form 
\begin{eqnarray} 
\label{NCSGCR} 
Y_i |\alpha, \bbeta, \phi_i &\sim & {\rm GC}\left(\alpha, \alpha\exp\left(\eta_i\right)\right),\quad i=1,\ldots,n \cr
\eta_i &=& \bx'_i\bbeta + \phi_i, 
\end{eqnarray} 
where $\alpha$ is the dispersion parameter, and $\bx_i$, $\bbeta$, and $\phi_i$ are the same as \eqref{sglmm1}. 
Here, our idea is to address both confounding and dispersion problems simultaneously to analyze count data in a unified Bayesian framework. For alleviating the spatial confounding, we will use the three approaches mentioned earlier in Section \ref{Sec2}.
As we described in Section \ref{Sec2}, developing the Bayesian GC model for both SPOCK and RHZ methods is relatively straightforward. For the SPOCK model, it suffices to obtain the new adjacency matrix using multiplying the $\bP^{\bot}$ by the available adjacency matrix and replacing it in the model \eqref{NCSGCR} \citep{Prates2019,Azevedo2020}. 
According to \cite{Dupont2020}, a spatial+ model arose from a partial TPS model. Therefore, the predictor in a spatial+ GC regression model would be defined by \eqref{spatial+3}. 
Implementing likelihood-based inferences in such models nearly always involves intractable integrals which is a severe computational difficulty, particularly by increasing the number of observations. Hence, we develop our proposed model in a Bayesian framework. 

\subsection{Bayesian inferences} 
For analysis of non-confounding spatial GC regression model, in a Bayesian framework, it is necessary to choose some suitable prior distributions for parameters of the model, $\alpha$, $\bbeta$, $\tau_\phi$, $\tau_x$, $f_{tps_x}(\bs)$, and $f^+_{tps}(\bs)$, that can reflect our prior beliefs about them. The vector of parameters $\bbeta$ is assumed to have independent zero-mean Gaussian priors with fixed variances. We determine a PC prior \citep{Nadifar2021,Simpson2017} for $\alpha$, and it is embedded in the INLA arsenal. We also consider PC priors for the precision parameters of the spatial effect and $\bx$'s error in the equation \eqref{spatial+2} \citep{Simpson2017,Gomez2020}. Finally, to develop a Bayesian version of the spatial+ model so that we can use the recommended INLA methodology, we consider a two-dimensional second-order random walk model (RW2D) for the TPS functions, $f_{tps_x}(\bs)$ and $f^+_{tps}(\bs)$, following \cite{Wang2012}. These prior distributions are flexible enough to represent the prior beliefs via the appropriate choice of their hyper-parameters. We suppose that the parameters are a priori independent, and by accepting them, the joint posterior density for SPOCK and RHZ models can be written as 
\begin{eqnarray} 
\label{bayesianGCSR} 
\pi(\alpha, \bbeta, \bPhi, \tau_{\phi} |\by)\propto \prod_{i=1}^{n} {\rm GC}\left(\alpha , \alpha\exp\left(\eta_i\right)\right)\pi(\bbeta)\pi(\alpha)\pi(\bPhi)\pi(\tau_{\phi}),
\end{eqnarray} 
and for spatial+ model is as follows: 
\begin{eqnarray} 
\label{bayesianGCS+} 
\pi(\alpha, \bbeta, f_{tps_x}(\bs), \tau_{x}, f^+_{tps}(\bs)|\by)&\propto &\prod_{i=1}^{n} \left\{{\rm GC}\left(\alpha , \alpha\exp\left(\eta_i\right)\right)\pi(\bbeta)\pi(\alpha)\right.\cr
&~~~&~~~\left. {\rm RW2D}(f_{tps_x}(\bs))\pi(\tau_{x}){\rm RW2D}(f^+_{tps}(\bs))\right\}. 
\end{eqnarray} 
The conventional approach to inference for the models \eqref{bayesianGCSR} and \eqref{bayesianGCS+} is based on MCMC sampling. It is well known, however, that MCMC methods have serious problems, regarding both convergence and computational time, when applied to such models \citep{Rue2009}. Particularly, the complexity of the proposed model for large spatial data could lead to several hours or even days of computing time to implement Bayesian inference via MCMC algorithms. To overcome this issue, \cite{Rue2009} introduced the INLA method that is a deterministic algorithm and provides accurate results in seconds or minutes. INLA combines Laplace approximations \citep{Tierney1986} and numerical integration in a very efficient manner to approximate posterior marginal distributions. Let $\btheta=(\alpha, \tau_{\phi}, \tau_x)^\prime$ denote the hyper-parameters of the models \eqref{bayesianGCSR} and \eqref{bayesianGCS+}. Let also $\bpsi$ denote the $\ell\times 1$ vector of latent variables that is $(\bbeta,\bPhi)^\prime$ in \eqref{bayesianGCSR} or $(\bbeta,\bPhi,f_{tps_x}(\bs),f^+_{tps}(\bs))^\prime$ in \eqref{bayesianGCS+}, where $\ell$ is the appropriate dimension depending on the related model. In practice, the primary interest lies in the marginal posterior distributions for  elements of the latent variables vector and hyper parameters, respectively, given by
\[\pi(\psi_j|\by)=\int\pi(\psi_j,\btheta|\by)d\btheta=\int\pi(\psi_j|\btheta,\by)\pi(\btheta|\by)d\btheta,~~j=1,\ldots,\ell,\] 
and 
\[\pi(\theta_k|\by)=\int\pi(\btheta|\by)d\btheta_{-k},~~k=1,2,3,\] 
where $\btheta_{-k}$ is equal to $\btheta$ with the eliminated $k$th element. The essential feature of INLA is to use this form to construct nested approximations 
\begin{eqnarray*} 
\tilde{\pi}(\psi_j|\by)&=&\int\tilde{\pi}(\psi_j|\btheta,\by)\tilde{\pi}(\btheta|\by)d\btheta, \\ 
\tilde{\pi}(\theta_k|\by)&=&\int\tilde{\pi}(\btheta|\by)d\btheta_{-k}, 
\end{eqnarray*} 
where Laplace approximation is applied to carry out the integrations required for evaluation of $\tilde{\pi}(\psi_j|\btheta,\by)$. 
A crucial success of INLA is its ability to compute model comparison criteria, such as deviance information criterion (DIC; \citeauthor{Spiegelhalter2002}, \citeyear{Spiegelhalter2002}) and Watanabe-Akaike information criterion (WAIC; \citeauthor{Watanabe2012}, \citeyear{Watanabe2012}; \citeauthor{Gelman2013}, \citeyear{Gelman2013}), and various predictive measures, e.g., conditional predictive ordinate (CPO; \citeauthor{Pettit1990}, \citeyear{Pettit1990}), to compare the complexity and fit of different possible models. Our proposed GC model has already implemented in the \texttt{R-INLA} package as a \texttt{family} argument with the name "\texttt{gammacount}". Further, both RHZ and SPOCK models are implemented in the \texttt{RASCO} R package in which the INLA could be applied.

\section{Experimental Assessment}\label{Sec4} 
We conducted a simulation study with two primary purposes: First, assessing the performance of our proposed model with different dispersion situations  in the presence of spatial confounding. To this end, we compared the model with some historical alternatives, the Poisson (Pois), negative binomial (NB), and generalized Poisson (GP) \citep{Zamani2012} models. Second, comparing the ability of proposed Bayesian non-confounding models, RHZ, SPOCK, and Spatial+ (denoted as S+), for alleviating spatial confounding. In our evaluation, we added two confounding models as well: the model with an ICAR effect given by (2) (denoted as a parametric spatial (PS) model), and the model with a non-parametric spatial effect, modeled by a RW2D process (denoted as a non-parametric spatial (NPS) model). Also, to evaluate both confounded and non-confounded covariates simultaneously in the same model, we brought up both of them in the linear predictor of the model. As a result, we deliberated the estimators of the covariate effects amongst different dispersed and non-dispersed count models. Finally, we controled the degree of spatial confounding according to \cite{Thaden2018} via variability on the covariates scale. We applied the following linear predictor to generate the data
\begin{eqnarray}\label{ysim} 
\eta_i &=& \beta_1x_{1i}+\beta_2x_{2i}+ \phi_i, ~~~~~i=1,\ldots, 192,\\ 
\label{xf}x_{2i} &=& -0.8\phi_i + e_x, ~~~~~~~~~~~~~~~~~~~ e_x\sim N(0,\tau_x), 
\end{eqnarray} 
  where $\bPhi = (\phi_1, \ldots , \phi_n)$ follows the ICAR model \eqref{CAR} with $\tau_\phi = 3.33$. 
The model used to generate the data is a GC regression model with the linear predictor \eqref{ysim} by considering $\beta_{1}=0.7$ and $\beta_{2}=-1$. 
Since the parameter $\alpha$ indicates the various dispersion conditions in the GC model, we considered a range of variations for this parameter in $\{0.5,1,1.3, 2\}$. It is useful to remind that $\alpha=1$, $\alpha>1$, and $\alpha<1$ are corresponding to equivalent-dispersion, under-dispersion, and over-dispersion, respectively. Furthermore, a GC model with $\alpha=1$ corresponds to a Poisson model. The covariate $x_1$ was simulated from the normal distribution, $N(0,0.5)$; covariate $x_2$ is confounded by spatial effect, $\bPhi$, and it was generated from \eqref{xf}. We considered three values for $\tau_x$ in $\{1, 4, 11\}$ to control the degree of spatial confounding. 
Furthermore, according to the Slovenia map, the number of regions was chosen to have similar conditions to real data. The number of replications was also fixed at $R=200$. 
As non-confounding models recover $\beta^{*} = \beta + (\bX^{'}\bX)^{-1}\bX^{'}\bPhi$ \citep{Prates2019}, we considered $\beta^{*}$ instead of $\beta$ in all computing measures for non-confounding models. 

We computed several measures to evaluate the performance of the models during all scenarios of dispersion and confounding degree as well. The efficiency of the fixed effect estimators is assessed by relative bias (RB) and squared error (SE), which are defined as  
$ \frac{\hat{\beta}}{\beta} - 1$ and $(\hat{\beta}-\beta)^2$, respectively. Note that $\beta$($\hat{\beta}$) should be replaced by $\beta^*$($\hat{\beta^{*}}$) for non-confounding models. We also calculated WAIC and DIC as model selection criteria and Log-score (LS; \citeauthor{Adrion2012}, \citeyear{Adrion2012}) and mean squared prediction error  (MSPE) of fitted values regarding prediction power. 
\begin{figure}[ppt]
\centerline{\includegraphics[width=500pt,height=15pc]{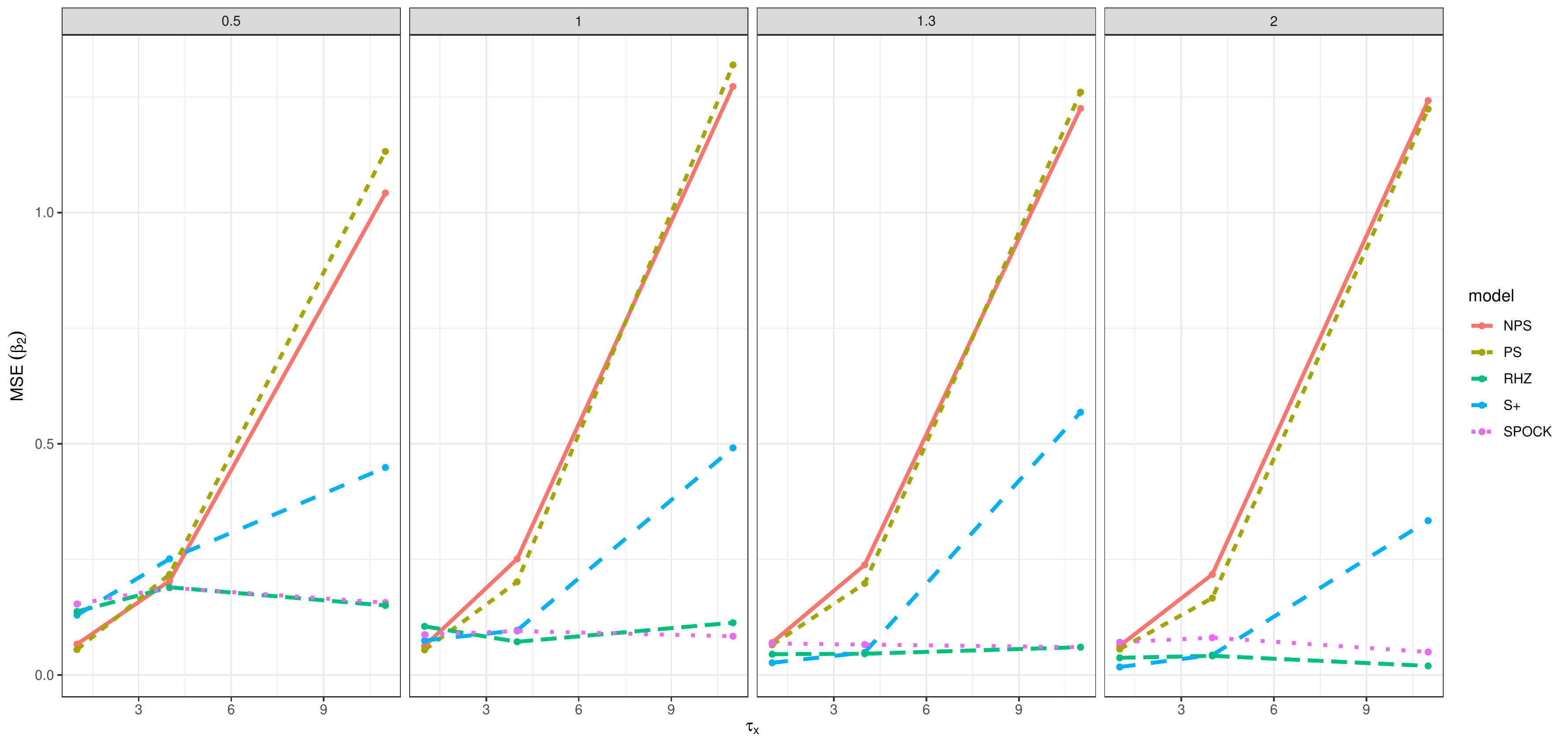}} 
\caption{MSE of estimated effect of confounded covariate, $\hat{\beta}_2$, for different scenarios: from the left over-dispersion (first column), equivalent-dispersion (second column), under-dispersion (third and forth columns) under different degrees of confounding.\label{fig-mse}}
\end{figure} 
 
\begin{figure}[ppt]
\begin{center} 
\begin{tabular}{cc} 
\includegraphics[width=230pt,height=6.8pc]{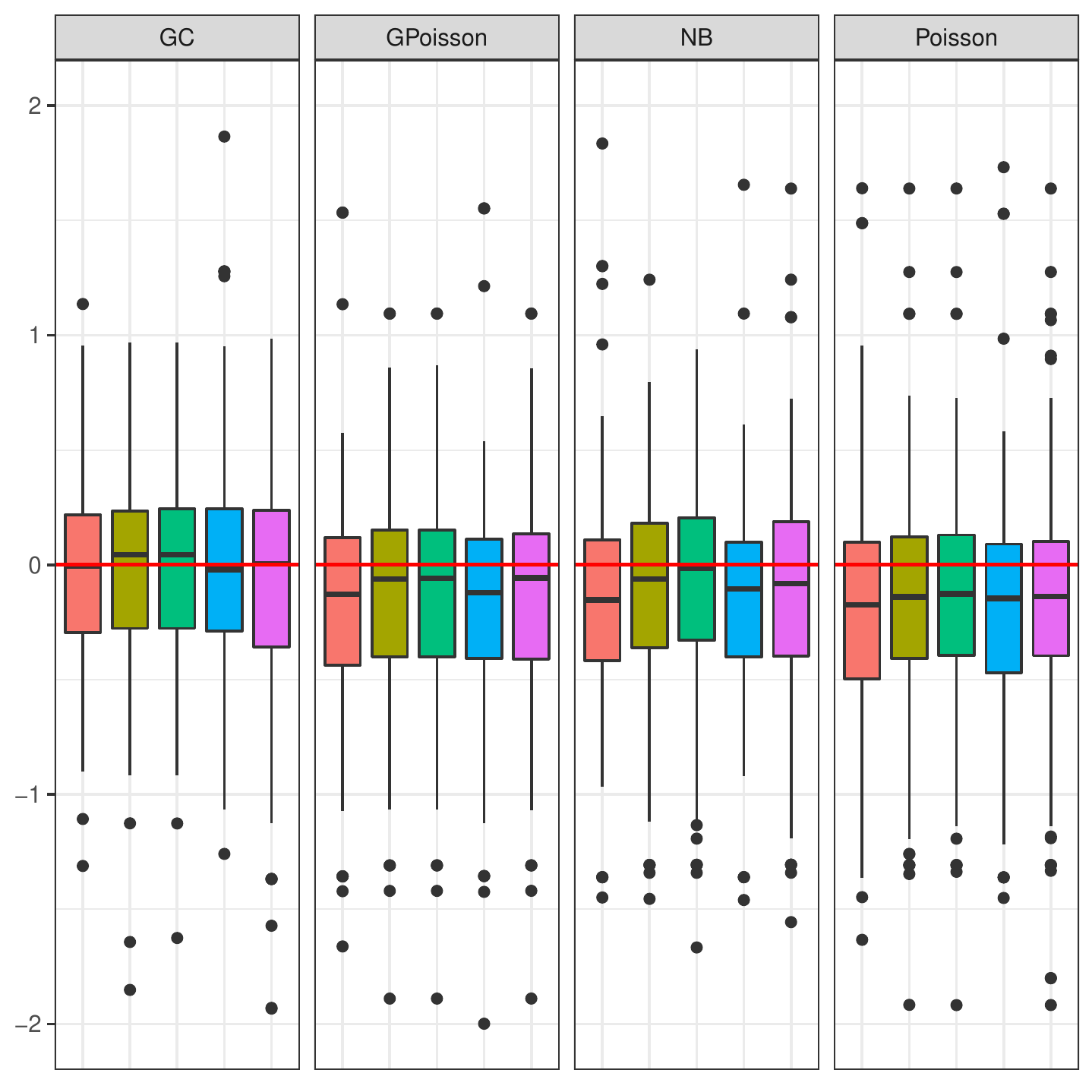} &
 \includegraphics[width=230pt,height=6.8pc]{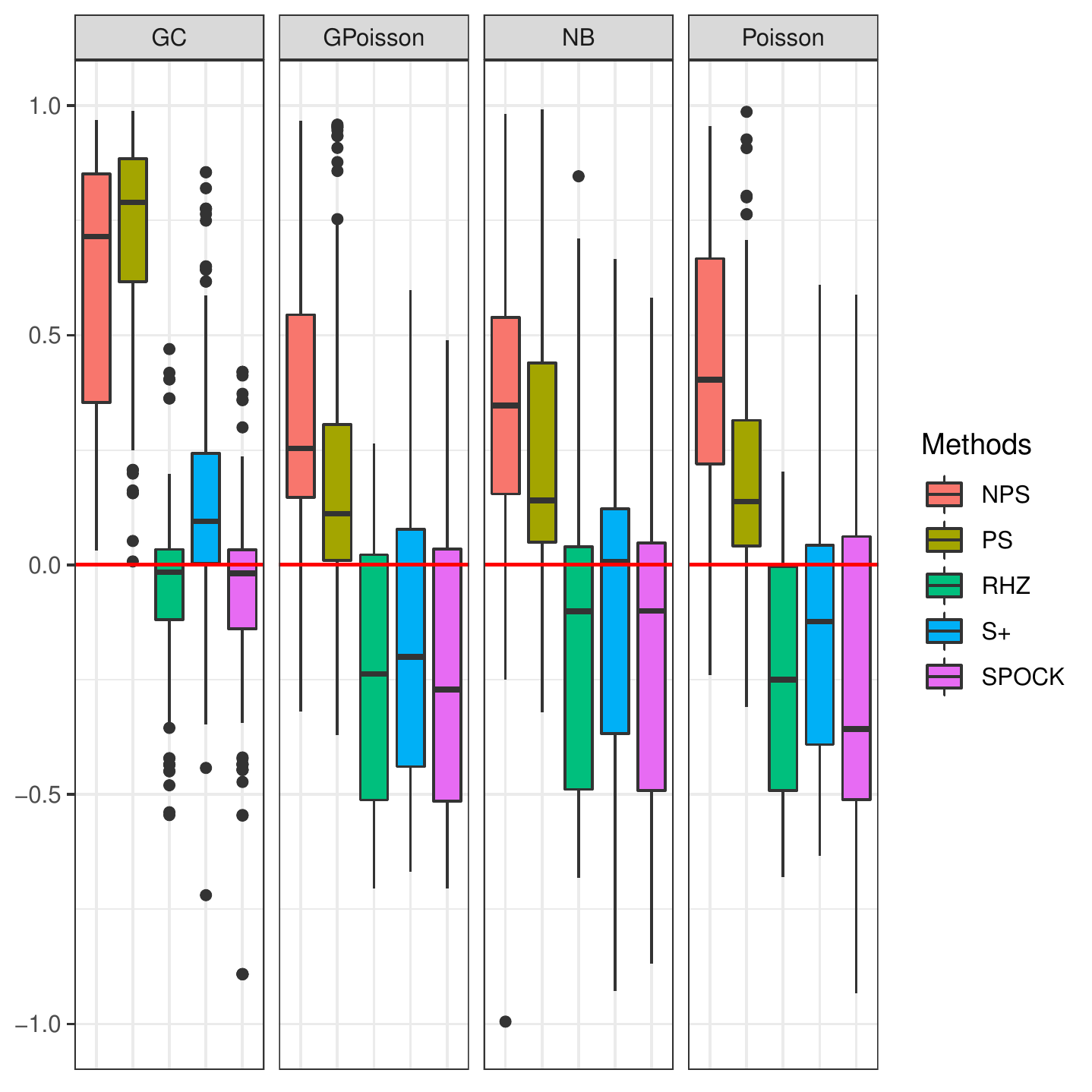}\\ 
\includegraphics[width=230pt,height=6.8pc]{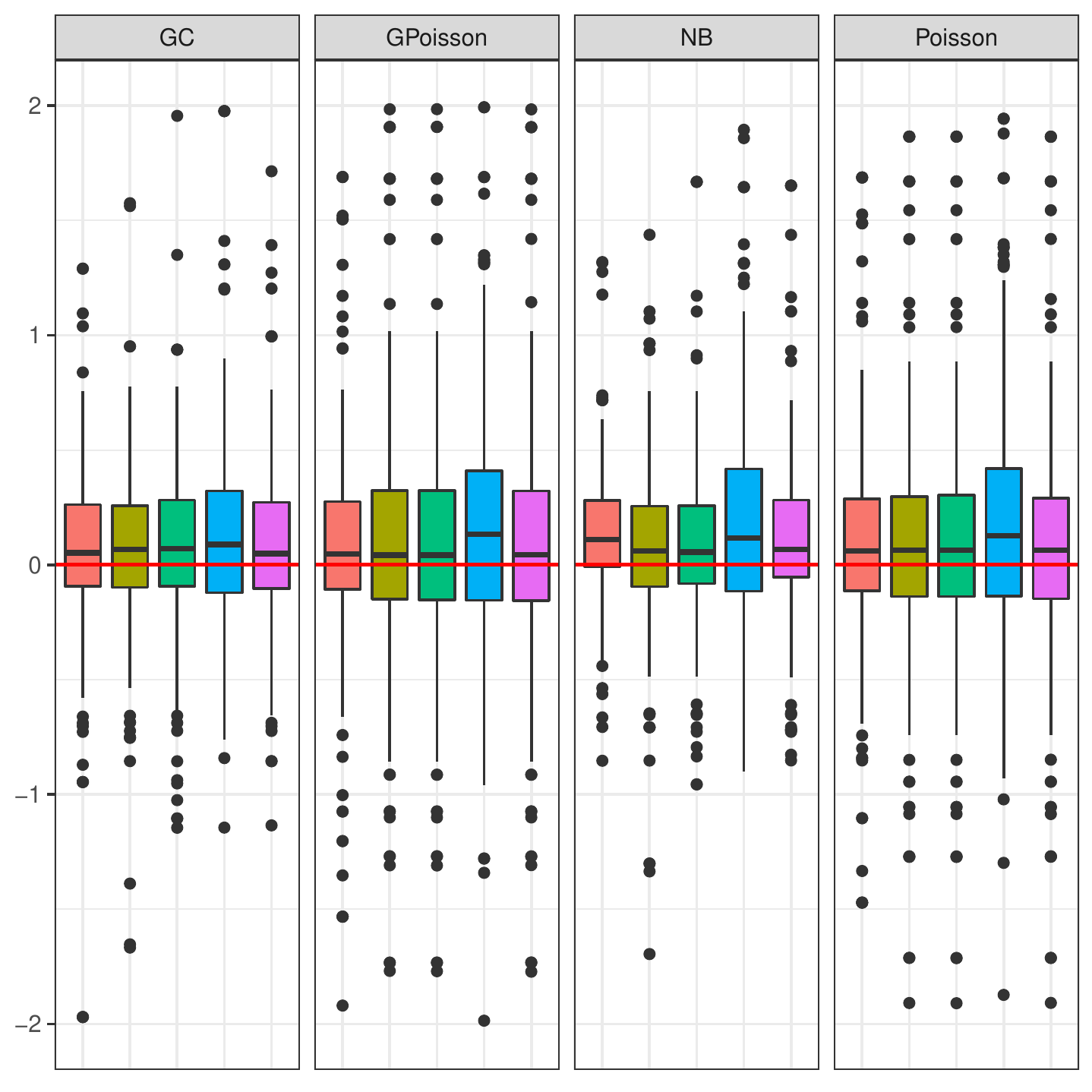} & 
\includegraphics[width=230pt,height=6.8pc]{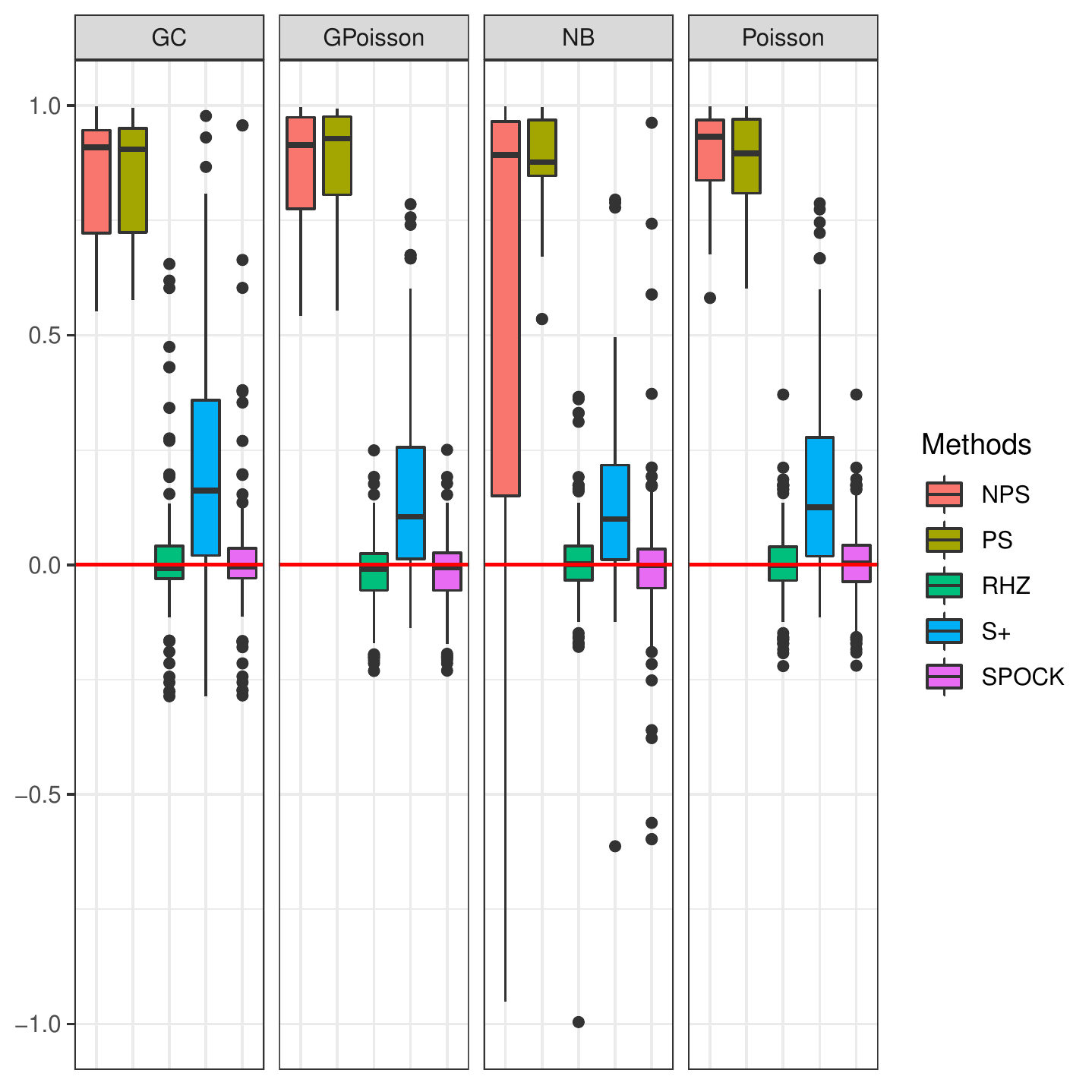}\\ 
\includegraphics[width=230pt,height=6.8pc]{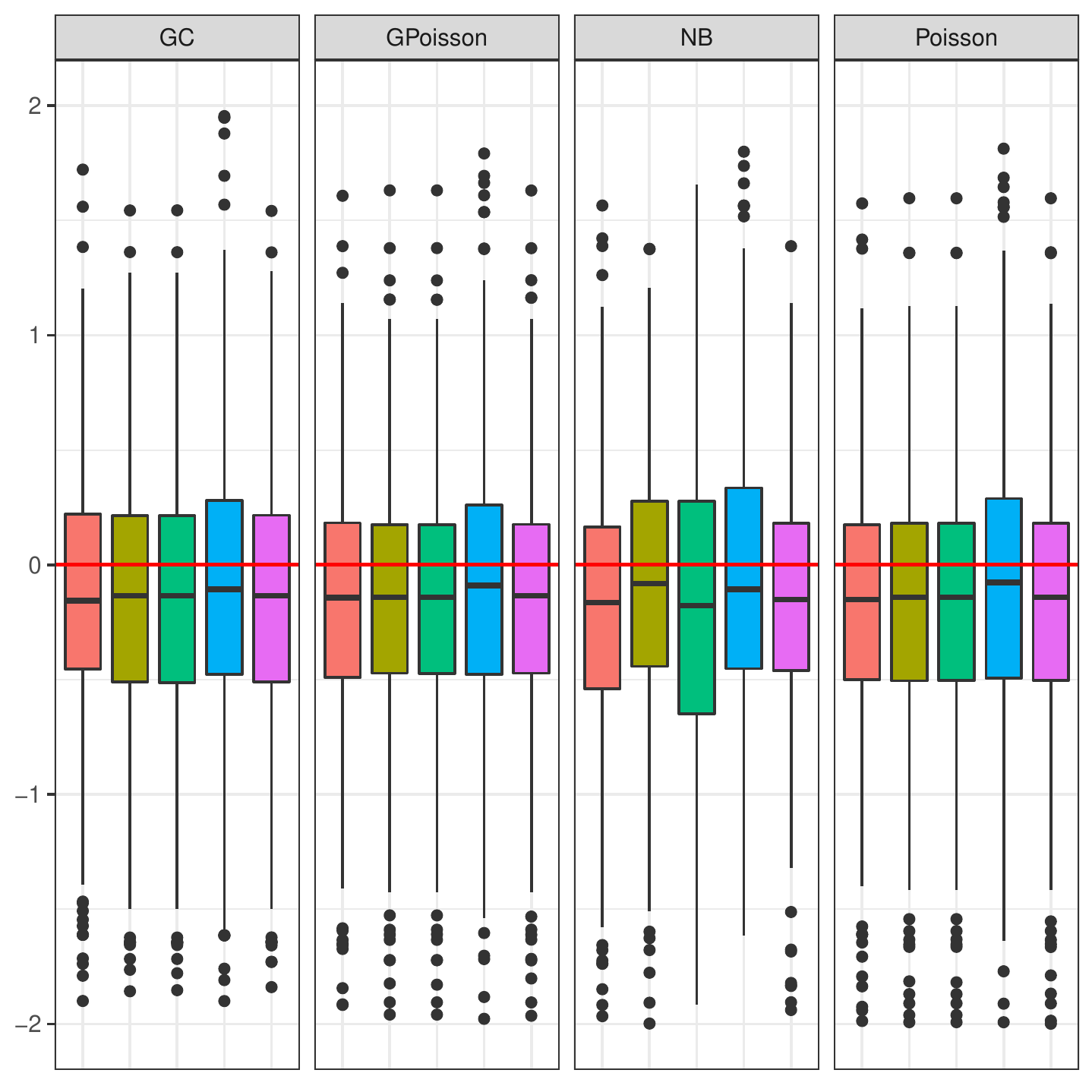} & 
\includegraphics[width=230pt,height=6.8pc]{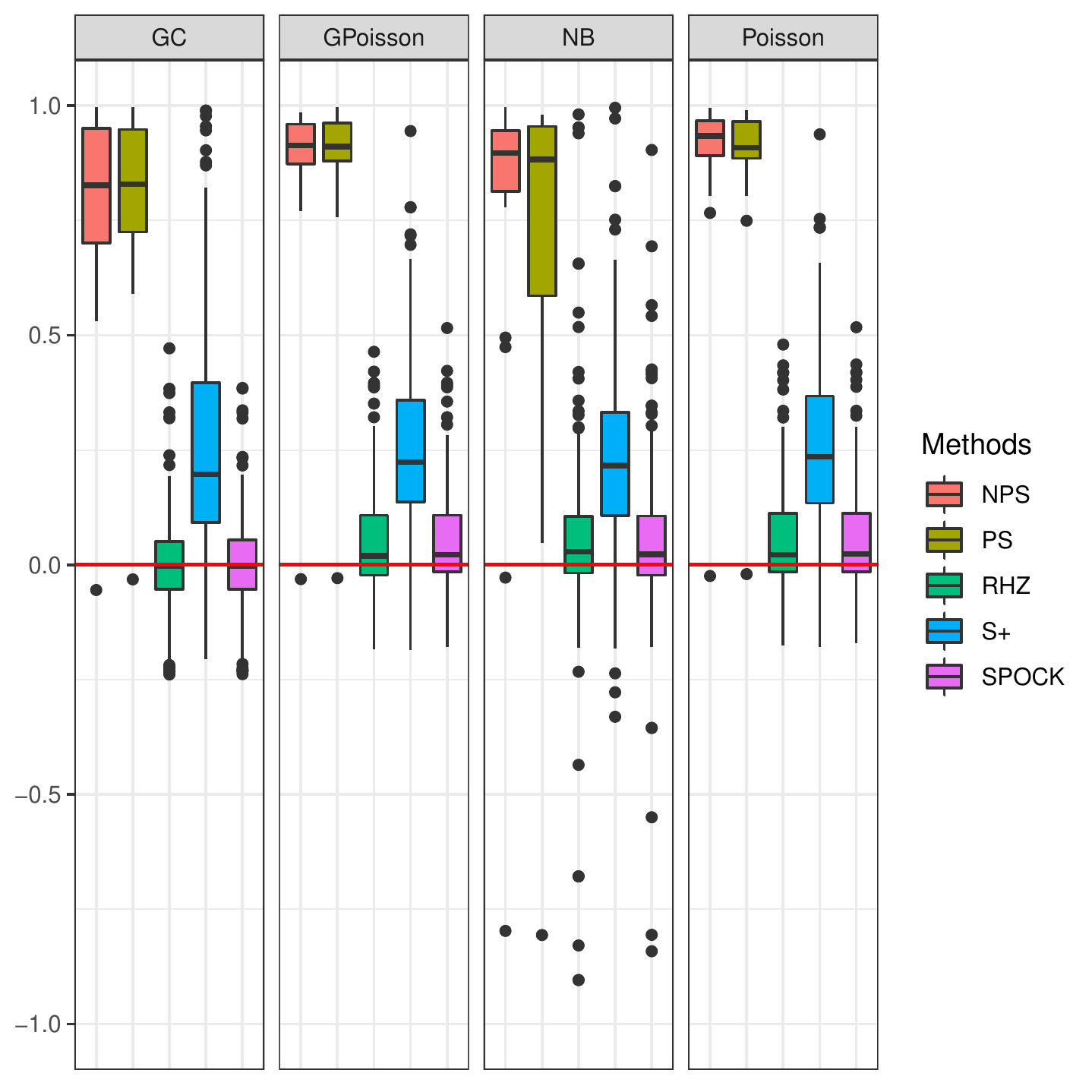}\\ 
\includegraphics[width=230pt,height=6.8pc]{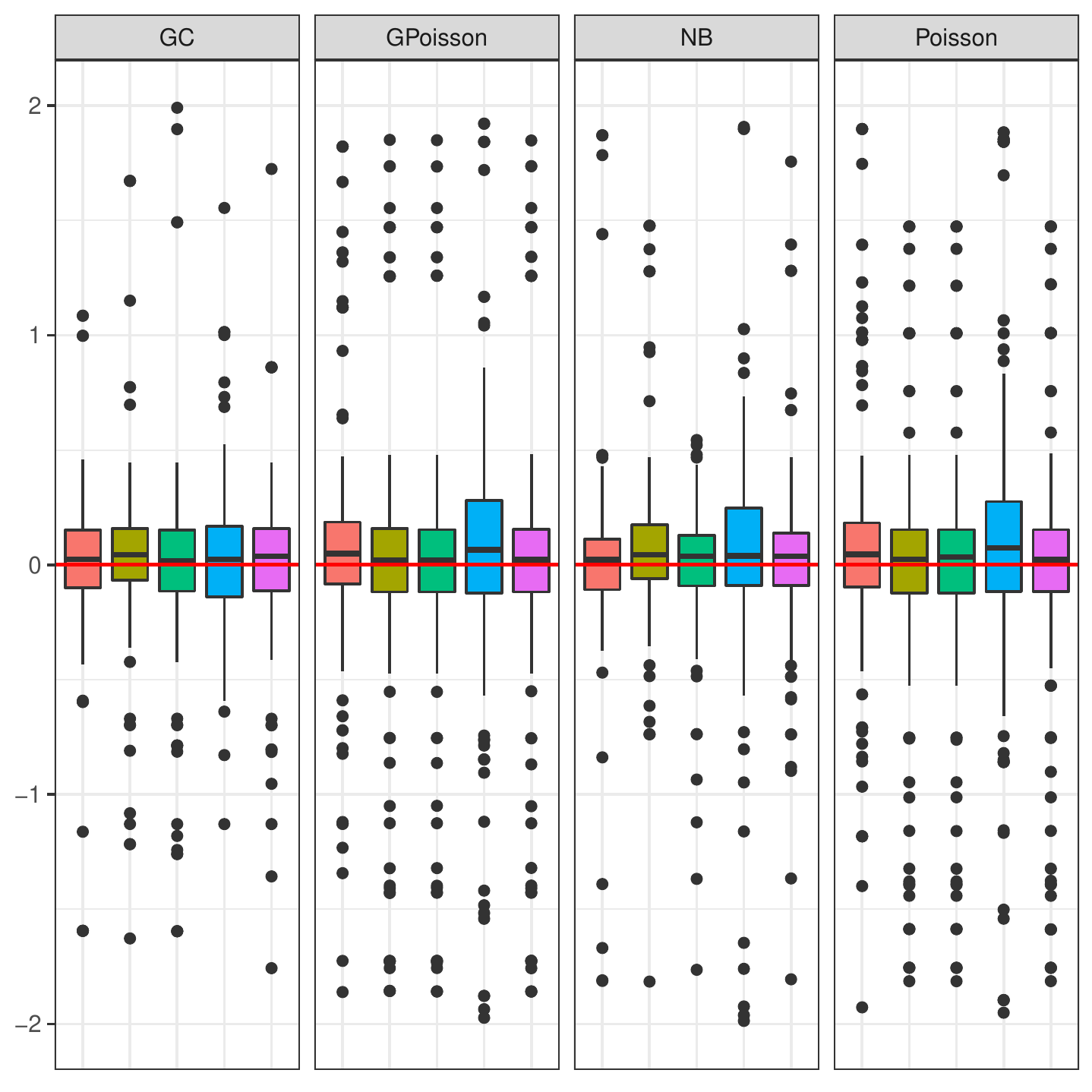} & 
\includegraphics[width=230pt,height=6.8pc]{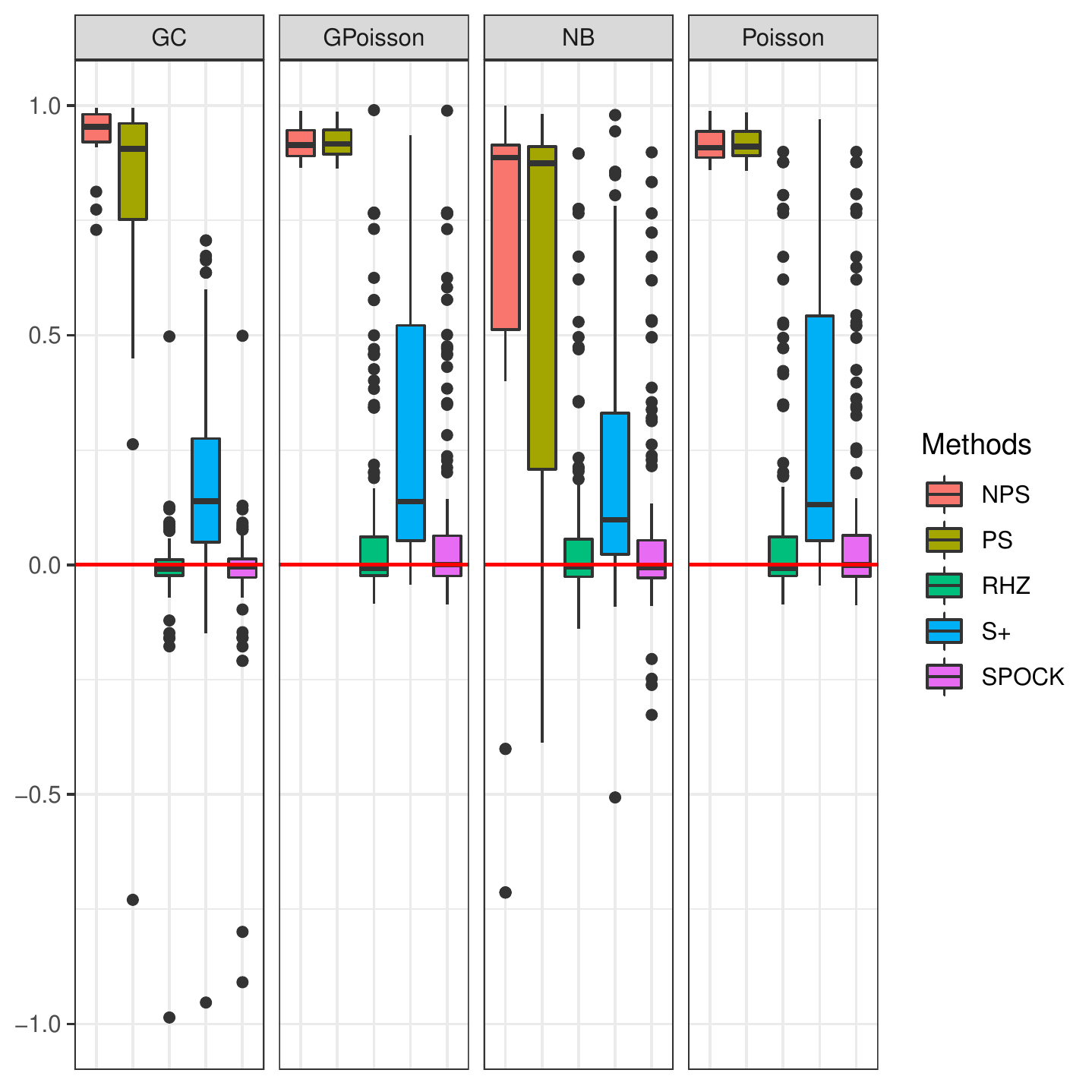}
\end{tabular} 
\end{center} 
\caption{Estimated relative bias of non-confounded covariate effect, $\beta_1$, (left column),  confounded covariate effect, $\beta_2$, (right column)  for each 
model with $\tau_x=11$ for scenarios: over-dispersion (first row), equivalent-dispersion (second row), and under-dispersion (third row; $\alpha=1.3$ and forth row; $\alpha=2$).\label{fig1}} 
\end{figure}

\begin{figure}[ppt] 
\begin{center} 
\begin{tabular}{cc} 
\includegraphics[width=230pt,height=6.8pc]{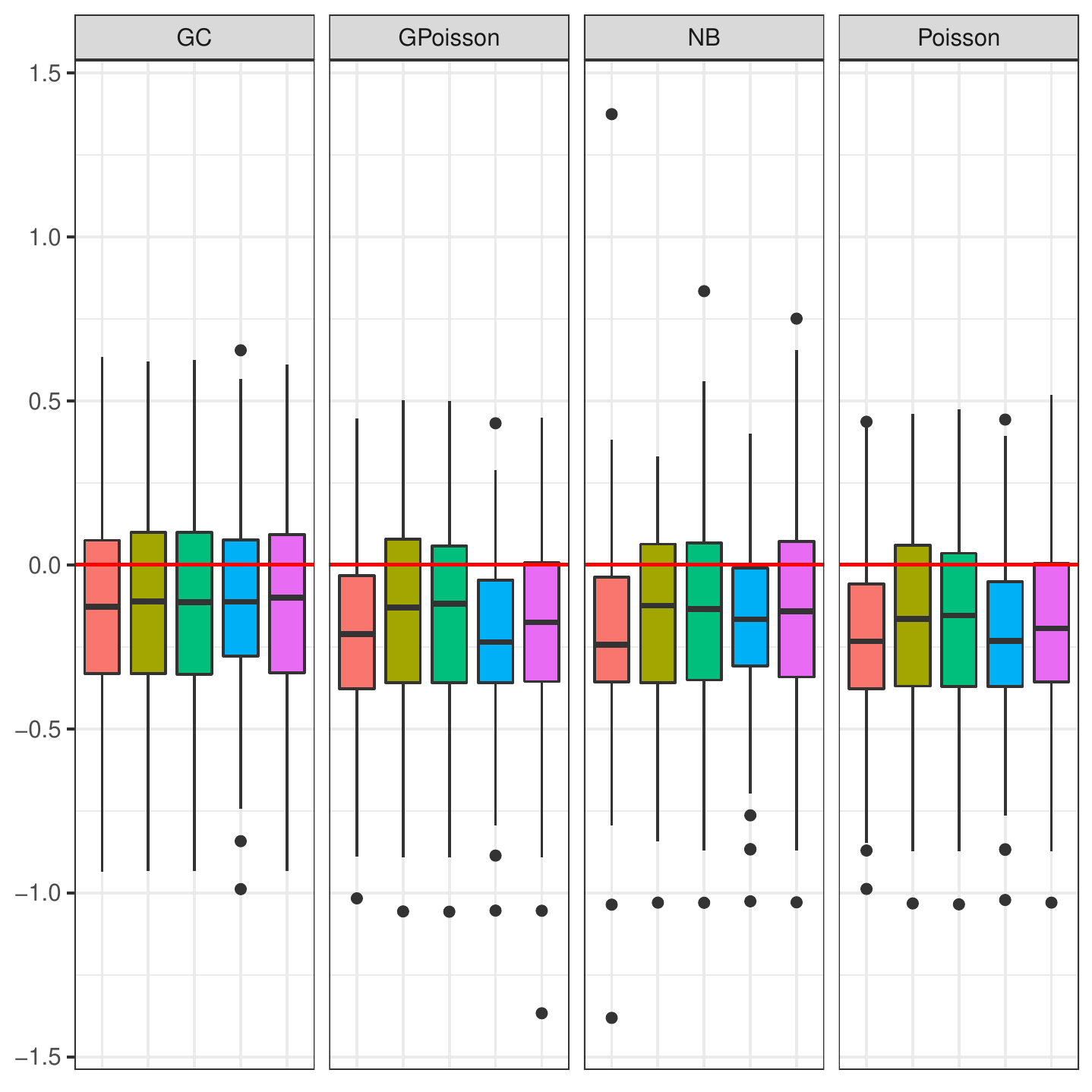} &
 \includegraphics[width=230pt,height=6.8pc]{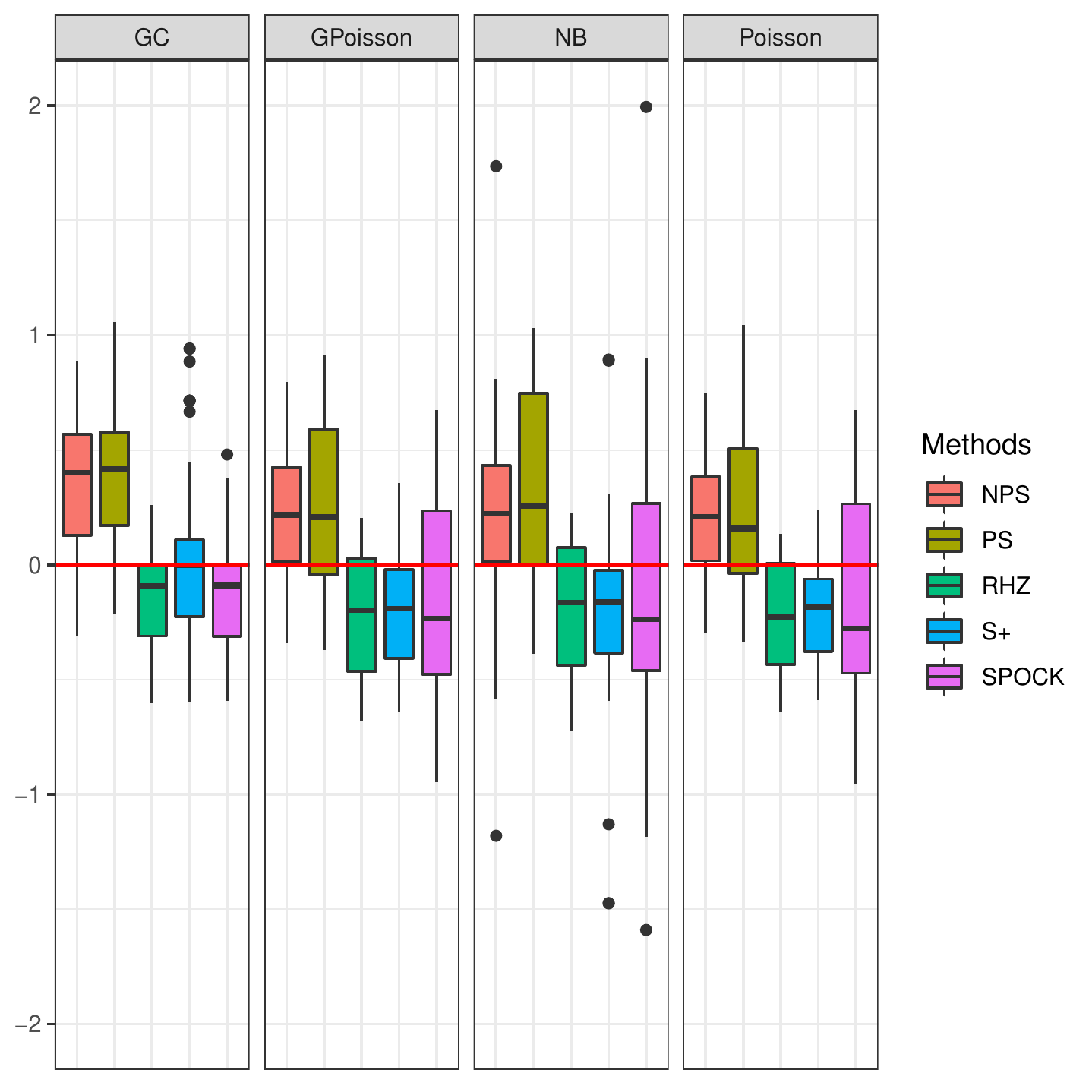}\\ 
\includegraphics[width=230pt,height=6.8pc]{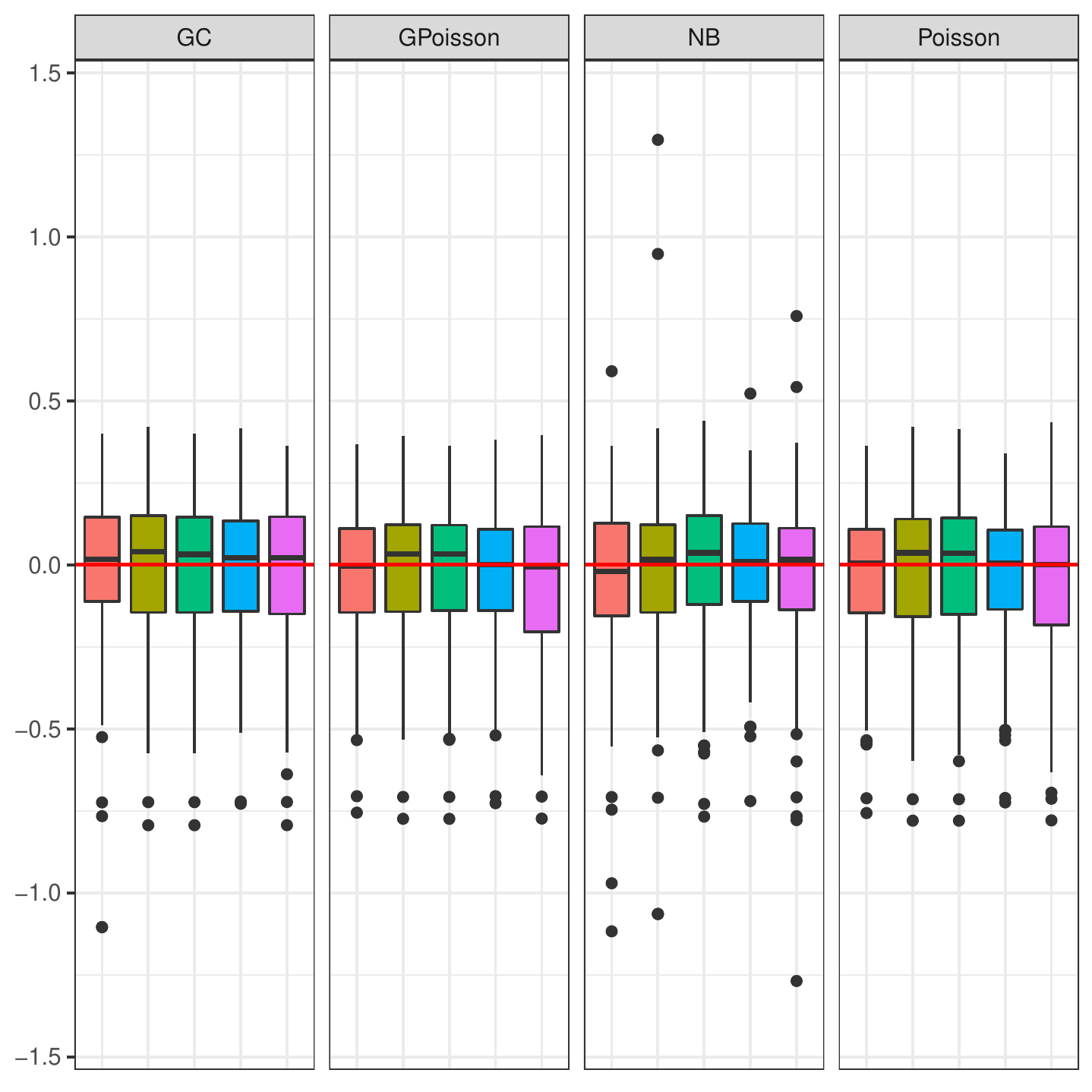} & 
\includegraphics[width=230pt,height=6.8pc]{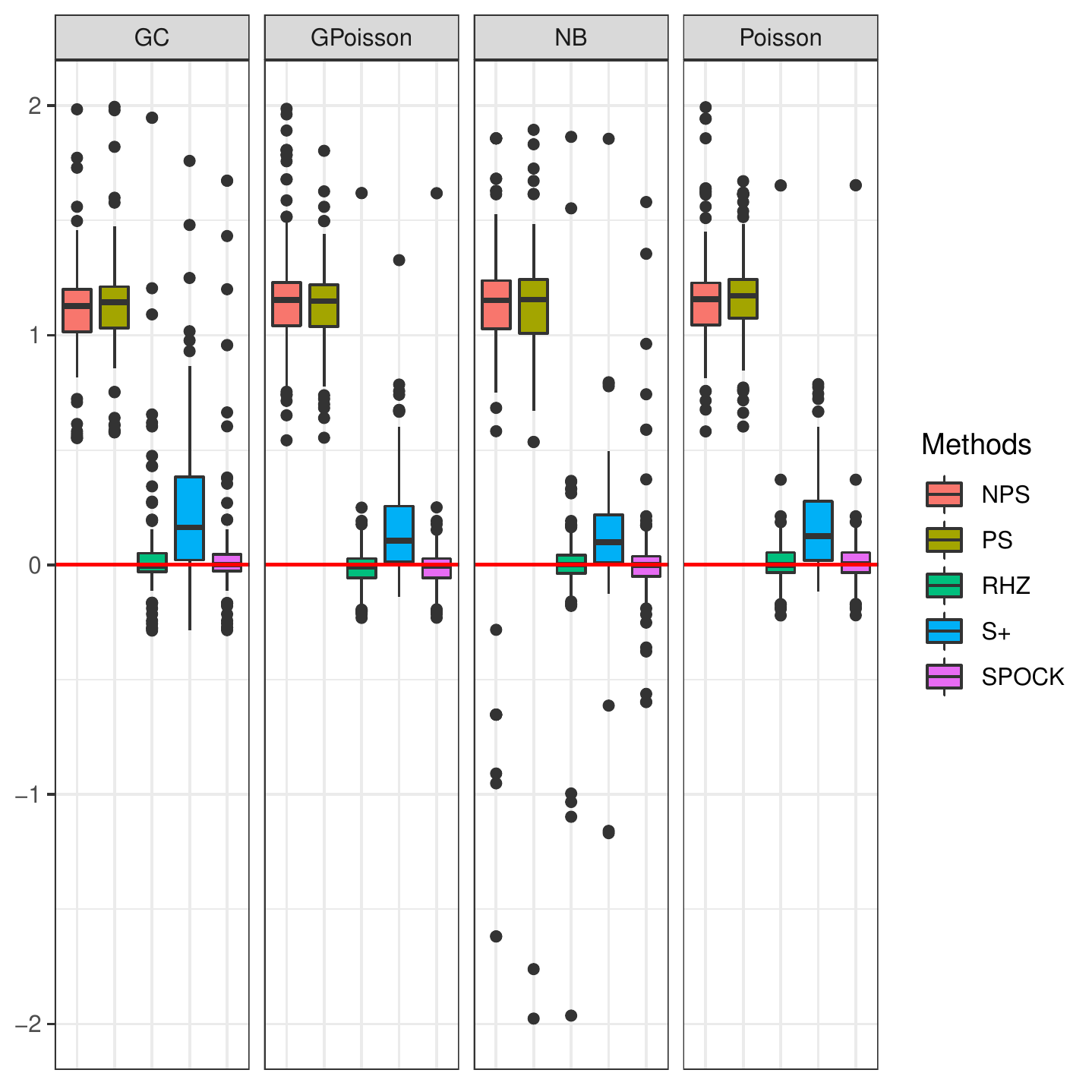}\\ 
\includegraphics[width=230pt,height=6.8pc]{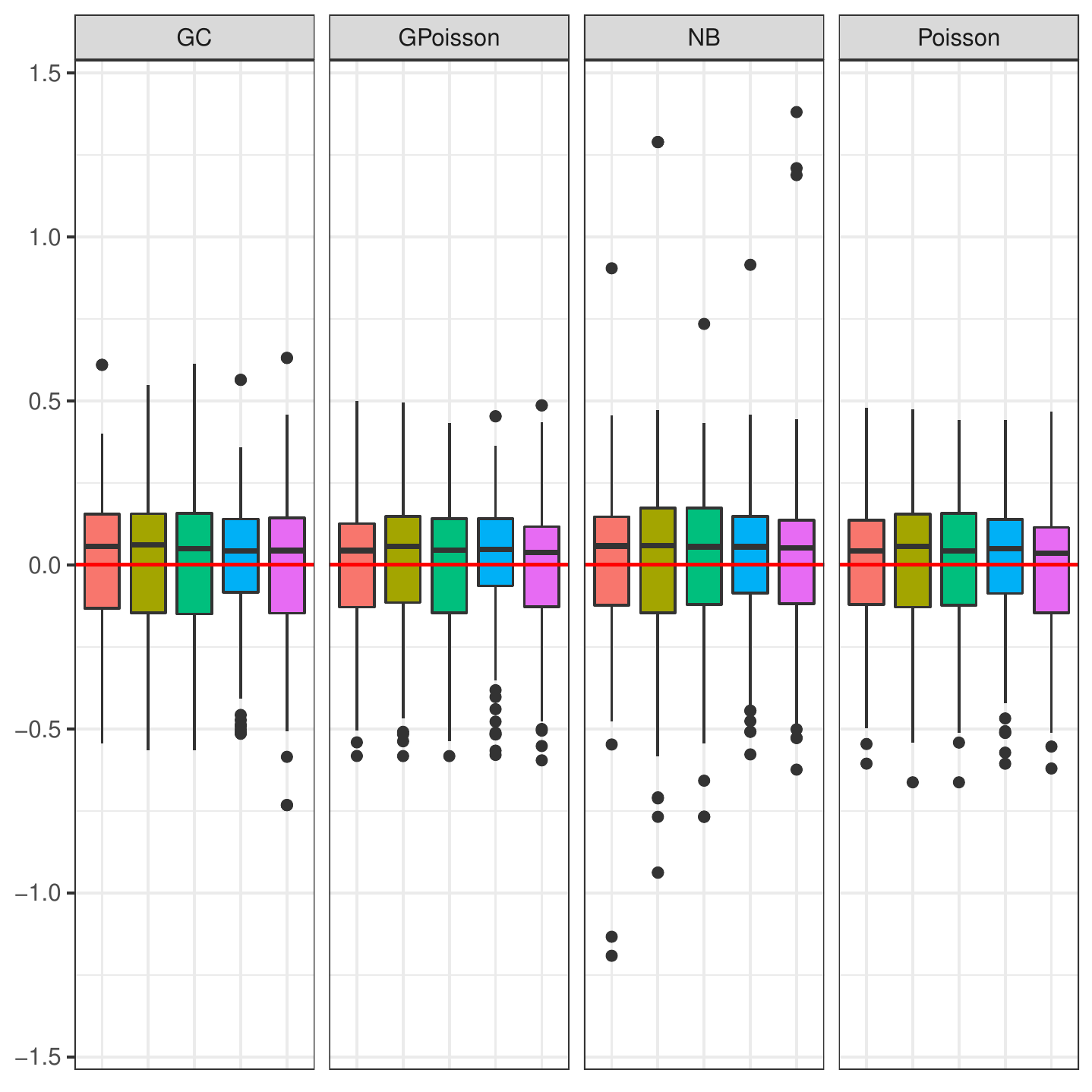} & 
\includegraphics[width=230pt,height=6.8pc]{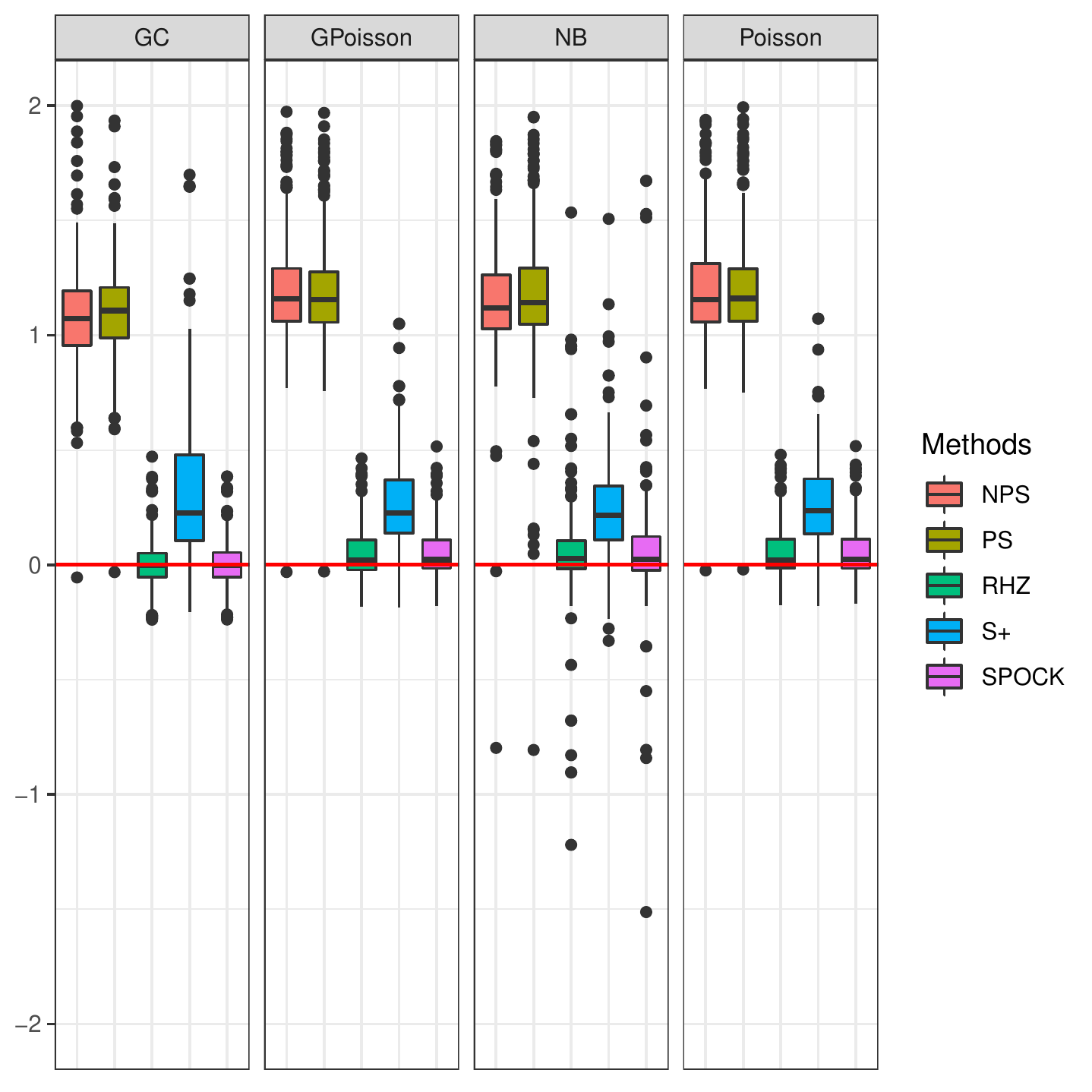} \\ 
\includegraphics[width=230pt,height=6.8pc]{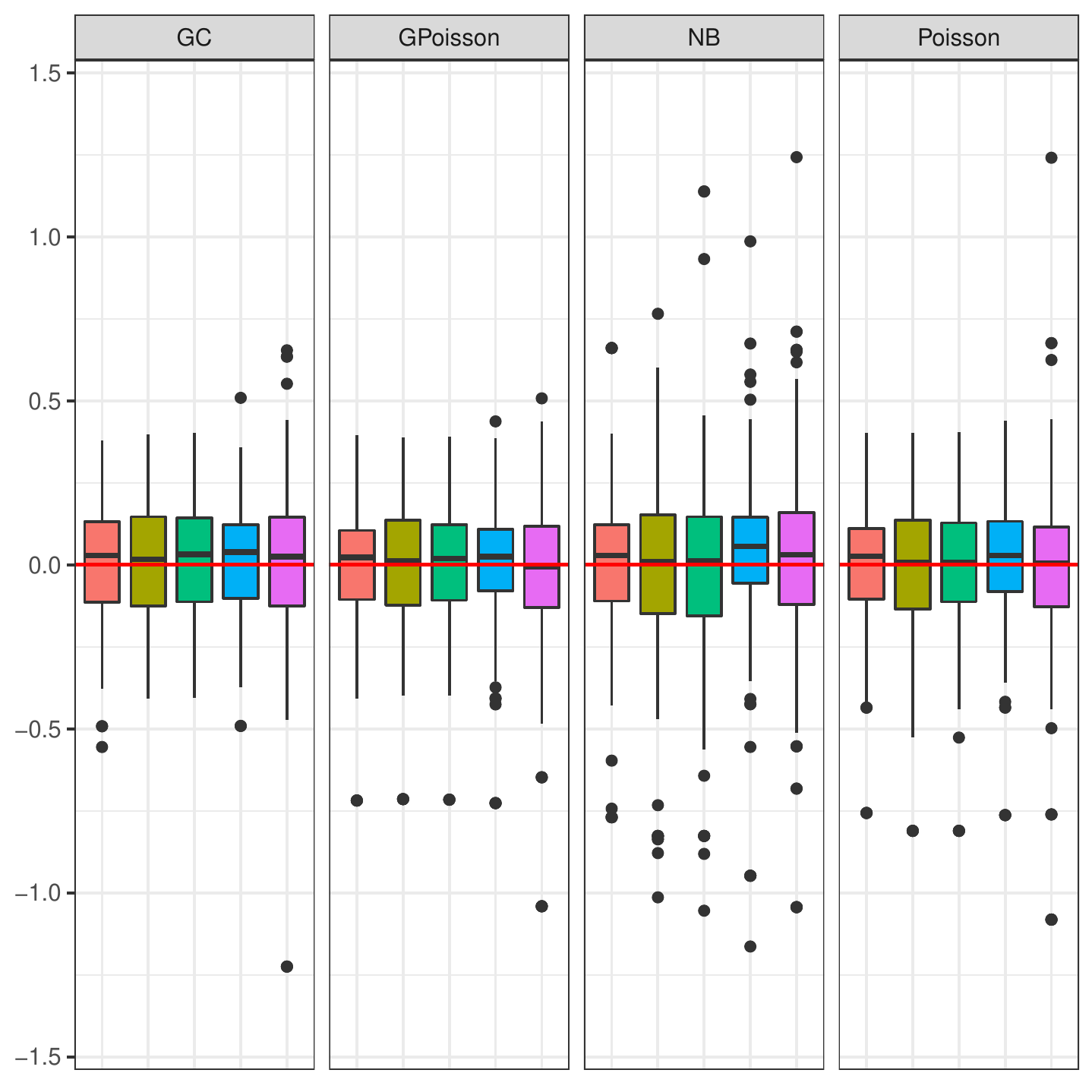} & 
\includegraphics[width=230pt,height=6.8pc]{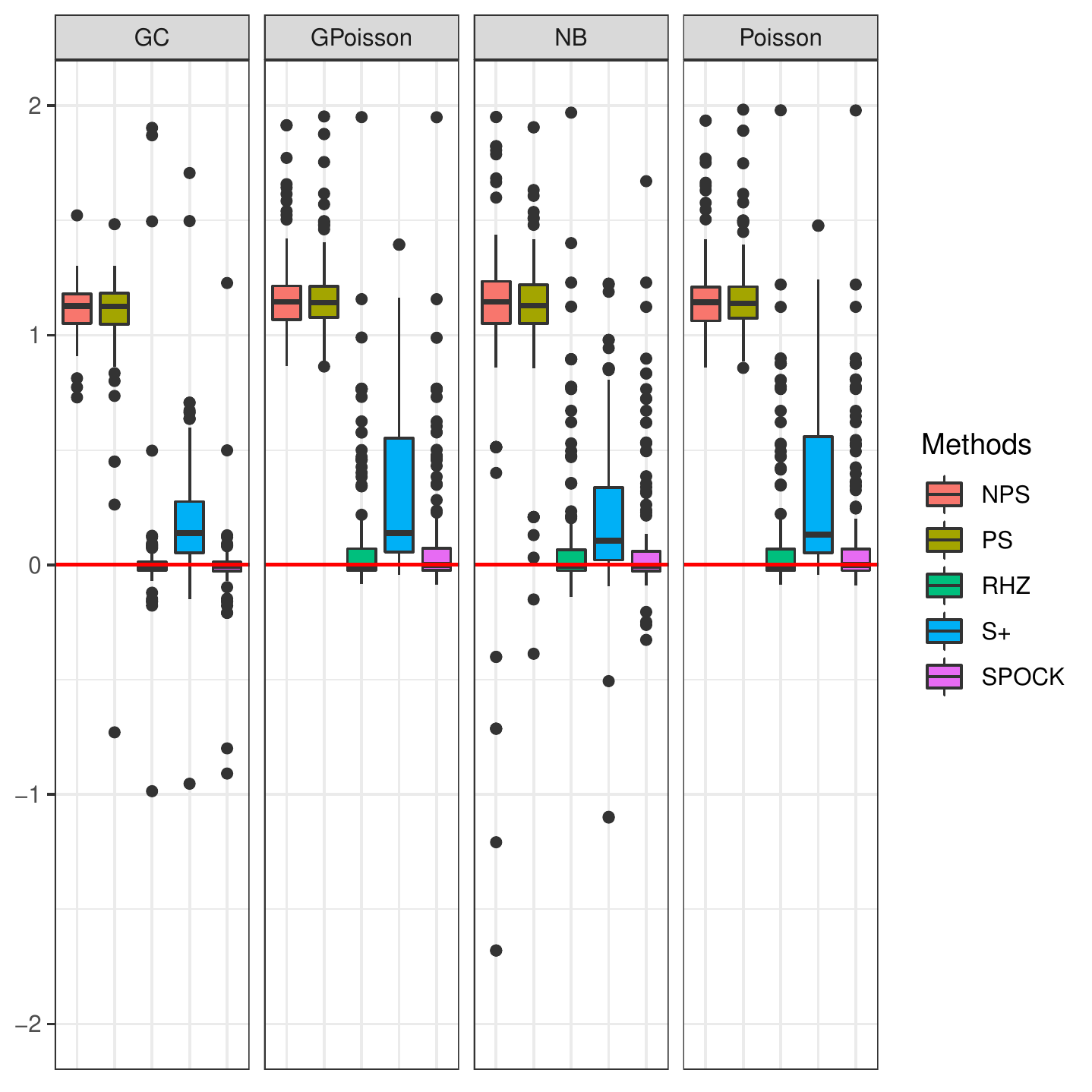}
\end{tabular} 
\end{center} 
\caption{Estimated relative bias of non-confounded covariate effect, $\beta_1$, (left column),  confounded covariate effect, $\beta_2$, (right column)  for each 
model with $\tau_x=4$ for scenarios: over-dispersion (first row), equivalent-dispersion (second row), and under-dispersion (third row; $\alpha=1.3$ and forth row; $\alpha=2$).\label{fig2}} 
\end{figure}

\begin{figure}[ppt]
\begin{center} 
\begin{tabular}{cc} 
\includegraphics[width=230pt,height=6.8pc]{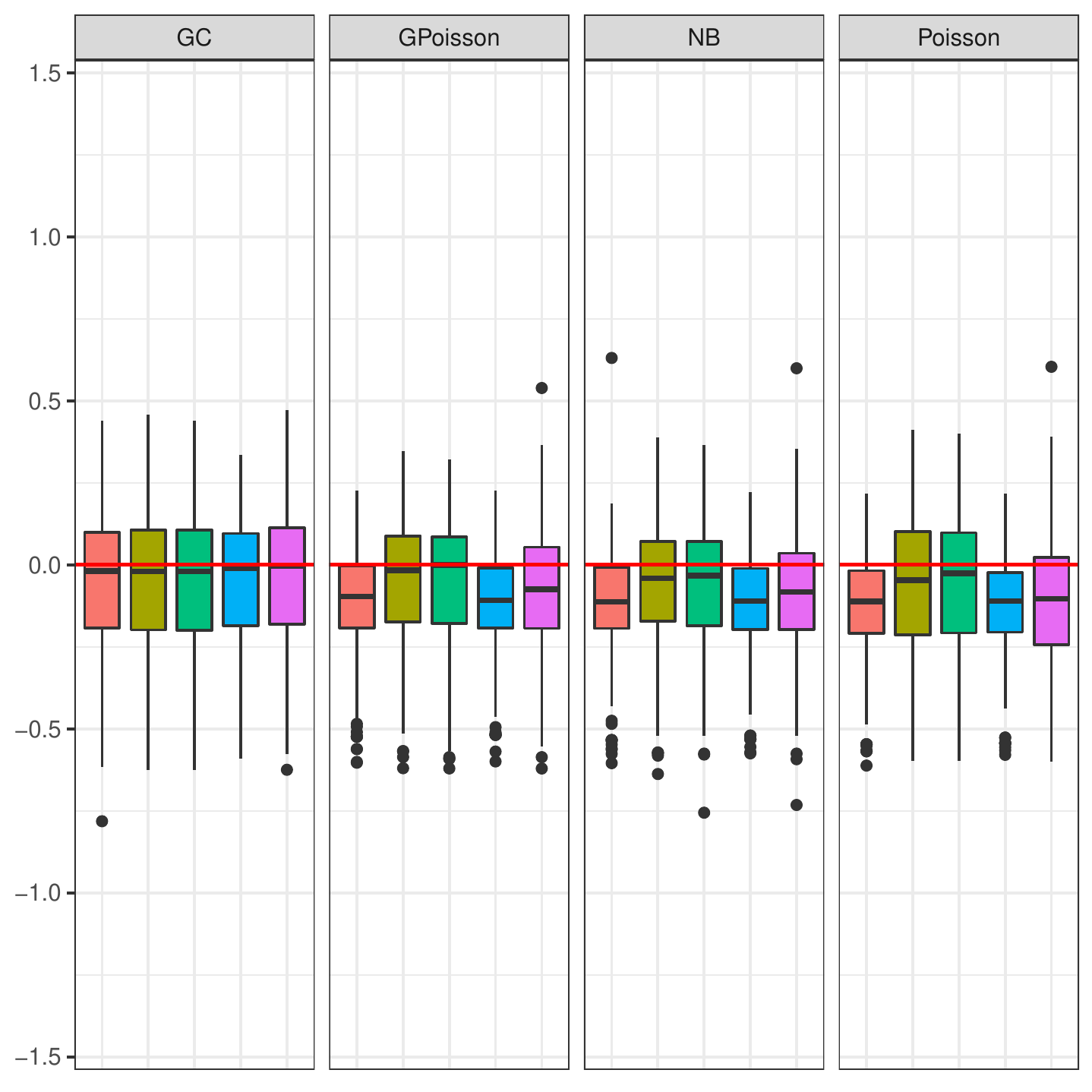} &
 \includegraphics[width=230pt,height=6.8pc]{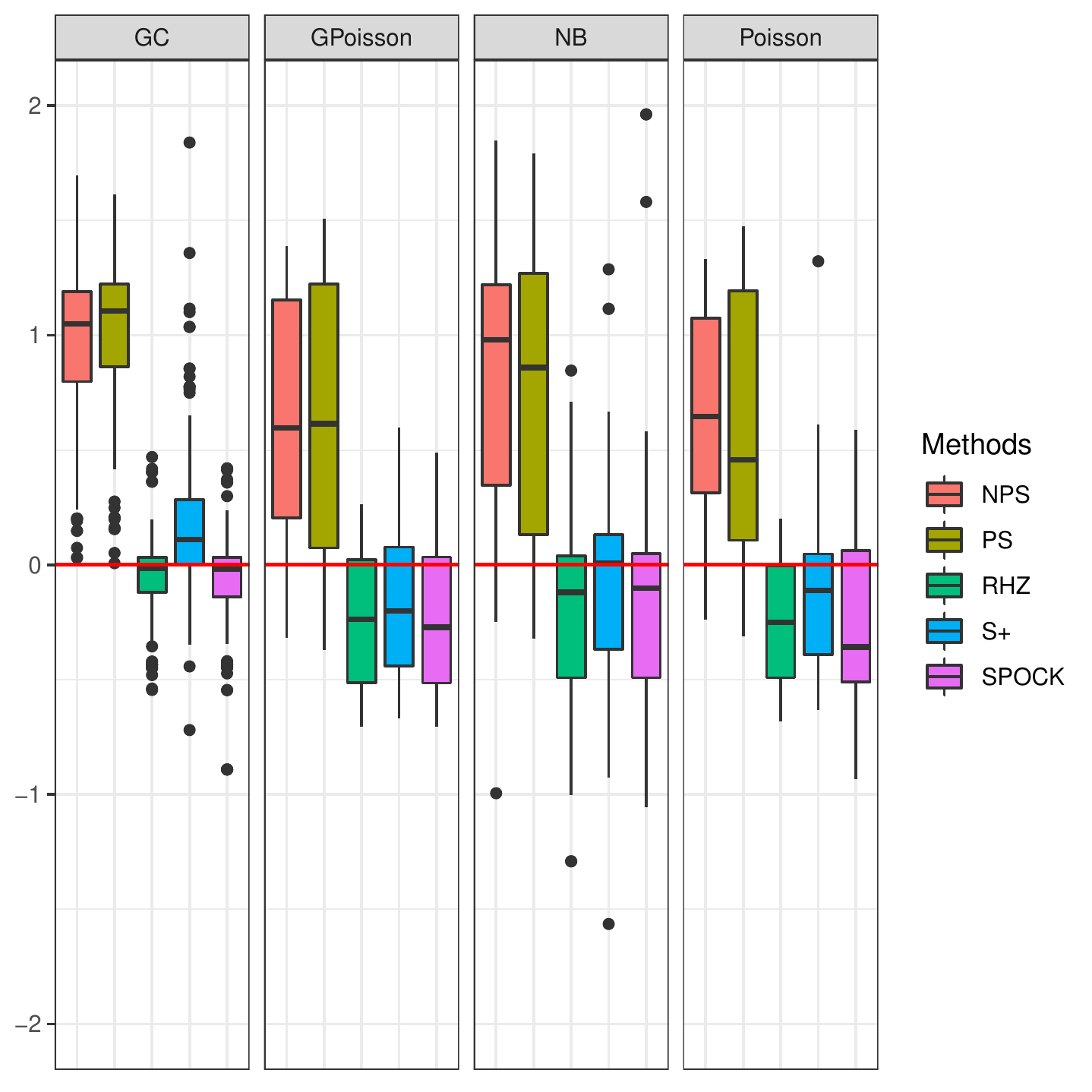}\\ 
\includegraphics[width=230pt,height=6.8pc]{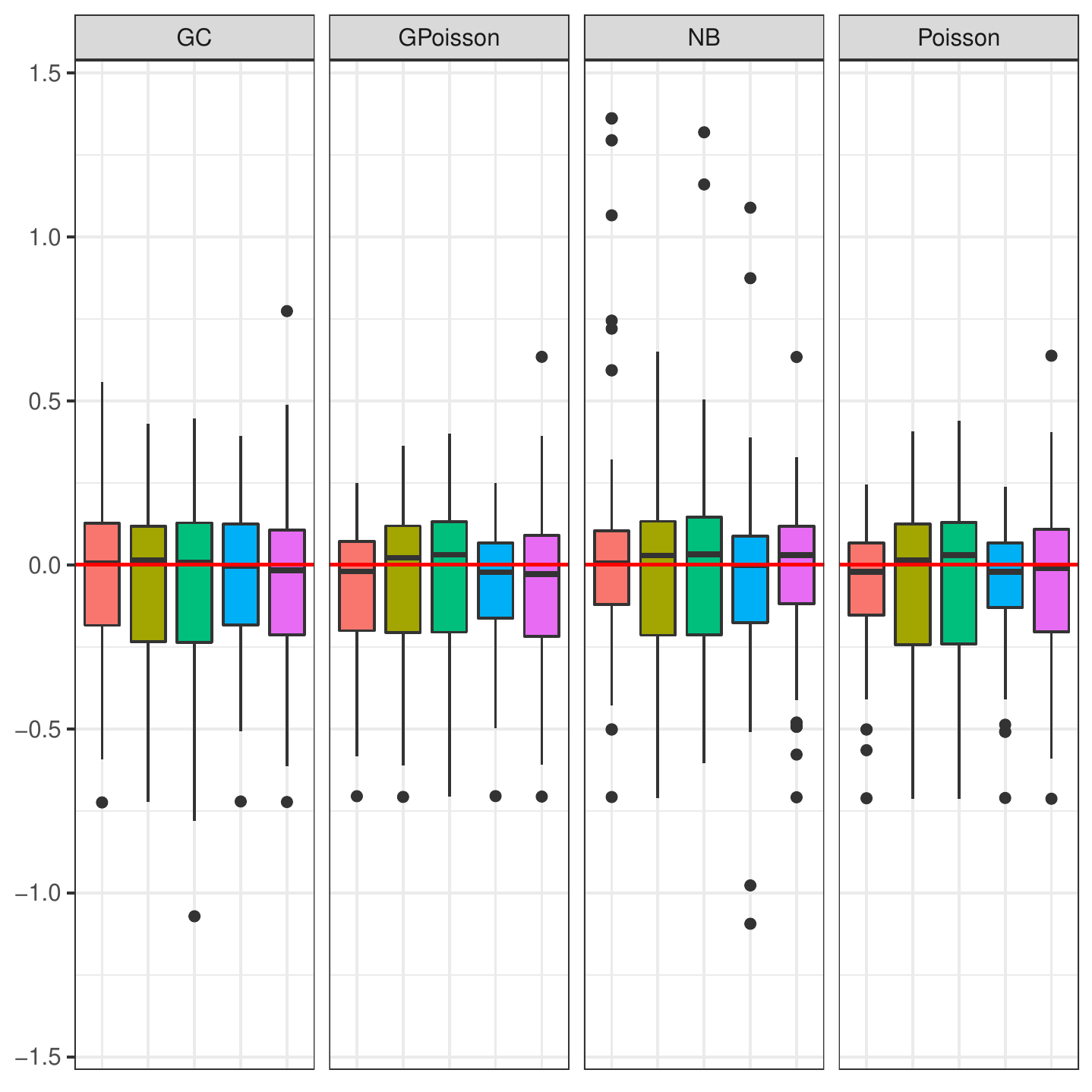} & 
\includegraphics[width=230pt,height=6.8pc]{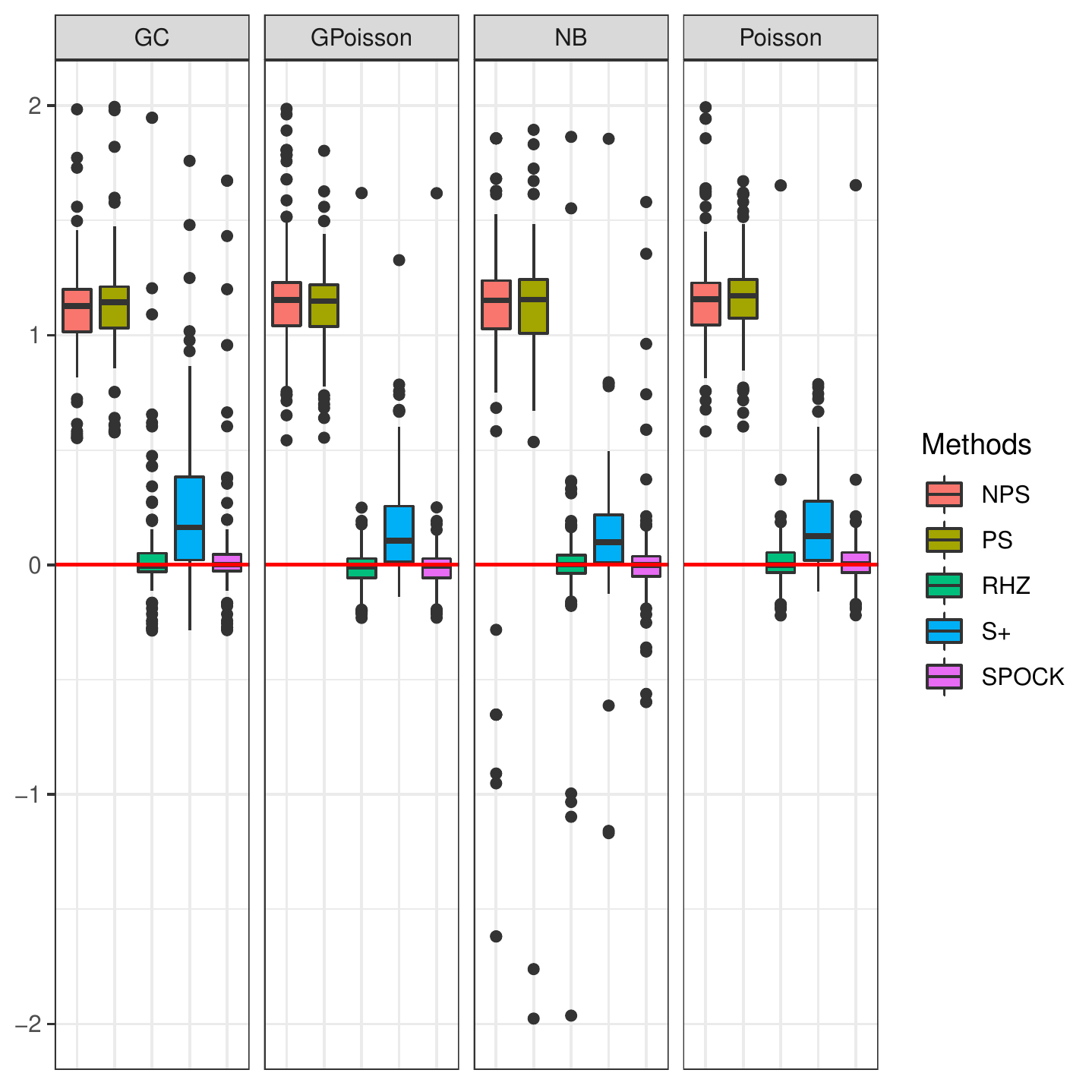}\\ 
\includegraphics[width=230pt,height=6.8pc]{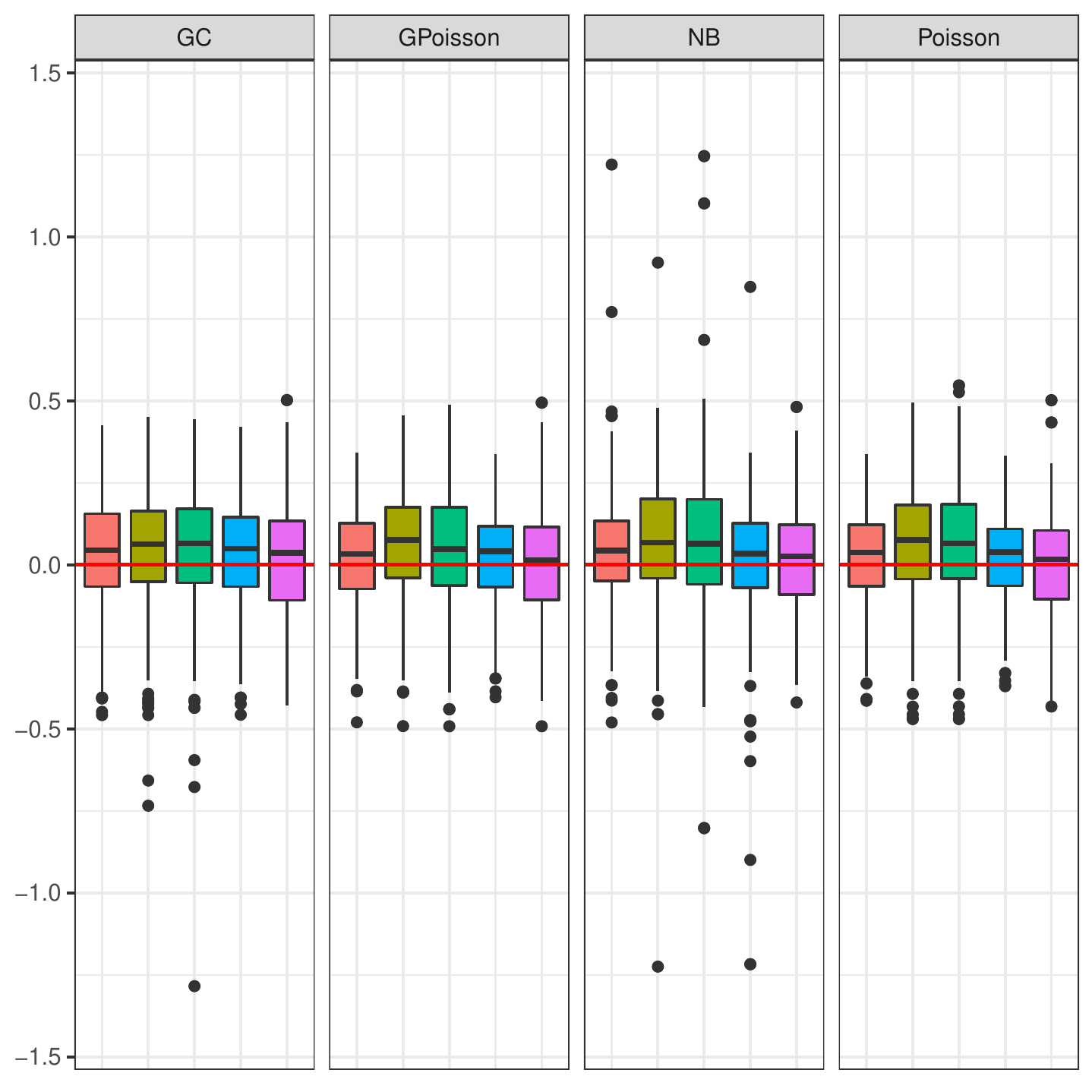} & 
\includegraphics[width=230pt,height=6.8pc]{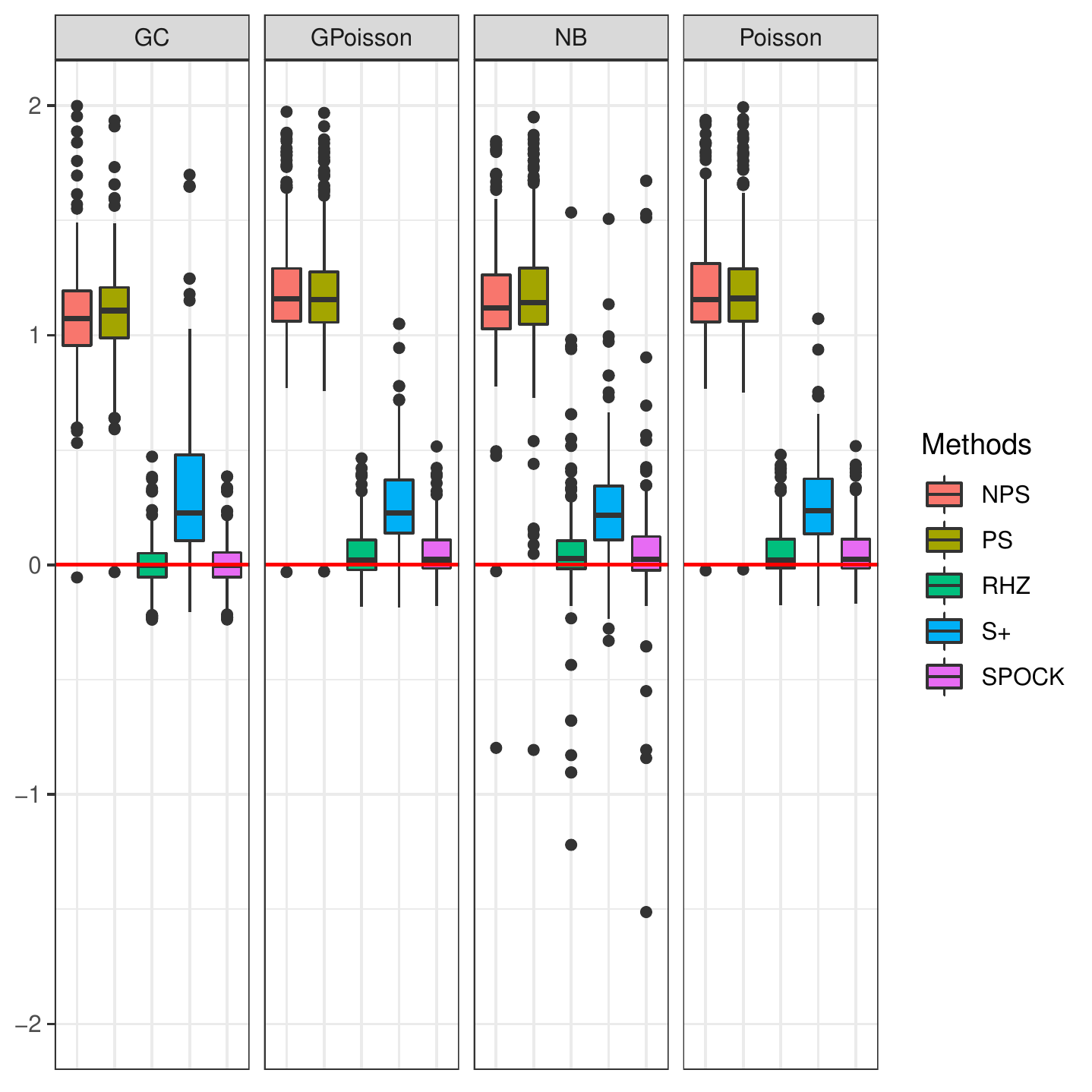}\\ 
\includegraphics[width=230pt,height=6.8pc]{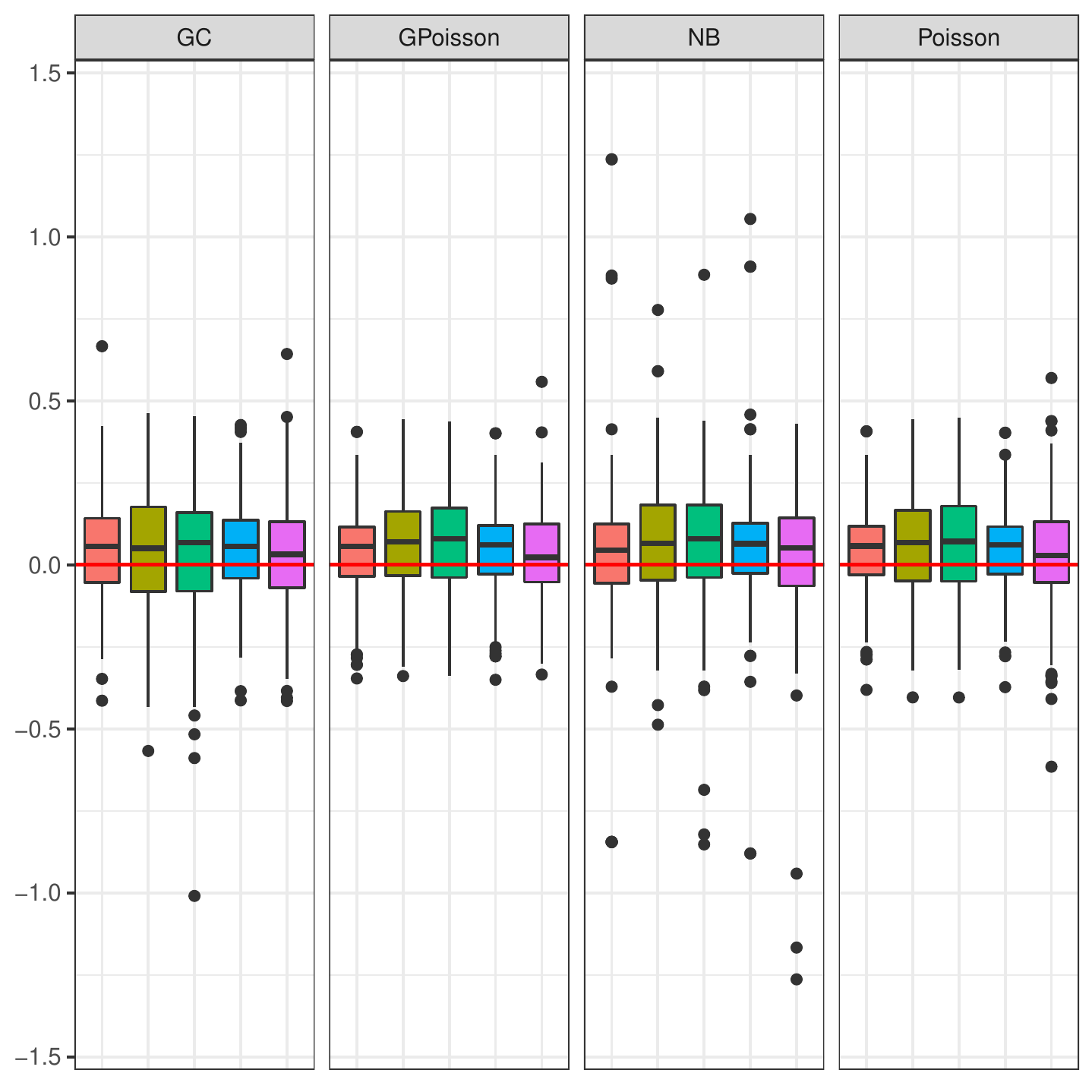} & 
\includegraphics[width=230pt,height=6.8pc]{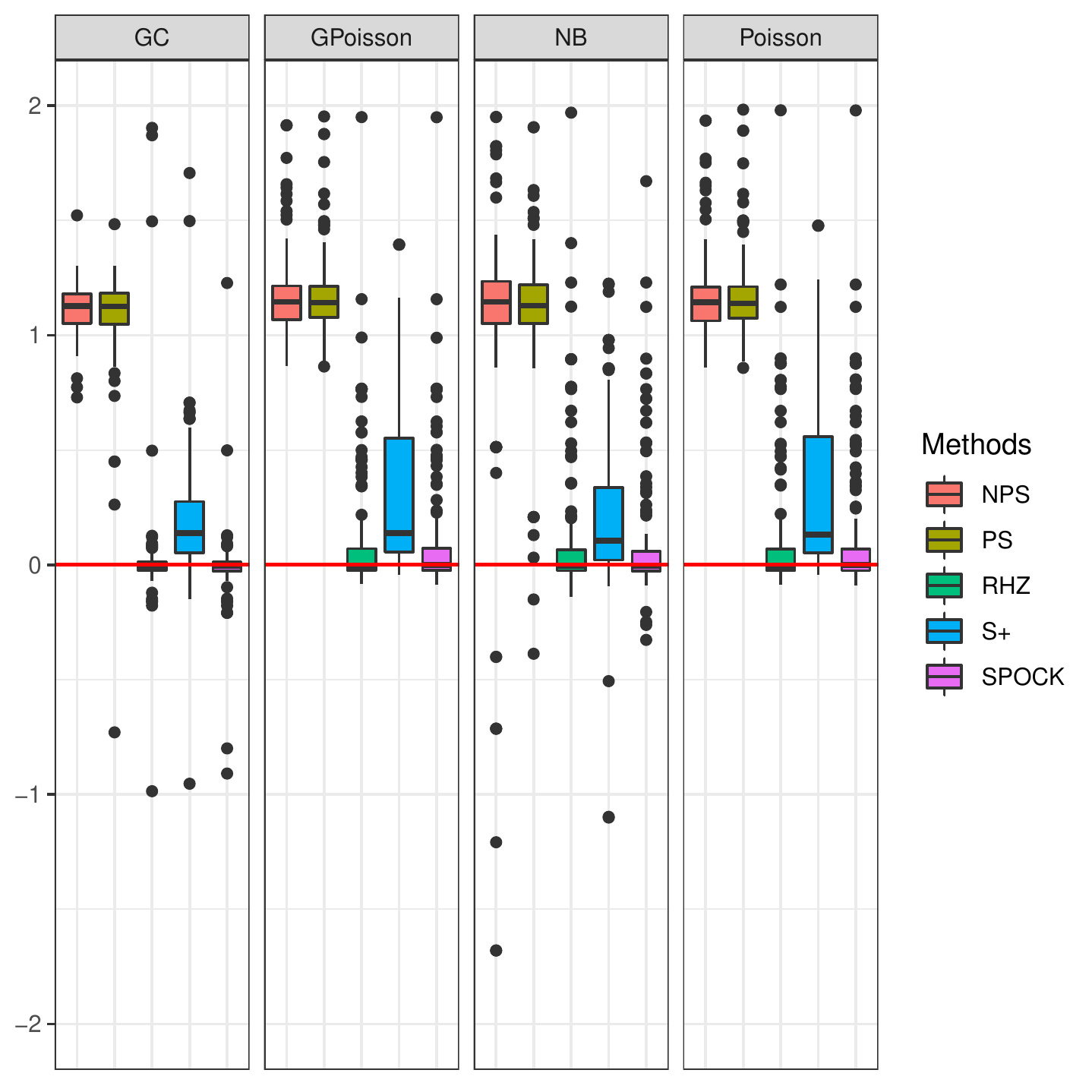}
\end{tabular} 
\end{center} 
\caption{Estimated relative bias of non-confounded covariate effect, $\beta_1$, (left column),  confounded covariate effect, $\beta_2$, (right column)  for each 
model with $\tau_x=1$ for scenarios: over-dispersion (first row), equivalent-dispersion (second row), and under-dispersion (third row; $\alpha=1.3$ and forth row; $\alpha=2$).\label{fig3}} 
\end{figure}

\begin{figure}[ppt] 
\begin{center} 
\begin{tabular}{cccc} 
\includegraphics[width=110pt,height=4.5pc]{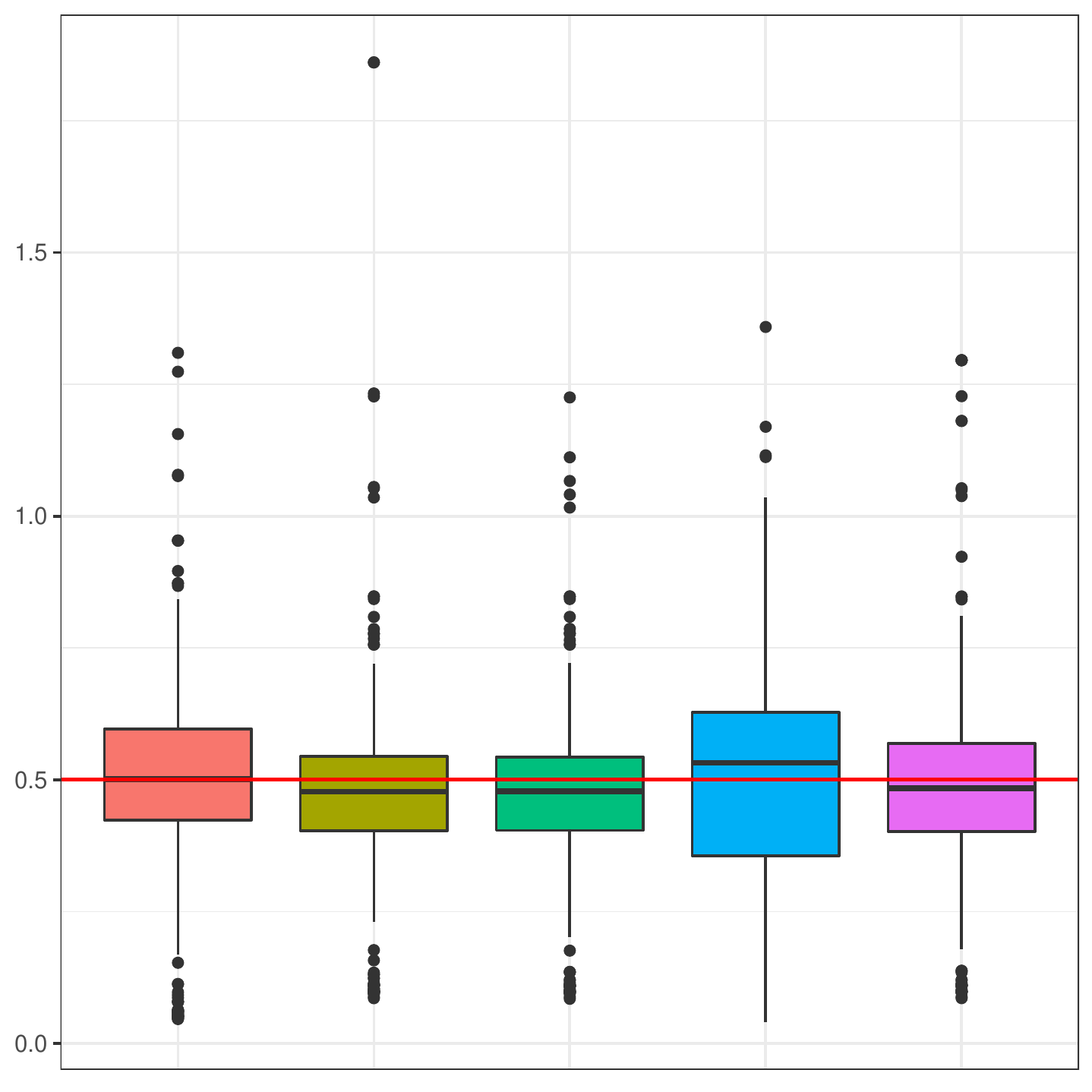} & 
\includegraphics[width=110pt,height=4.5pc]{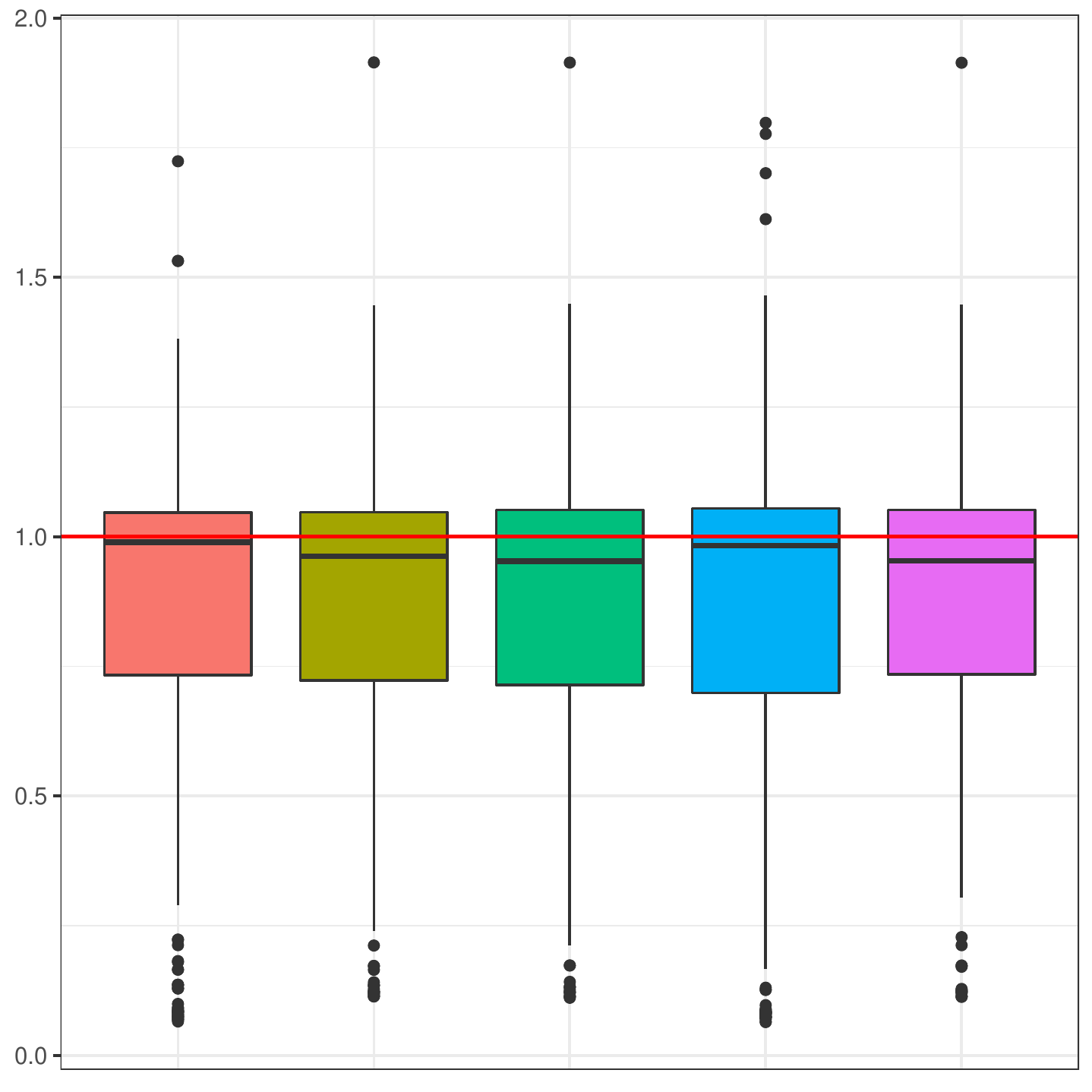}&
\includegraphics[width=110pt,height=4.5pc]{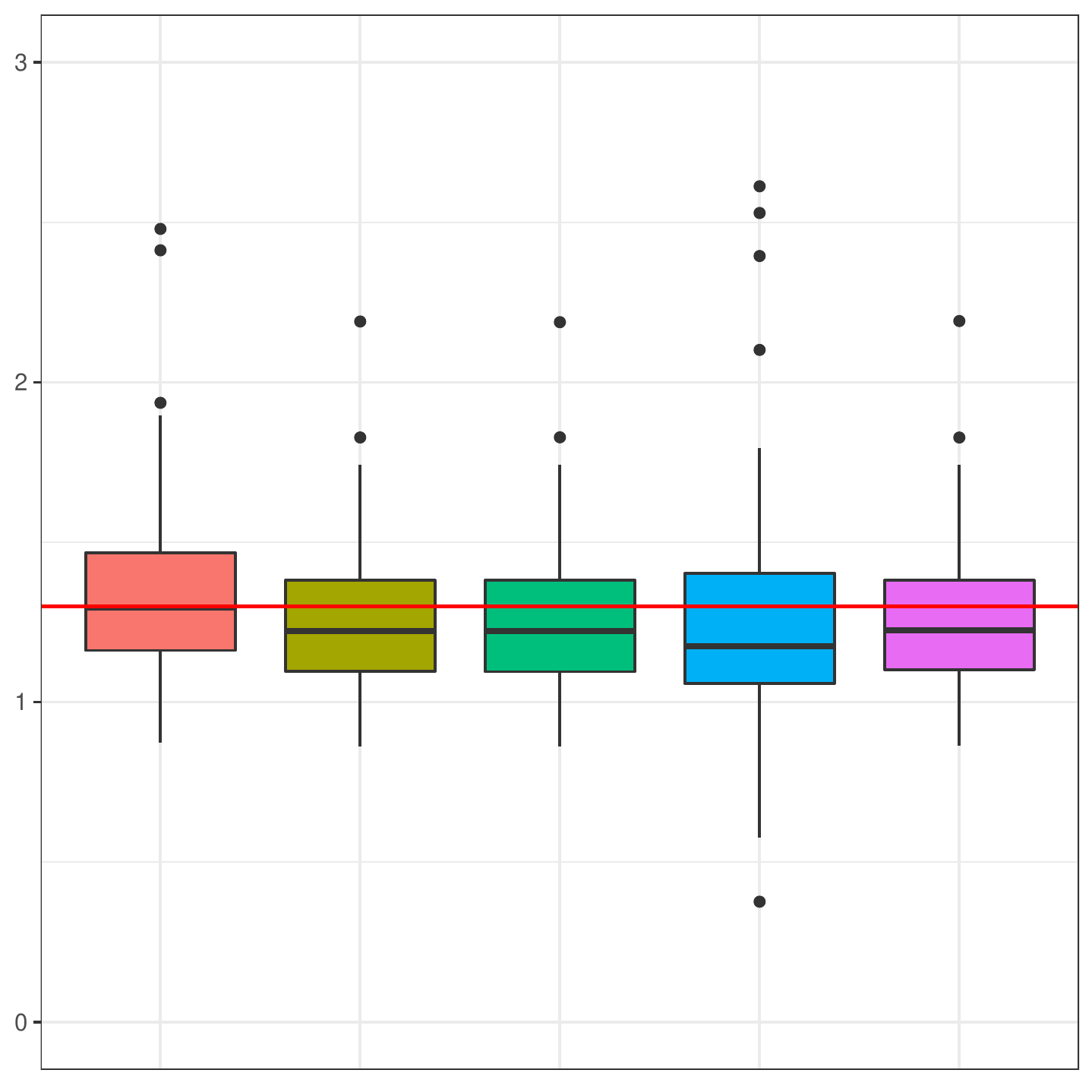}&
\includegraphics[width=110pt,height=4.5pc]{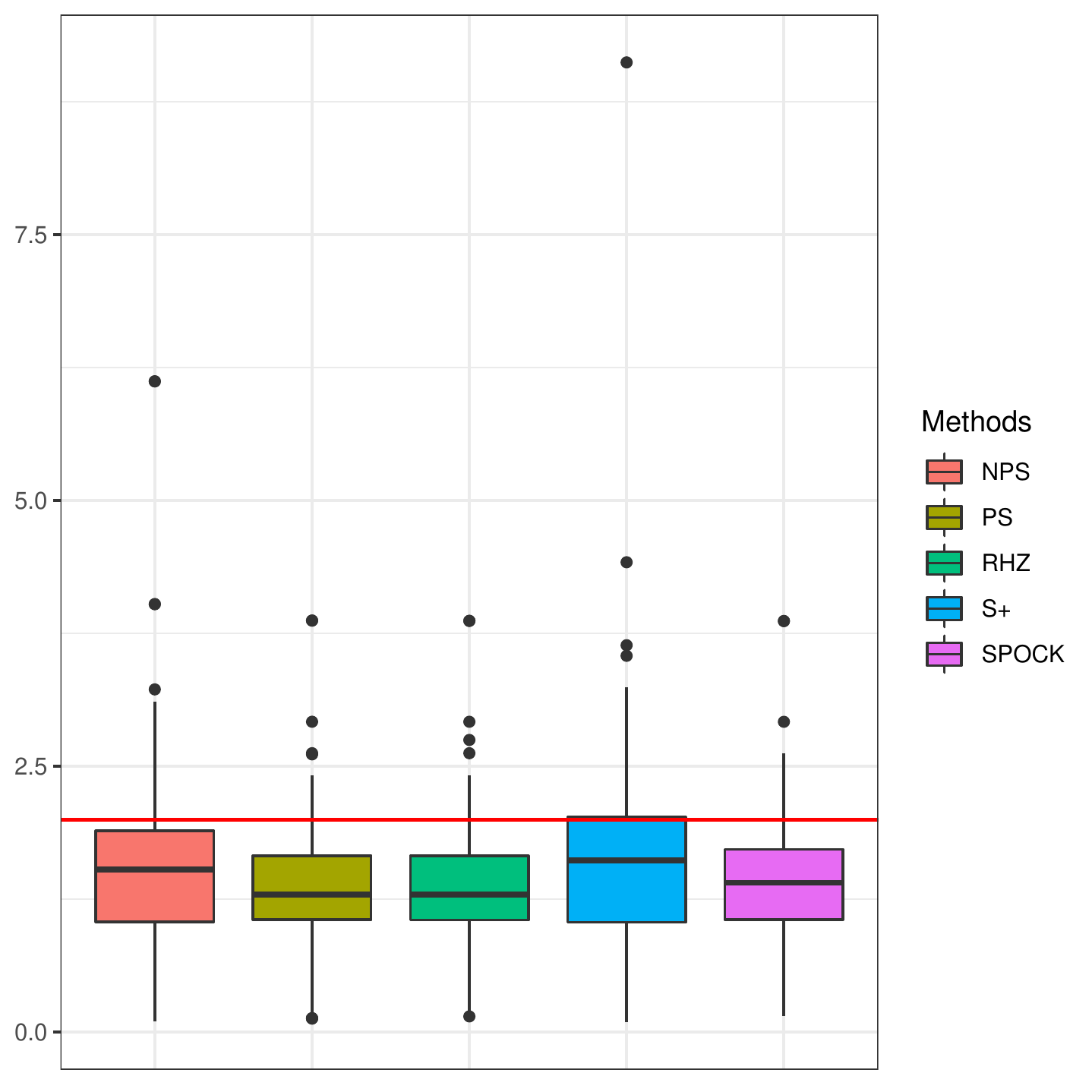} \\
\includegraphics[width=110pt,height=4.5pc]{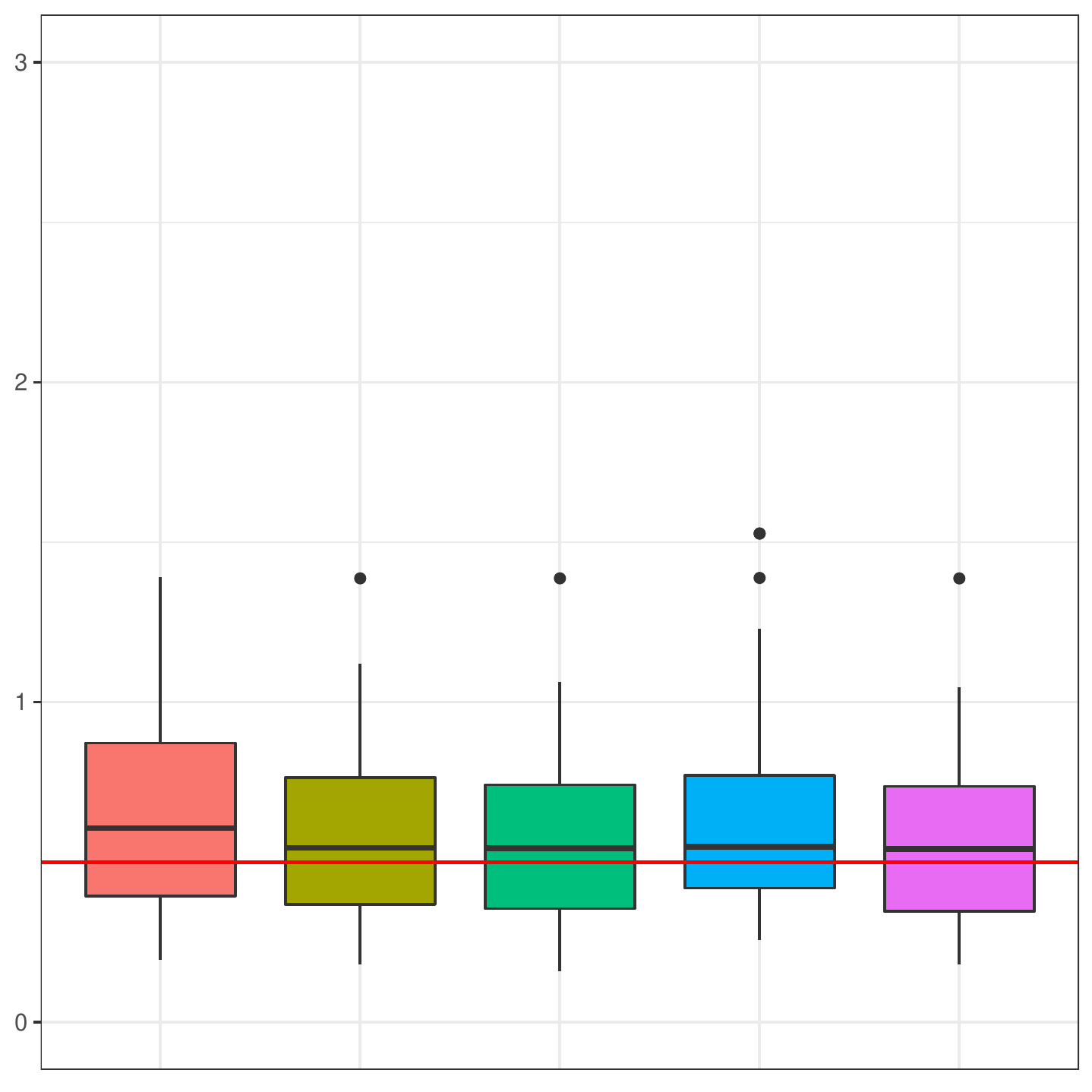} & 
\includegraphics[width=110pt,height=4.5pc]{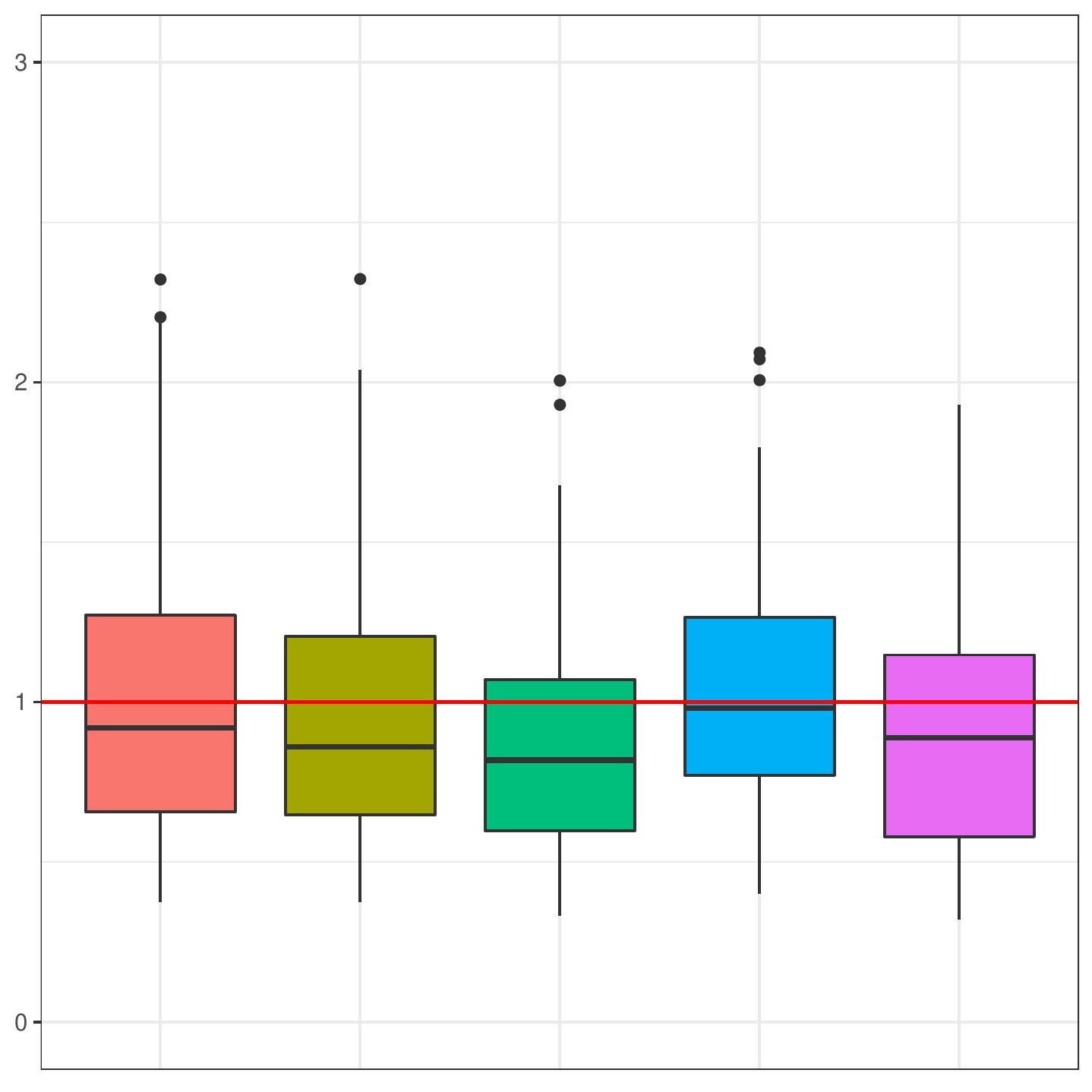}&
\includegraphics[width=110pt,height=4.5pc]{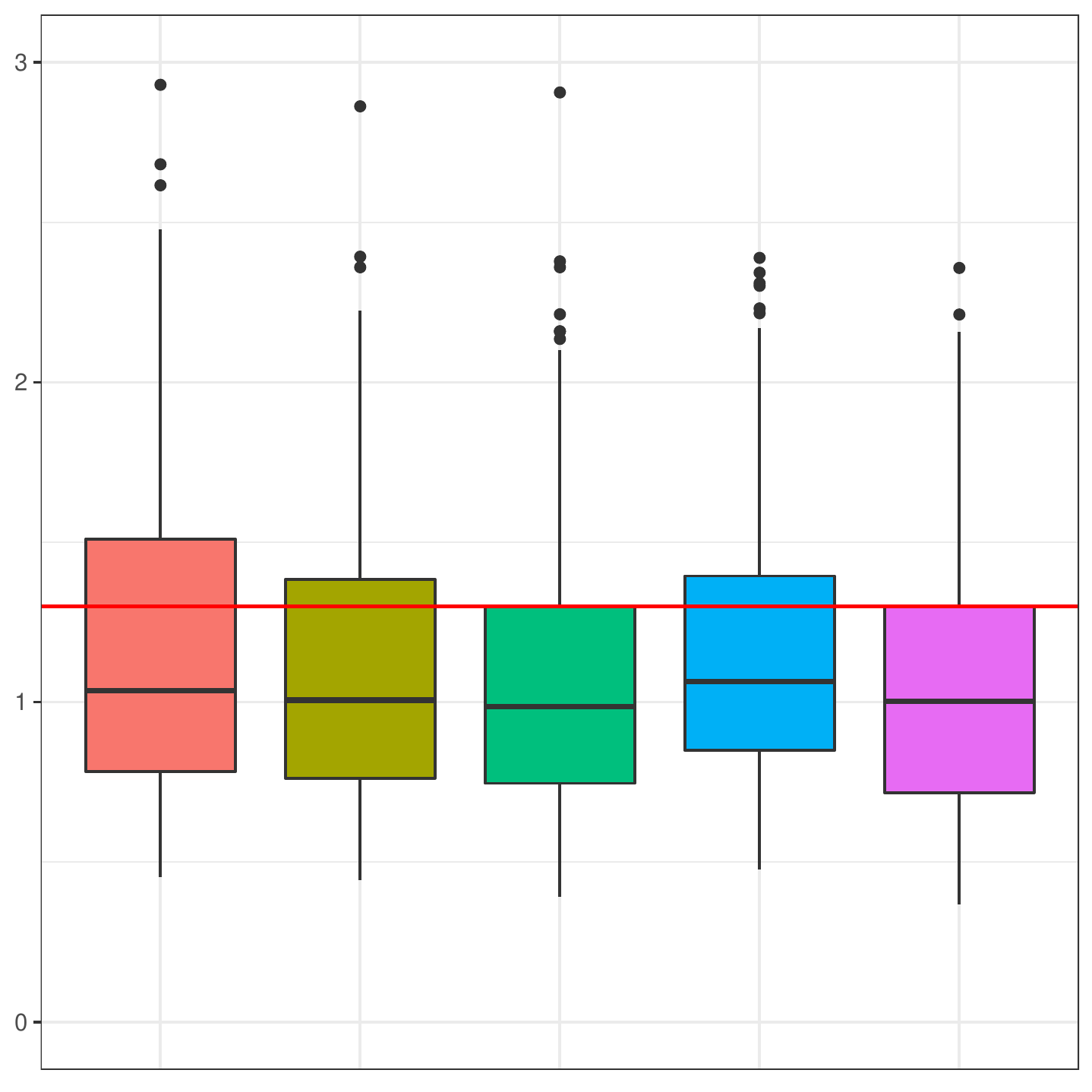} &
\includegraphics[width=110pt,height=4.5pc]{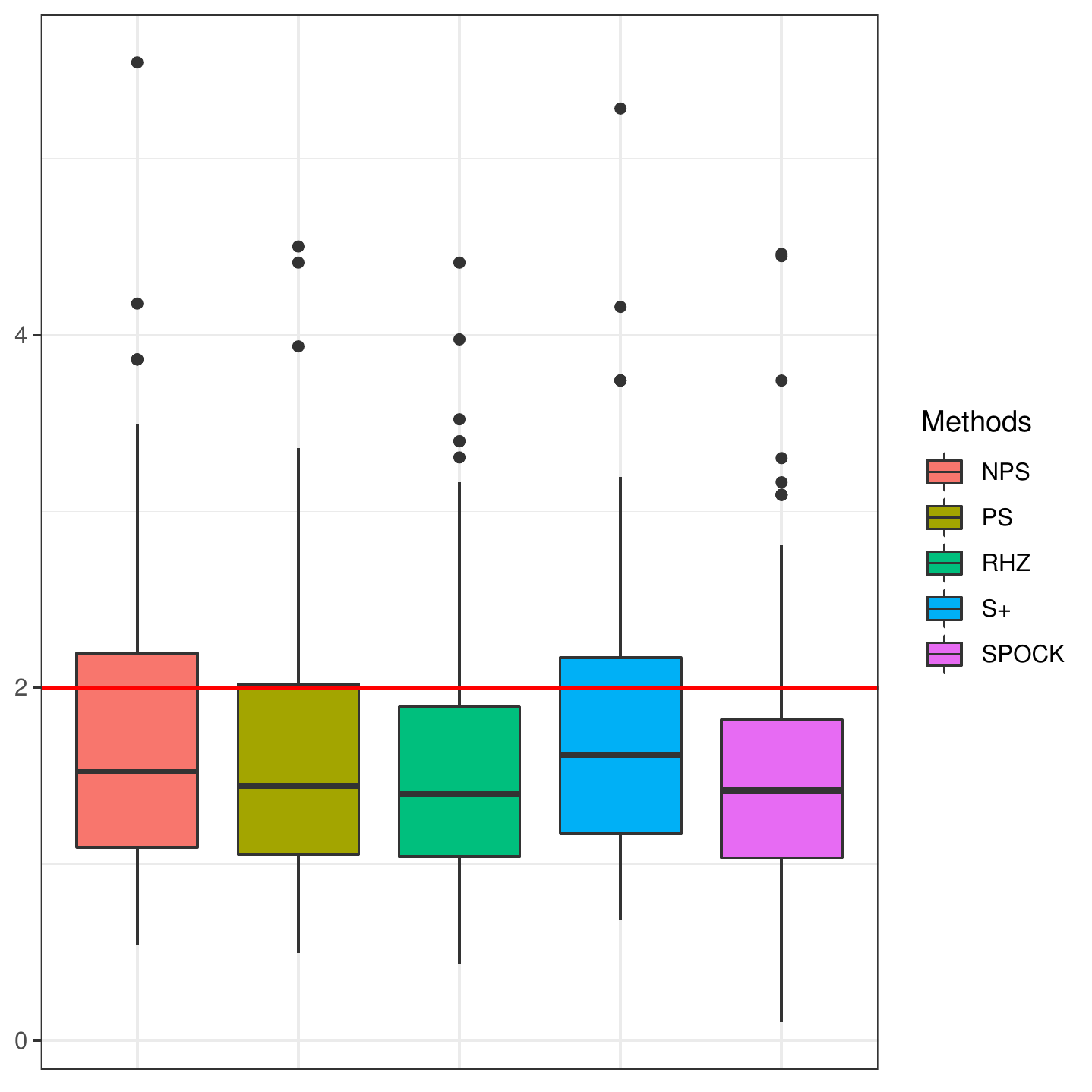} \\
\includegraphics[width=110pt,height=4.5pc]{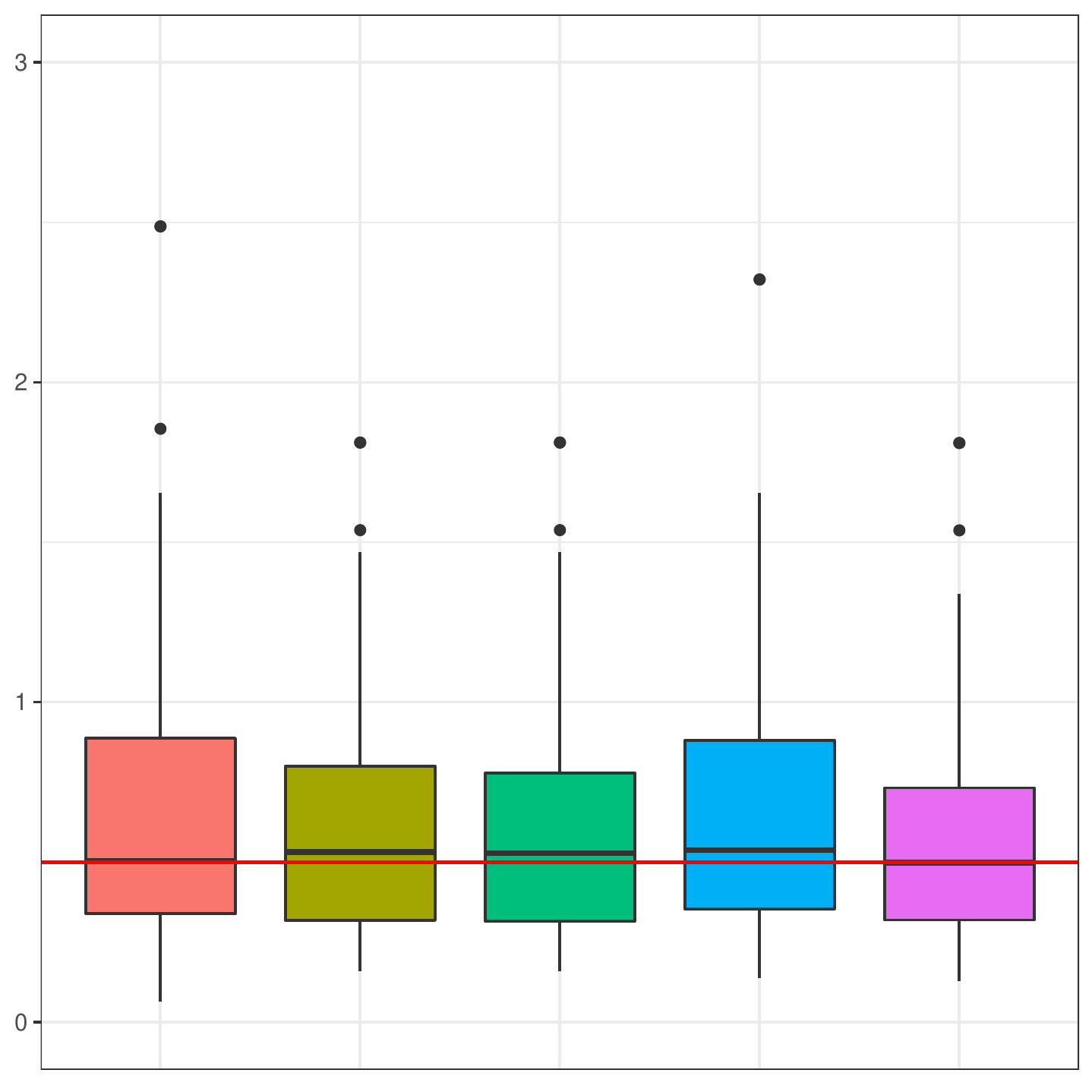} & 
\includegraphics[width=110pt,height=4.5pc]{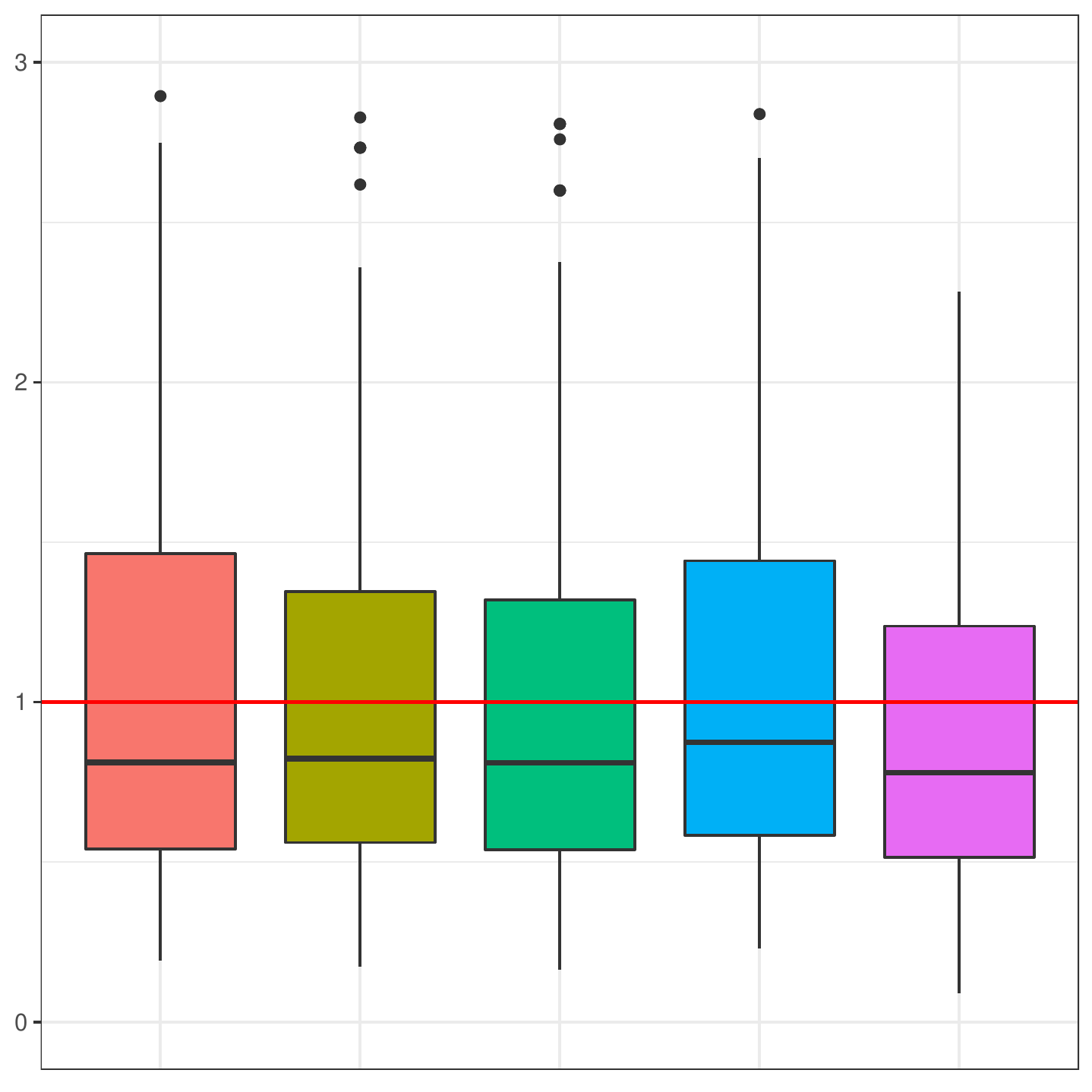}&
\includegraphics[width=110pt,height=4.5pc]{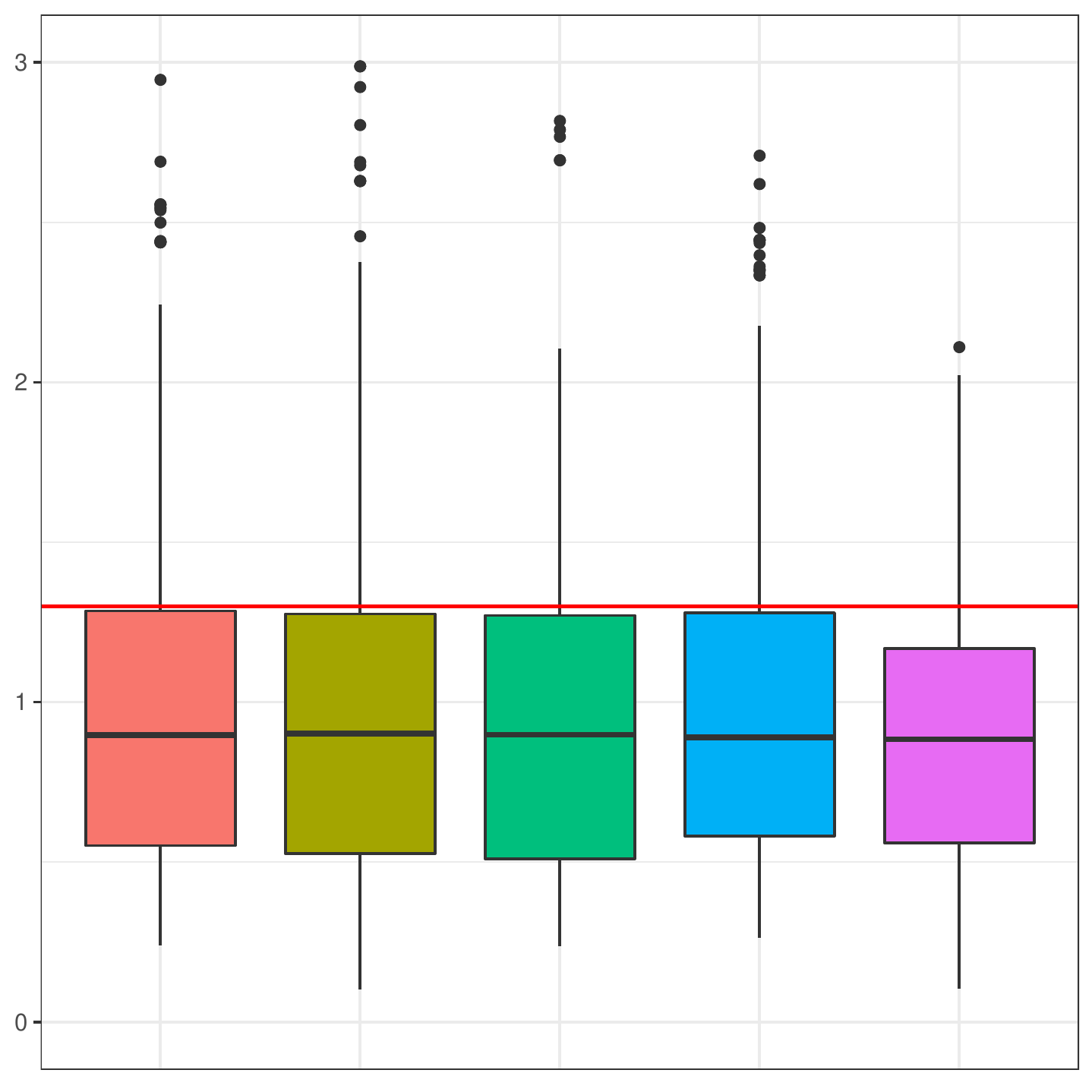}&
\includegraphics[width=110pt,height=4.5pc]{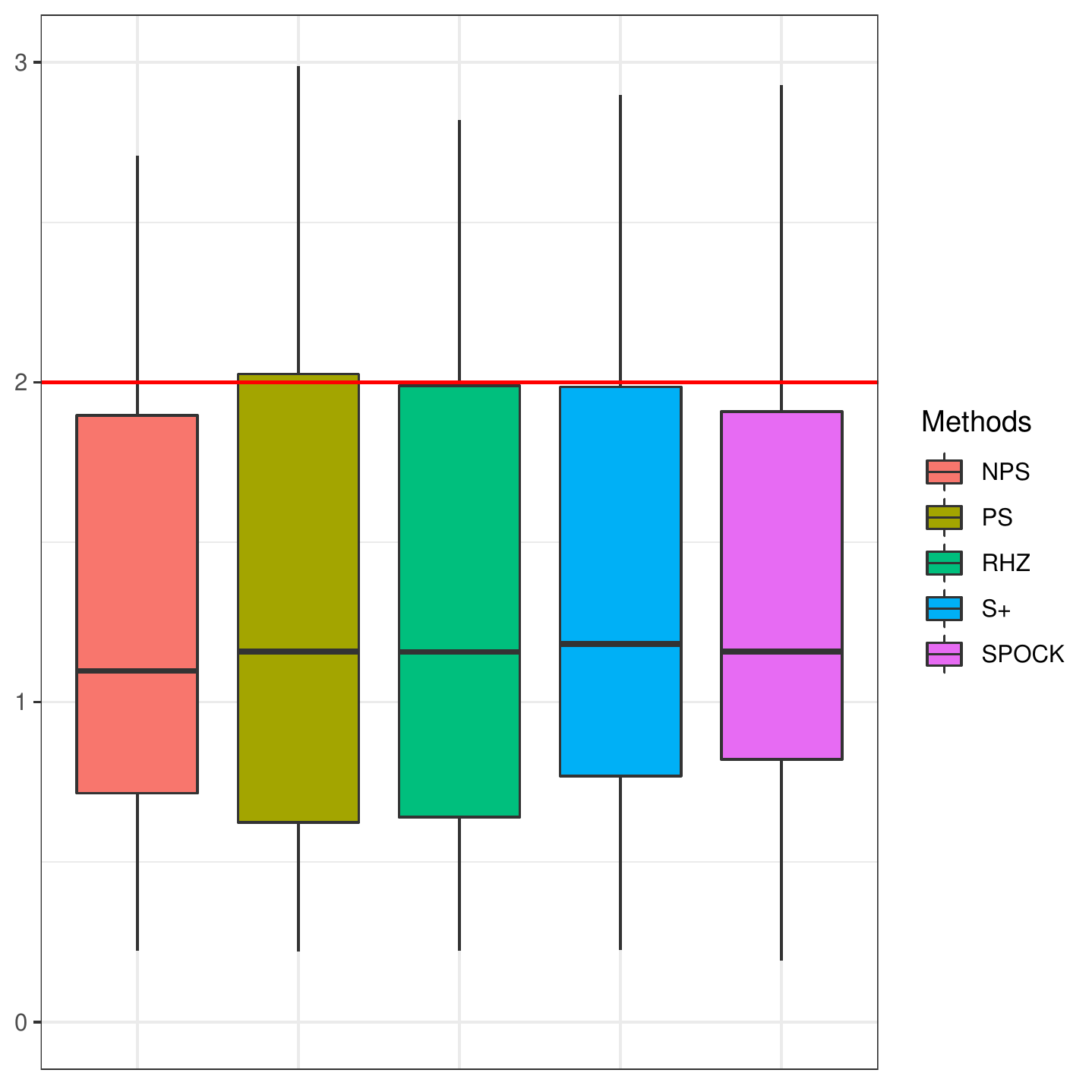}  \\
\end{tabular} 
\end{center} 
\caption{ Estimation of dispersion parameter, $\alpha$,  using the GC distribution for different spatial models and dispersion scenarios, from the left side: over-dispersion (first column), equivalent-dispersion (second column), under-dispersion ($\alpha=1.3$; third column) and under-dispersion ($\alpha=2$; forth column) and degree of confounding scenarios: high degree (first row), mild degree (second row) and low degree (third row).\label{fig4}} 
\end{figure}

We assess the impact of confounding degree by looking at numerous precision combinations, $\tau_x$, in \eqref{xf} on the covariate effect estimation. If $\tau_x$ is small, the covariate involves much beyond the spatial effect; there is poor confounding \citep{Thaden2018}. Figure \ref{fig-mse} visualizes the MSE attitude in a confounding setting. The MSE of  estimators except RHZ and SPOCK models increases by multiplying the confounding degree in all dispersion scenes. However, if $\tau_x$ gets small, spatial confounding is less problematic and all models behave similarly. Moreover, the SPOCK and RHZ models have the least MSE and the same appearance for all confounding degrees, displaying the best and stable to confounding degrees. The NPS and PS models have the same performance, and further, the S+ model performs better than those. Although NPS and PS models perform best in an over-dispersion situation with less confounded, they are worst in almost all scenarios. Our results for the spatial+ model are similar to \cite{Thaden2018}. 
Boxplots in Figures \ref{fig1} to \ref{fig3} visualize the relative bias of covariates effects for all models and all three dispersion cases under different levels of confounding. These figures show that, in general, all models have the same performance to resume the effect of the non-confounded covariate, $\beta_1$. Moreover, 
the GC model has a relatively better performance than the alternative models, especially in comparison to the RHZ and SPOCK models. Indeed, the differences in recovering the actual effect of the confounded covariate, $\beta_2$, are more evident in favor of the GC model and non-equivalent dispersions. As we expected, when the degree of confounding is small, the differences are less appealing.  
Moreover, Figure \ref{fig4} displays boxplots for estimation of dispersion parameter of the GC distribution, $\alpha$, for all proposed models. It shows that there is no essential difference between the models. However, we could see that the accuracy of the estimates decreases as the degree of confounding increases. The estimates for the under-dispersion situation are also not accurate, compared to over- or equivalent-dispersion. 

Table \ref{Tab2} presents the MSE and coverage rate (CR) of fixed effects for all confounding scenarios for the GC model when the nominal rate is 95$\%$. As the results show, by increasing the confounding degree, MSE values become greater, but the coverage rate is almost the same for all scenarios. Moreover, coverage rate values of the confounded covariate effect, $\beta_2$, vary between 0.4 and 0.94. We can see the same results for the other alternative count models in Tables \ref{Tab4}  to \ref{Tab6} in the Appendix. We also computed MSPE of fitted values and SE of estimates for $\beta_2$; their boxplots are given in Figures \ref{fig8} to \ref{fig10} in the appendix. The MSPE of fitted values represents that the S+ model performs better than others. The SE values also show that the GC model is the best, especially in the over-dispersion case. Furthermore, Figures \ref{fig11} to \ref{fig13} (see the appendix) present boxplots of DIC, WAIC and LS criteria. The DIC values show the GC model  has a relative superiority over competing models, and also LS criterion assesses the NB model has weak performance in the prediction. 
  
\begin{table}[ppt]
\begin{footnotesize}
\centering\caption{\label{Tab2} MSE and coverage rate for a nominal rate of 95$\%$ for all scenarios for the GC model.}
\begin{tabular}{llllllllllllll}
  \hline
$\alpha$ &Model &\multicolumn{6}{c}{$\beta_1$}&\multicolumn{6}{c}{$\beta_2$}\\
   \cmidrule[0.7pt](lr{0.125em}){3-8}\cmidrule[0.7pt](lr{0.125em}){9-14}
   &&\multicolumn{6}{c}{$\tau_x$}&\multicolumn{6}{c}{$\tau_x$}\\
  &  &\multicolumn{2}{c}{11}&\multicolumn{2}{c}{4}&\multicolumn{2}{c}{1}&\multicolumn{2}{c}{11}&\multicolumn{2}{c}{4}&\multicolumn{2}{c}{1}\\
  \cmidrule[0.7pt](lr{0.125em}){3-4}\cmidrule[0.7pt](lr{0.125em}){5-6}  \cmidrule[0.7pt](lr{0.125em}){7-8}\cmidrule[0.7pt](lr{0.125em}){9-10}  \cmidrule[0.7pt](lr{0.125em}){11-12}\cmidrule[0.7pt](lr{0.125em}){13-14}
 && MSE & CR & MSE & CR & MSE& CR & MSE & CR & MSE & CR & MSE & CR \\ 
  \hline
&NPS & 0.13 & 0.98 & 0.05 & 0.90 & 0.03 & 0.98 & 1.24 & 0.73 & 0.20 & 0.57 & 1.04 & 0.73 \\ 
 & S+ & 0.11 & 0.97 & 0.04 & 0.97 & 0.02 & 0.97 & 0.33 & 0.59 & 0.25 & 0.70 & 0.45 & 0.59 \\ 
0.5  &PS & 0.16 & 0.94 & 0.05 & 0.90 & 0.03 & 0.94 & 1.22 & 0.74 & 0.22 & 0.57 & 1.13 & 0.74 \\ 
 & SPOCK & 0.13 & 0.95 & 0.05 & 0.92 & 0.03 & 0.95 & 0.05 & 0.74 & 0.19 & 0.62 & 0.16 & 0.74 \\ 
 & RHZ & 0.11 & 0.97 & 0.05 & 0.90 & 0.03 & 0.97 & 0.02 & 0.76 & 0.19 & 0.58 & 0.15 & 0.76 \\ 
  [3mm]
 & NPS & 0.20 & 0.96 & 0.03 & 0.96 & 0.02 & 0.96 & 1.27 & 0.66 & 1.27 & 0.66 & 1.27 & 0.66 \\ 
 & S+ & 0.26 & 0.94 & 0.02 & 0.94 & 0.02 & 0.94 & 0.49 & 0.49 & 0.49 & 0.49 & 0.49 & 0.49 \\ 
1 & PS & 0.10 & 0.96 & 0.03 & 0.96 & 0.03 & 0.96 & 1.32 & 0.68 & 1.32 & 0.68 & 1.32 & 0.68 \\ 
  &SPOCK & 0.15 & 0.96 & 0.05 & 0.96 & 0.03 & 0.96 & 0.08 & 0.74 & 0.08 & 0.74 & 0.08 & 0.74 \\
  &  RHZ & 0.14 & 0.96 & 0.03 & 0.96 & 0.03 & 0.96 & 0.11 & 0.71 & 0.11 & 0.71 & 0.11 & 0.71 \\ 
  [3mm] 
 & NPS & 0.25 & 0.94 & 0.02 & 0.94 & 0.02 & 0.94 & 1.22 & 0.94 & 1.22 & 0.94 & 1.22 & 0.94 \\ 
 & S+ & 0.36 & 0.94 & 0.02 & 0.94 & 0.01 & 0.94 & 0.57 & 0.56 & 0.57 & 0.56 & 0.57 & 0.56 \\ 
1.3 & PS & 0.26 & 0.92 & 0.02 & 0.92 & 0.03 & 0.92 & 1.26 & 0.92 & 1.26 & 0.92 & 1.26 & 0.92 \\ 
 & SPOCK & 0.26 & 0.92 & 0.03 & 0.92 & 0.02 & 0.92 & 0.06 & 0.93 & 0.06 & 0.93 & 0.06 & 0.93 \\ 
 & RHZ & 0.26 & 0.92 & 0.03 & 0.92 & 0.02 & 0.92 & 0.06 & 0.90 & 0.06 & 0.90 & 0.06 & 0.90 \\ 
  [3mm]
 & NPS & 0.18 & 0.94 & 0.02 & 0.94 & 0.01 & 0.94 & 1.24 & 0.68 & 1.24 & 0.68 & 1.24 & 0.68 \\ 
 & S+ & 0.22 & 0.94 & 0.02 & 0.94 & 0.01 & 0.94 & 0.33 & 0.40 & 0.33 & 0.40 & 0.33 & 0.40 \\ 
2 & PS & 0.23 & 0.93 & 0.02 & 0.93 & 0.02 & 0.93 & 1.22 & 0.76 & 1.22 & 0.76 & 1.22 & 0.76 \\ 
 & SPOCK & 0.21 & 0.94 & 0.04 & 0.94 & 0.01 & 0.94 & 0.05 & 0.80 & 0.05 & 0.80 & 0.05 & 0.80 \\ 
 & RHZ & 0.22 & 0.94 & 0.02 & 0.94 & 0.02 & 0.94 & 0.02 & 0.76 & 0.02 & 0.76 & 0.02 & 0.76 \\ 
   \hline
\end{tabular}
\end{footnotesize}
\end{table}
\section{Data analysis: Slovenia stomach cancer}\label{Sec5} 
\begin{figure}[ppt] 
\begin{center} 
\begin{tabular}{cc} 
\includegraphics[width=180pt,height=7pc]{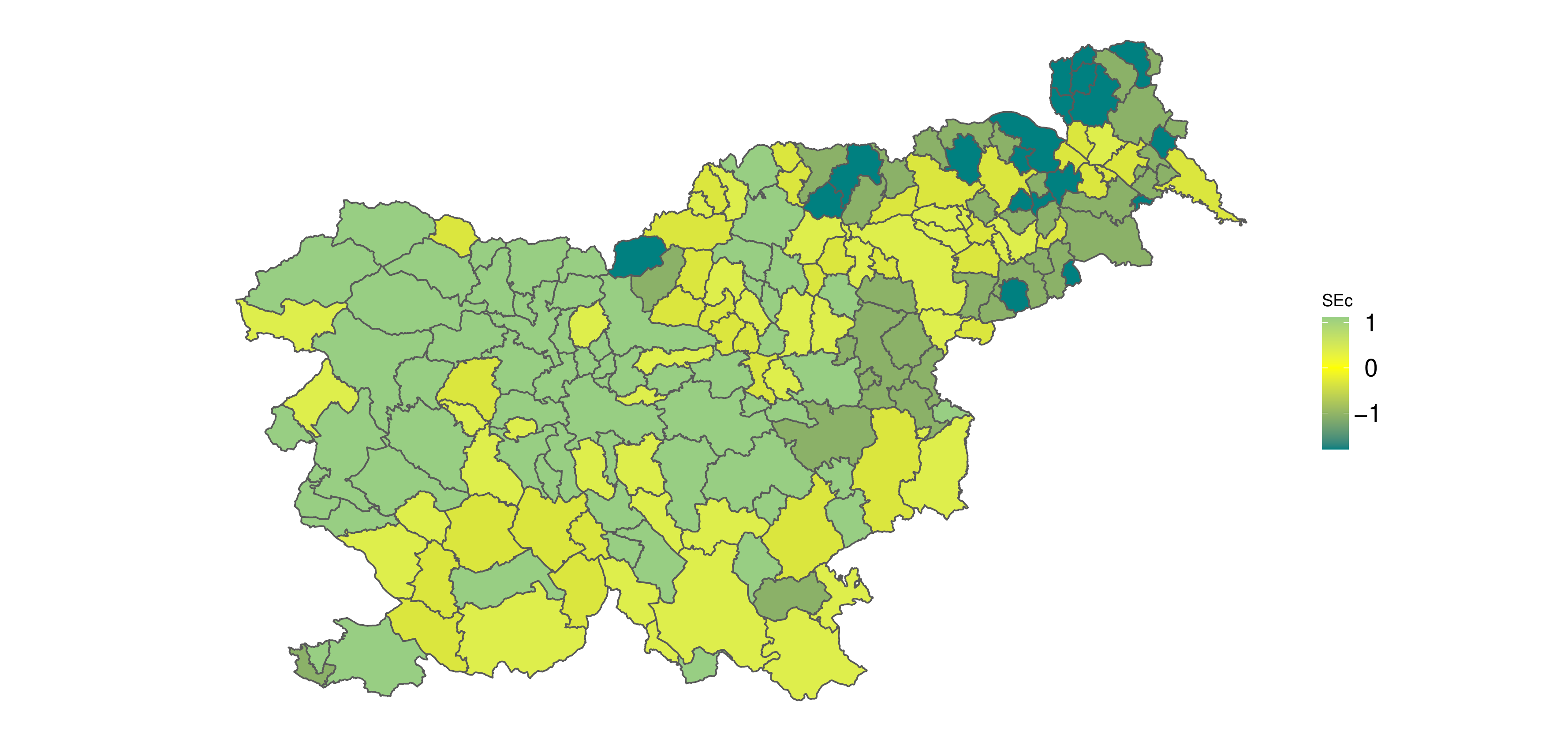} & \includegraphics[width=180pt,height=7pc]{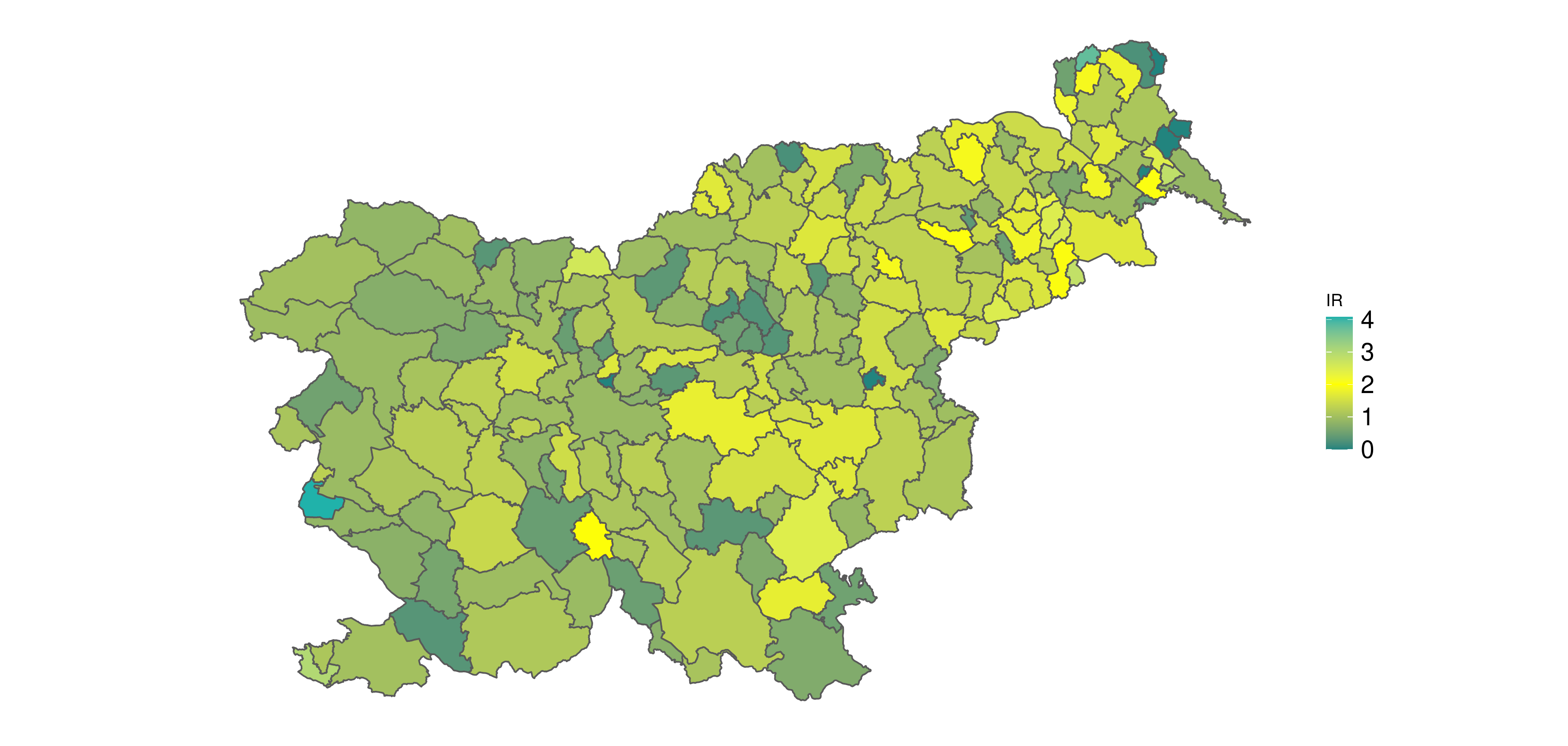} 
\end{tabular} 
\end{center} 
\caption{Slovenia municipality's incidence ratio (IR, left panel) and centered socioeconomic status (SEc, right panel).\label{fig5}} 
\end{figure}

In this example, we remark analyzing the relevance amongst socioeconomic factors and stomach cancer incidence in Slovenia for the years 1995-2001, in 192 municipalities \citep{Zadnik2006}. In this study, the objective variable is the ratio of observed to the expected number of cases $y_i/{\rm E}_i$, where $i$ indexes municipalities. Each region's socioeconomic status is placed into five arranged classes by Slovenia's Institute of Macroeconomic Analysis and Department \citep{Reich2006}. Figure \ref{fig5}, shows the incidence ratio (IR), $y_i/{\rm E}_i$, and the centered version of the covariate socioeconomic status (SEc). Both IR and SEc represent strong spatial patterns. Clearly, there is a negative association between IR and SEc. We considered the following linear predictor for analyzing these data \begin{eqnarray} 
\label{data} 
\eta_i = \log({\rm E}_i) + \beta_0 + \beta_{{\rm SEc}} {\rm SEc}_i + \phi_i, ~~~~~~~ i= 1,\ldots, 192 
\end{eqnarray} 
where $\beta_0$, $\beta_{{\rm SEc}}$, ${\rm E}_i$ and $\phi_i$ denote  the intercept,  the fixed effect coefficient for covariate SEc, the offset and the spatial effect, respectively.
Similar to the simulation study, we examined the same models for controlling dispersion behavior, including GC, GP, Pois, and NB models, to analyze the data.  

\begin{figure}[ppt]
\centerline{\includegraphics[width=350pt,height=20pc]{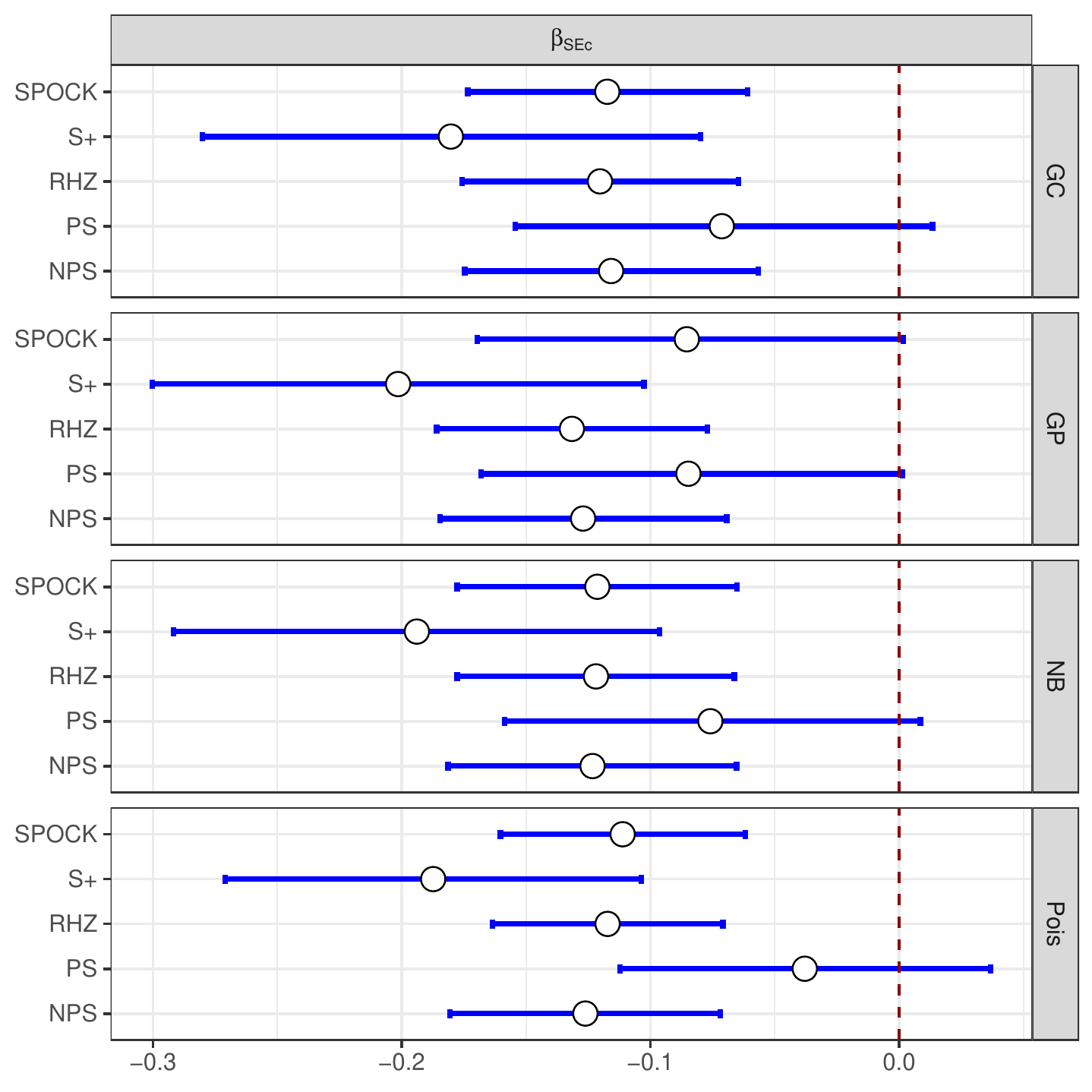}} 
\caption{Slovenia data: Posterior mean estimates (point) and 95\% HPD intervals (lines) of $\beta_{{\rm SEc}}$, 
the effect of socioeconomic status, for different count models, with different approaches to deal with spatial confounding.\label{fig6}} 
\end{figure}

\begin{figure}[ppt] 
\begin{center} 
\begin{tabular}{ccccc}
\includegraphics[width=100pt,height=9pc]{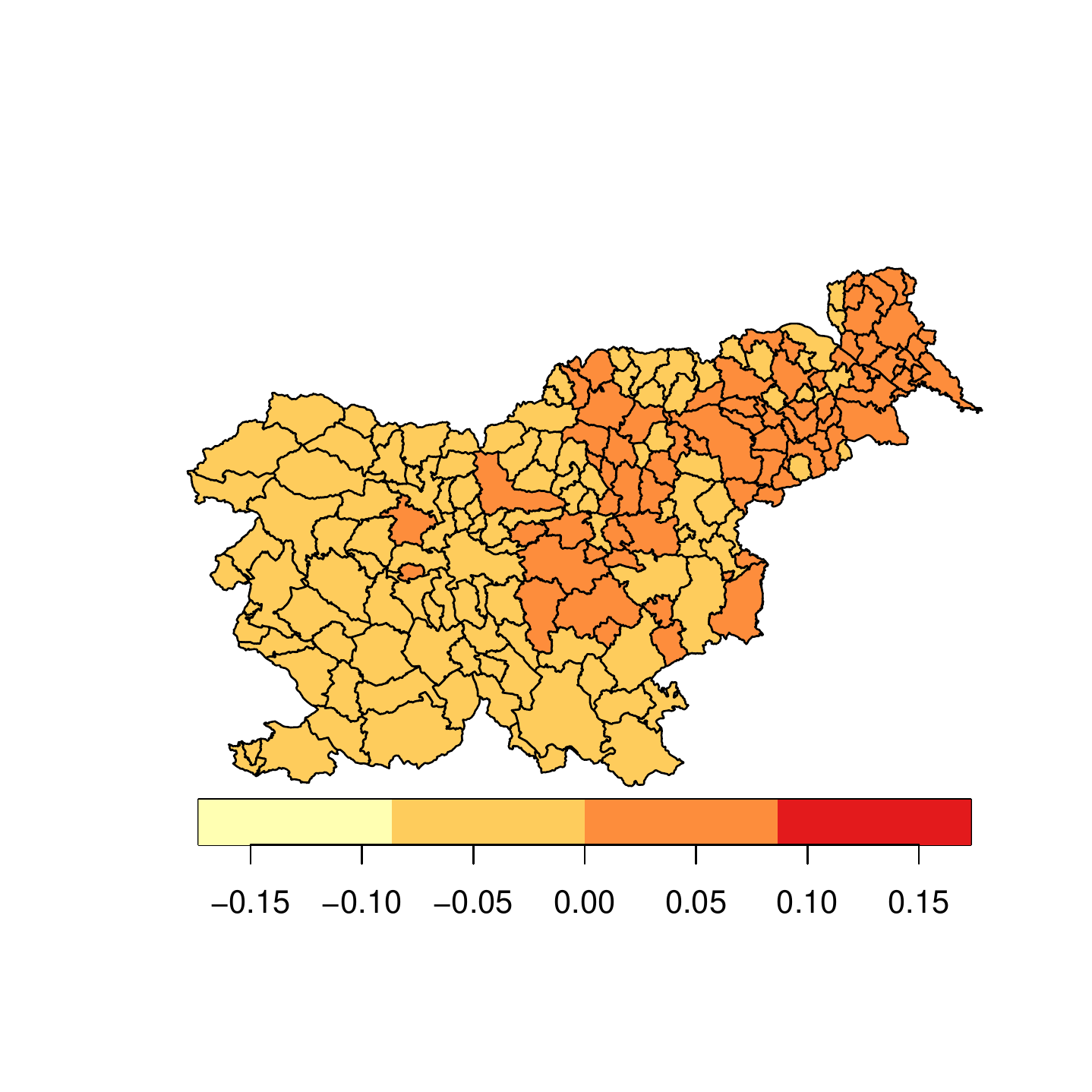} & 
 \includegraphics[width=100pt,height=9pc]{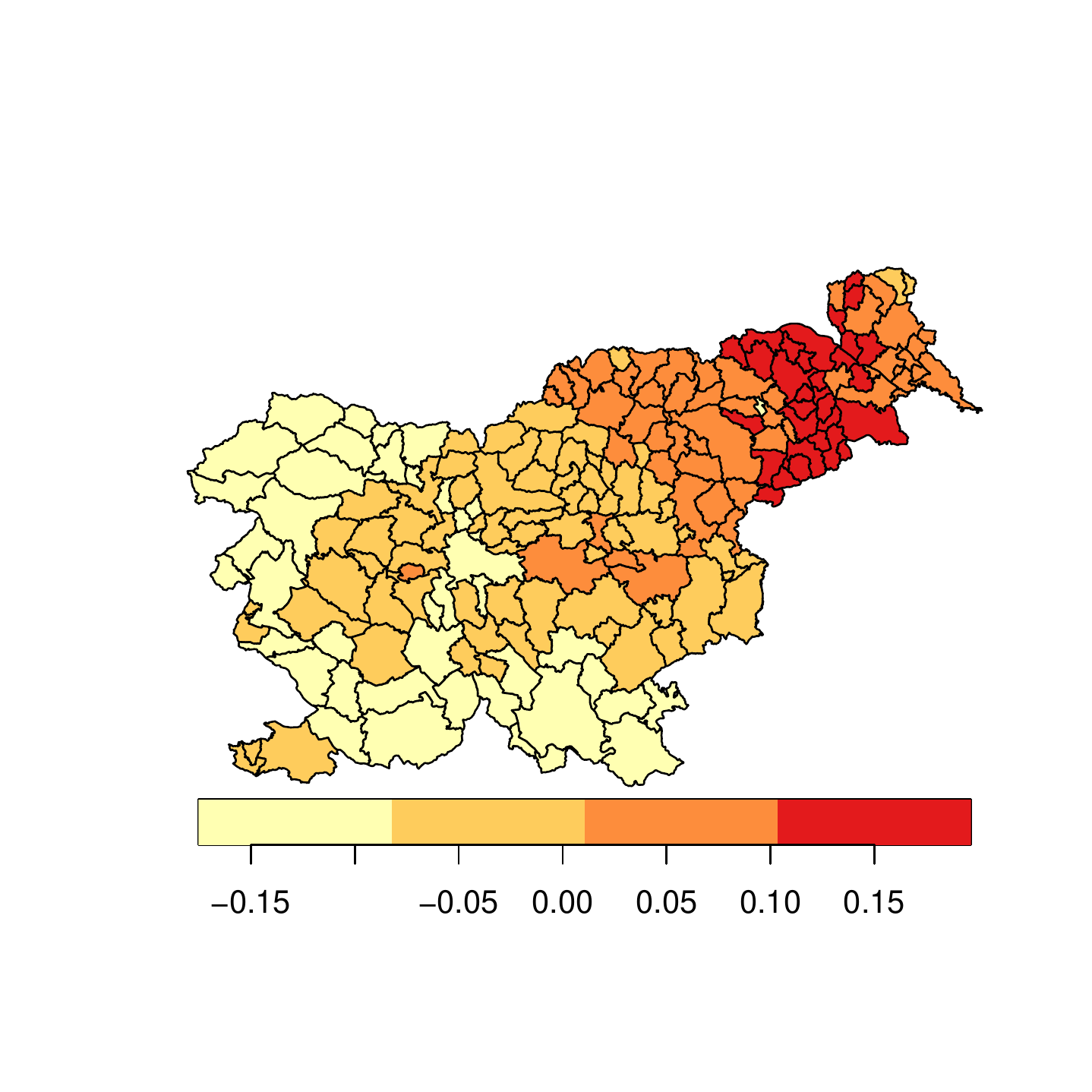} &  
\includegraphics[width=100pt,height=9pc]{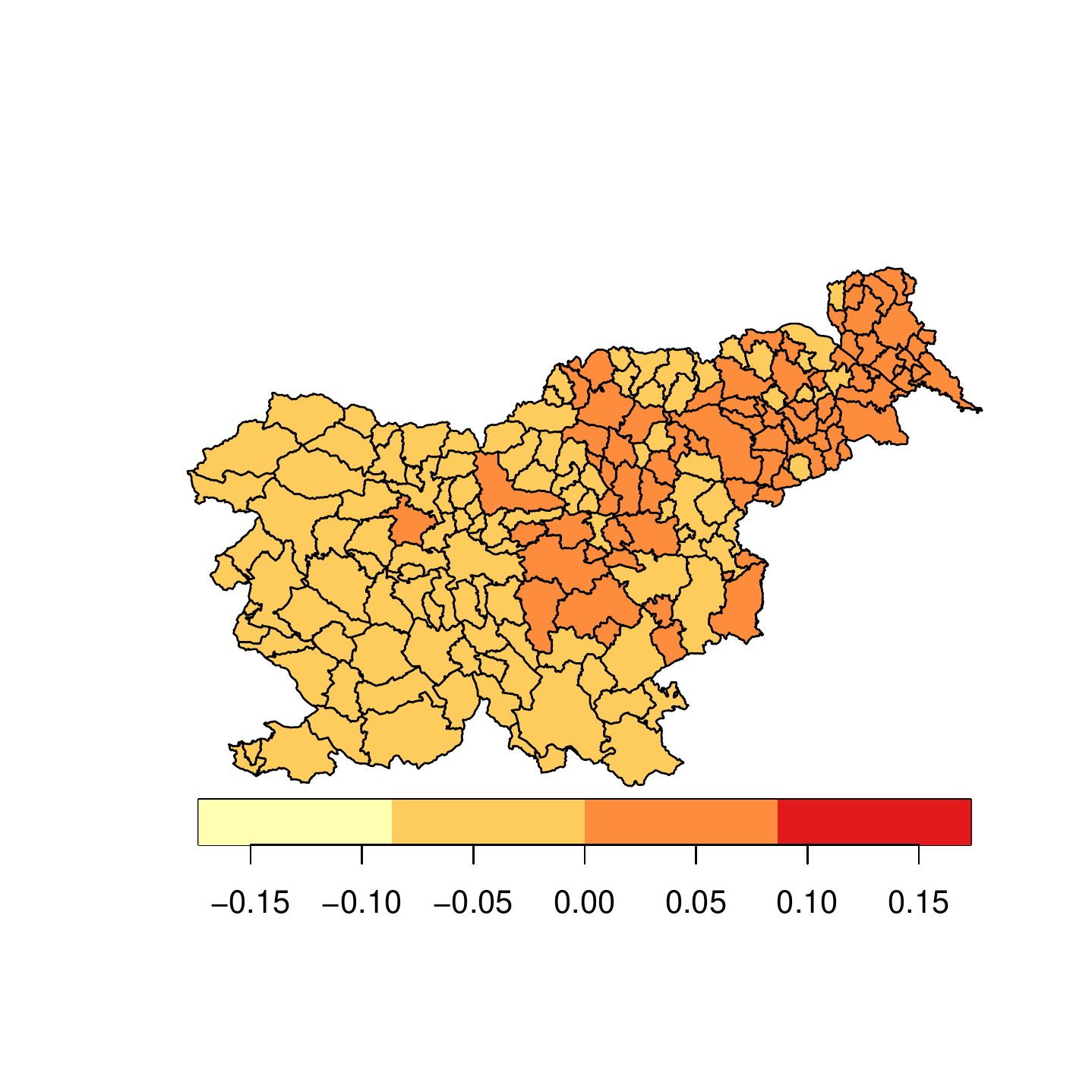} & 
\includegraphics[width=100pt,height=9pc]{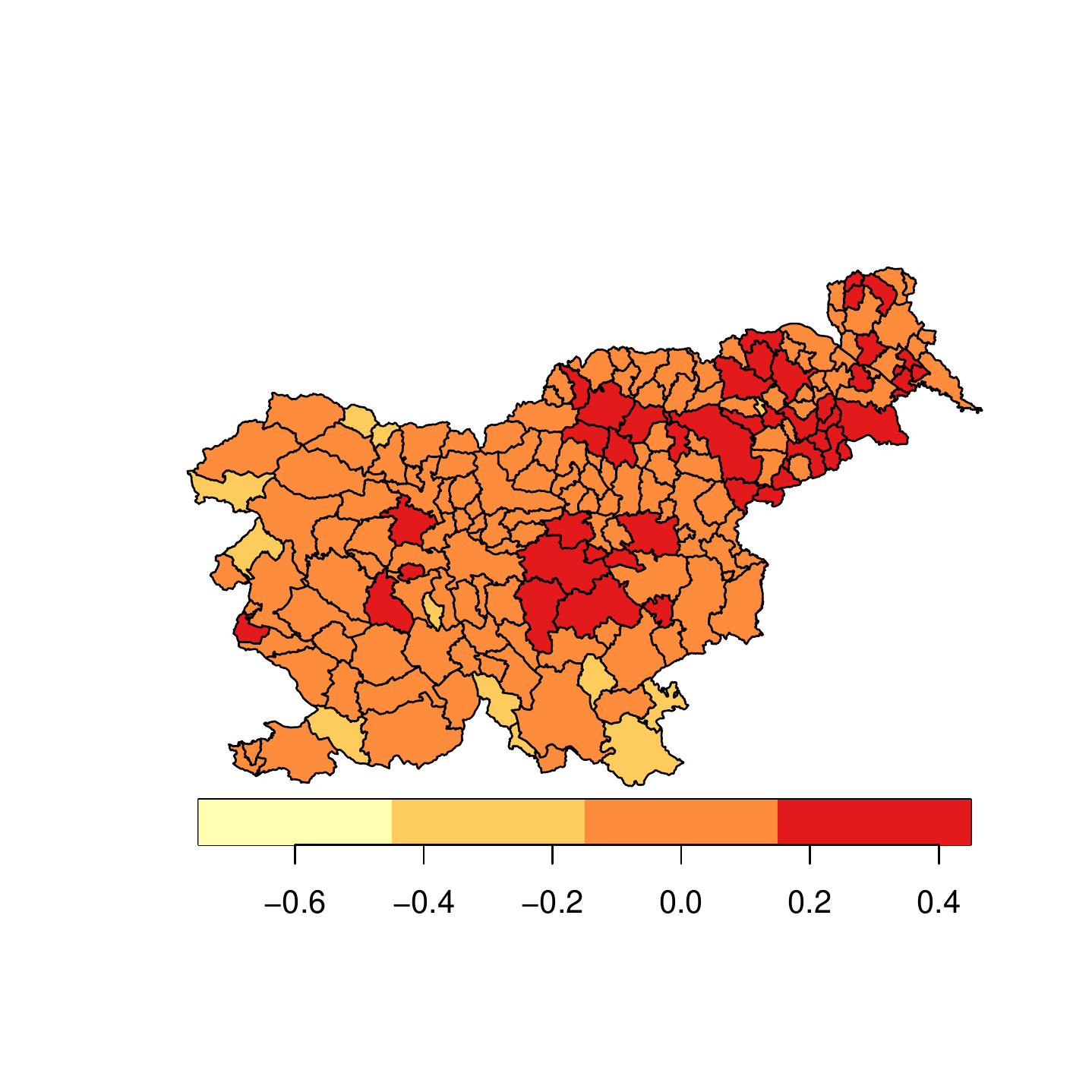}\\
\includegraphics[width=100pt,height=9pc]{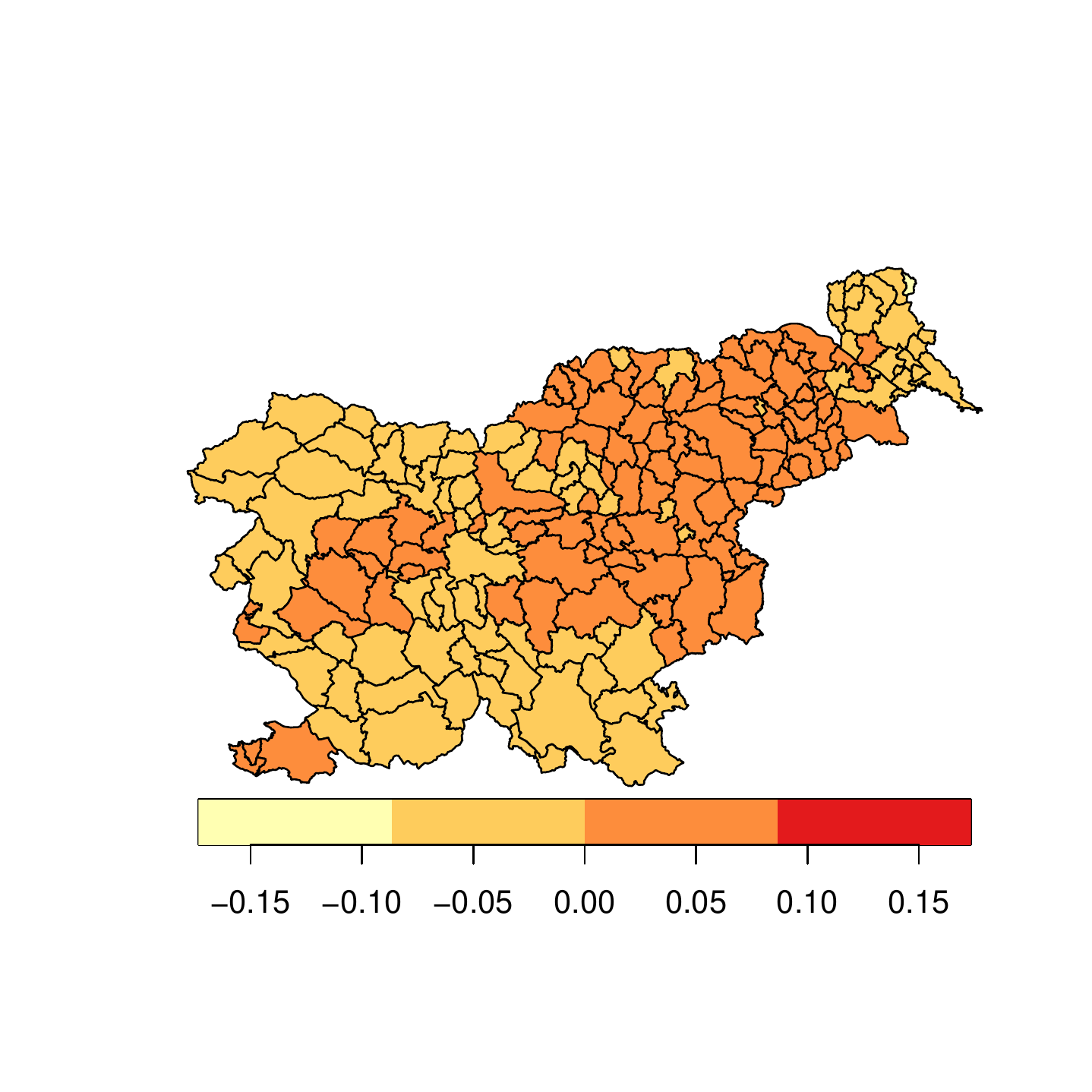} & 
 \includegraphics[width=100pt,height=9pc]{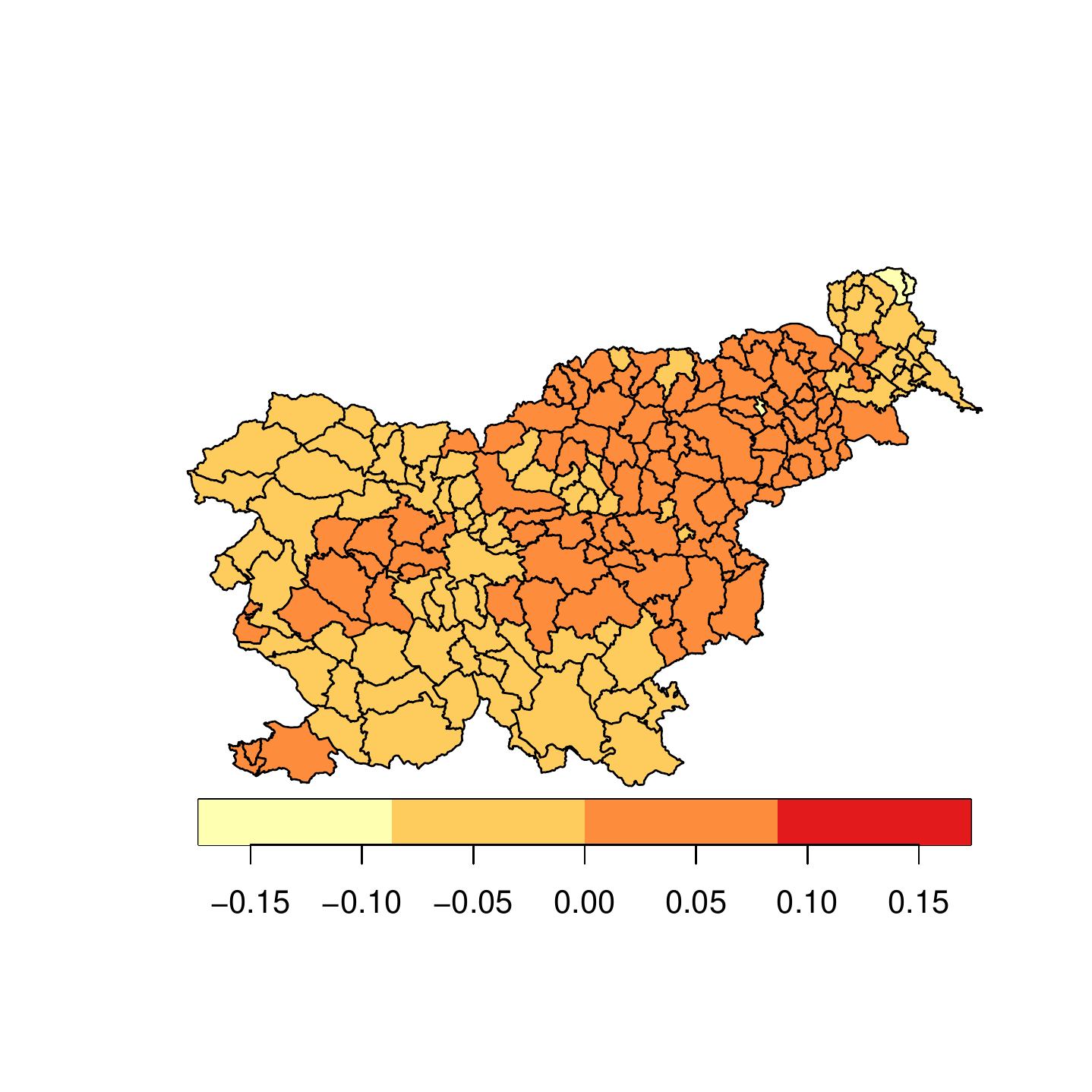} &  
\includegraphics[width=100pt,height=9pc]{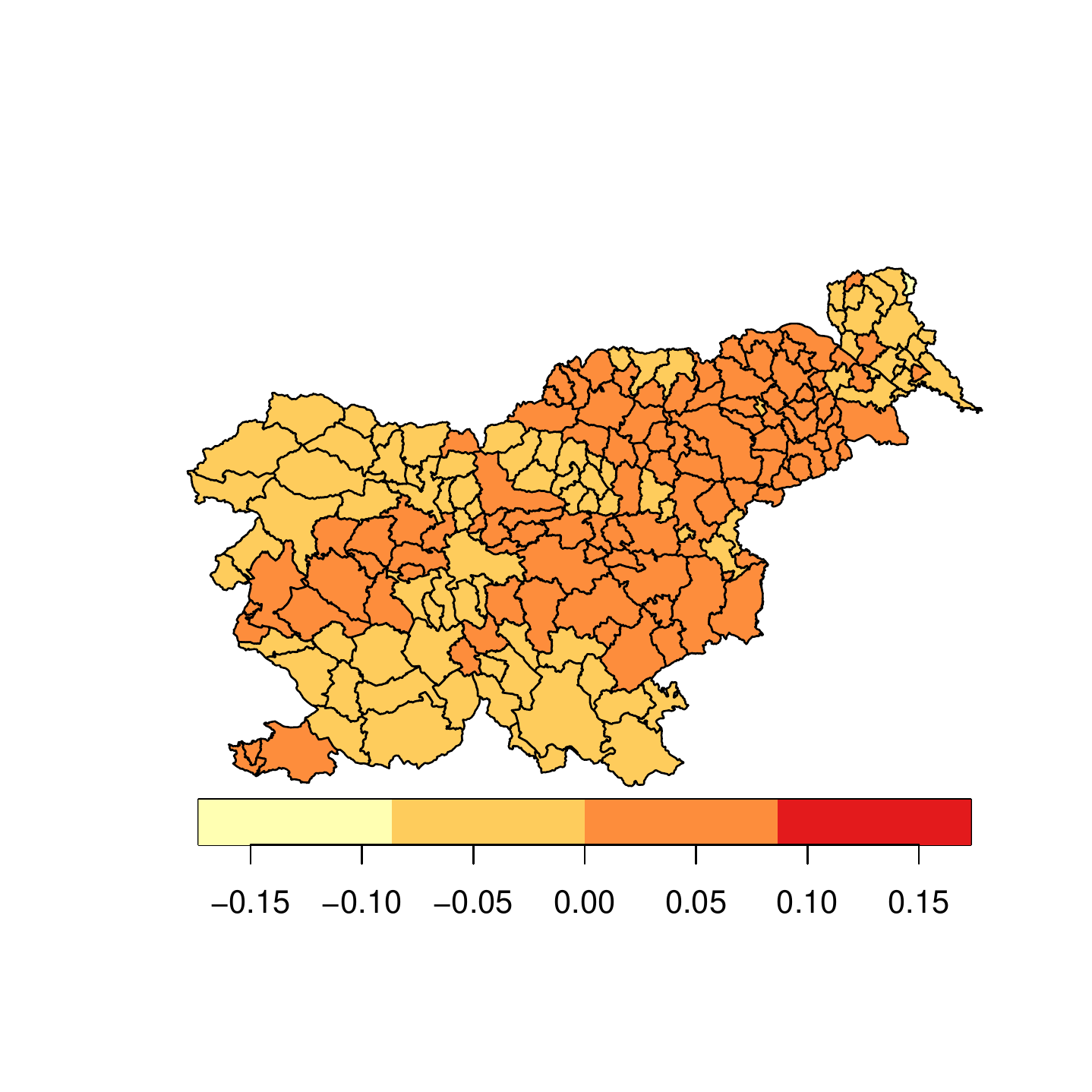} & 
\includegraphics[width=100pt,height=9pc]{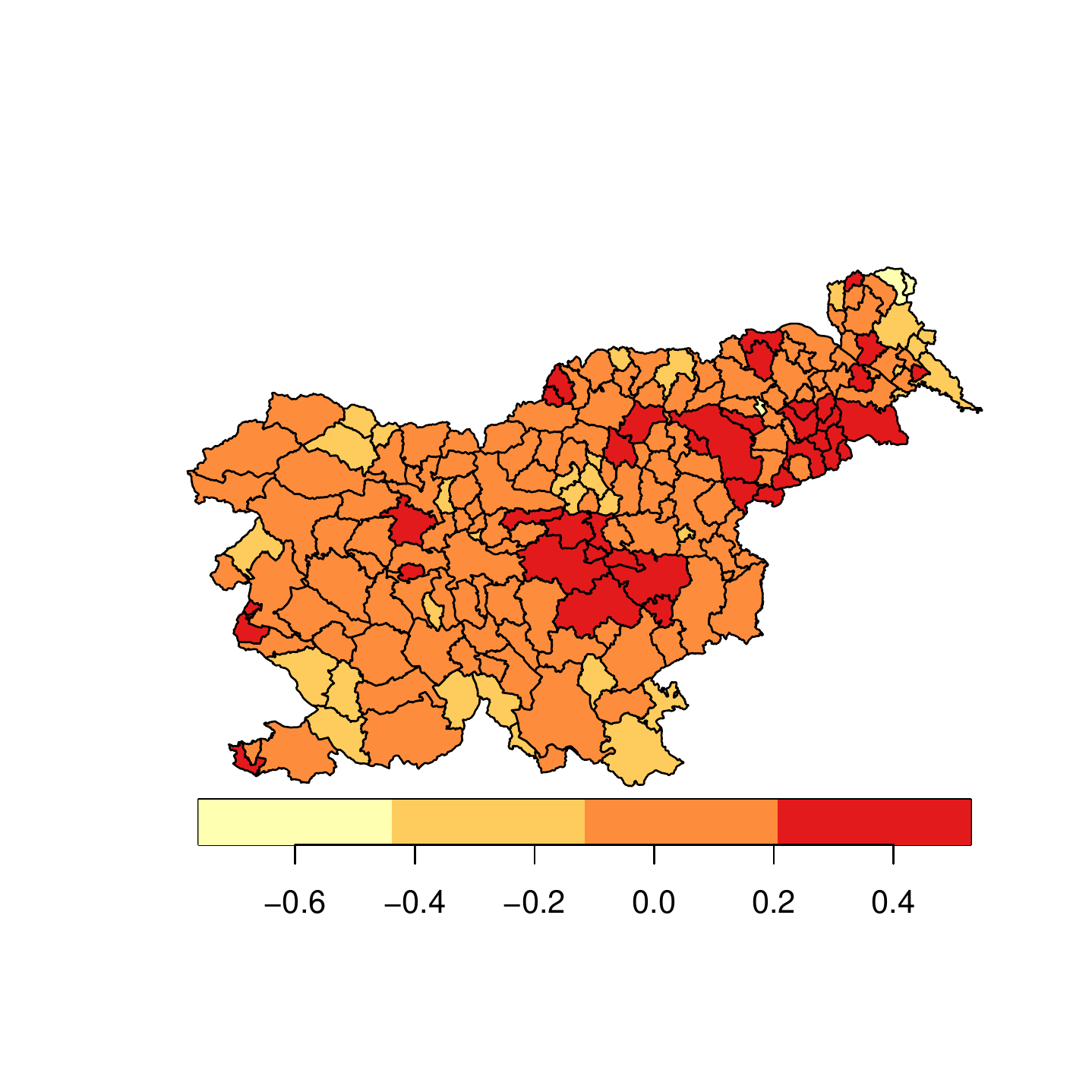}\\
\includegraphics[width=100pt,height=9pc]{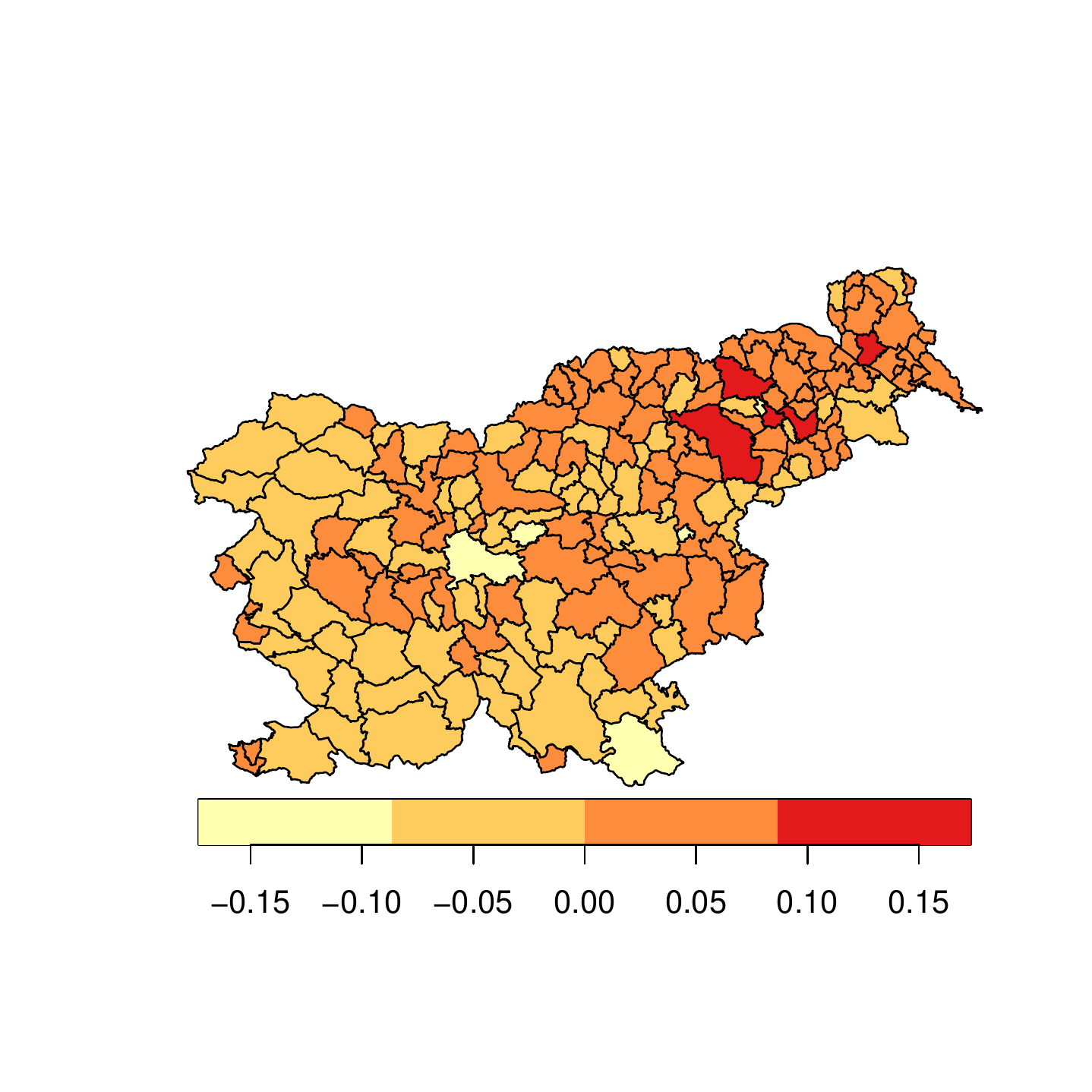} & 
 \includegraphics[width=100pt,height=9pc]{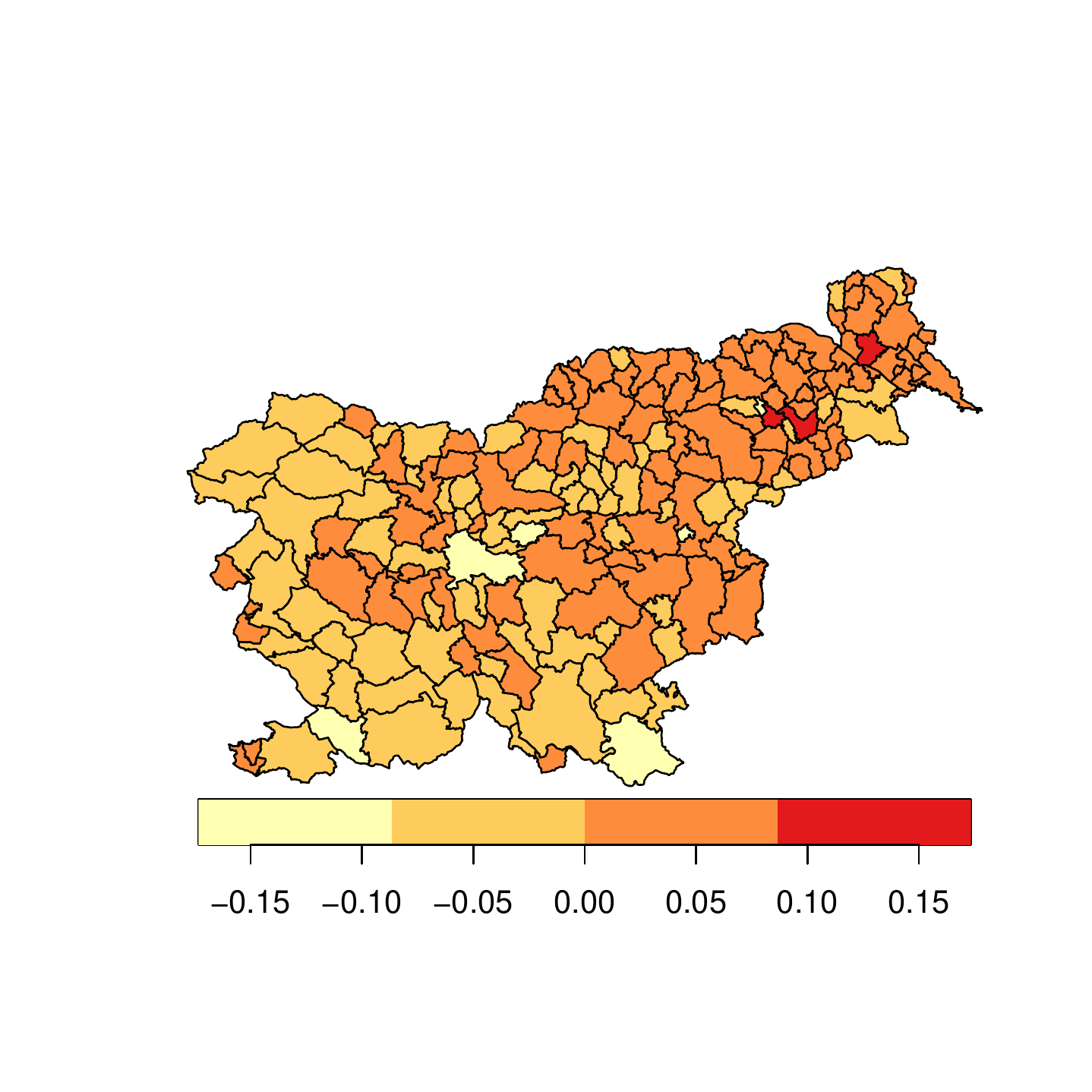} &  
\includegraphics[width=100pt,height=9pc]{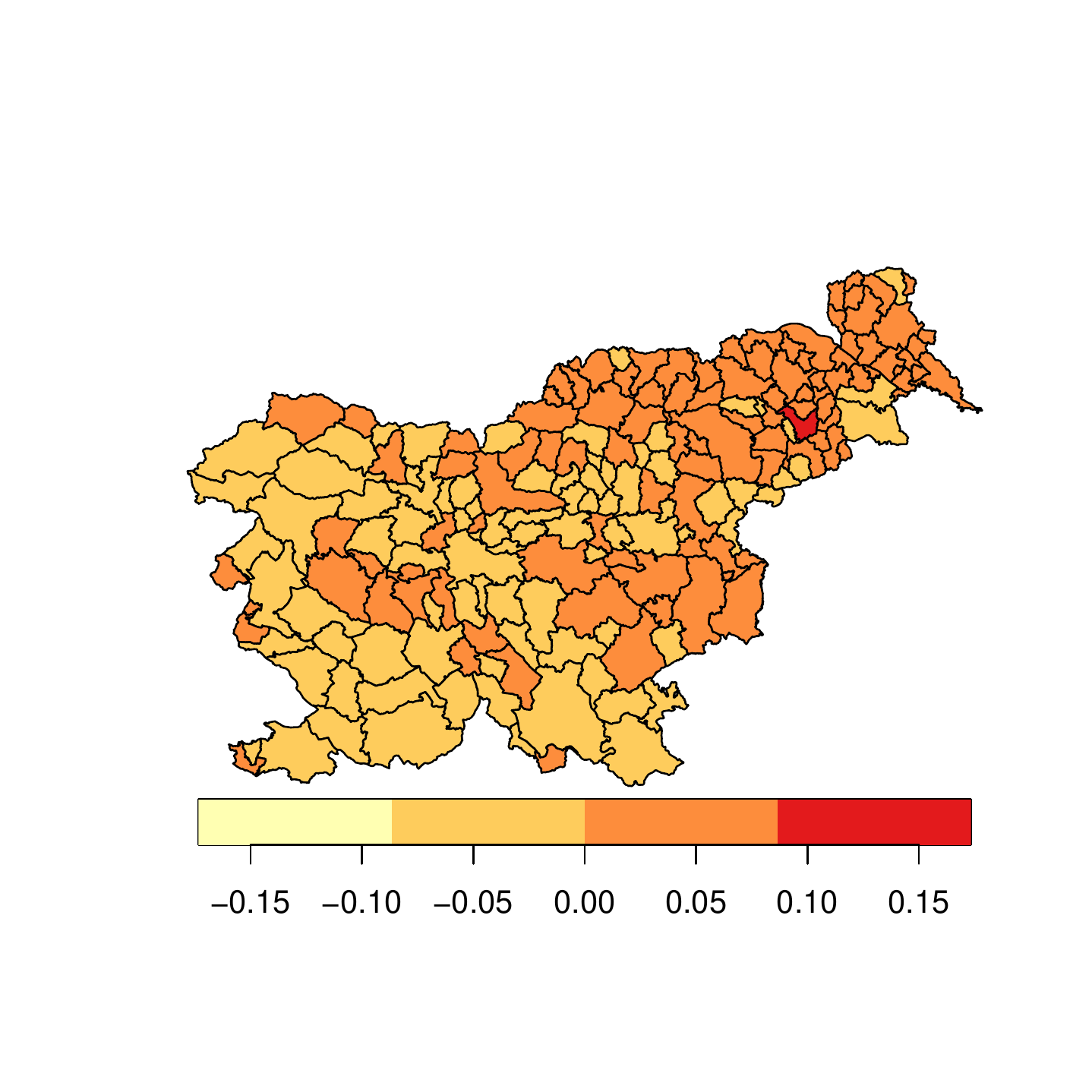} & 
\includegraphics[width=100pt,height=9pc]{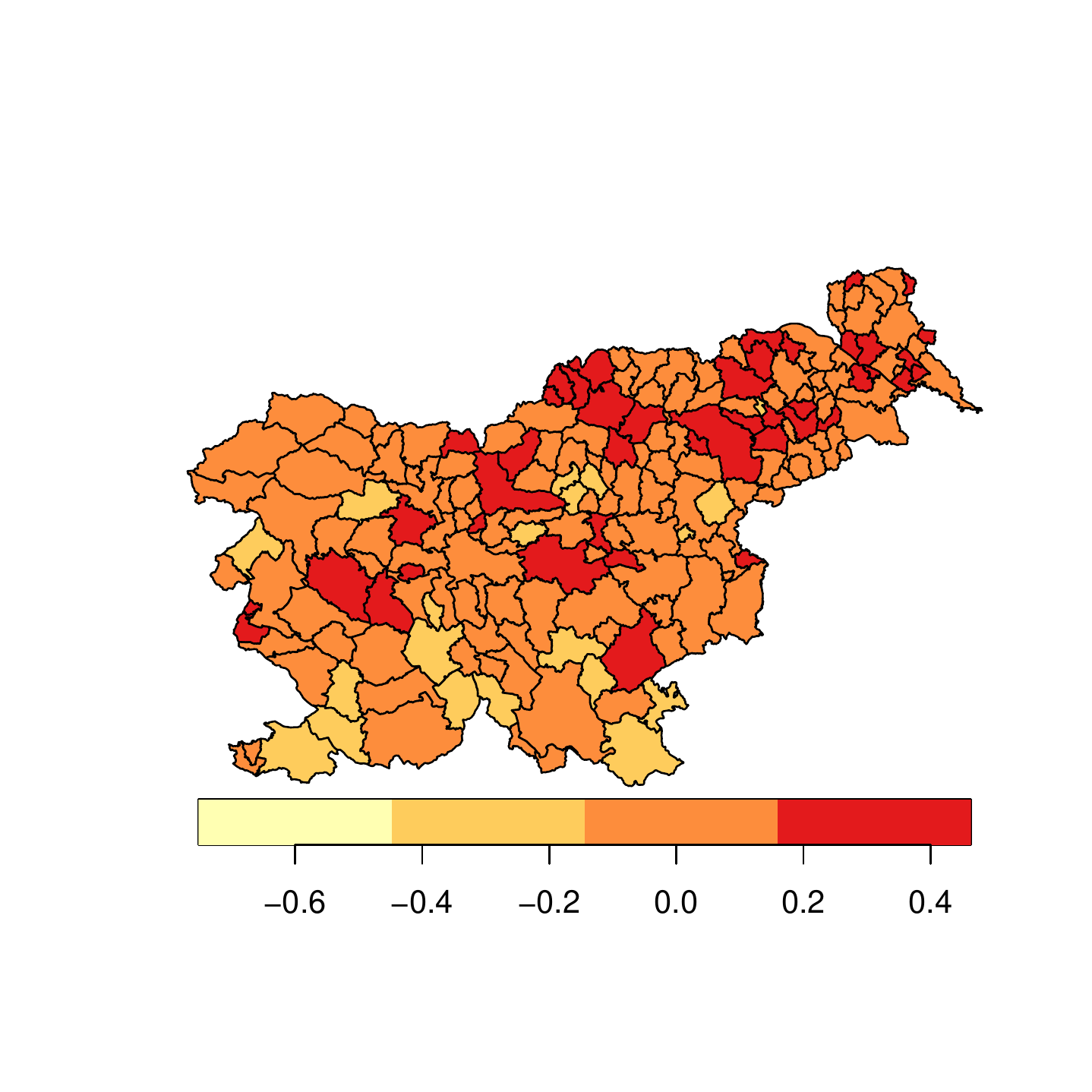}
\end{tabular} 
\end{center} 
\caption{\label{fig7} Slovenia data: Posterior mean estimates of spatial random effects in, from  left to right, the GC (first column), GP (second column), NB (third column) and Pois (forth column) models using the SPOCK (first row), RHZ (second row) and spatial+ (third row) confounding approaches.} 
\end{figure}

To illustrate the confounding problem, we represented the forest plots of the estimated SEc effect in Figure \ref{fig6}. The figure shows the 95\% HPD (highest posterior density) interval and the posterior mean estimate displayed by a white point for each model. The HPD interval of SEc effect for PS model under all count families, and SPOCK model under the GP count model contain zero, which implies socioeconomic status is not significantly related to stomach cancer incidence. However, according to the Figure \ref{fig5} and other previous studies \citep{Azevedo2020}, this result is questionable despite the acceptability of the model selection criteria, reported in Table \ref{Tab1}. 
It shows the posterior inferences for the precision parameter of the spatial models and dispersion parameter in the GC, GP, and NB models. The dispersion has its definition according to the properties of distributions. Moreover, we announced some model selection criteria, including DIC, WAIC, and LS in the table. As we mentioned, the estimated effect for SEc in the PS and SPOCK models under the GP count model is unreliable, and hence, we ignore these models in our final analysis. We first note that, as the estimate of $\alpha$ in the GC model shows, data are over-dispersed.
\begin{table}[ppt] 
\begin{scriptsize}
\centering\caption{\label{Tab1}Slovenia data: Extracted posterior inferences (mean and 95\% HPD intervals) for dispersion and spatial precision parameters, and model selection criteria for competing count models, with different approaches to deal with spatial confounding.} 
\begin{tabular}{lllllllll} 
\hline 
Family & Model & Dispersion  (HPD I.) & $\tau$ (HPD I.) & DIC & WAIC & LS \\ 
\hline  
&NPS & -- & 9.664 ( 3.154 , 18.801) & 986.119 & 1127.282 & 201.121 \\ 
& PS & -- &8.675 ( 3.599 , 15.289 )& 957.279 & 1088.933 & 189.031\\ 
Pois & S+ & -- & 8.821 ( 3.099 , 16.588) & 985.543 & 1126.277 & 199.960\\ 
& SPOCK & -- &9.999 ( 2.883 , 21.720) & 979.249 & 1119.057 &  199.109\\ 
& RHZ & -- & 7.352 ( 3.201 , 12.695) & 959.421 & 1089.960 & 188.435\\  
[3mm]
&NPS & 0.557 ( 0.428 , 0.690) & 62.909 ( 0.597 , 176.558) & 876.579 & 1104.910 &  184.215 \\ 
&PS & 0.588 ( 0.437 , 0.744 )& 94.840 ( 0.224 , 309.855) & 882.319 & 1099.192 & 167.958\\  
GC&S+ &  0.553 ( 0.423 , 0.688) & 51.464 ( 2.916 , 142.476 )& 876.937 & 1106.775 &  184.322\\ 
&SPOCK & 0.543 ( 0.419 , 0.670 )& 444.300 ( 0.295 , 2943.855) & 869.303 & 1103.444 &  182.639\\ 
&RHZ & 0.554 ( 0.422 , 0.691) & 32299.160 ( 0.142 , 434891.734 ) & 873.371 & 1102.739 & 181.415 \\  
[3mm]
&NPS &  0.353 ( 0.202 , 0.516) & 60.044 ( 3.977 , 167.526) & 1009.811 & 1104.976 &  171.661\\ 
&PS & 0.331 ( 0.176 , 0.515) & 90.889 ( 0.737 , 278.509) & 1000.085 & 1100.244 & 171.751\\ 
GP&S+ & 0.358 ( 0.206 , 0.522) & 52.300 ( 4.167 , 141.545) & 1011.858 & 1106.766 & 171.118\\ 
&SPOCK & 0.336 ( 0.183 , 0.520) & 98.043 ( 0.558 , 311.268) & 1000.075 & 1100.295 & 171.942\\  
&RHZ & 0.372 ( 0.209 , 0.559) & 279.099 ( 2.843 , 1756.817) & 1008.031 & 1102.975 & 171.018\\  
[3mm]
&NPS &24.502 ( 10.950 , 41.140) & 84.513 ( 0.213 , 251.111) & 988.772 & 1112.214 & 178.460 \\ 
&PS&  26.938 ( 10.678 , 46.849) & 156.977 ( 0.107 , 525.500) & 980.796 & 1106.019 &  224.306 \\ 
NB&S+ & 23.885 ( 10.646 , 39.974) & 77.092 ( 0.419 , 226.399) & 990.710 & 1114.120 &   177.949\\ 
&SPOCK & 24.017 ( 11.176 , 39.609) & 669.238 ( 0.117 , 4793.692) & 986.611 & 1109.724 &  178.746\\ 
 &RHZ & 22.552 ( 10.838 , 36.344) & 5014.970 ( 0.138 , 257858.881) & 986.533 & 1109.660 &  178.624\\ 
\hline 
\end{tabular} 
\end{scriptsize}
\end{table} 
The given results in Table \ref{Tab1} show that generally, the RHZ model has the best performance based on WAIC and LS and the SPOCK model based on DIC, ignoring the PS model. Furthermore, the GC model is the best-selected model based on DIC criterion; however, the GP model is preferable based on LS and Poisson model based on WAIC. Figure \ref{fig7} displays the estimated spatial effect map for all models, using the SPOCK, RHZ and S+ approaches. The recommended over-dispersed models exhibit the same spatial pattern, but the Poisson model shows more dependence in some regions. Moreover, the estimated precision parameter under RHZ and SPOCK models for Poisson is $\tau = 7.35, 9.99$, while for GC, GP, and NB are $32299.1,444.3$, $279.1, 98.04$, and $5014.97, 669.24$, respectively. Hence, the Poisson model seems to overfit the spatial effect; this could be due to ignoring the over-dispersion of the data that the Poisson model is unable to take into account.
\section{Conclusions}\label{Sec6} 
There are limitations in fitting spatial models with a lack of consideration about the relationship between fixed and random effects, called spatial confounding. Spatial confounding can signify inaccurate inferences about effective covariates in the model. Moreover, count data usually have various dispersion levels and, consequently, this inherent property of counts should be included in the model. In this paper, we presented a hierarchical Bayesian approach for modeling non-confounding spatially dispersed counts by mixing the non-confounding proposed models with the renewal theory that relates gamma distribution for waiting times between events and the distribution of the counts. Our proposed model framework flexibly allows various dispersions from under-dispersion to over-dispersion and considers the spatial confounding. We proposed to use the INLA method of \cite{Rue2009} as an efficient statistical tool for model fitting and inference in a Bayesian spatial GC model. Further, we extended the spatial+ model \eqref{bayesianGCS+} in a Bayesian framework. 
The results of the simulation study showed that our proposed methodology could handle both dispersion and confounding problems in spatial modeling.

\bibliographystyle{apalike}
\bibliography{gc}%

\newpage
\section*{Appendix}

 \begin{figure}[h!]
\begin{center} 
\begin{tabular}{cc} 
 \includegraphics[width=230pt,height=6.8pc]{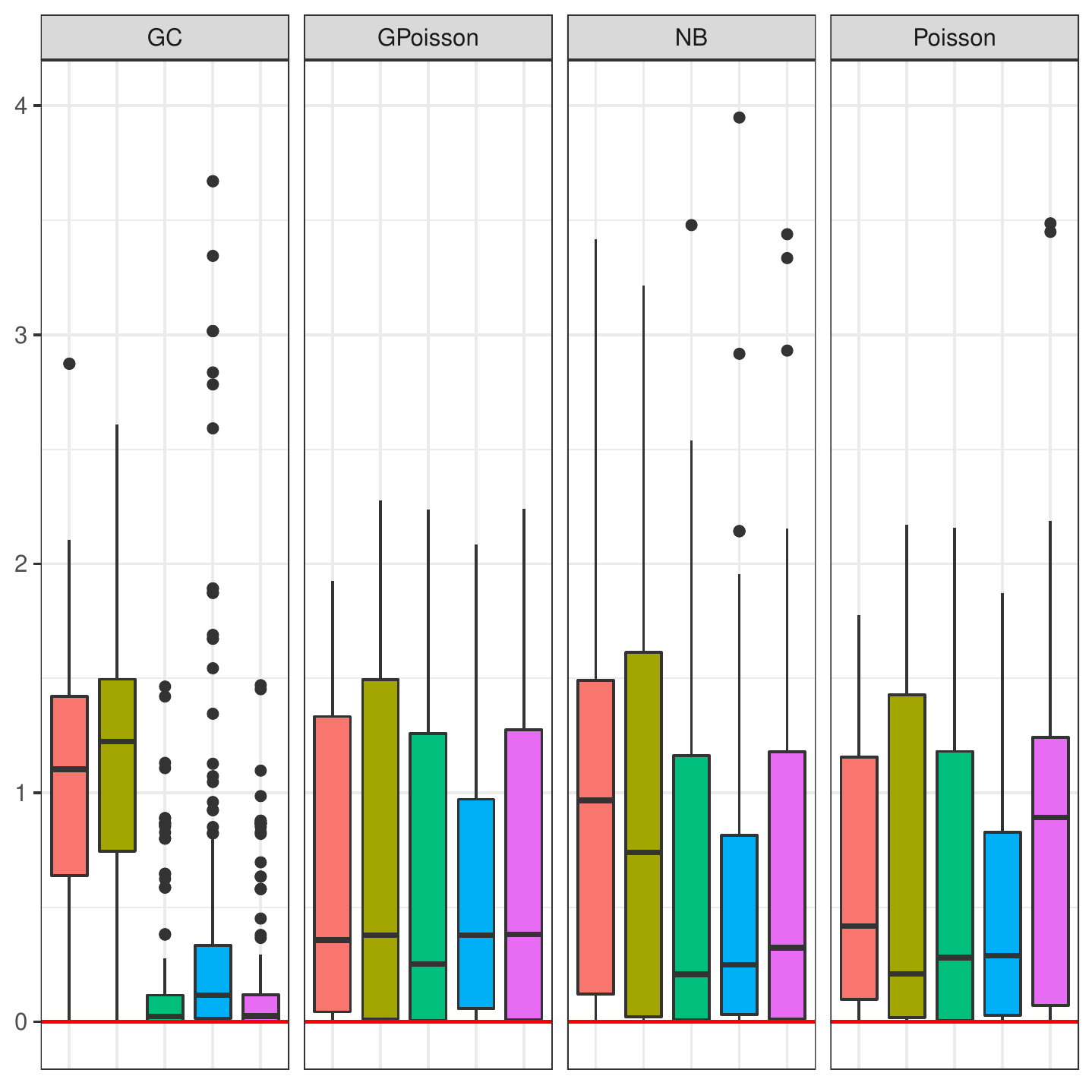}&
 \includegraphics[width=230pt,height=6.8pc]{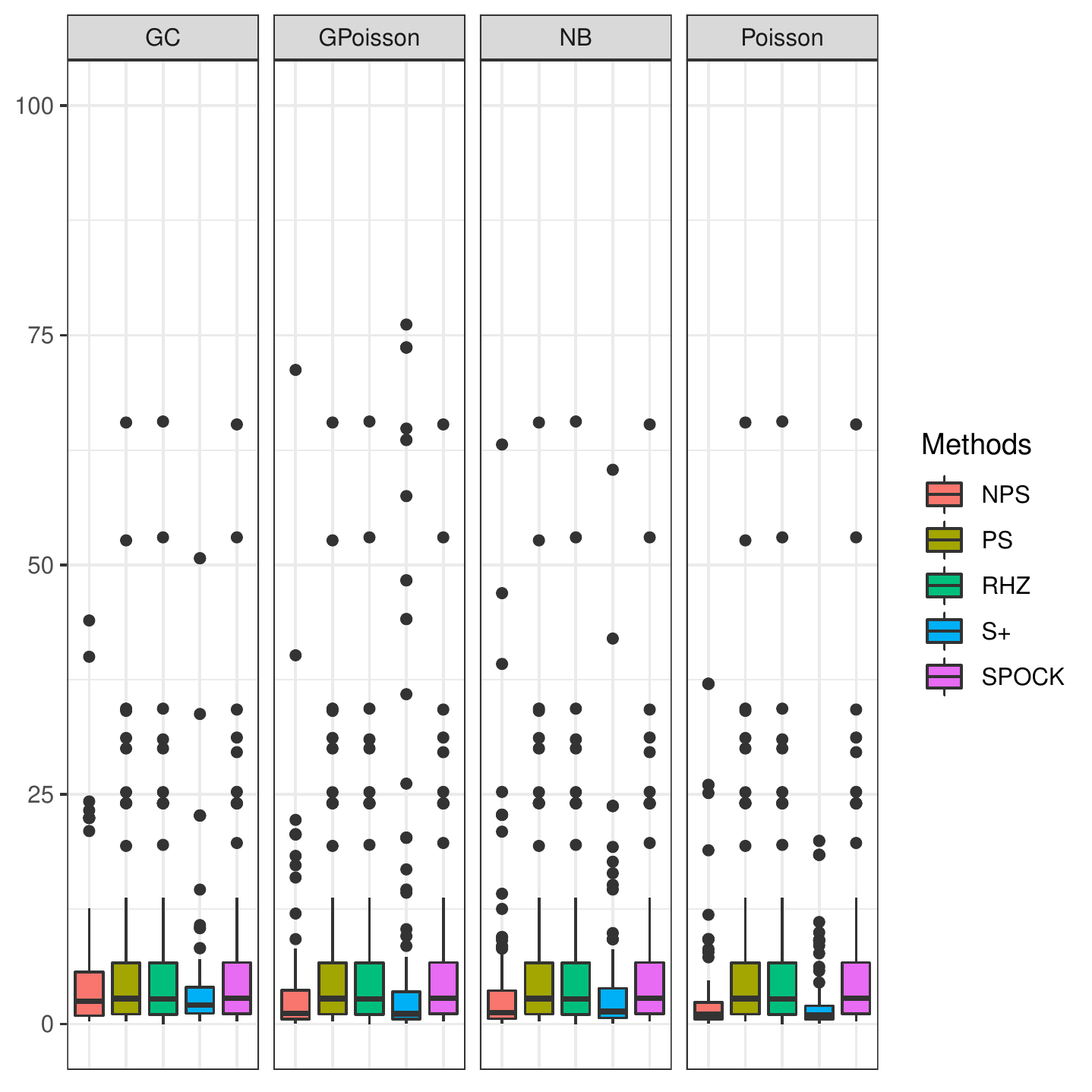}\\ 
\includegraphics[width=230pt,height=6.8pc]{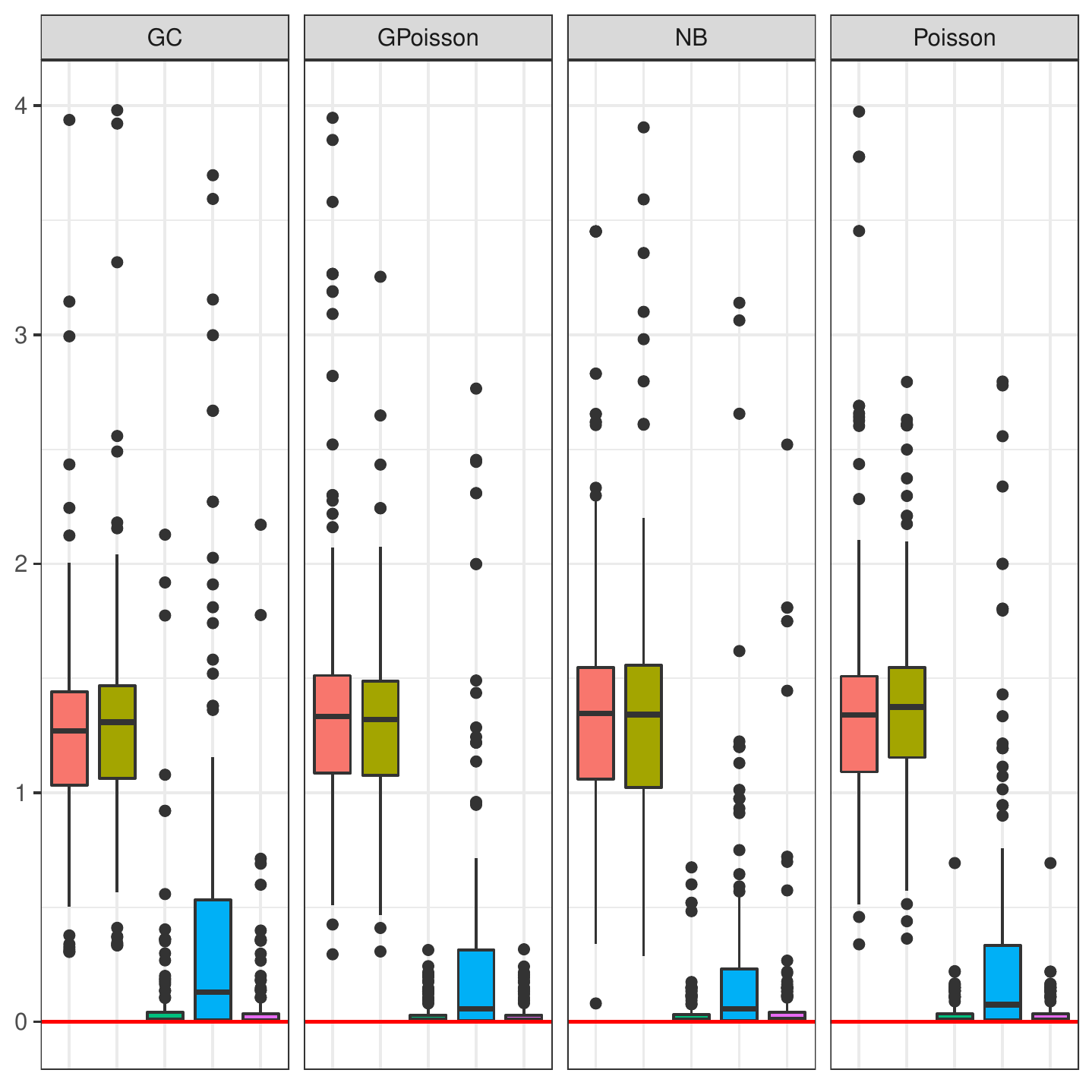}&
\includegraphics[width=230pt,height=6.8pc]{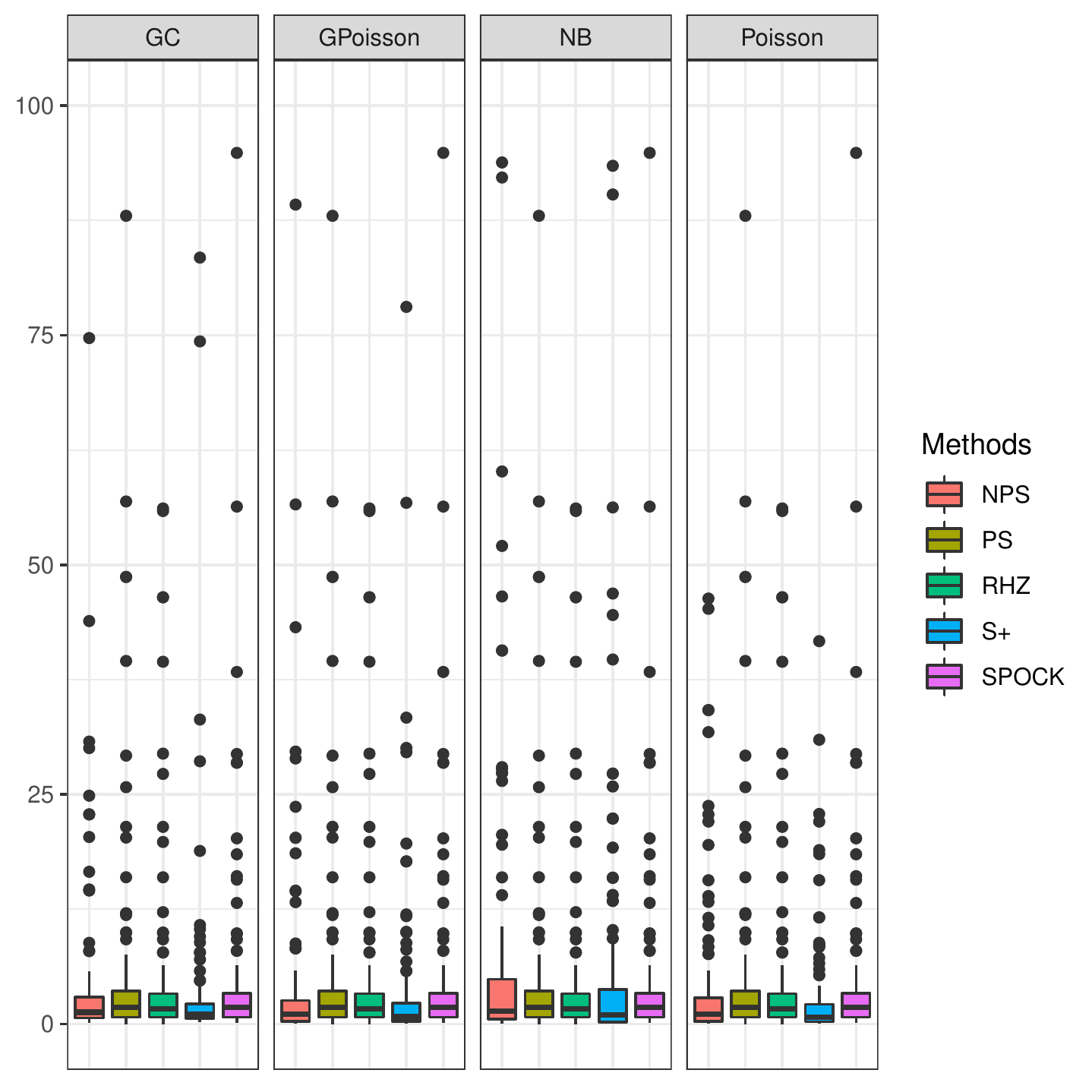}\\ 
\includegraphics[width=230pt,height=6.8pc]{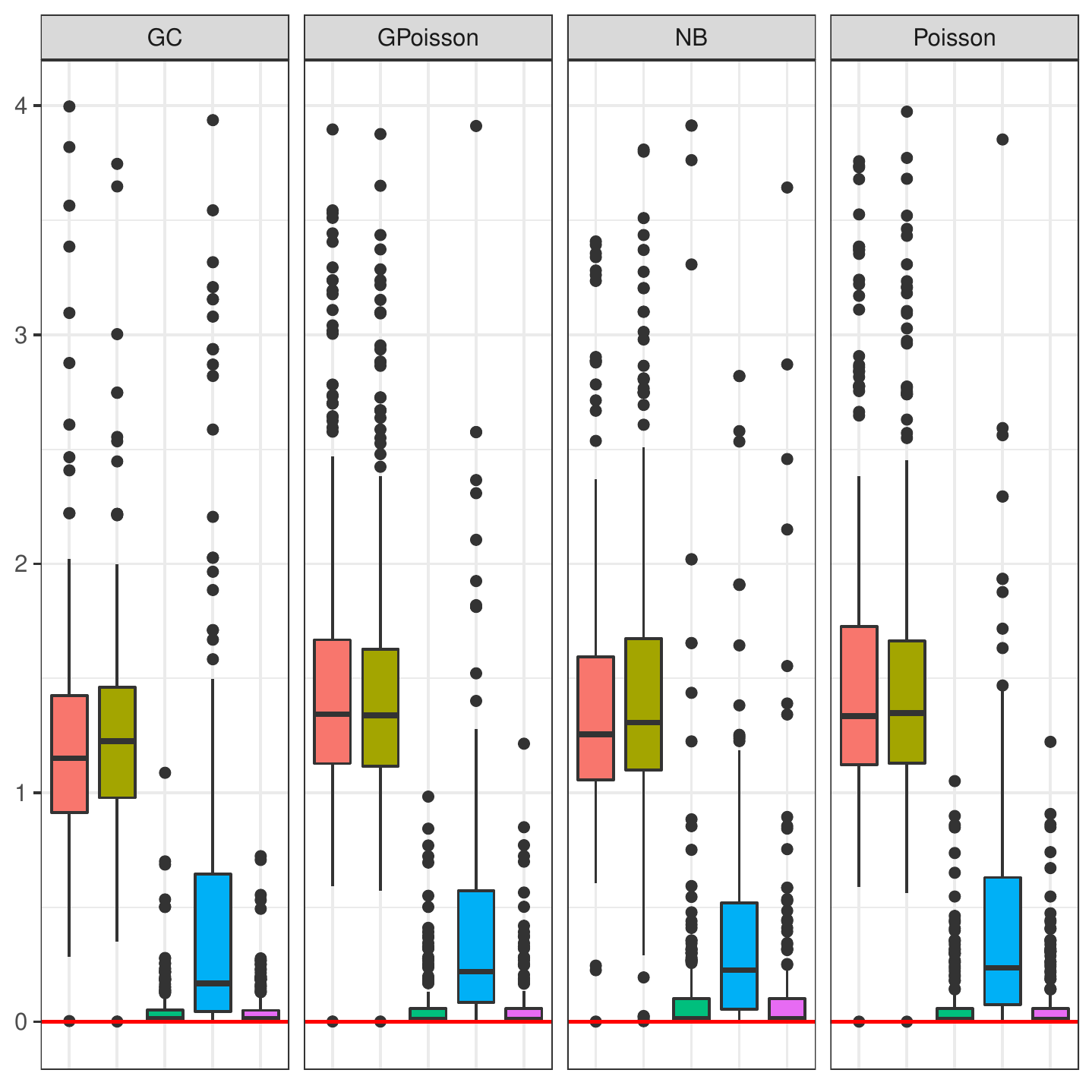}&
\includegraphics[width=230pt,height=6.8pc]{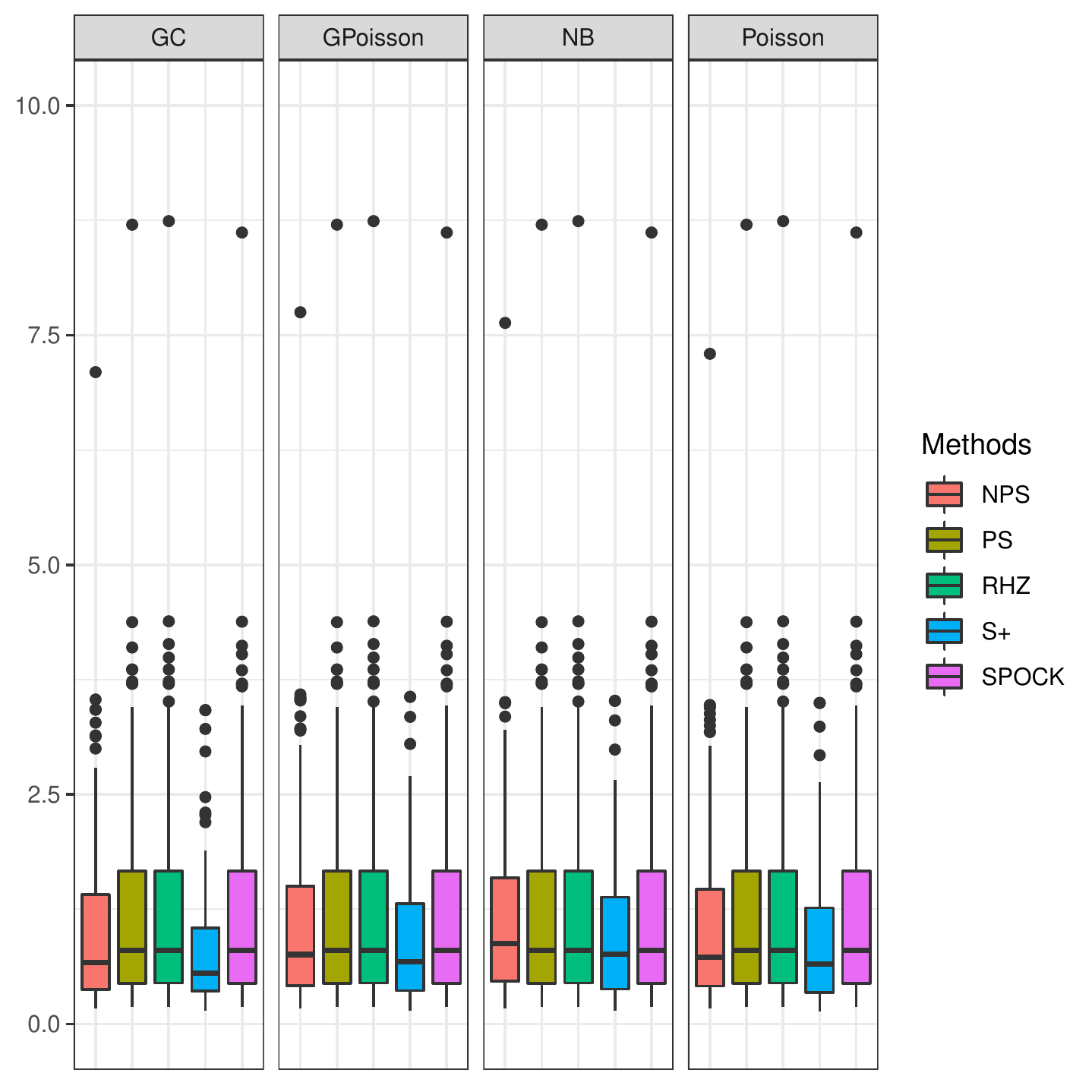}\\ 
\includegraphics[width=230pt,height=6.8pc]{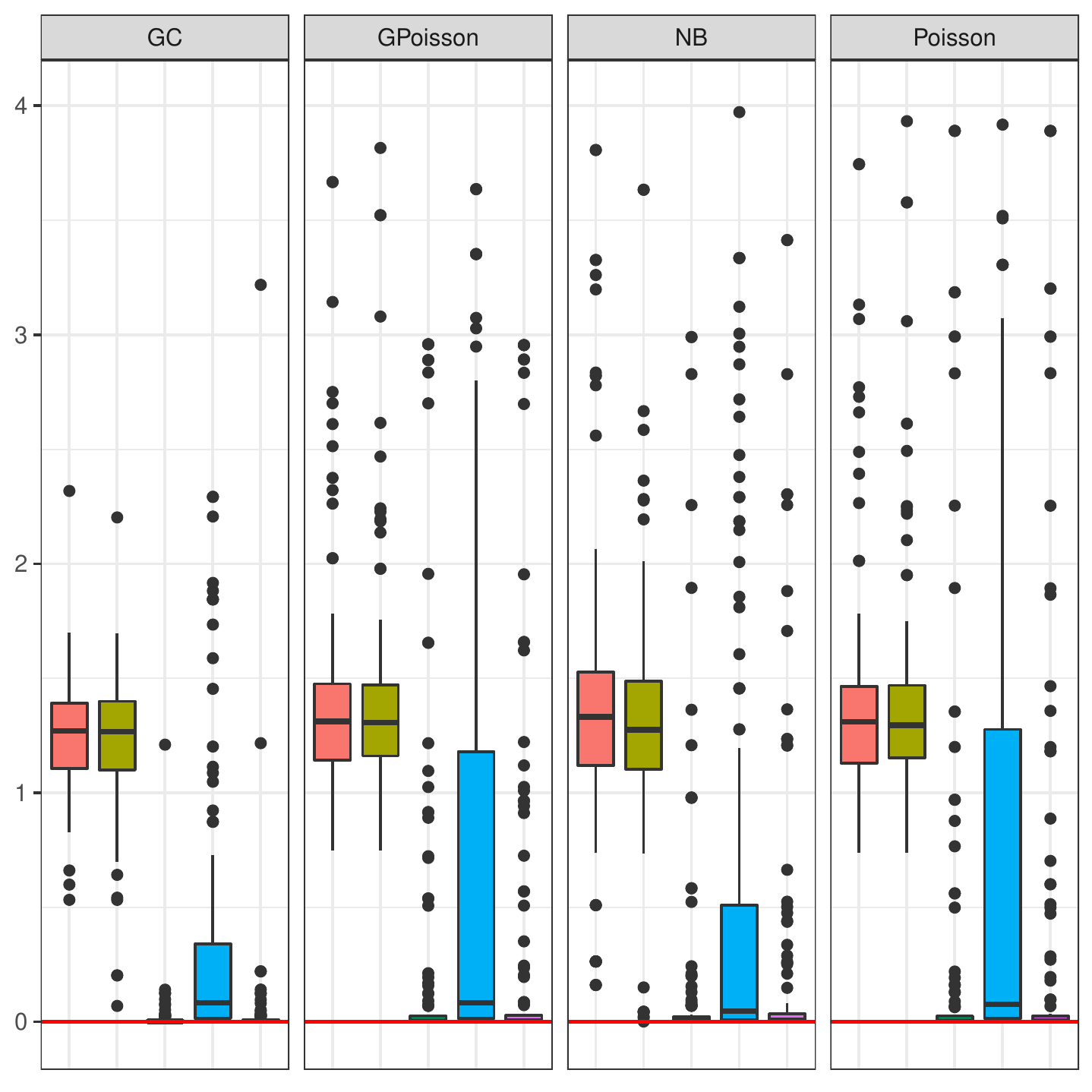}&
\includegraphics[width=230pt,height=6.8pc]{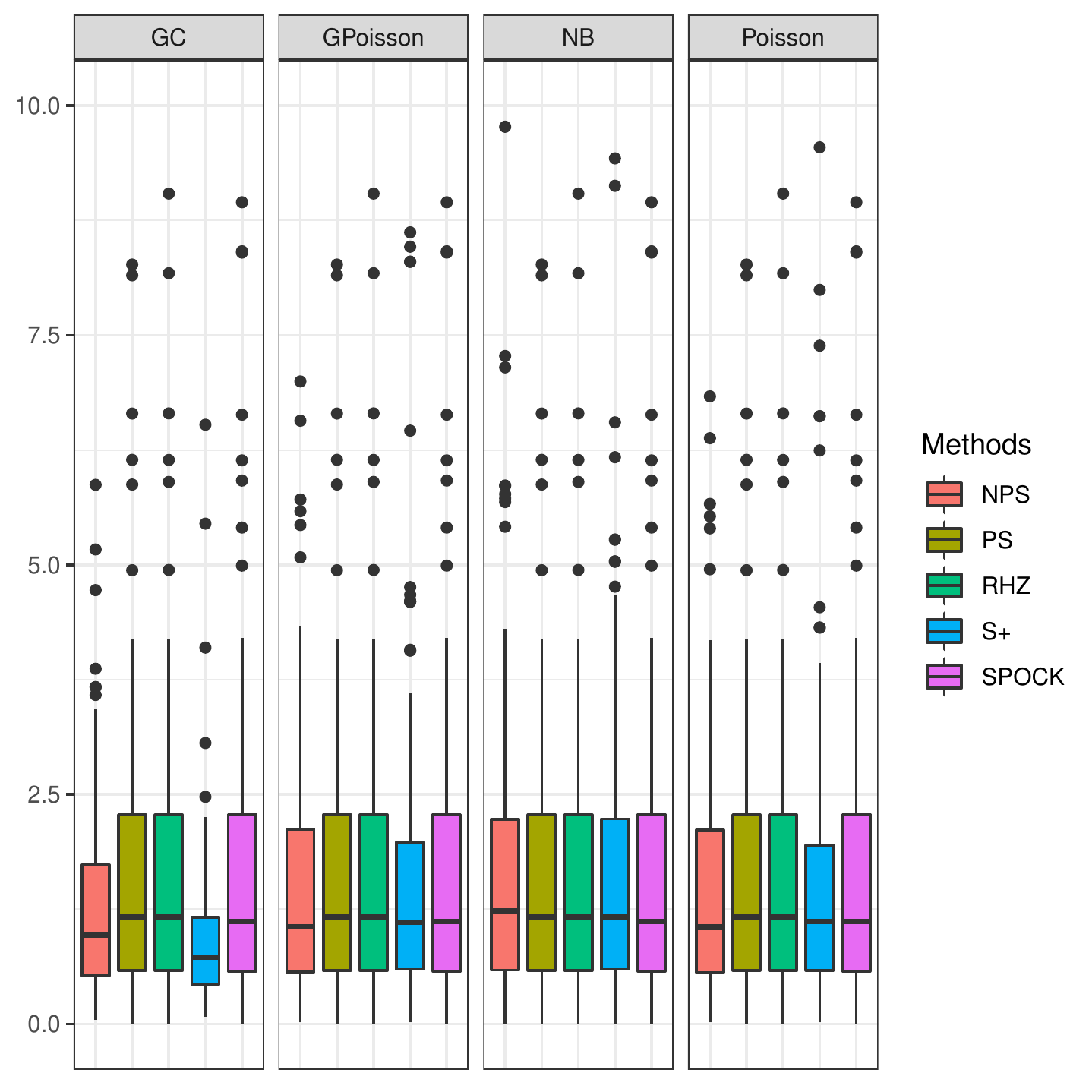} 
\end{tabular} 
\end{center} 
\caption{SE of confounded covariate effect, $\beta_2$, (left column) and 
MSPE of fitted values (right column) for each 
model with $\tau_x=11$ for scenarios: over-dispersion (first row), equivalent-dispersion (second row), under-dispersion (third row; $\alpha=1.3$) and under-dispersion (forth row; $\alpha=2$).\label{fig8}} 
\end{figure} 
\begin{figure}[h!] 
\begin{center} 
\begin{tabular}{cc} 
 \includegraphics[width=230pt,height=6.8pc]{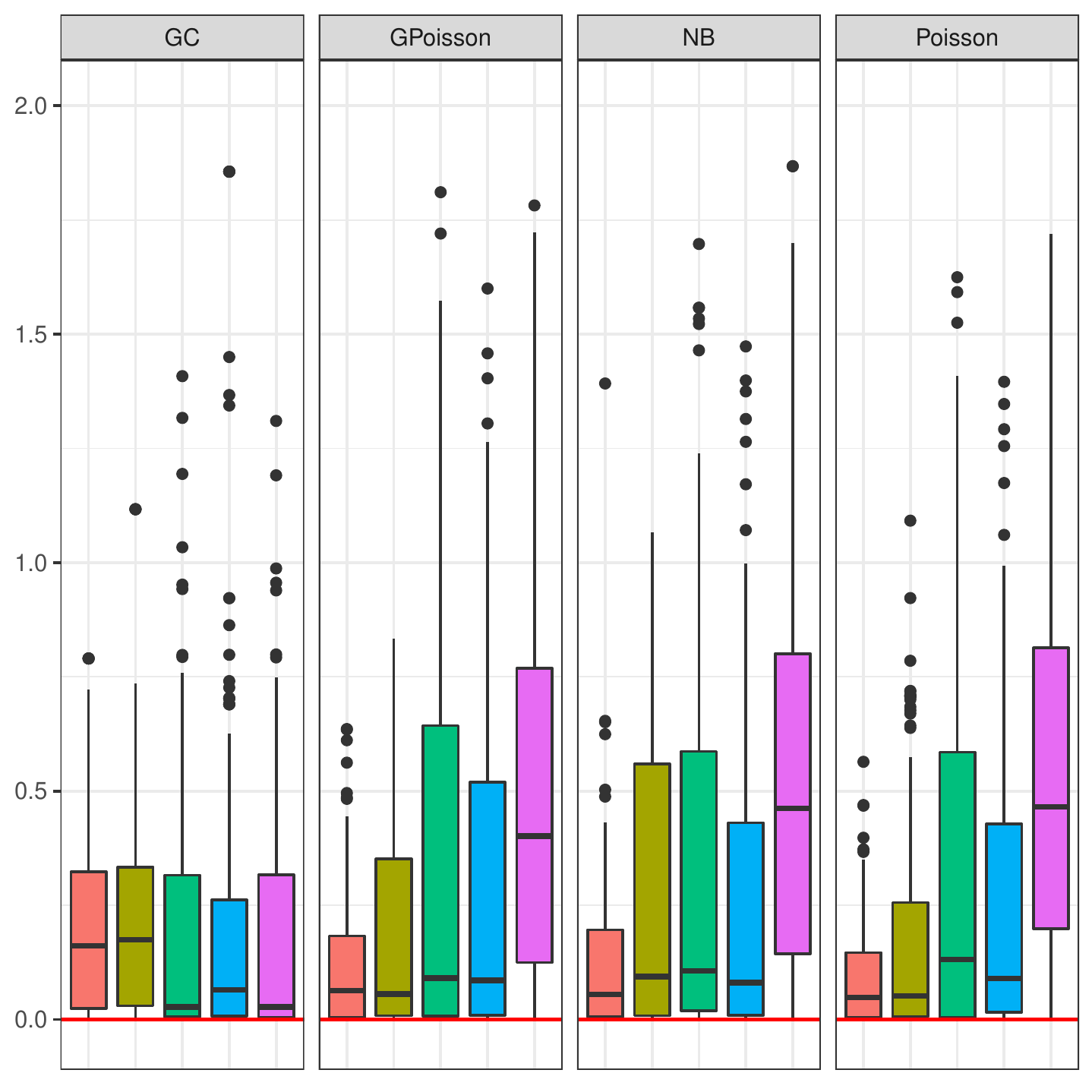}&
 \includegraphics[width=230pt,height=6.8pc]{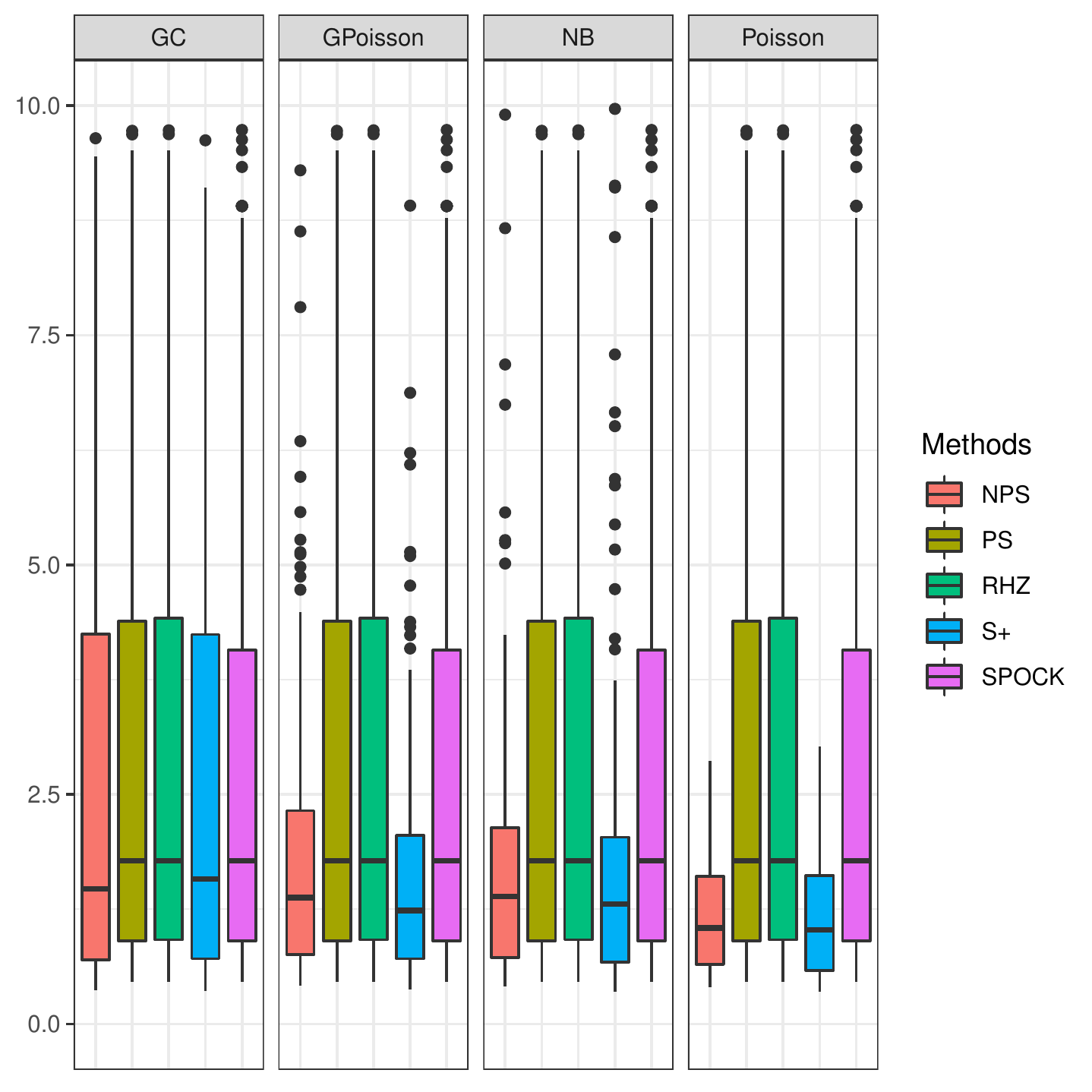}\\ 
\includegraphics[width=230pt,height=6.8pc]{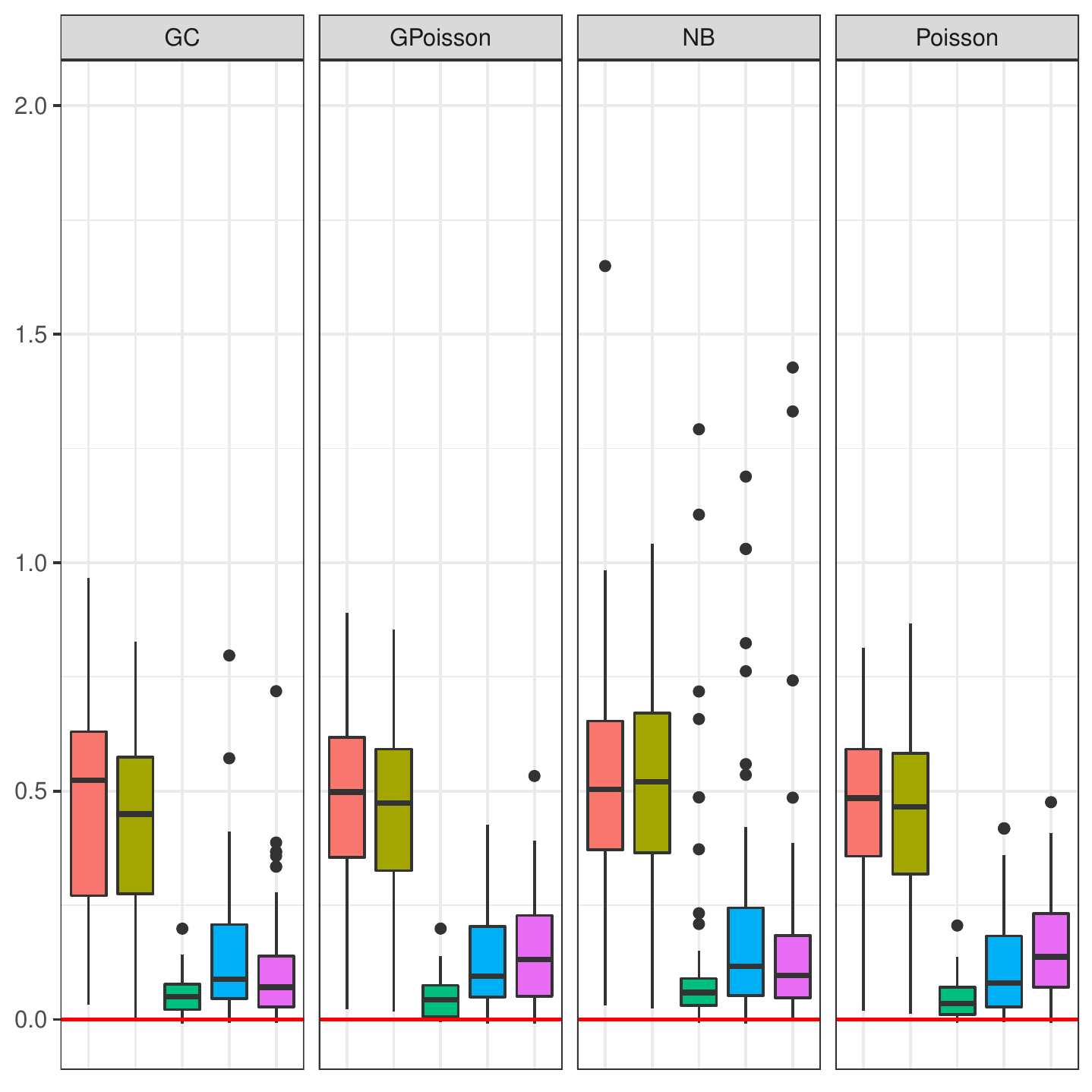}&
\includegraphics[width=230pt,height=6.8pc]{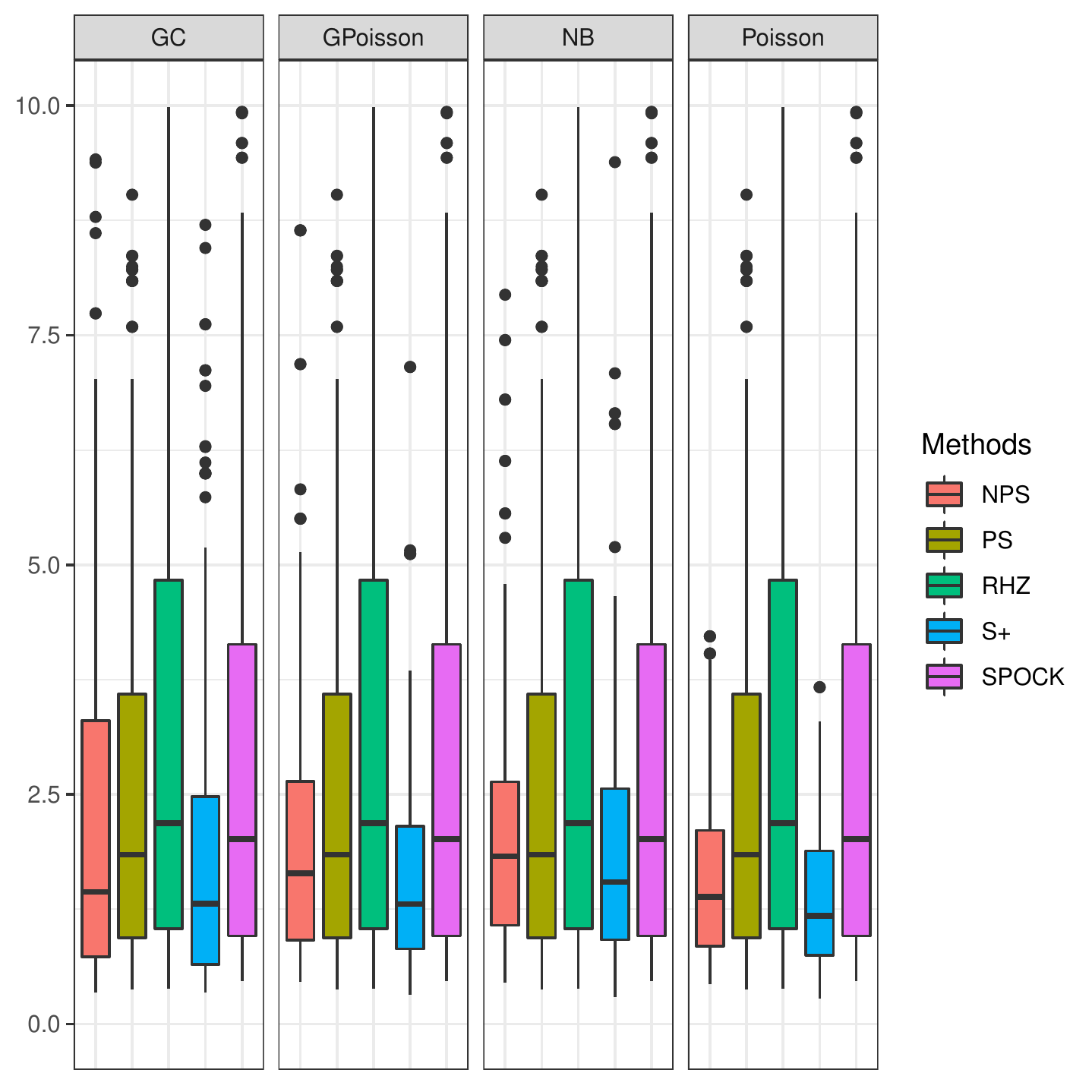}\\ 
\includegraphics[width=230pt,height=6.8pc]{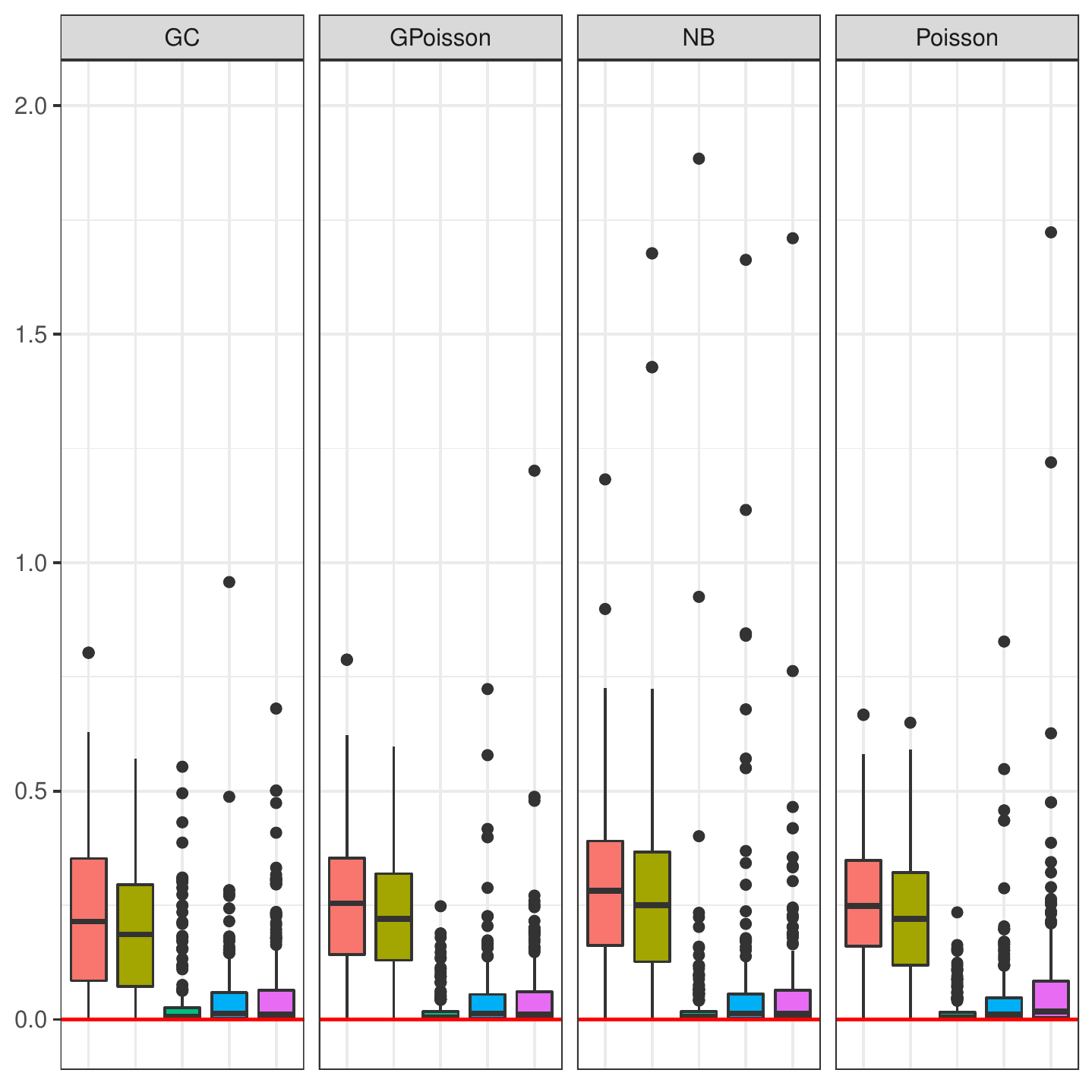}&
\includegraphics[width=230pt,height=6.8pc]{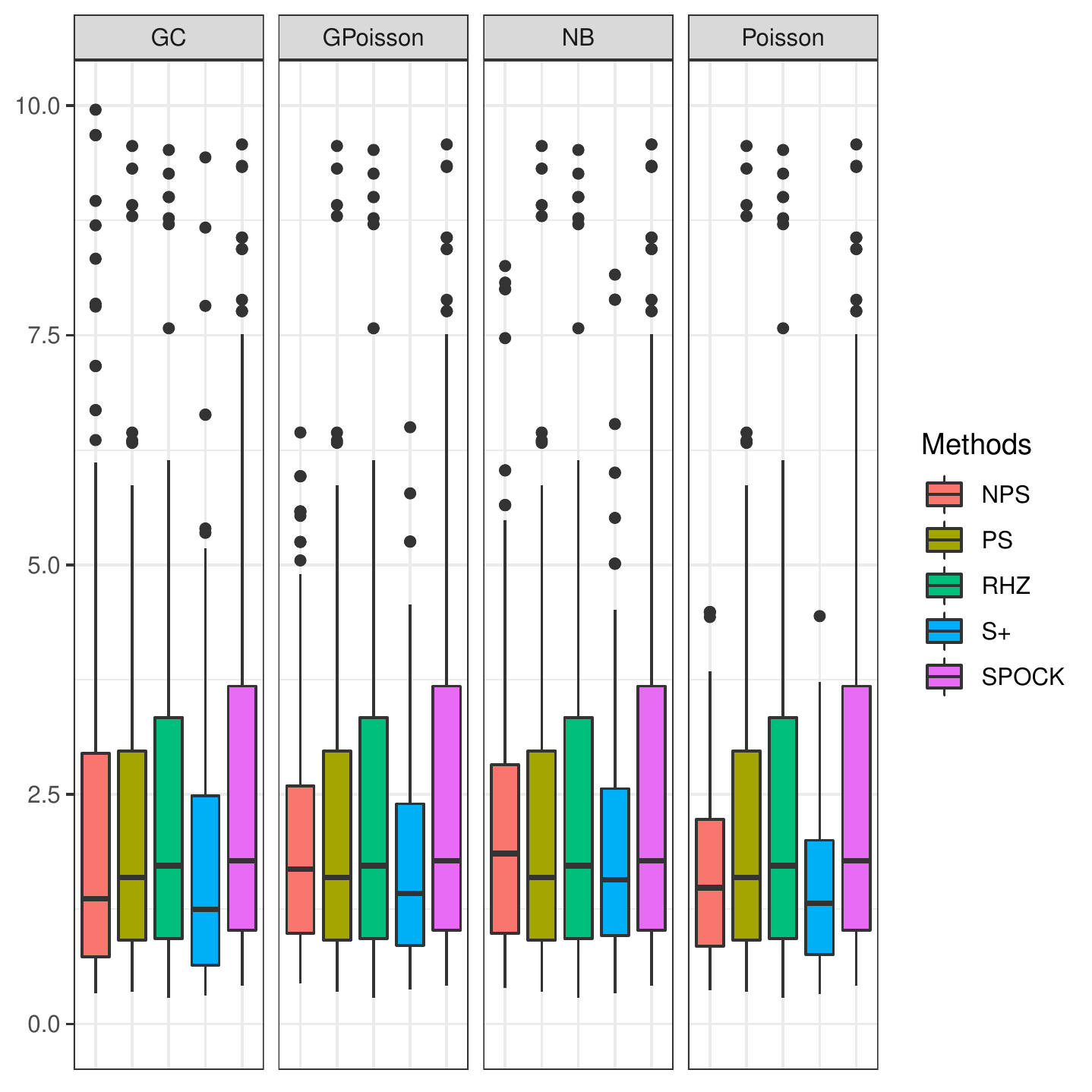} \\ 
\includegraphics[width=230pt,height=6.8pc]{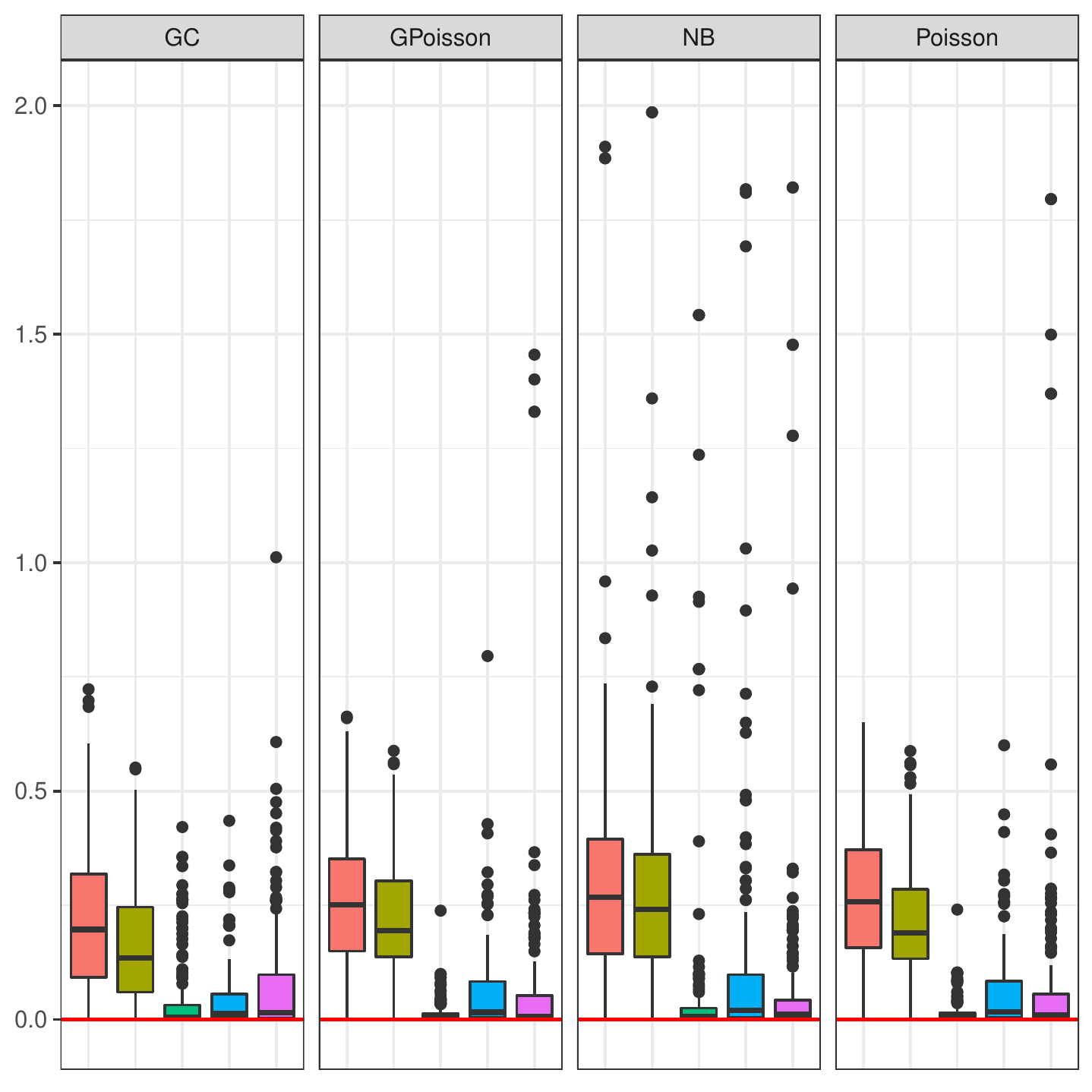}&
\includegraphics[width=230pt,height=6.8pc]{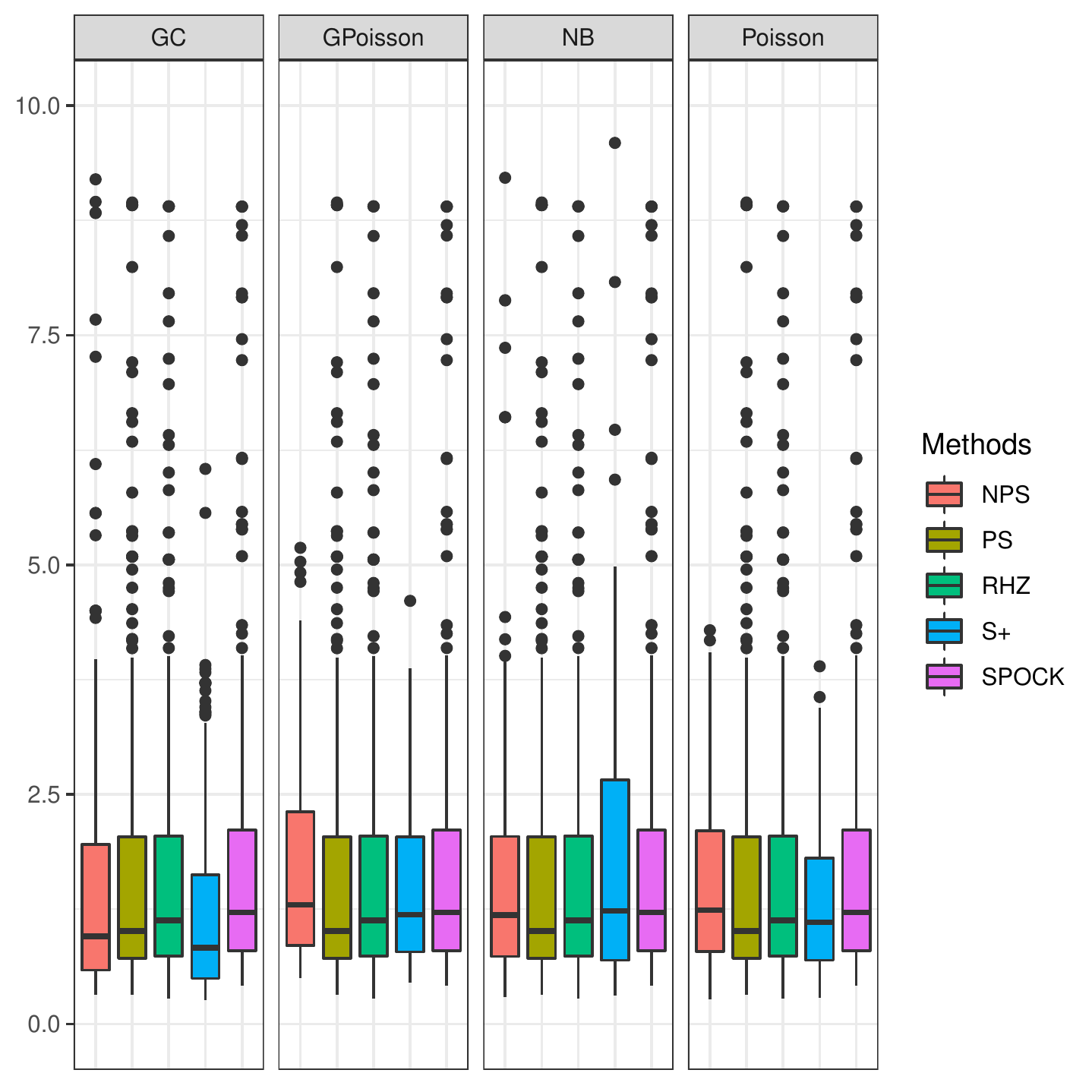}
\end{tabular} 
\end{center} 
\caption{SE of confounded covariate effect, $\beta_2$, (left column) and 
MSPE of fitted values (right column) for each 
model with $\tau_x=4$ for scenarios: over-dispersion (first row), equivalent-dispersion (second row), under-dispersion (third row; $\alpha=1.3$) and under-dispersion (forth row; $\alpha=2$).\label{fig9}} 
\end{figure} 
\begin{figure}[ppt]
\begin{center} 
\begin{tabular}{cc} 
 \includegraphics[width=230pt,height=6.8pc]{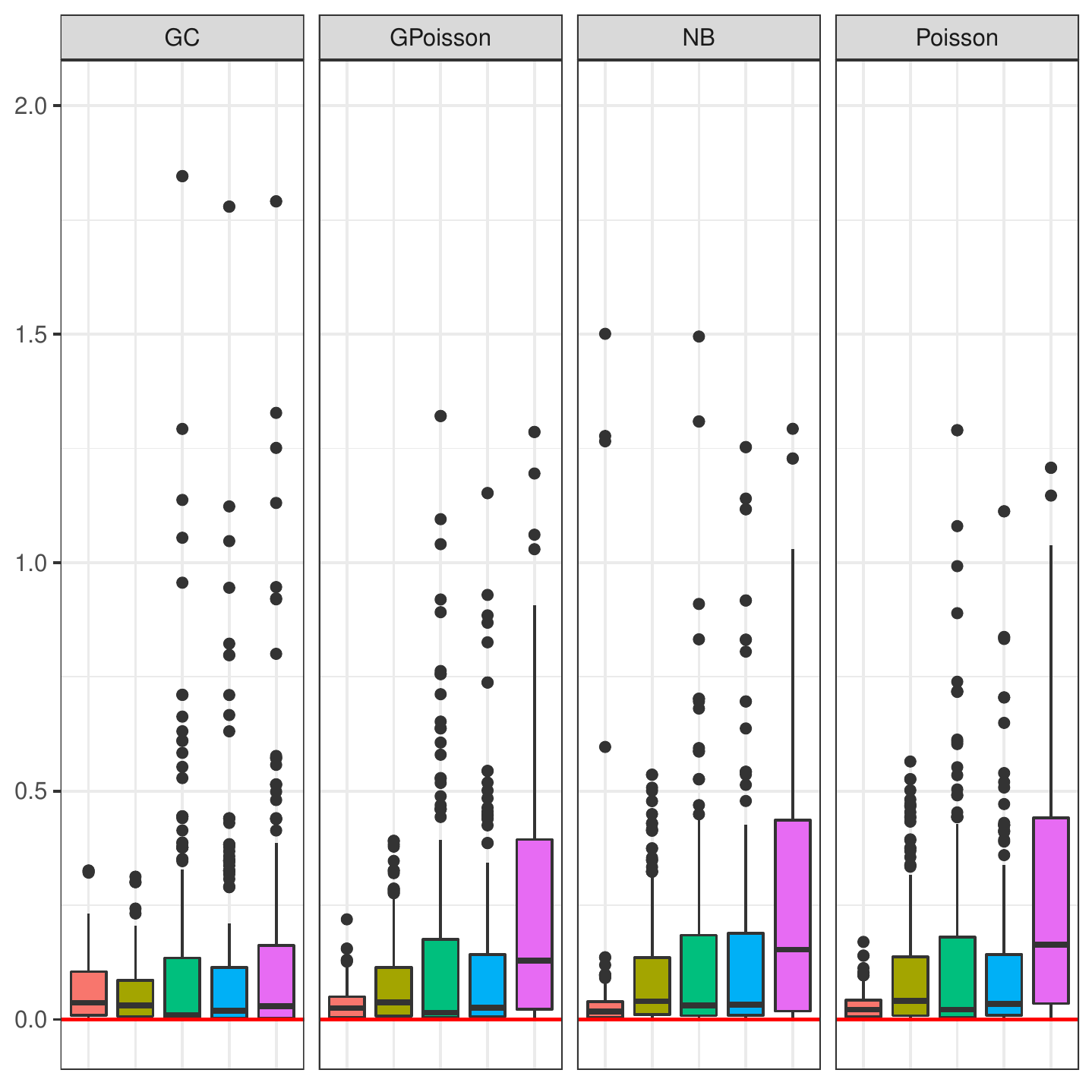}&
 \includegraphics[width=230pt,height=6.8pc]{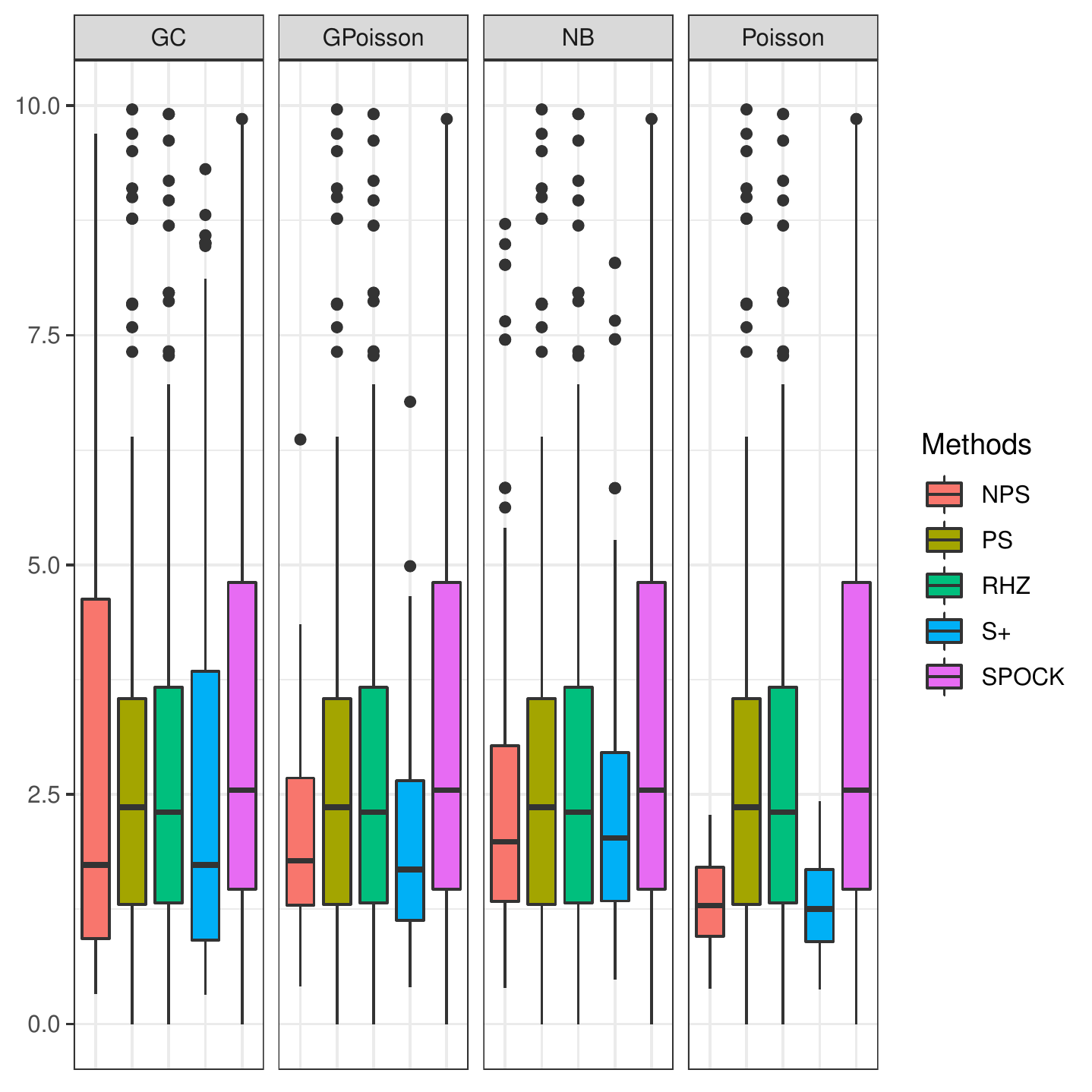}\\ 
\includegraphics[width=230pt,height=6.8pc]{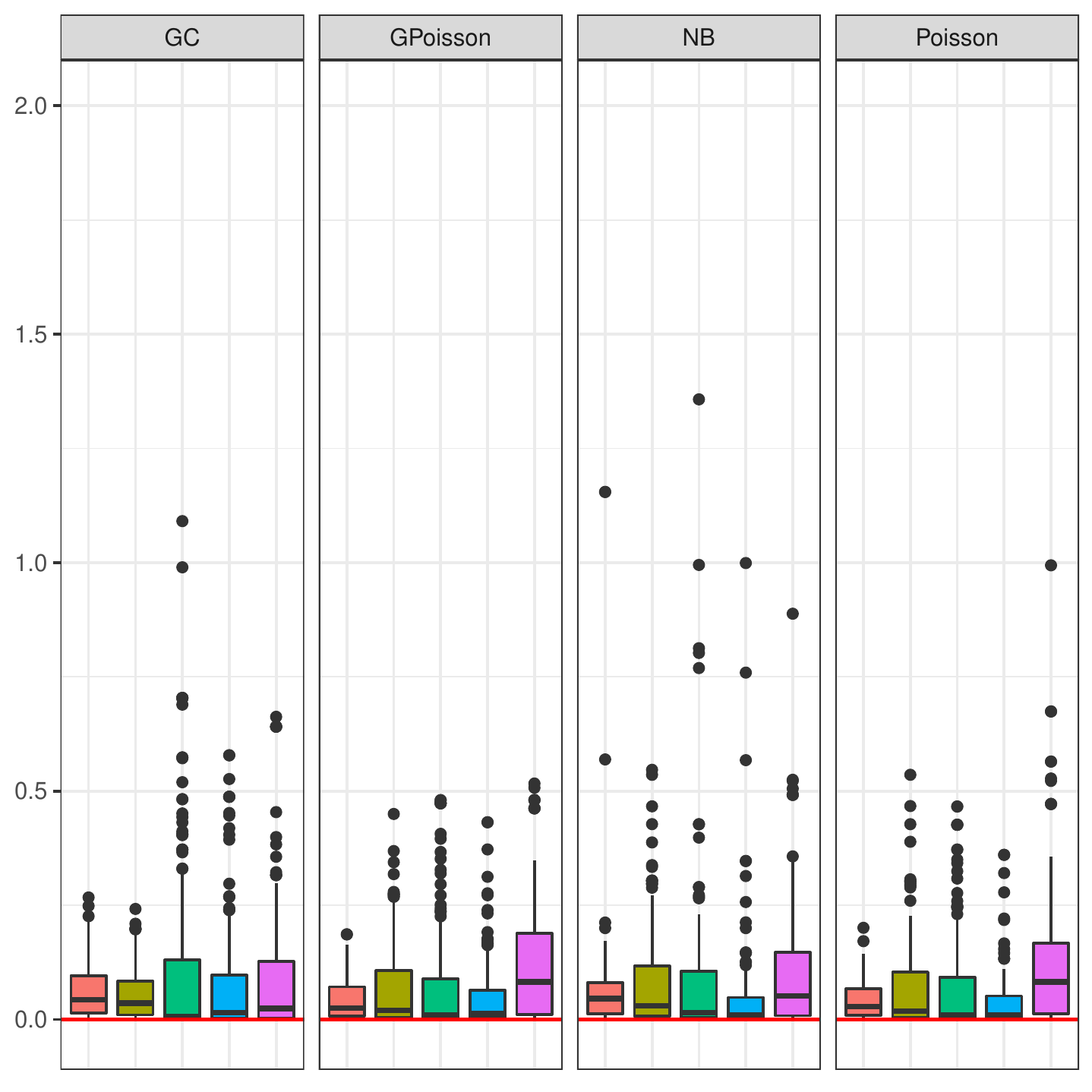}&
\includegraphics[width=230pt,height=6.8pc]{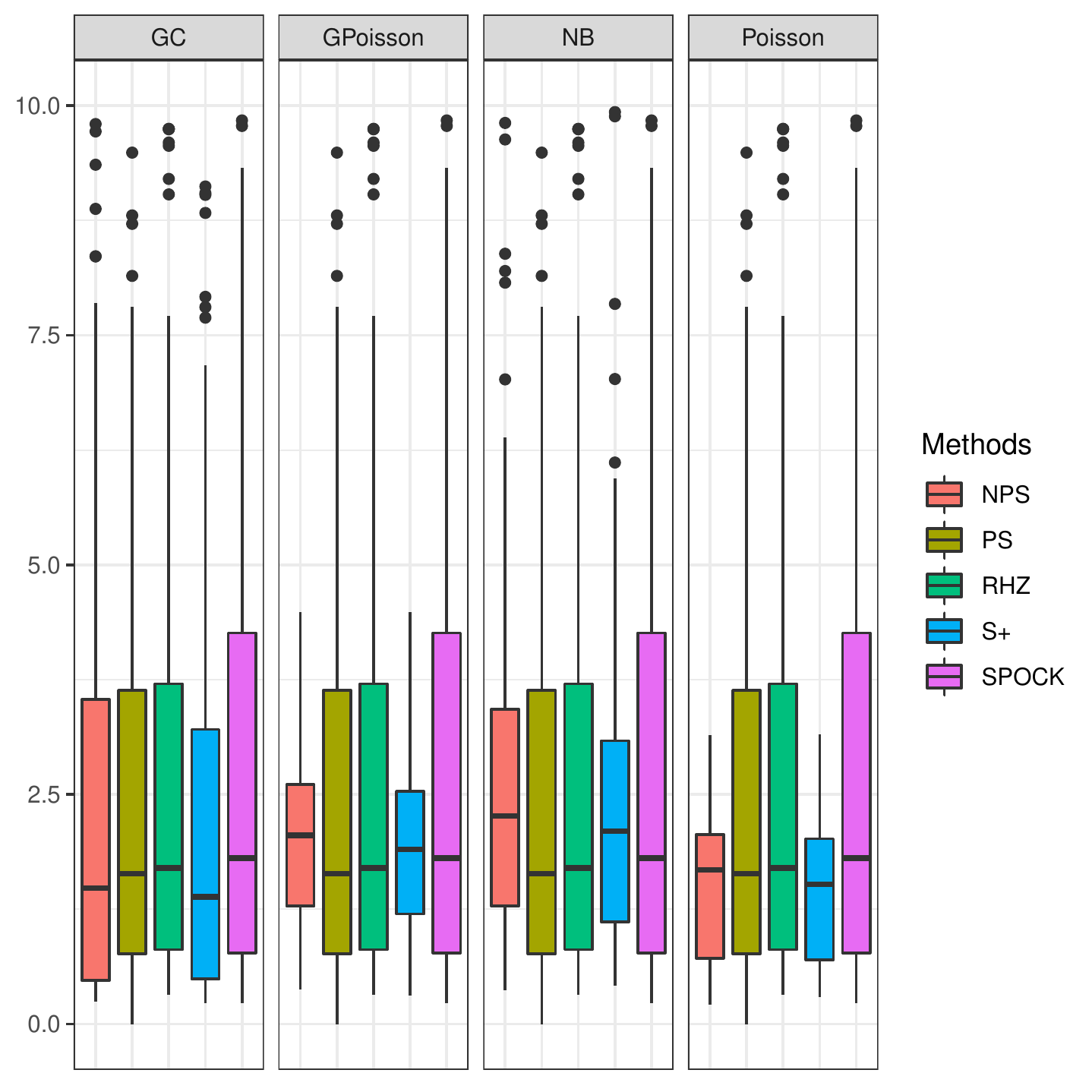}\\ 
\includegraphics[width=230pt,height=6.8pc]{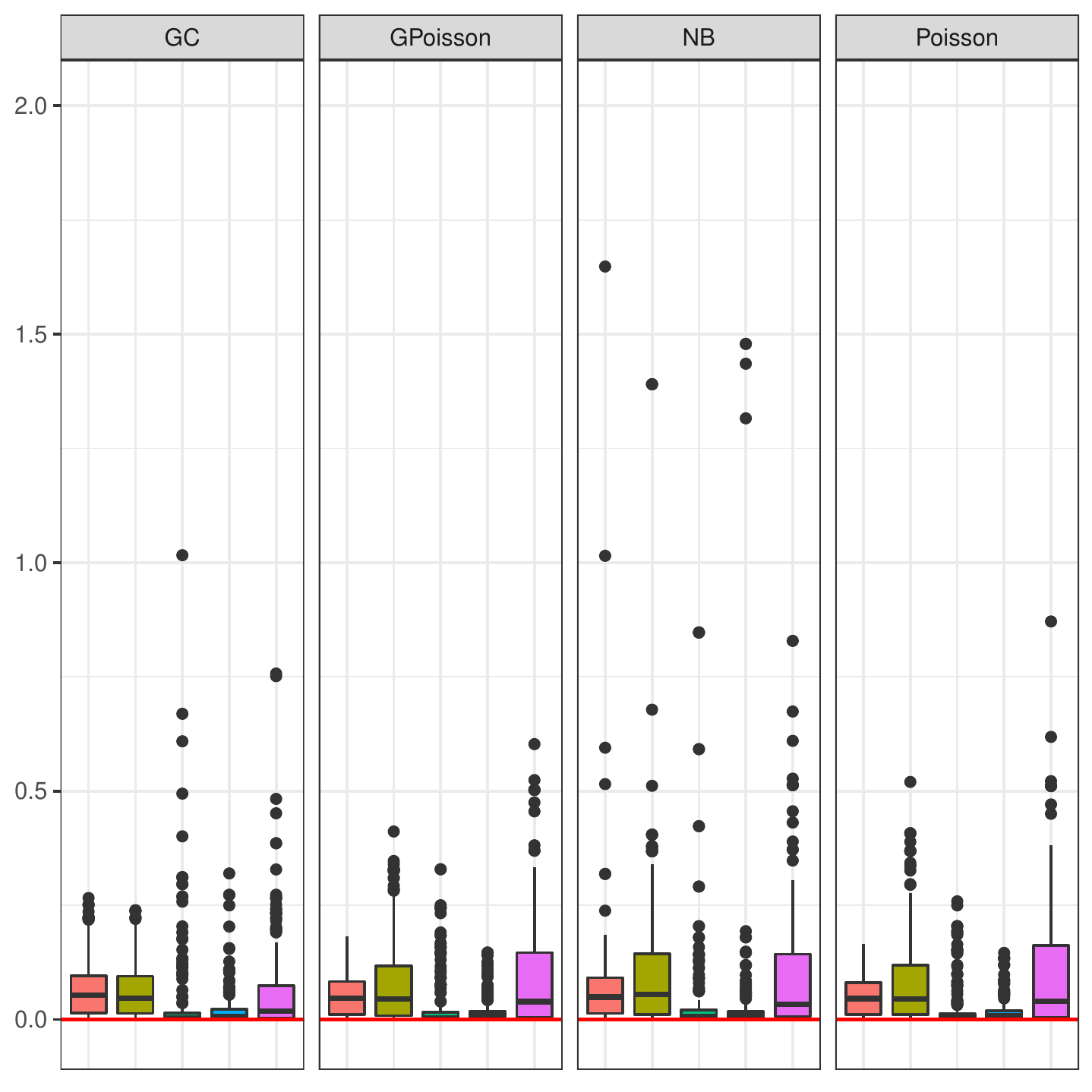}&
\includegraphics[width=230pt,height=6.8pc]{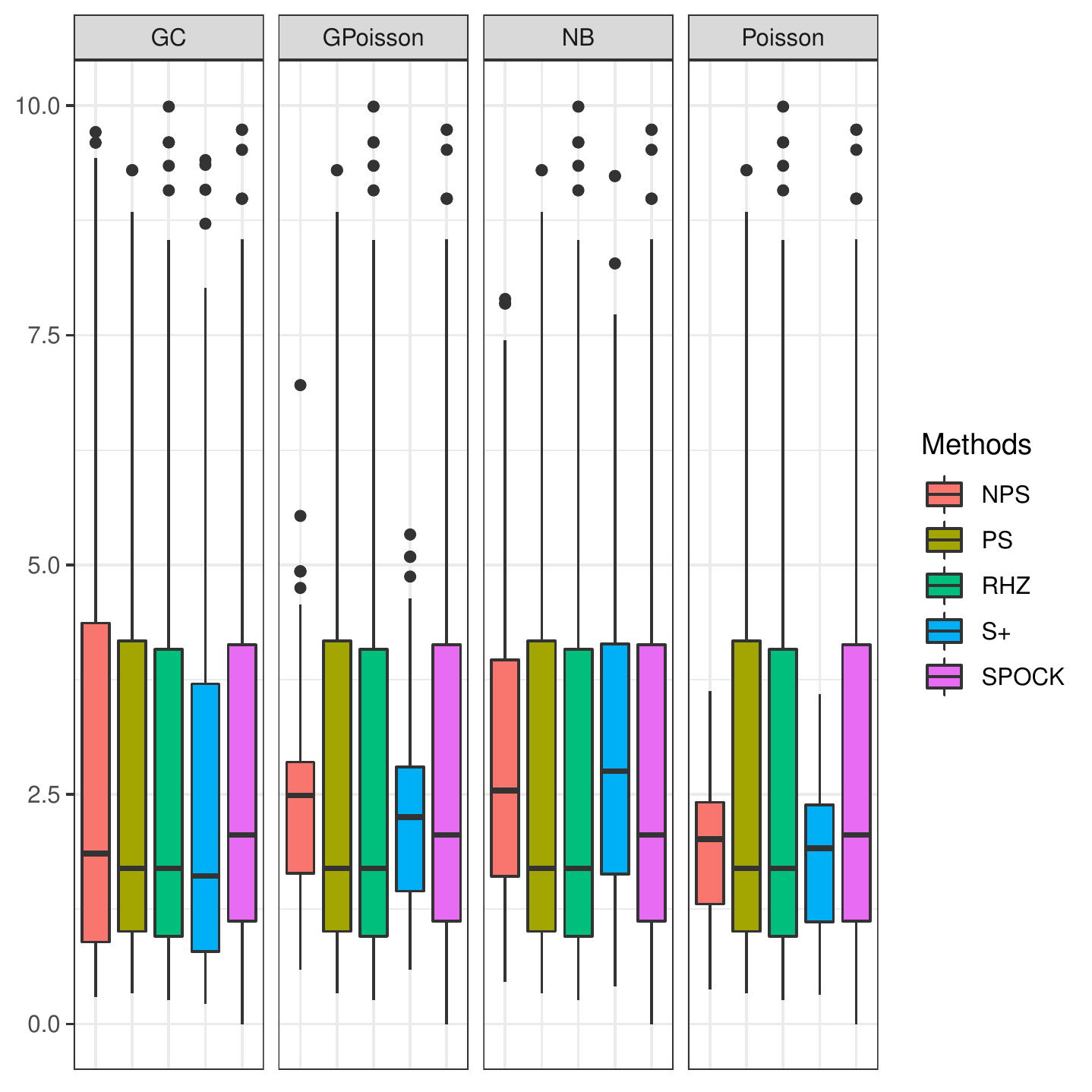}\\ 
\includegraphics[width=230pt,height=6.8pc]{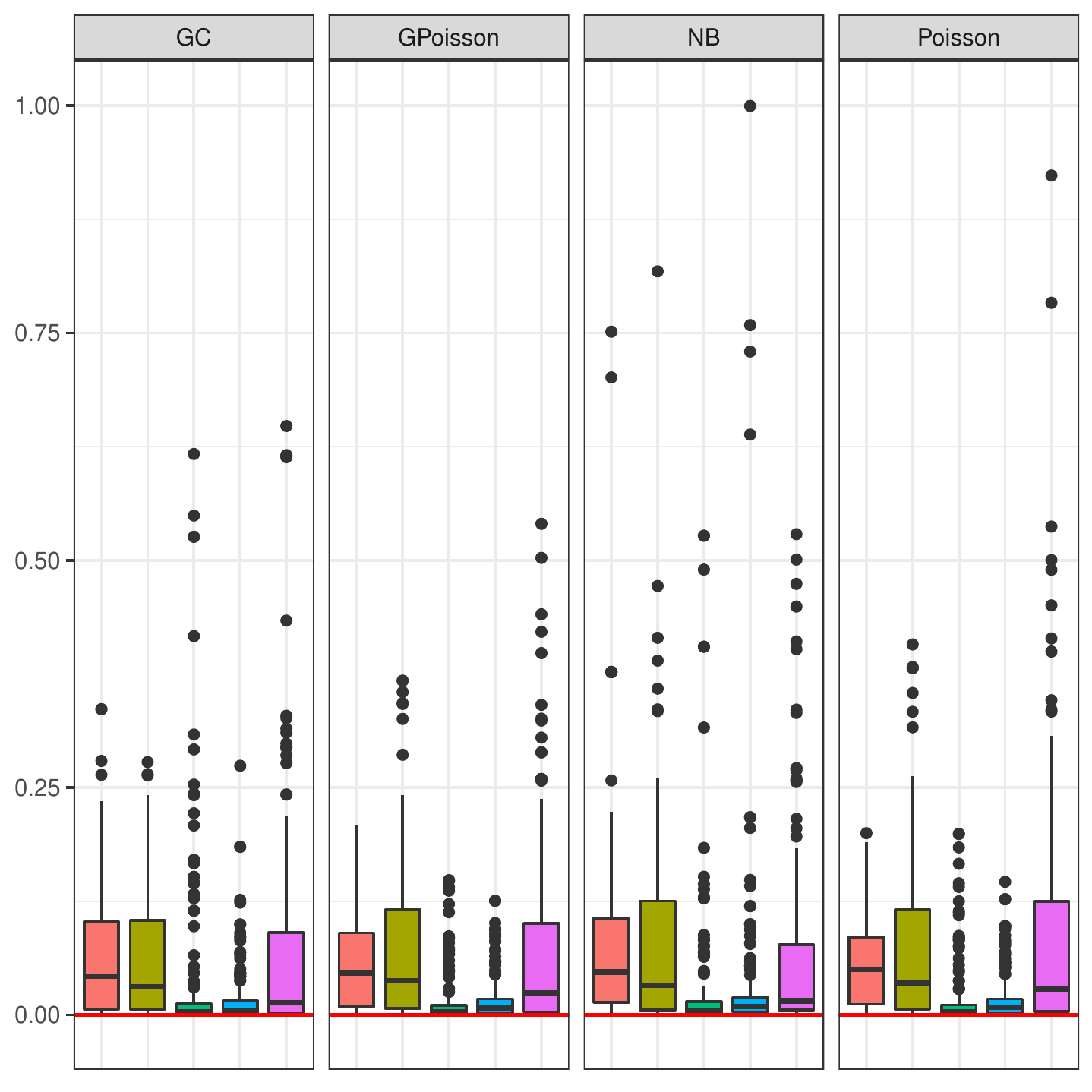}&
\includegraphics[width=230pt,height=6.8pc]{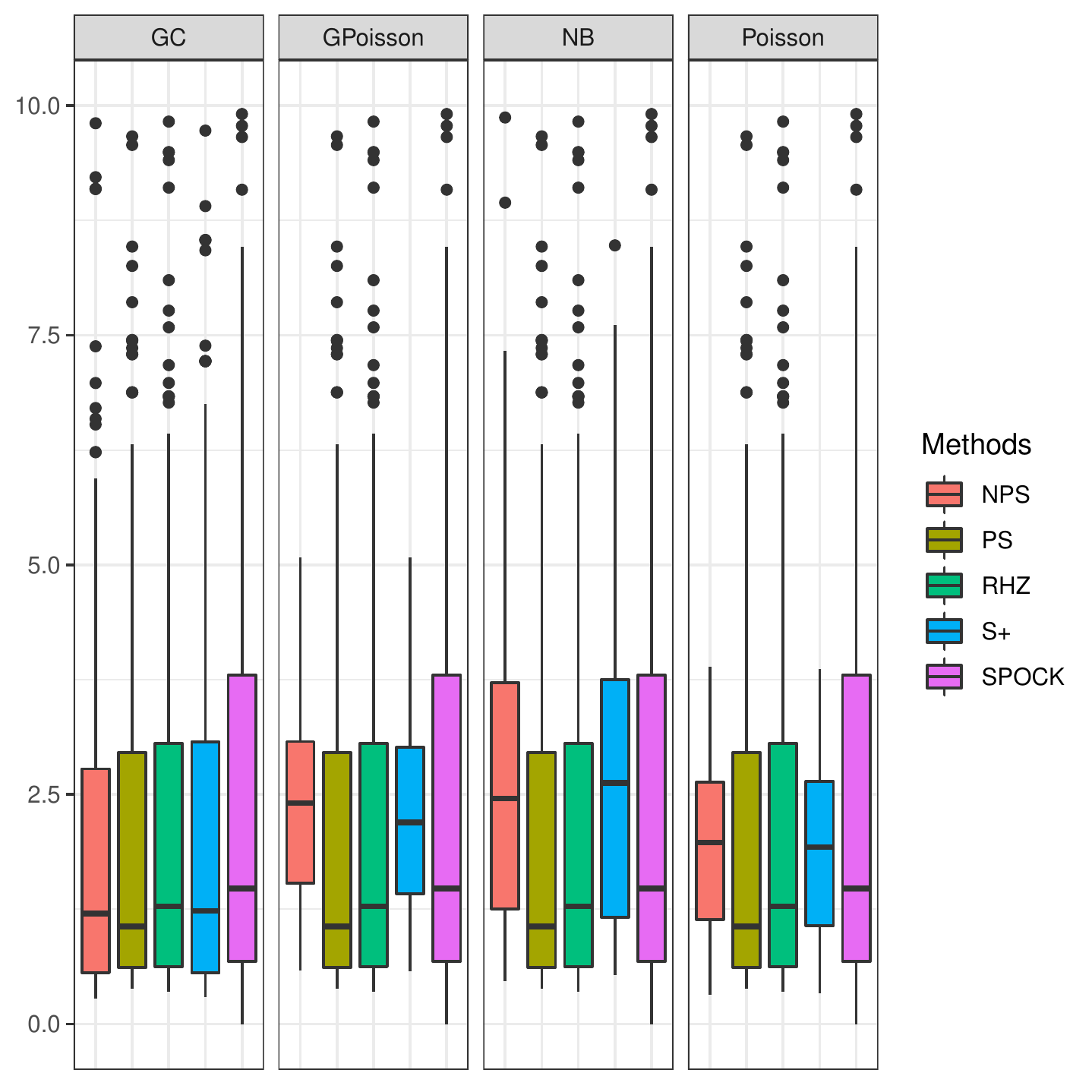} 
\end{tabular} 
\end{center} 
\caption{SE of confounded covariate effect, $\beta_2$, (left column) and 
MSPE of fitted values (right column) for each 
model with $\tau_x=1$ for scenarios: over-dispersion (first row), equivalent-dispersion (second row), under-dispersion (third row; $\alpha=1.3$) and under-dispersion (forth row; $\alpha=2$).\label{fig10}} 
\end{figure} 
\begin{figure}[ppt]
\begin{center} 
\begin{tabular}{ccc} 
\includegraphics[width=155pt,height=6.8pc]{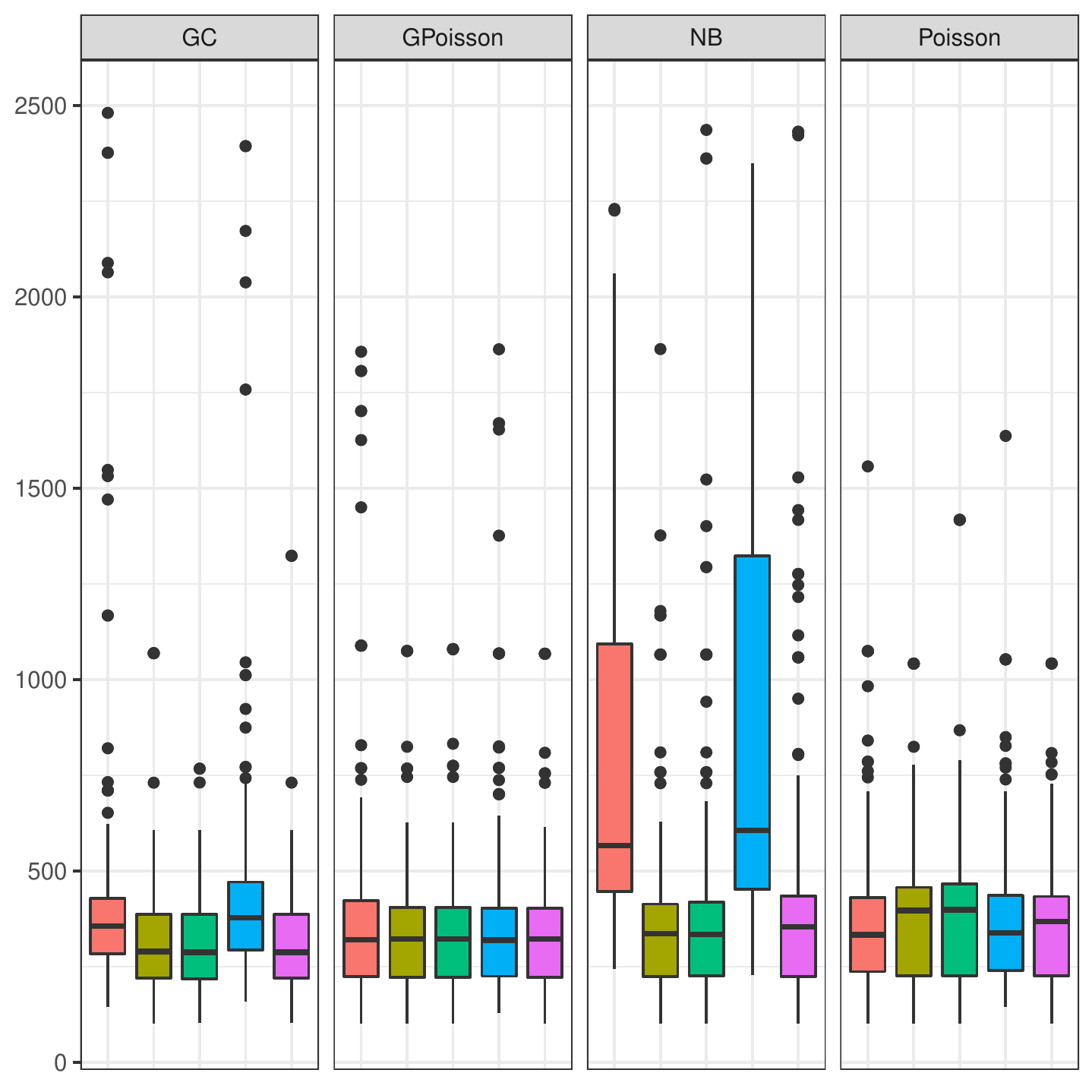} &
 \includegraphics[width=155pt,height=6.8pc]{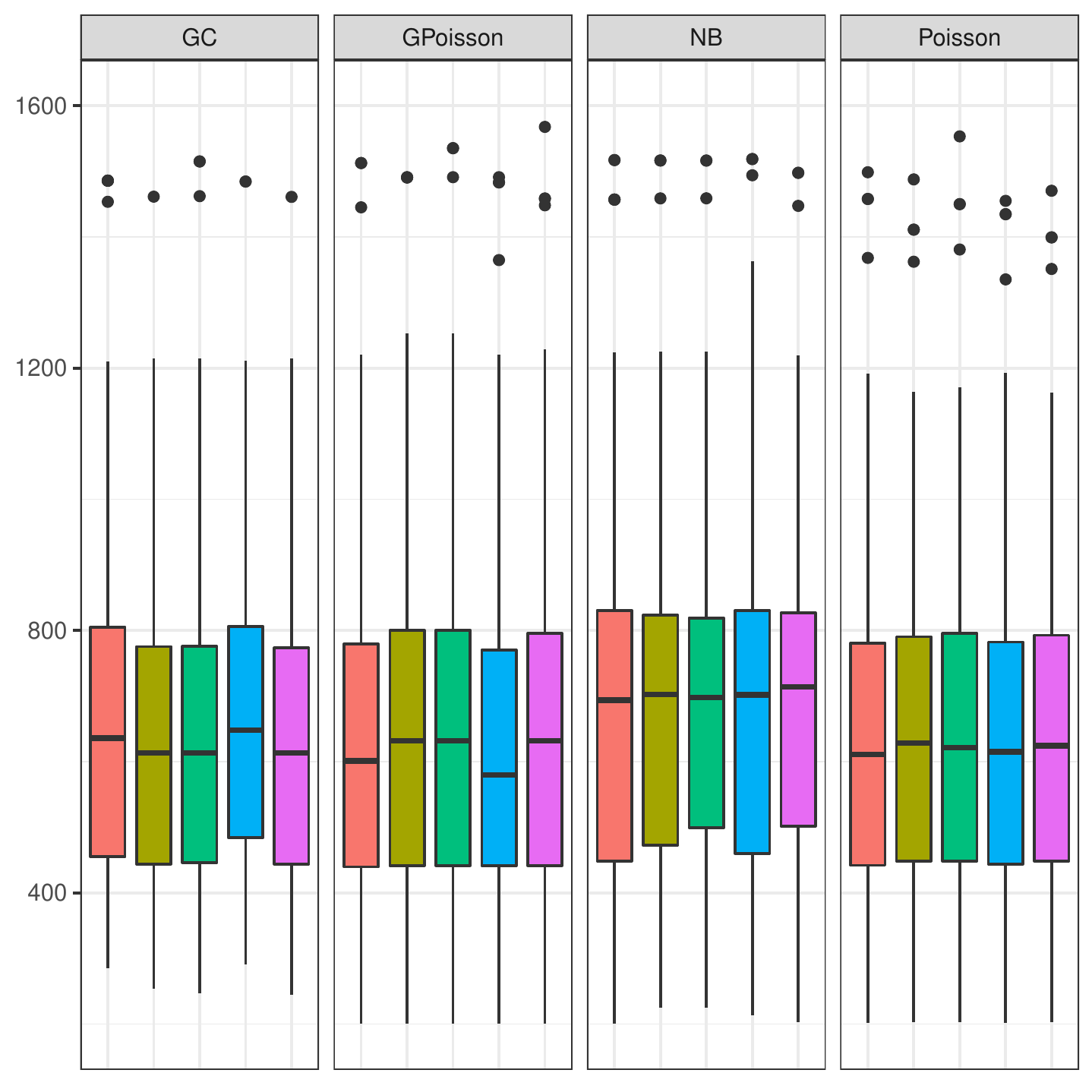}&
 \includegraphics[width=155pt,height=6.8pc]{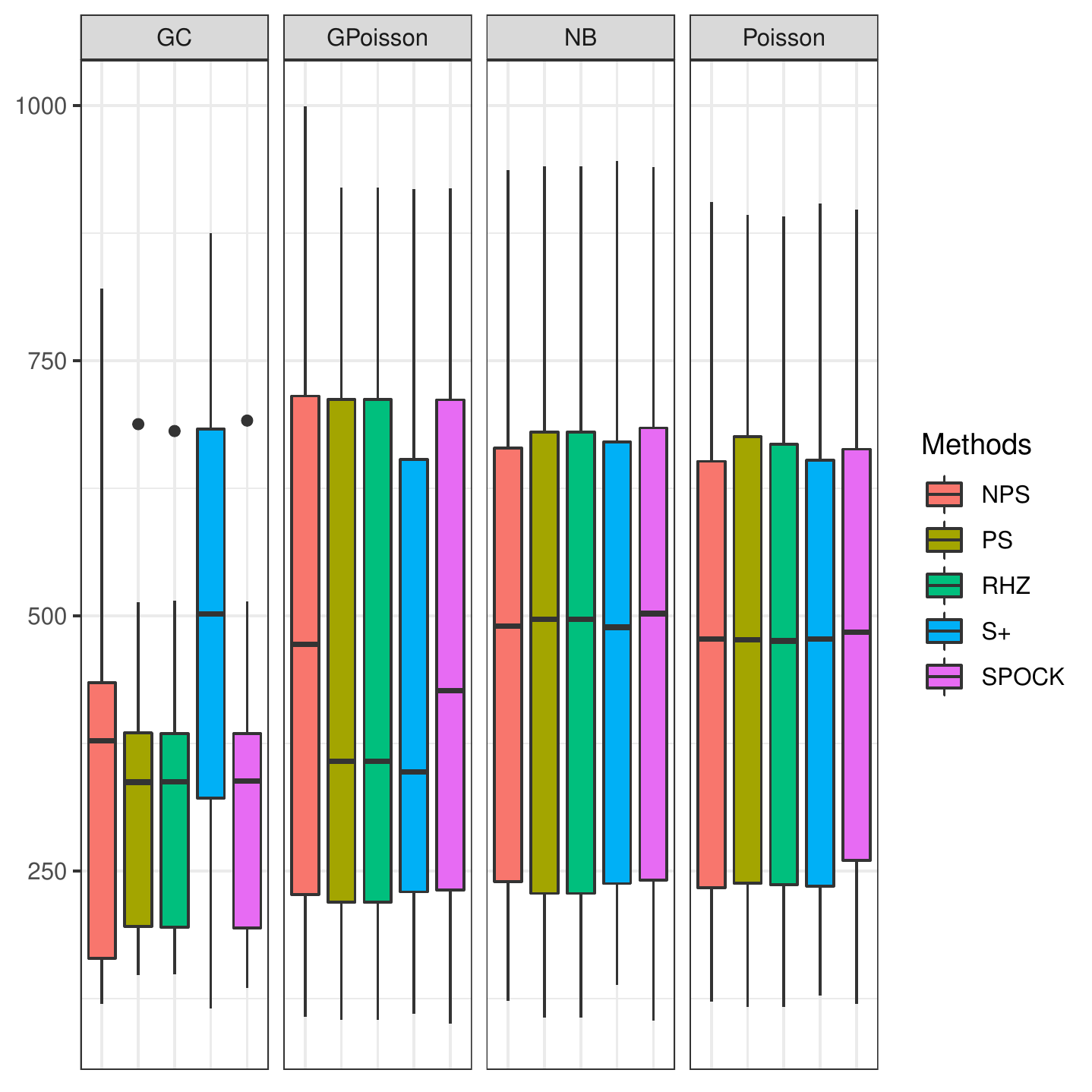}\\ 
\includegraphics[width=155pt,height=6.8pc]{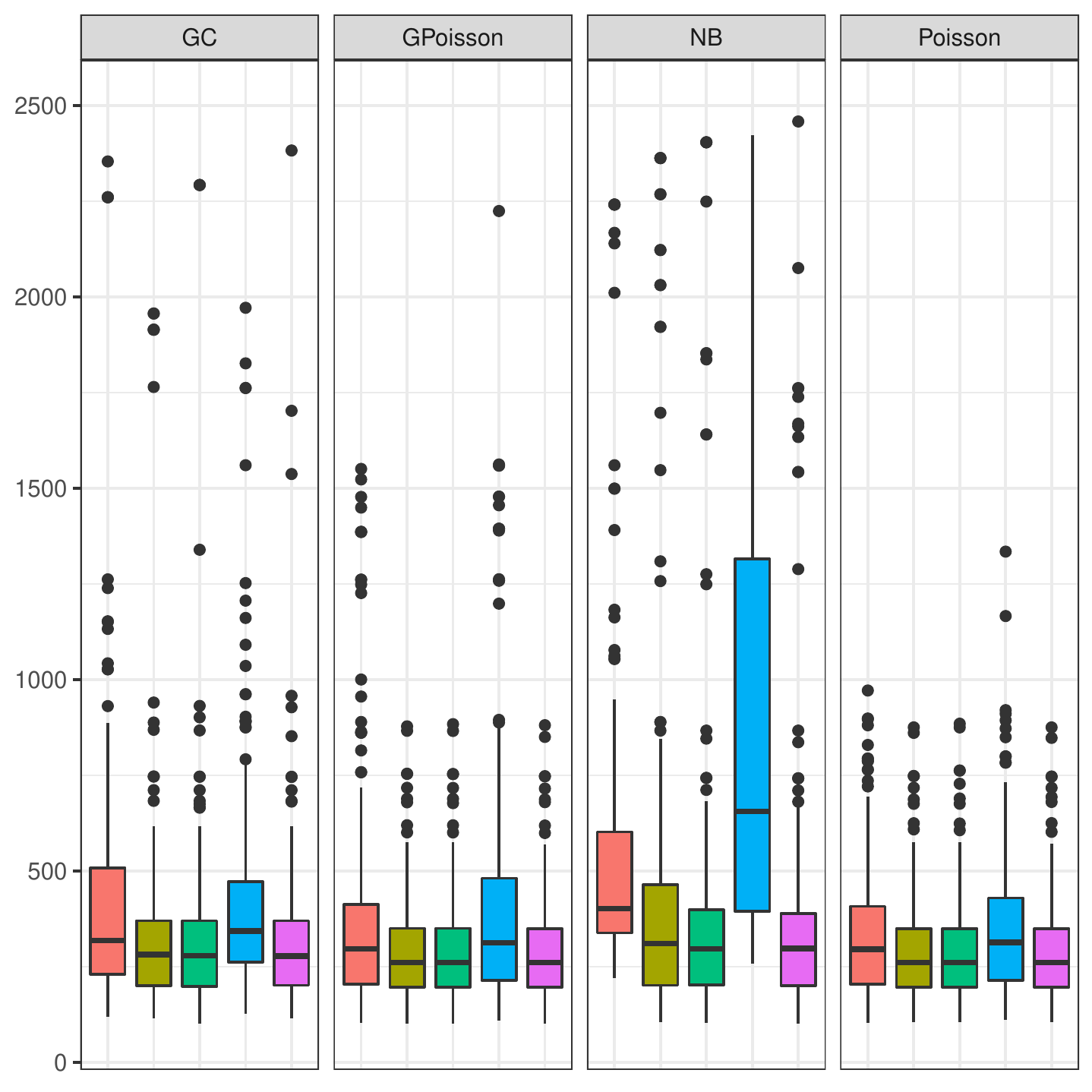} & 
\includegraphics[width=155pt,height=6.8pc]{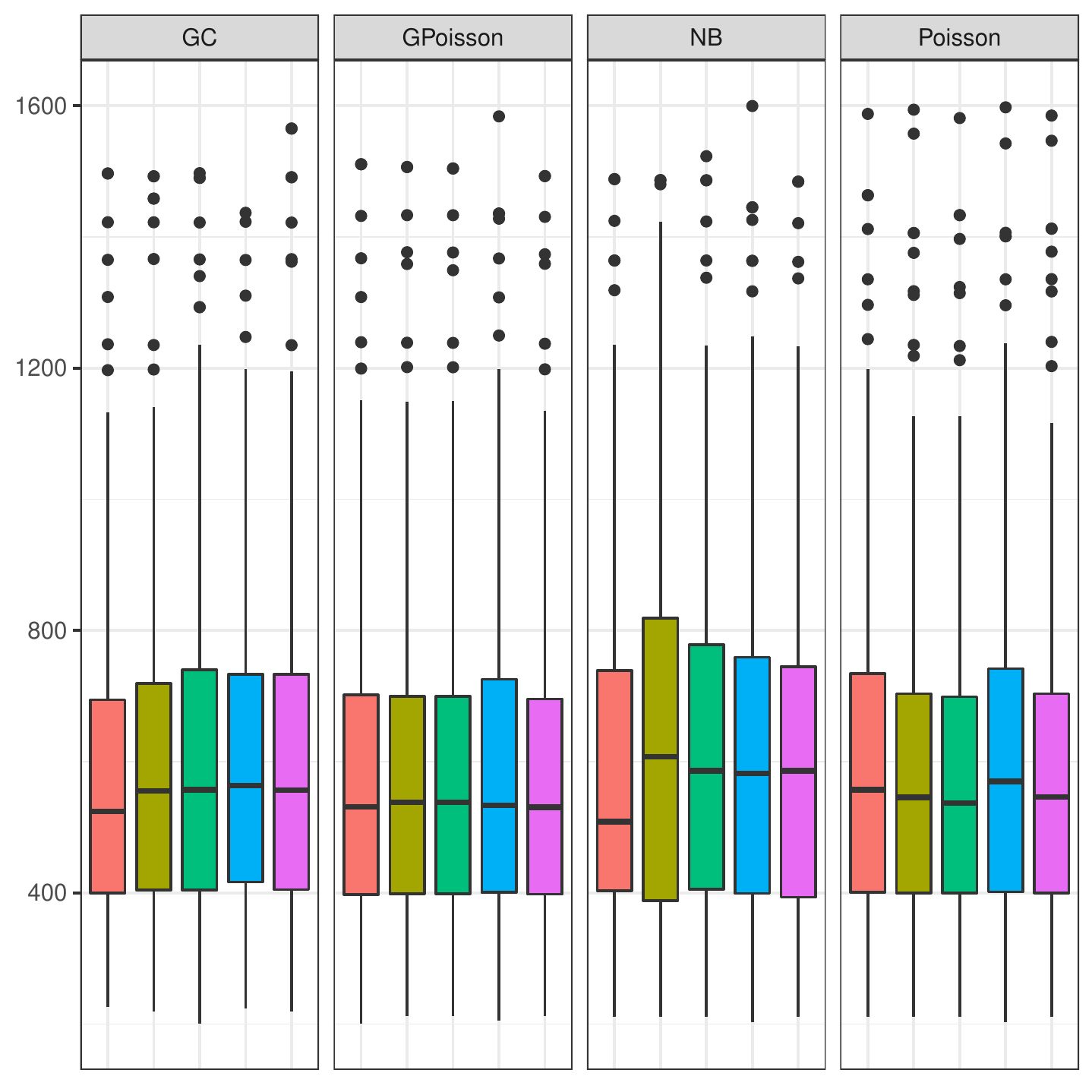}&
\includegraphics[width=155pt,height=6.8pc]{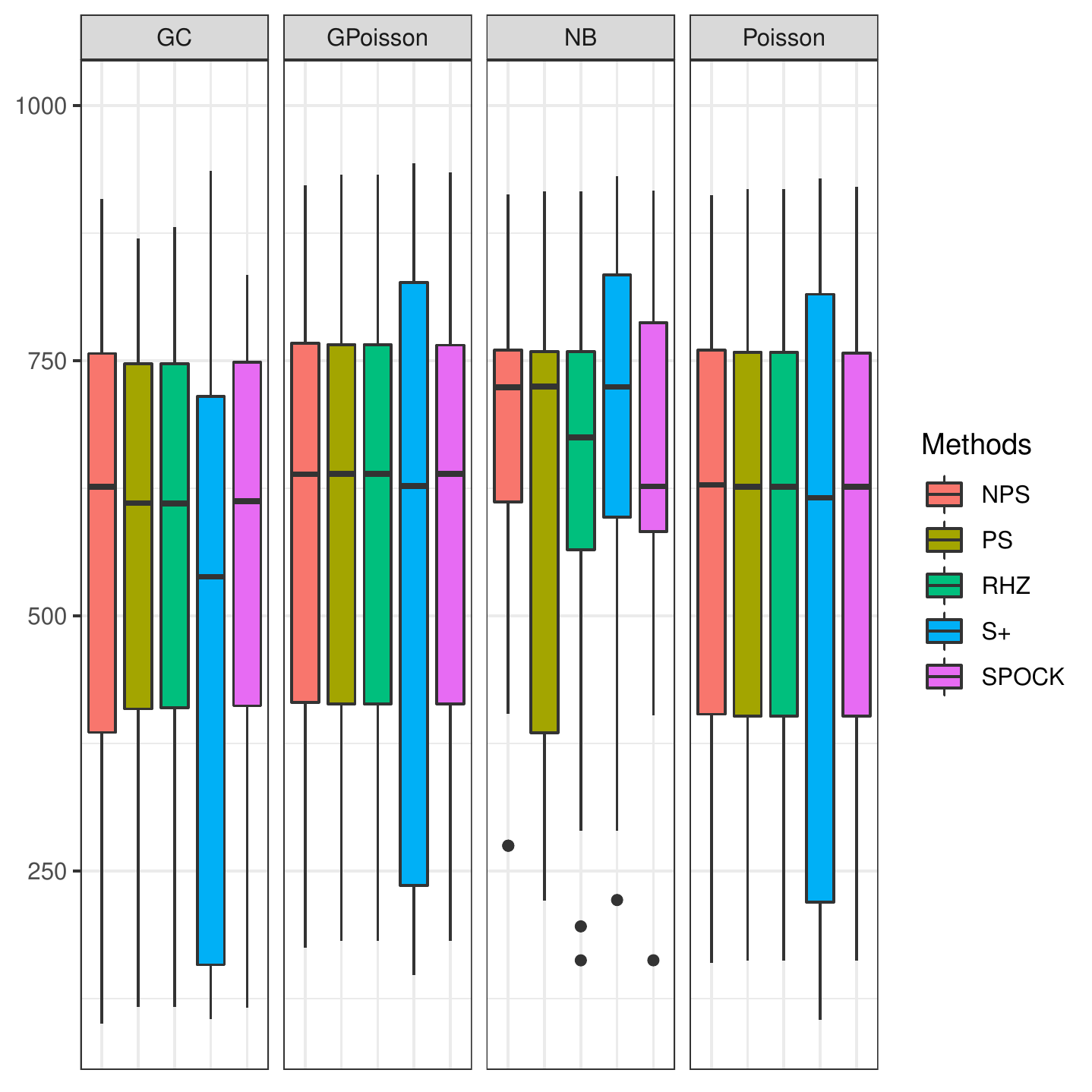}\\ 
\includegraphics[width=155pt,height=6.8pc]{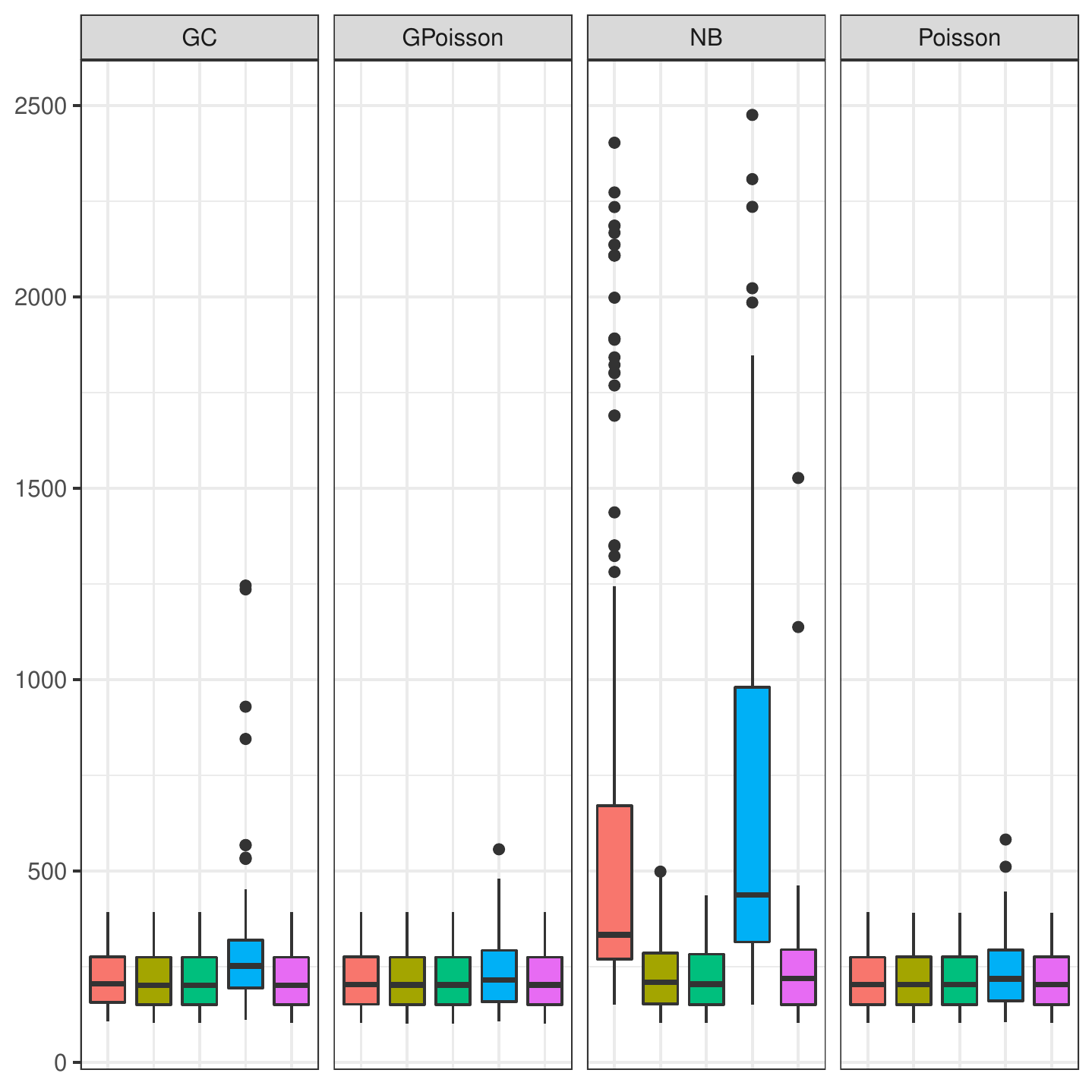} & 
\includegraphics[width=155pt,height=6.8pc]{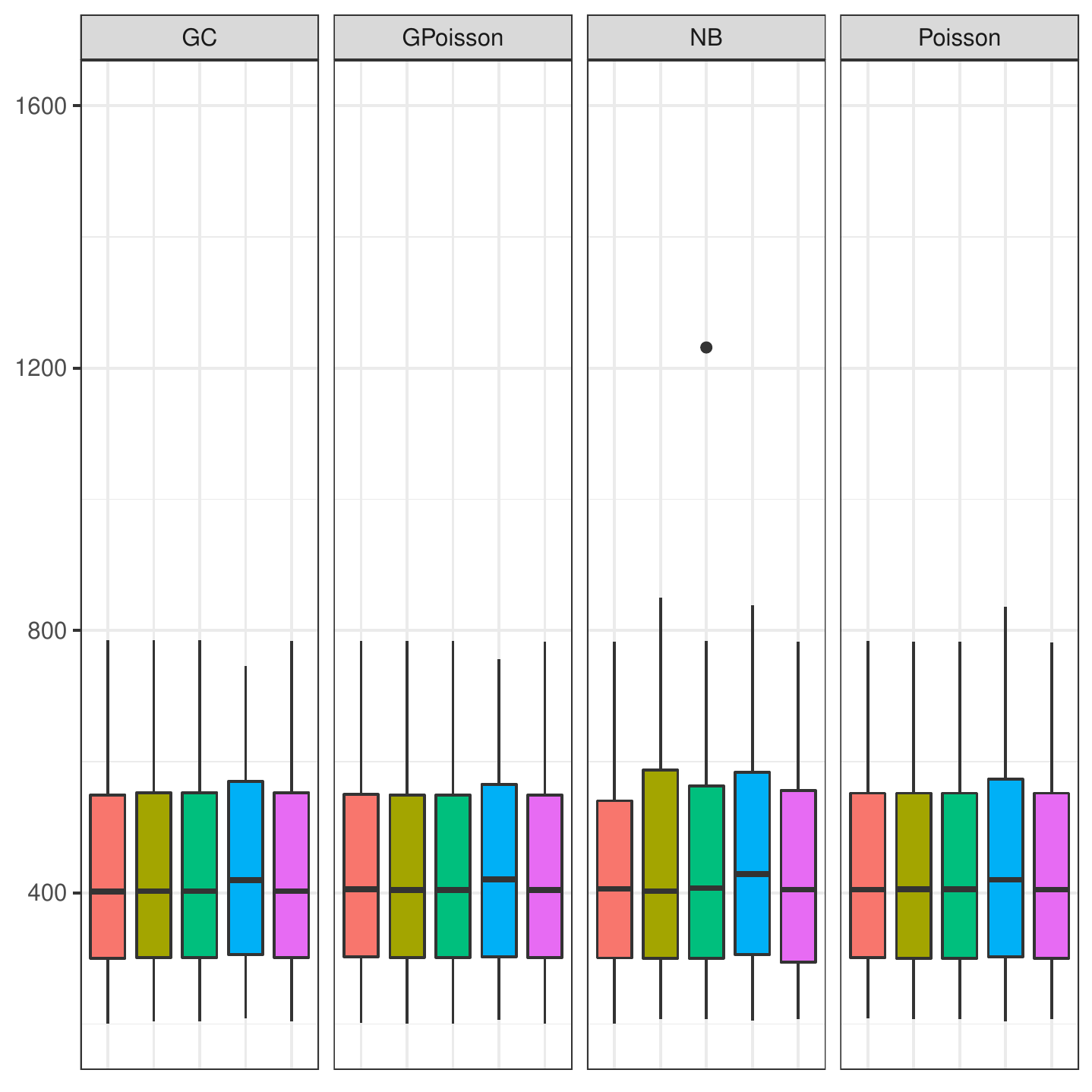}&
\includegraphics[width=155pt,height=6.8pc]{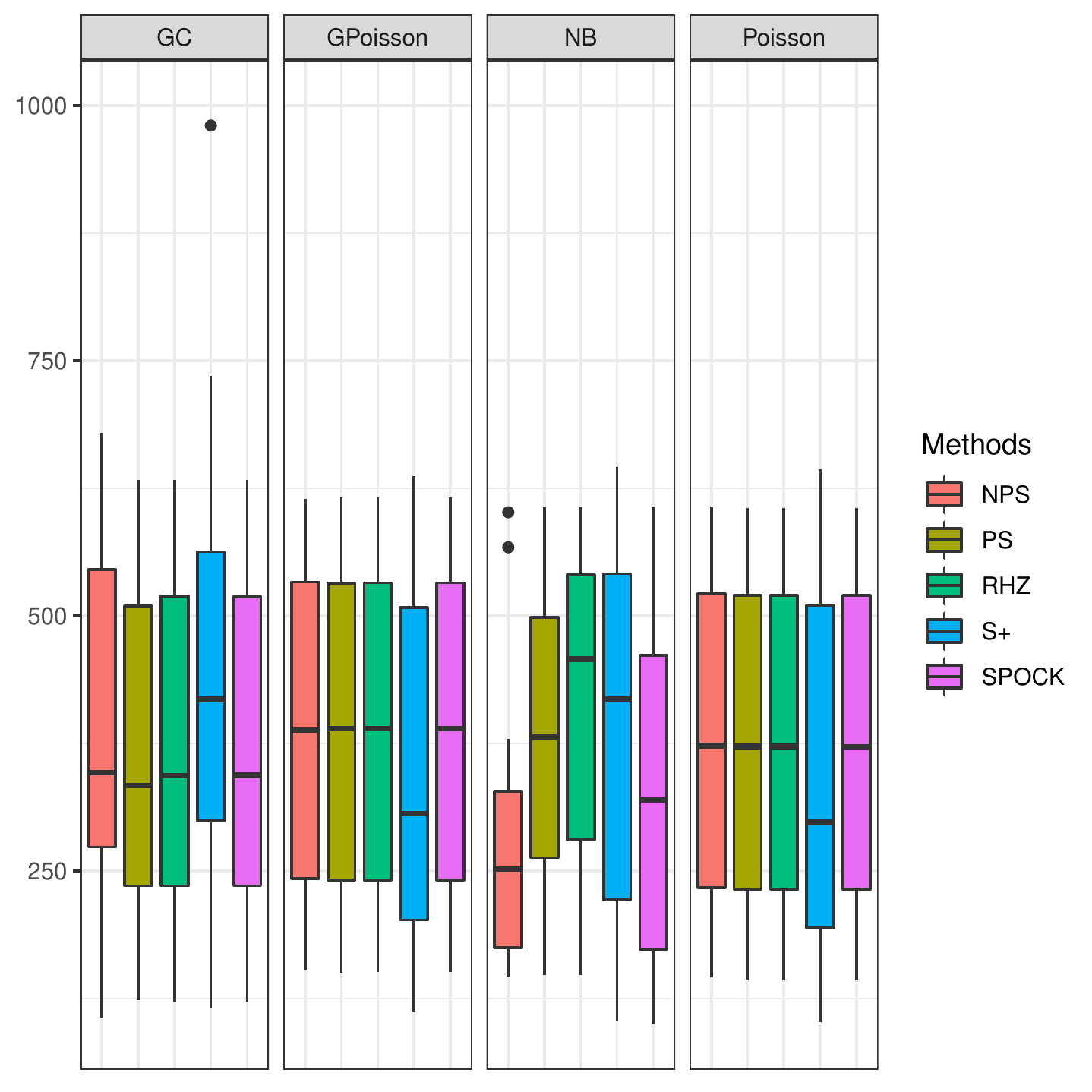}\\ 
\includegraphics[width=155pt,height=6.8pc]{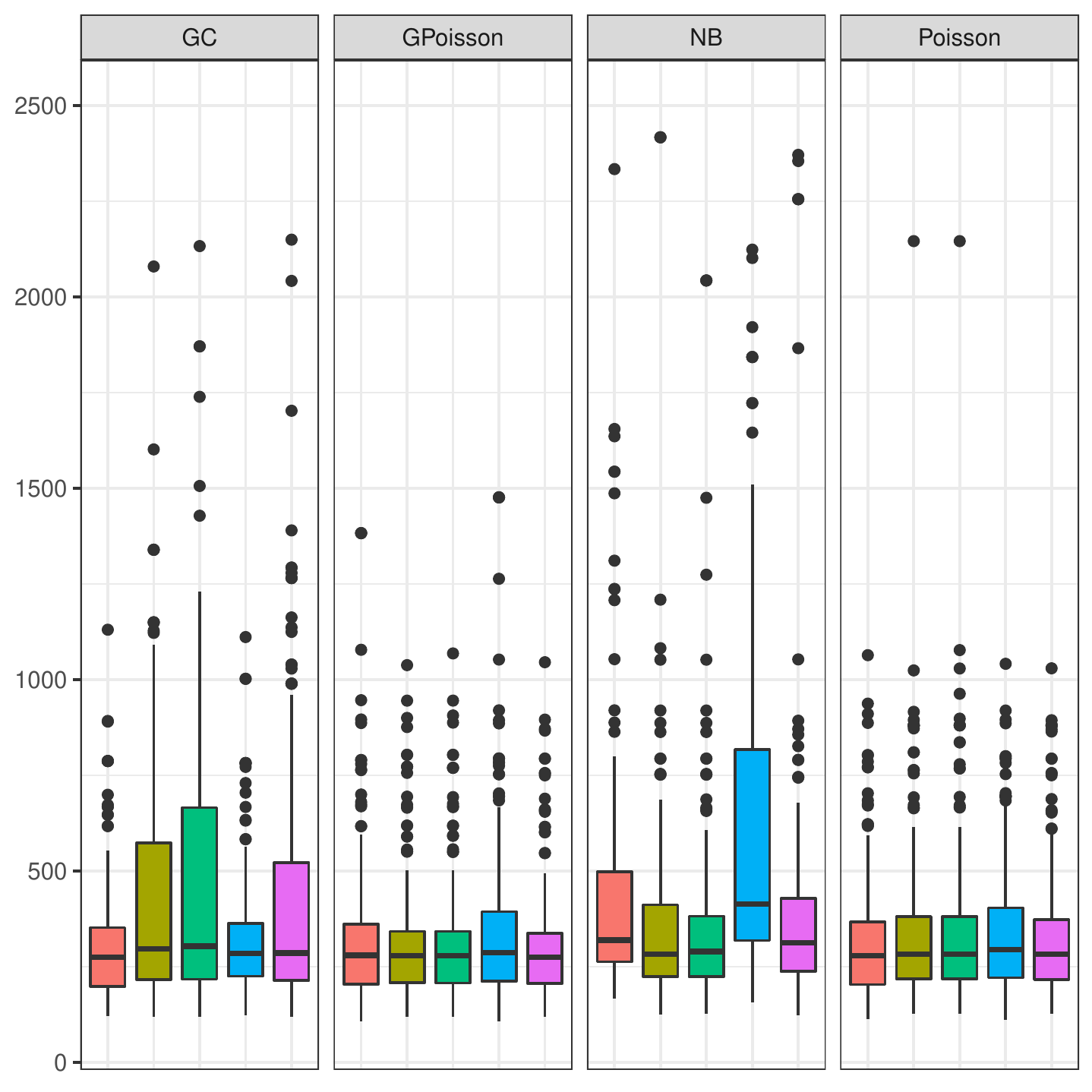} & 
\includegraphics[width=155pt,height=6.8pc]{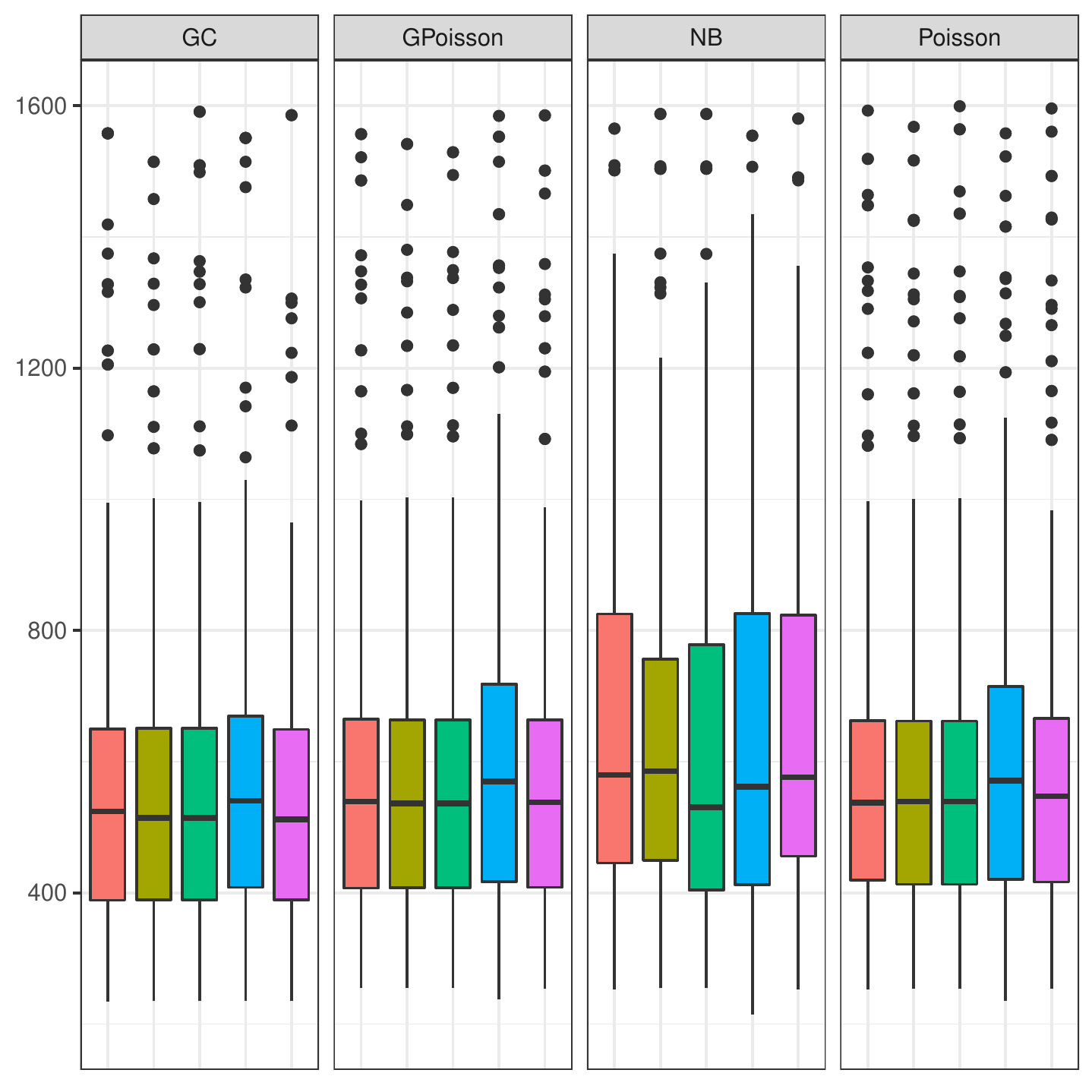}&
\includegraphics[width=155pt,height=6.8pc]{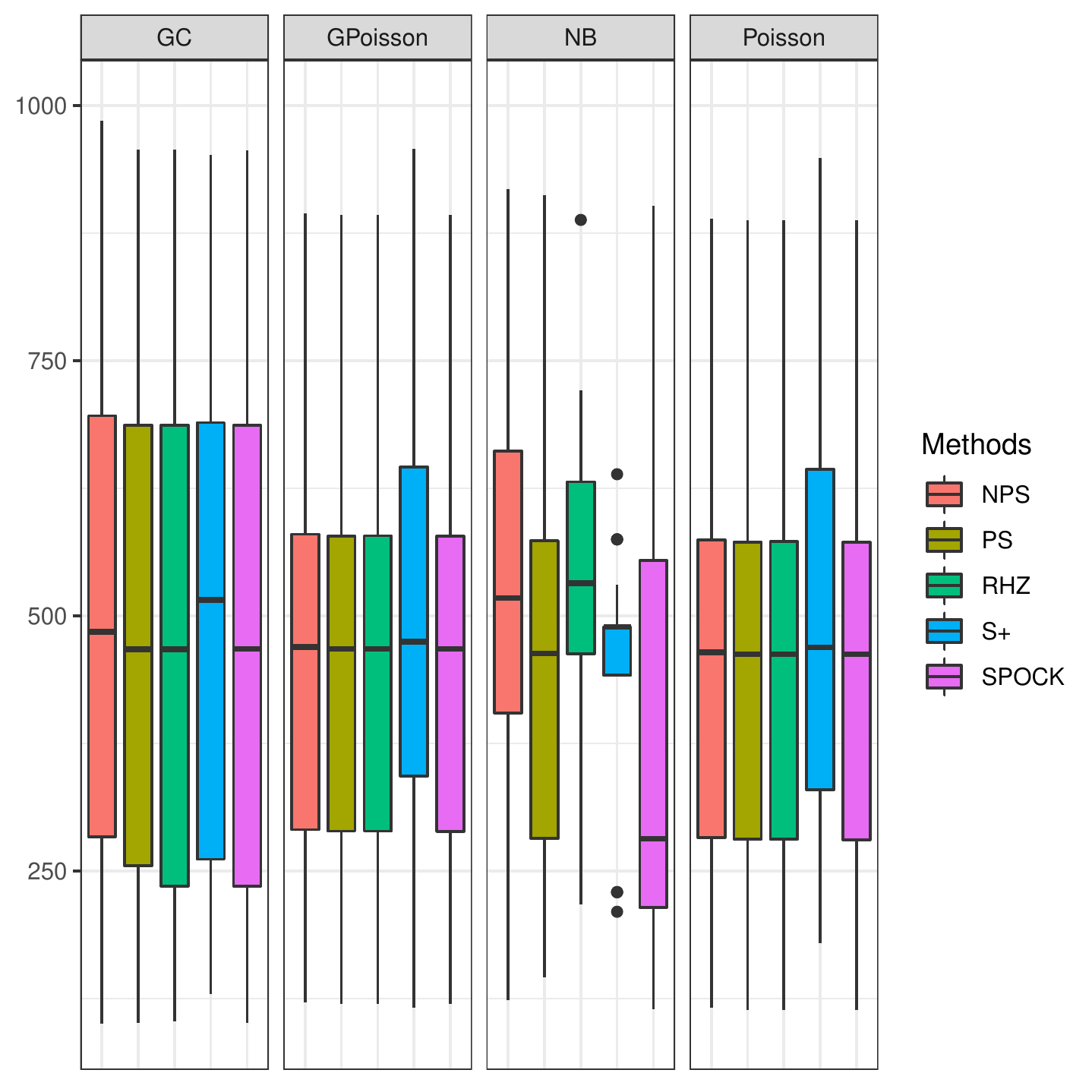} 
\end{tabular} 
\end{center} 
\caption{LS (left column), WAIC (middle column) and 
DIC (right column) for each 
model with $\tau_x=11$ for scenarios: over-dispersion (first row), equivalent-dispersion (second row), under-dispersion (third row; $\alpha=1.3$) and under-dispersion (forth row; $\alpha=2$).\label{fig11}} 
\end{figure} 

\begin{figure}[ppt] 
\begin{center} 
\begin{tabular}{ccc} 
\includegraphics[width=155pt,height=6.8pc]{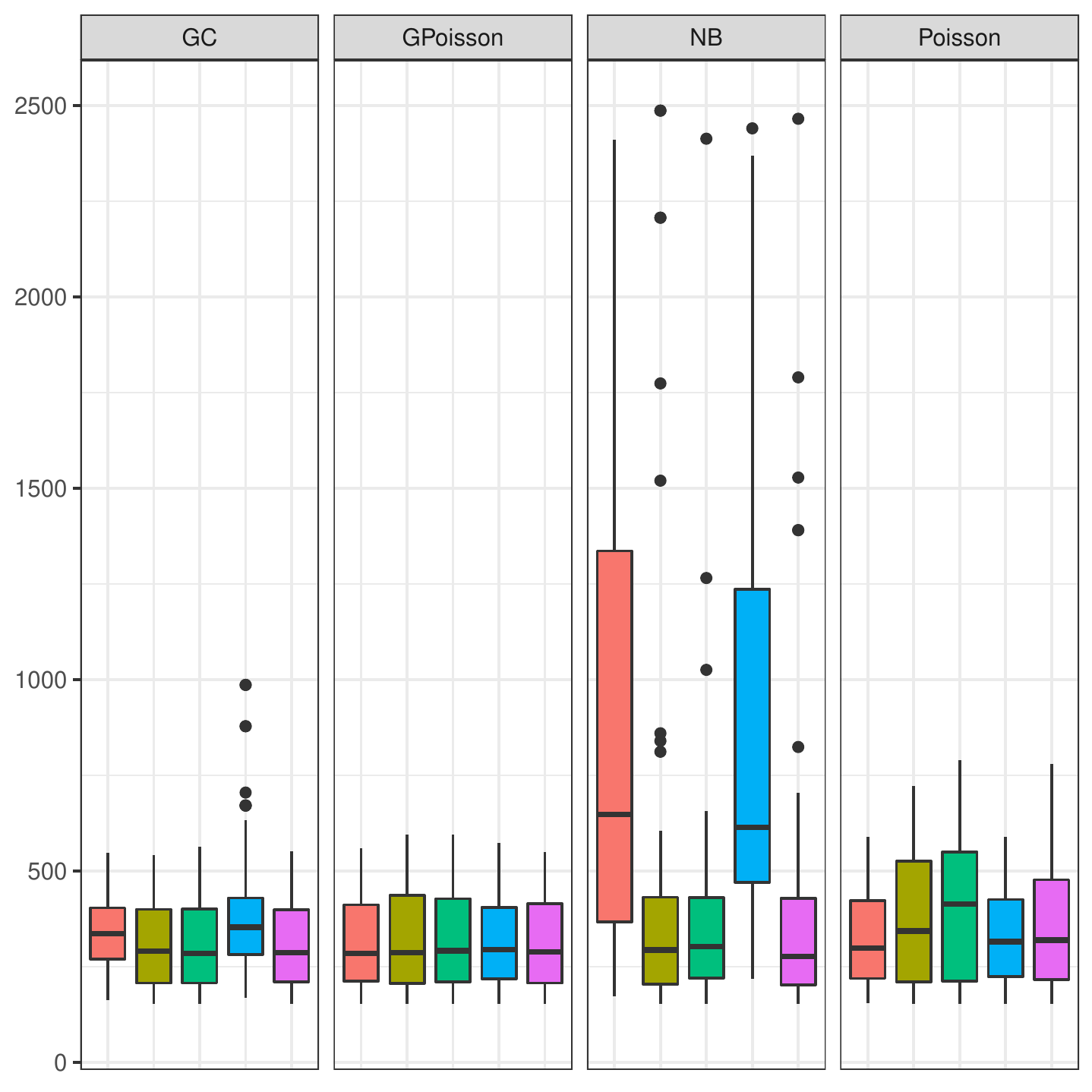} &
 \includegraphics[width=155pt,height=6.8pc]{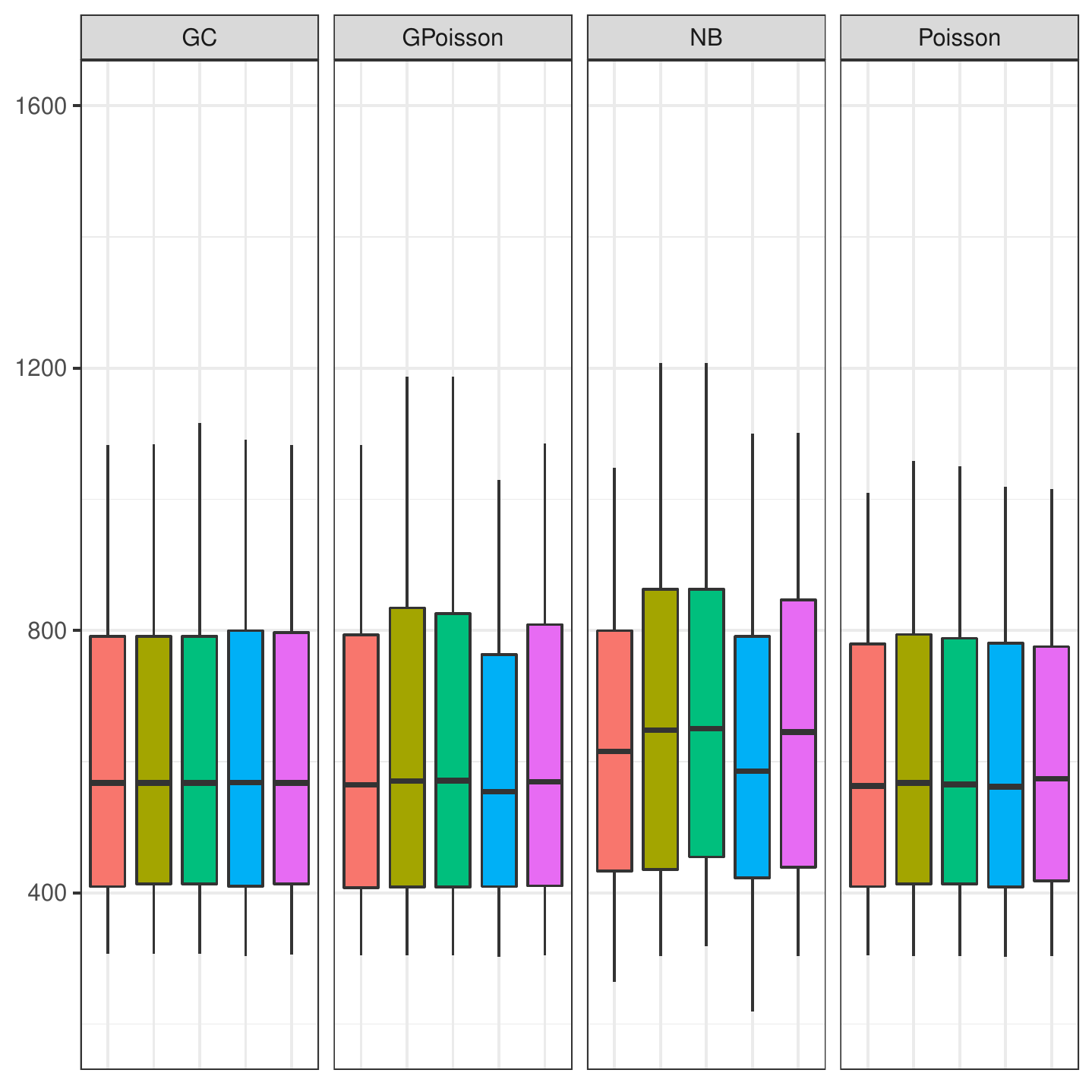}&
 \includegraphics[width=155pt,height=6.8pc]{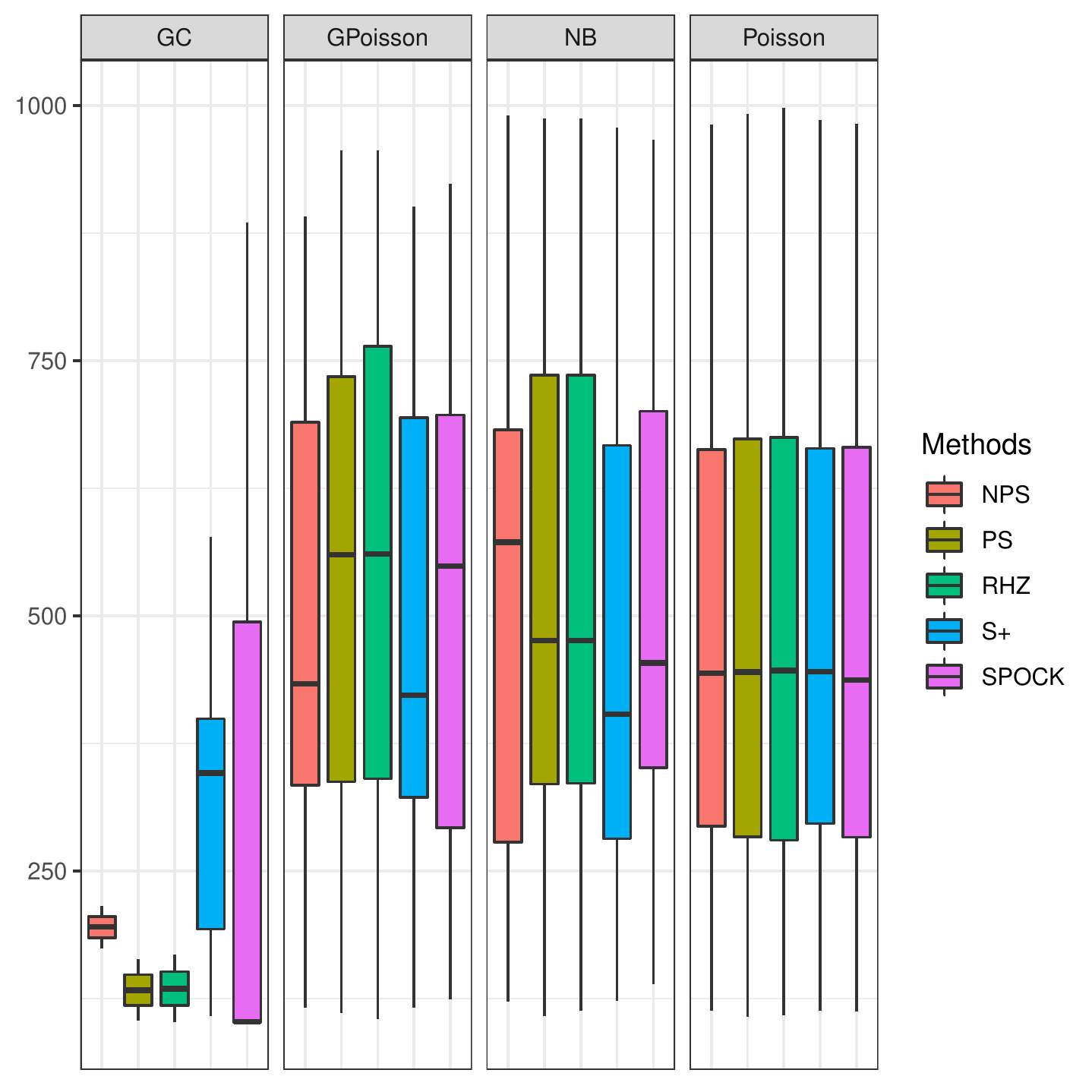}\\ 
\includegraphics[width=155pt,height=6.8pc]{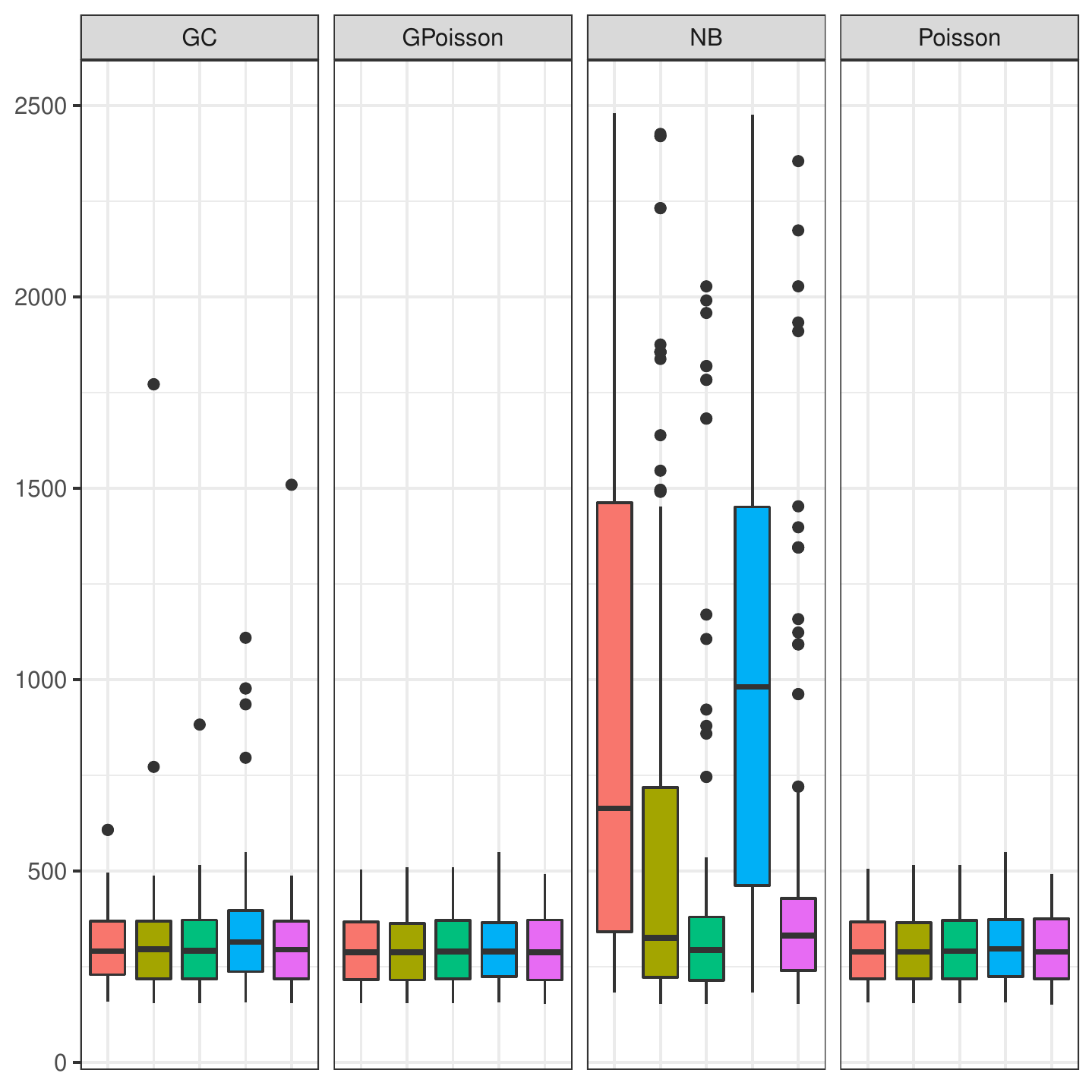} & 
\includegraphics[width=155pt,height=6.8pc]{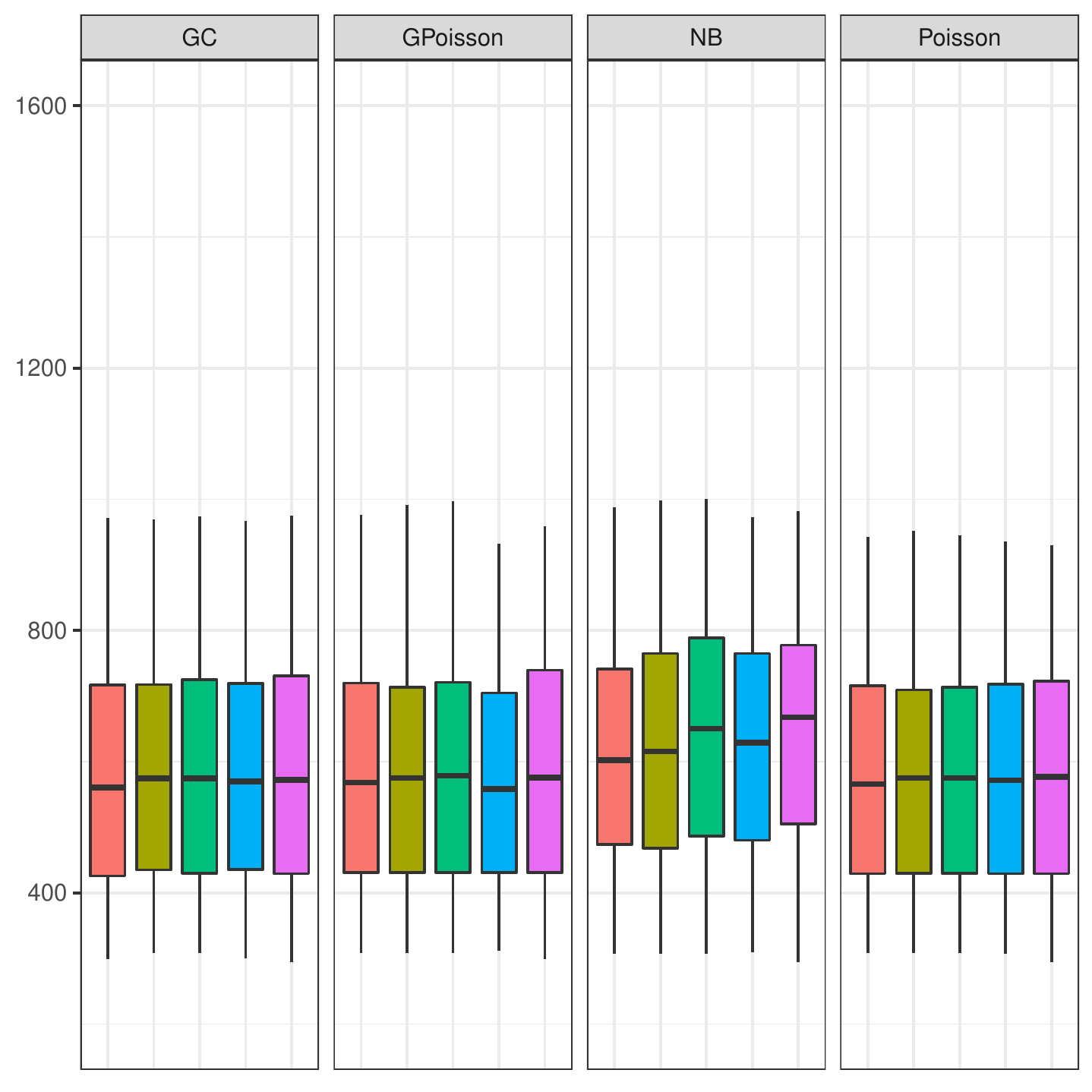}&
\includegraphics[width=155pt,height=6.8pc]{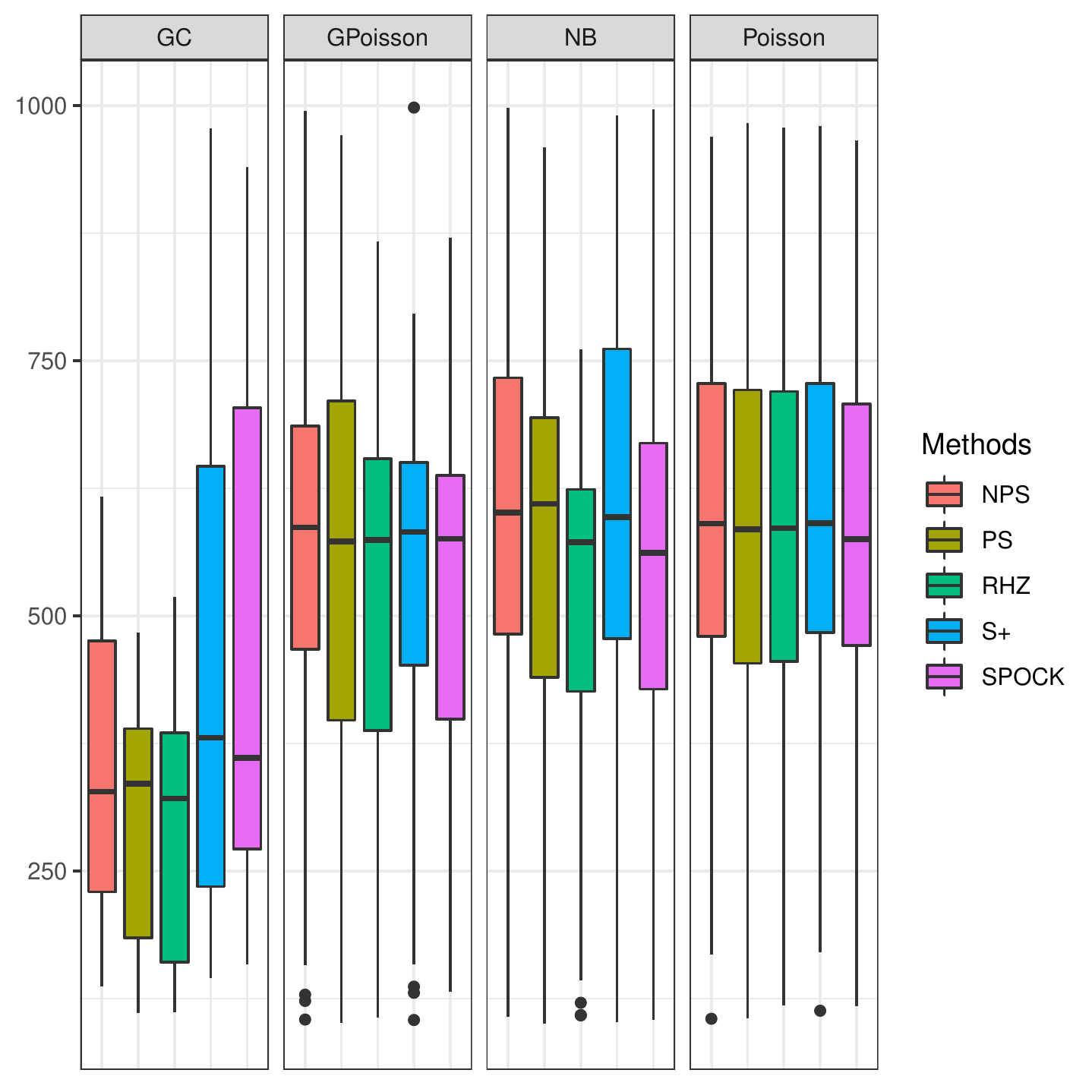}\\ 
\includegraphics[width=155pt,height=6.8pc]{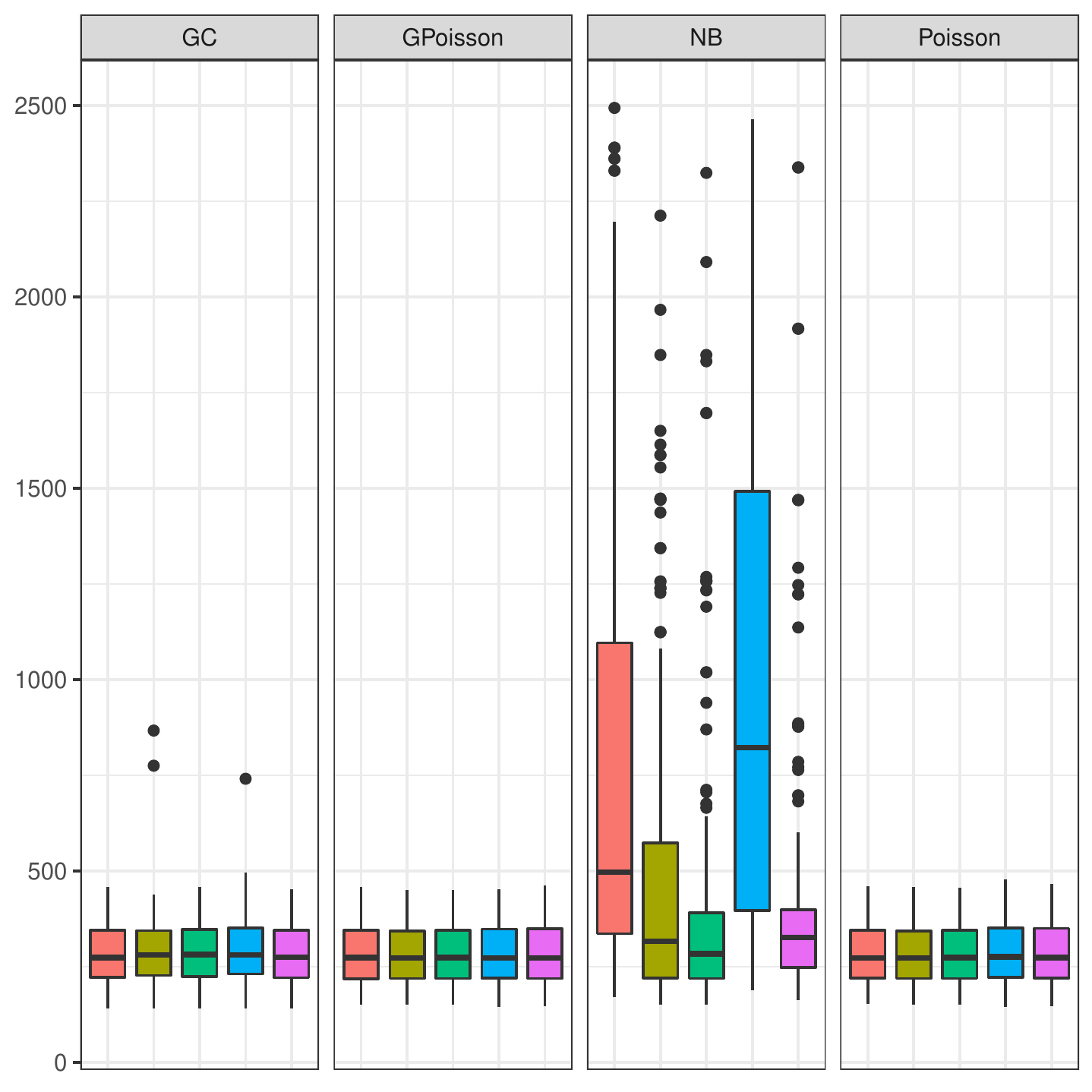} & 
\includegraphics[width=155pt,height=6.8pc]{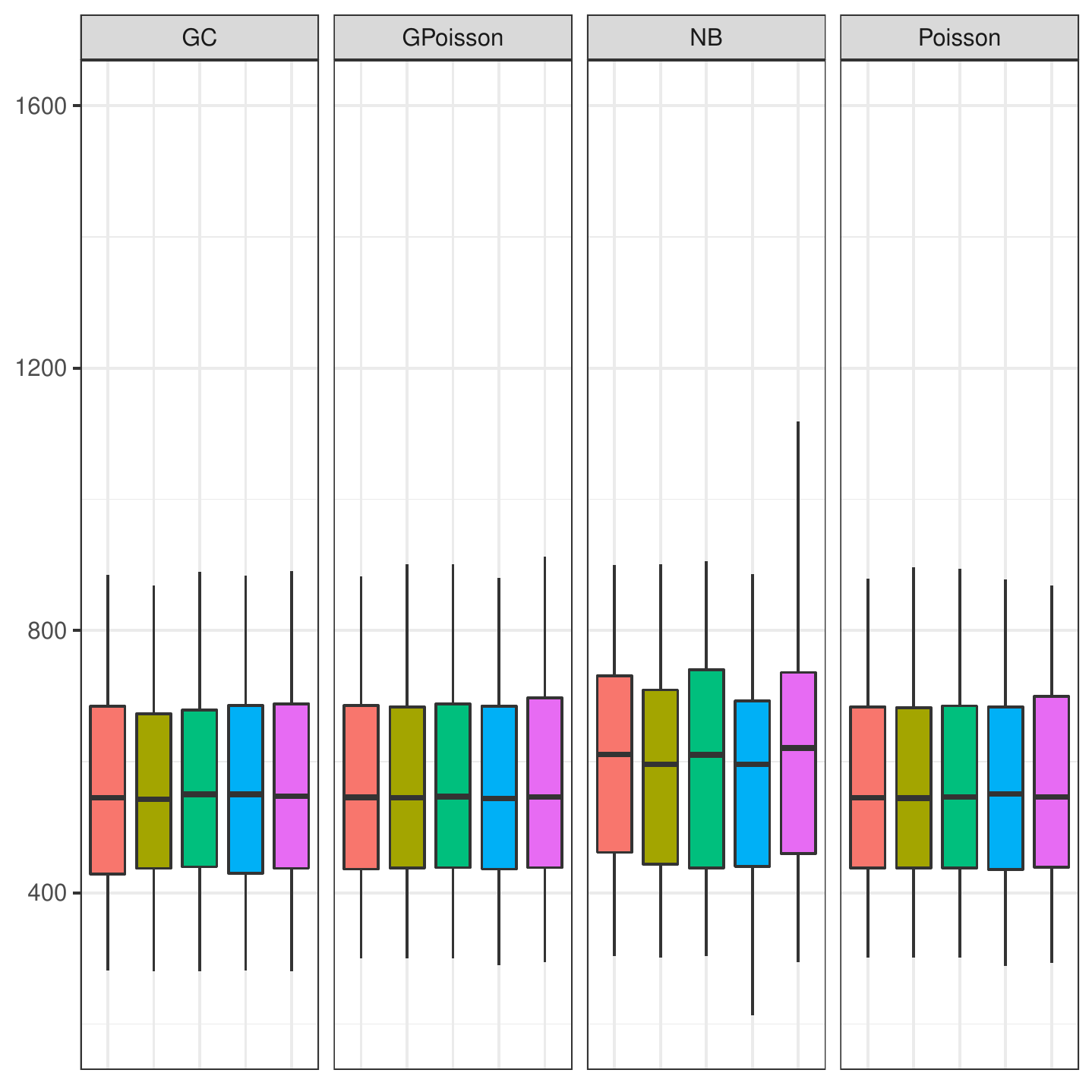}&
\includegraphics[width=155pt,height=6.8pc]{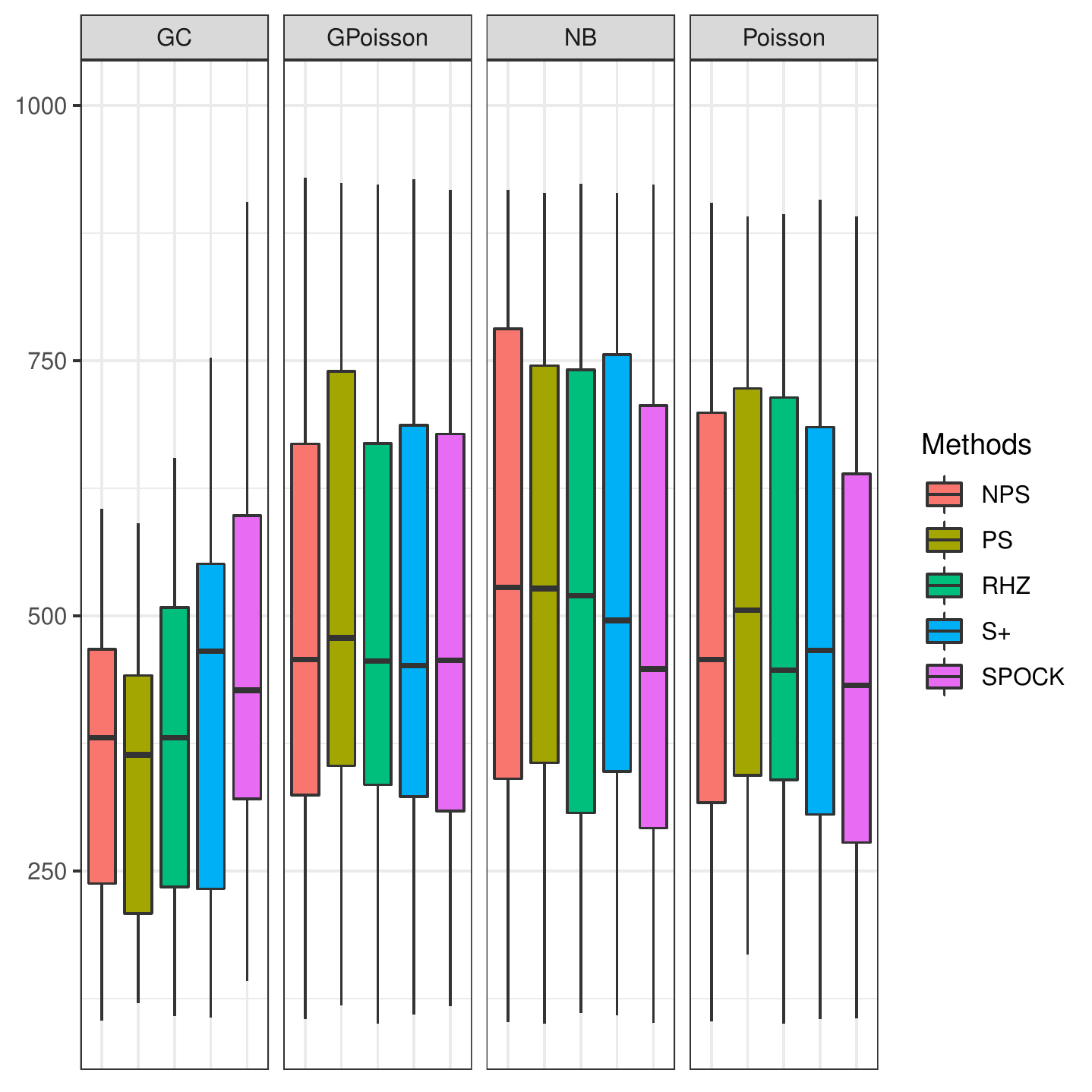}\\ 
\includegraphics[width=155pt,height=6.8pc]{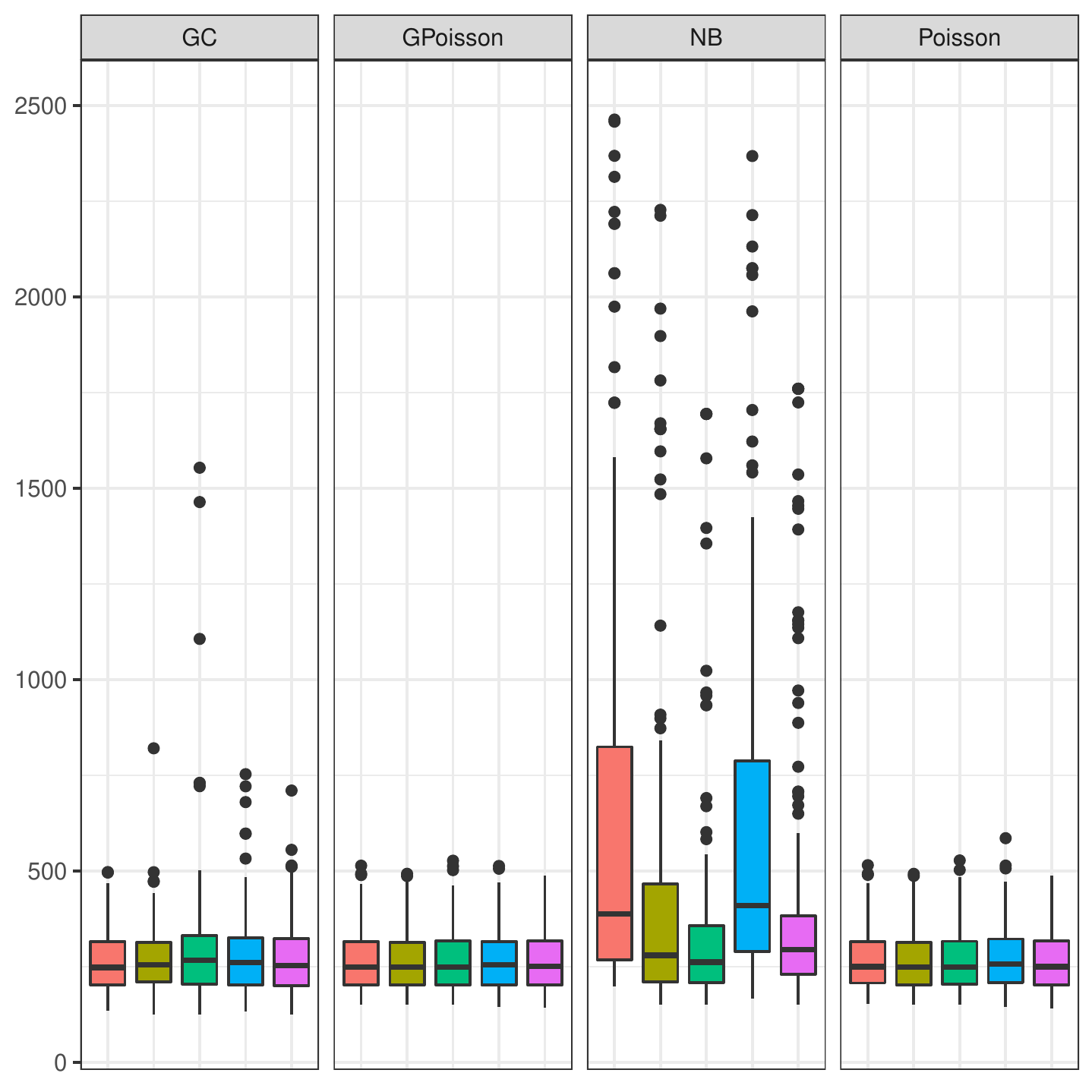} & 
\includegraphics[width=155pt,height=6.8pc]{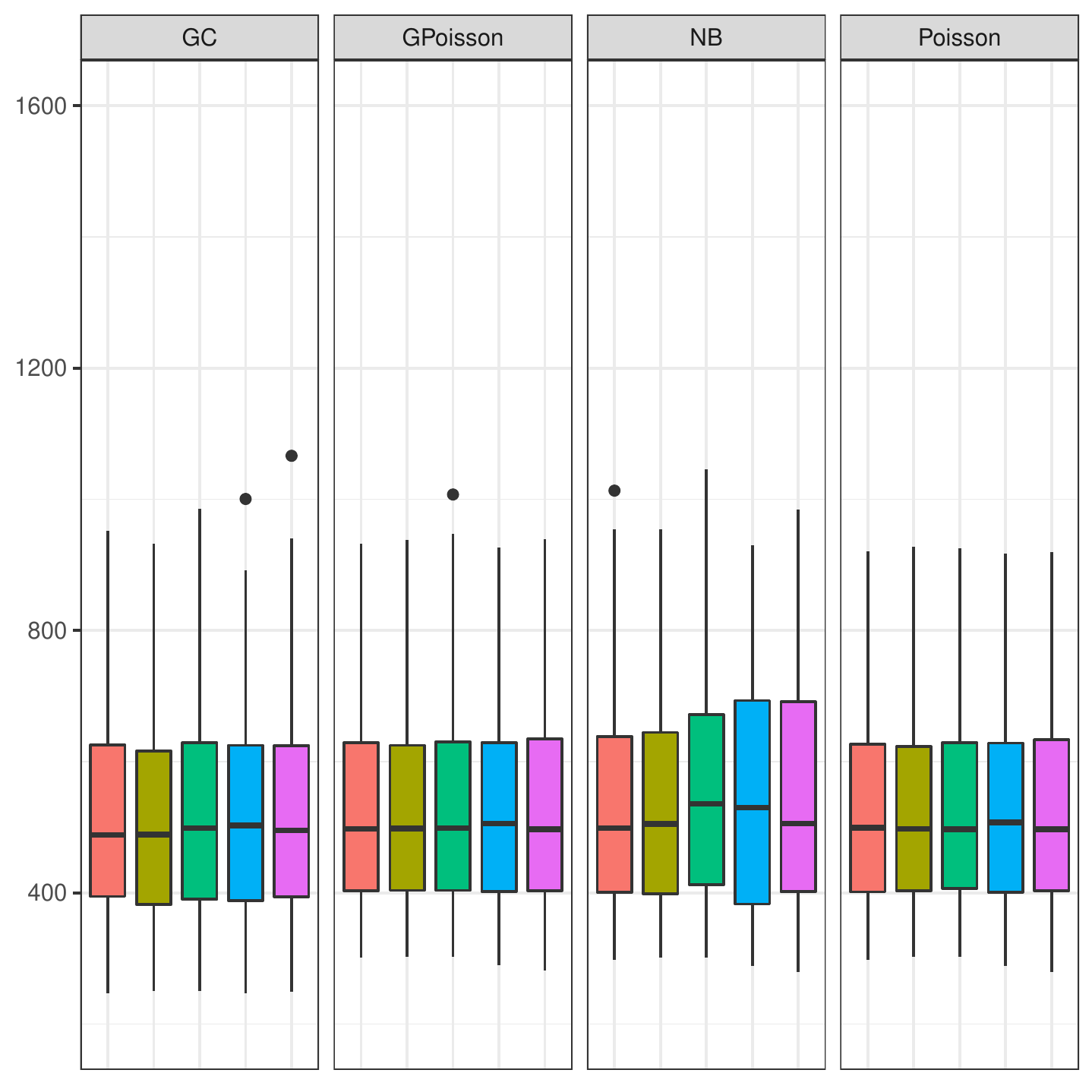}&
\includegraphics[width=155pt,height=6.8pc]{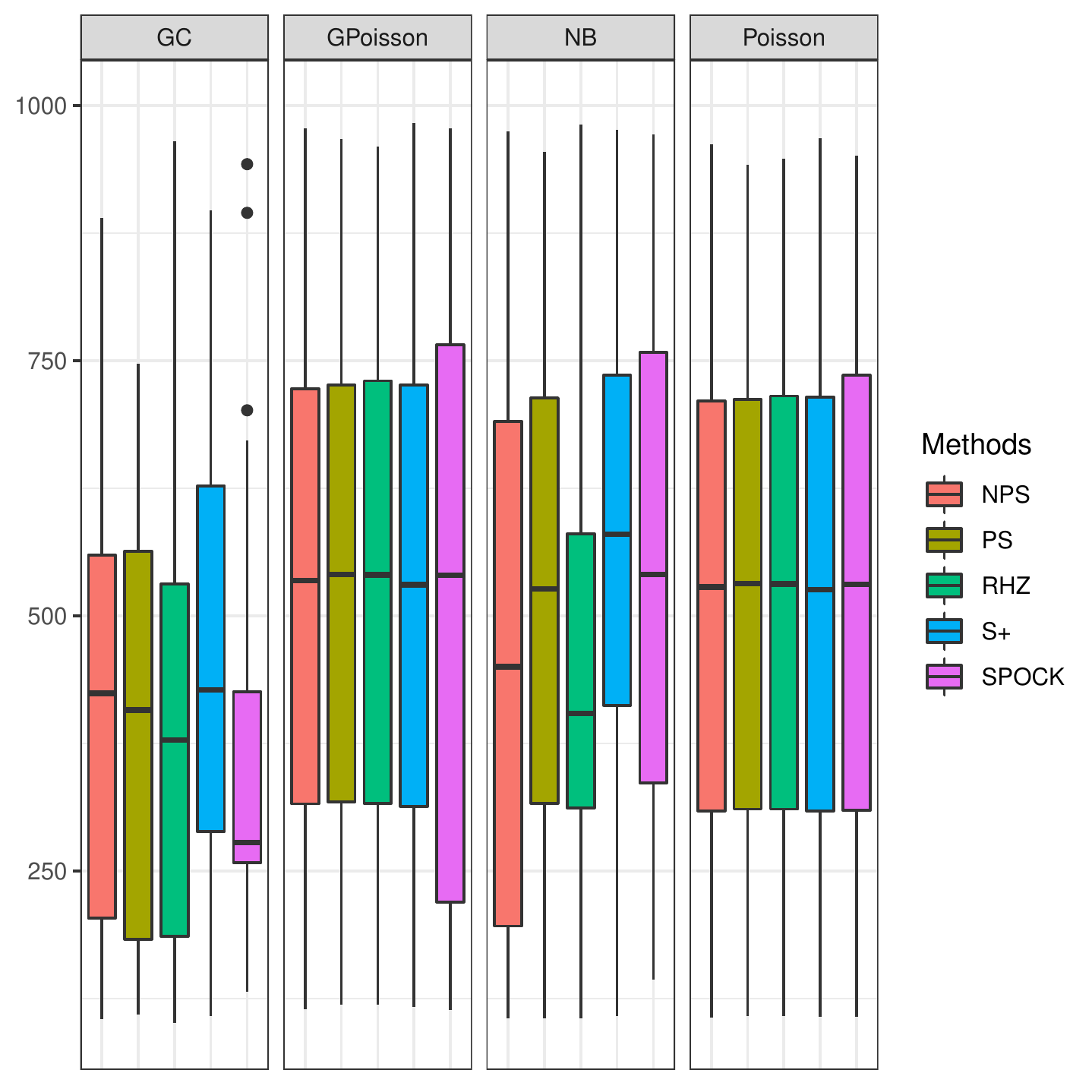} 
\end{tabular} 
\end{center} 
\caption{LS (left column), WAIC (middle column) and 
DIC (right column) for each 
model with $\tau_x=4$ for scenarios: over-dispersion (first row), equivalent-dispersion (second row), under-dispersion (third row; $\alpha=1.3$) and under-dispersion (forth row; $\alpha=2$).\label{fig12}} 
\end{figure} 
\begin{figure}[ppt]
\begin{center} 
\begin{tabular}{ccc} 
\includegraphics[width=155pt,height=6.8pc]{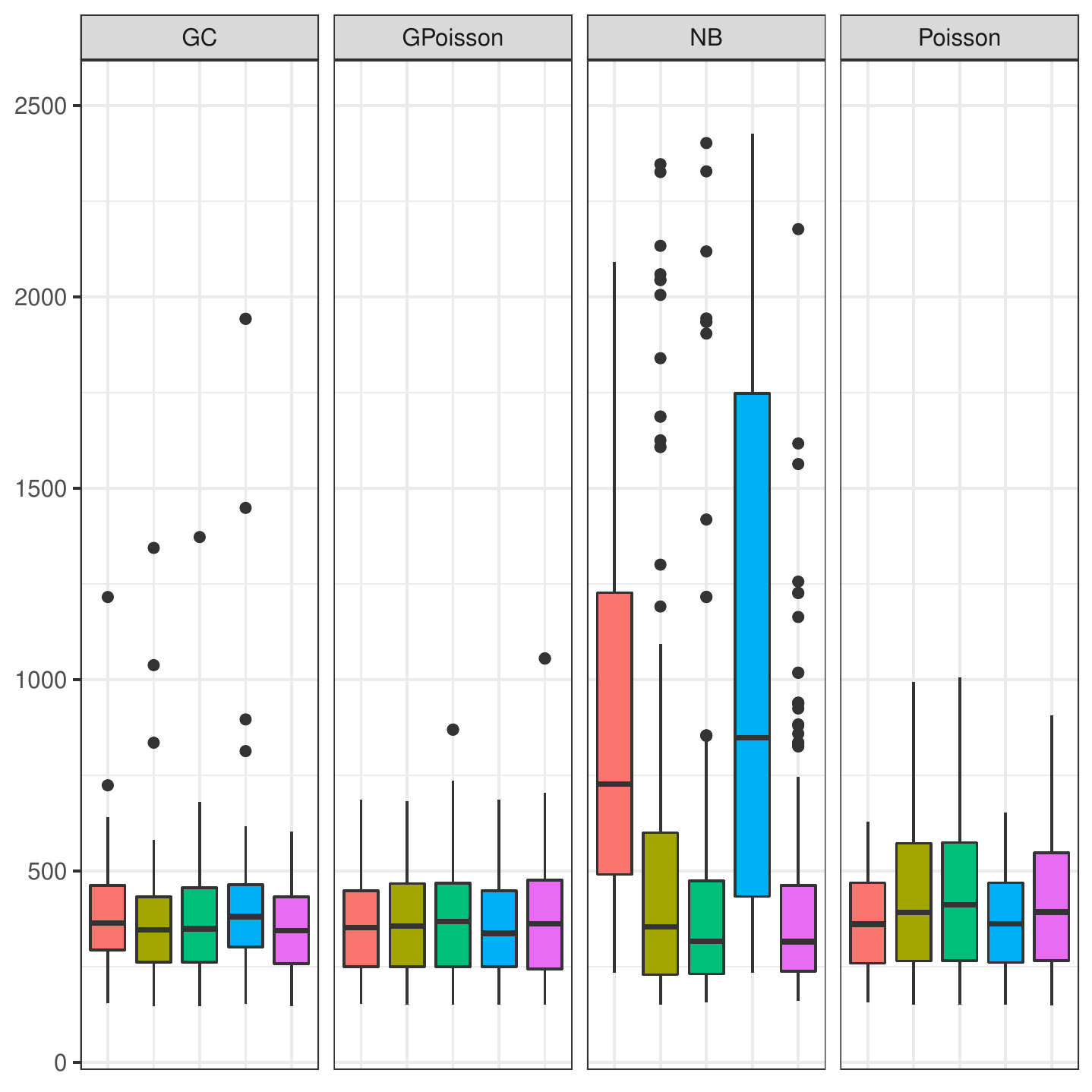} &
 \includegraphics[width=155pt,height=6.8pc]{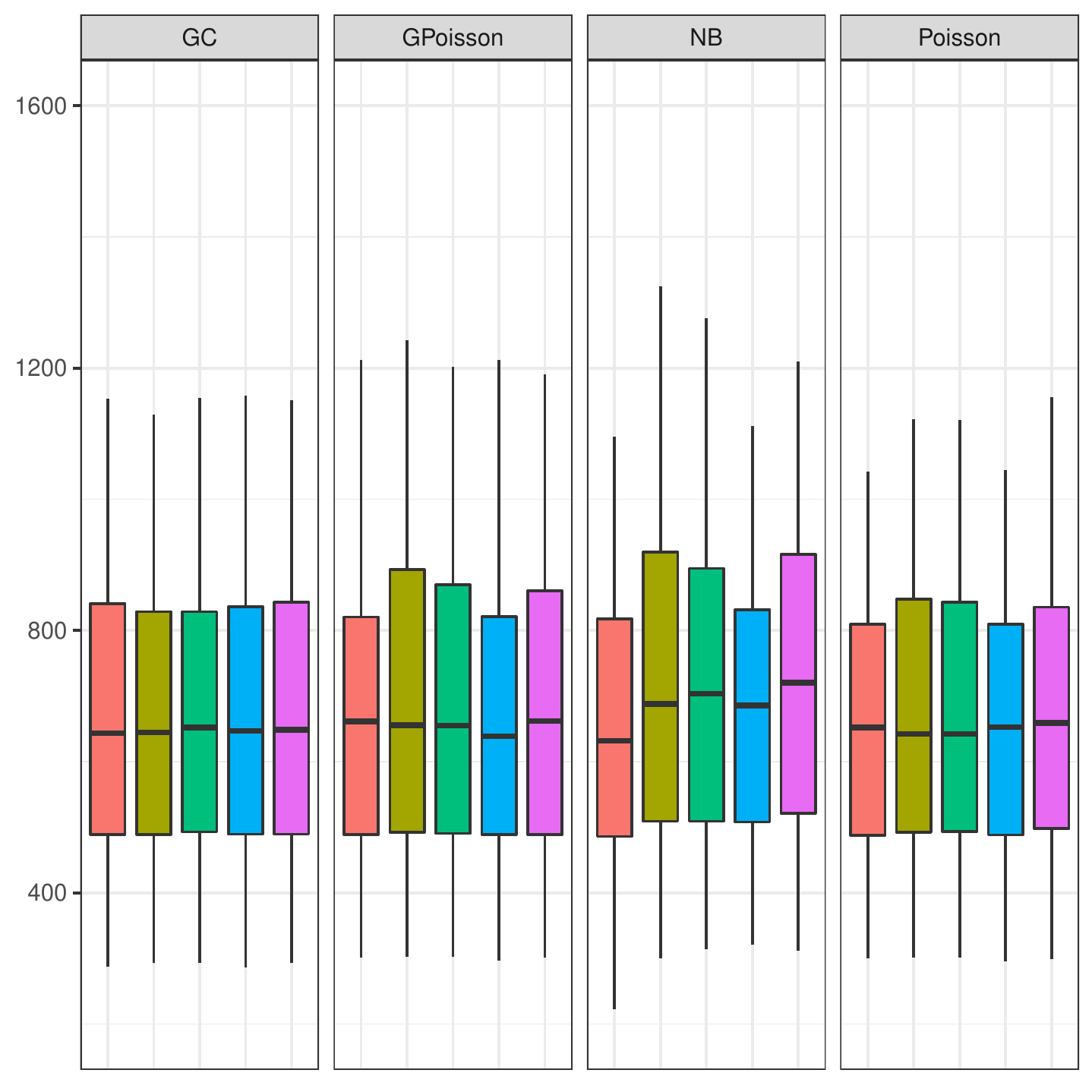}&
 \includegraphics[width=155pt,height=6.8pc]{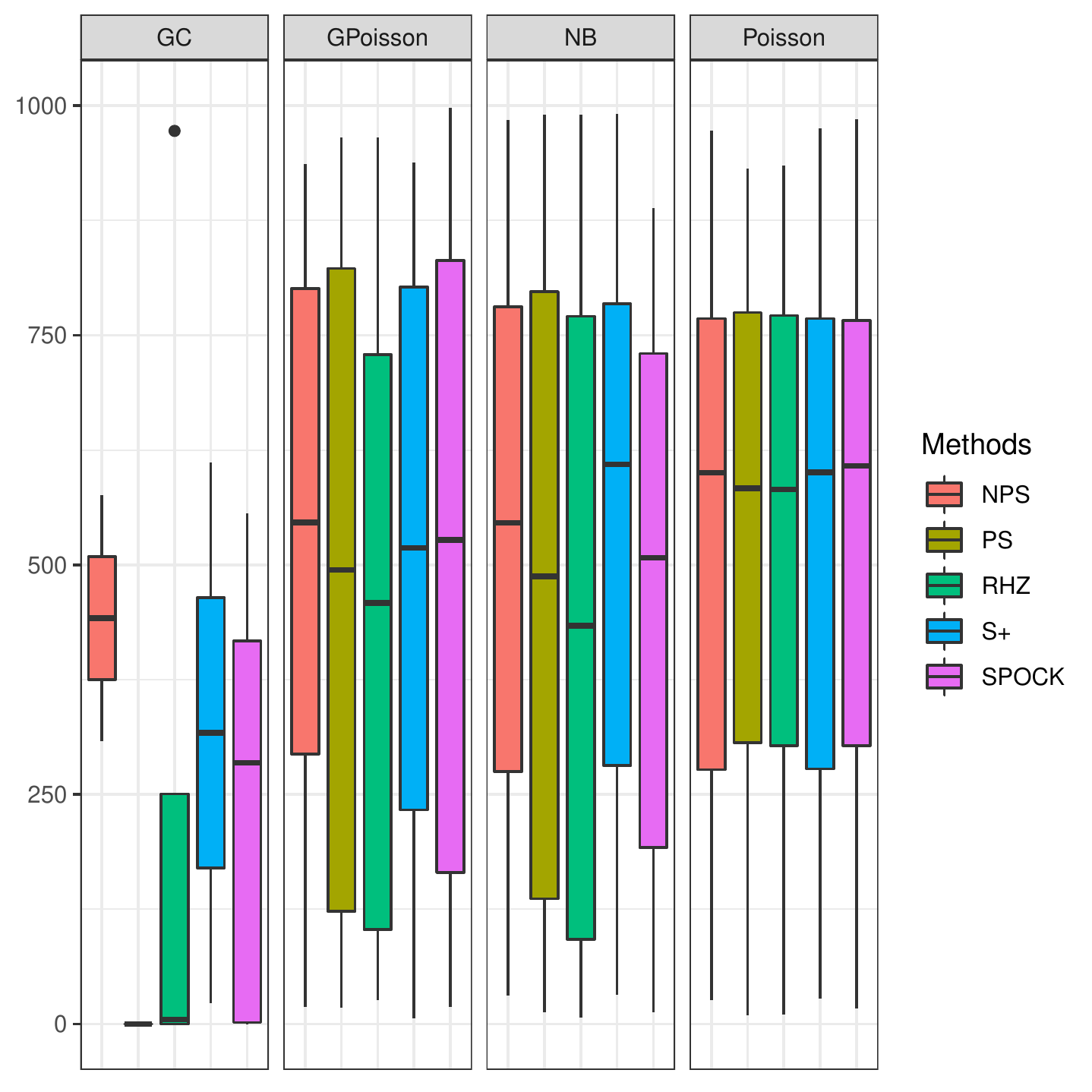}\\ 
\includegraphics[width=155pt,height=6.8pc]{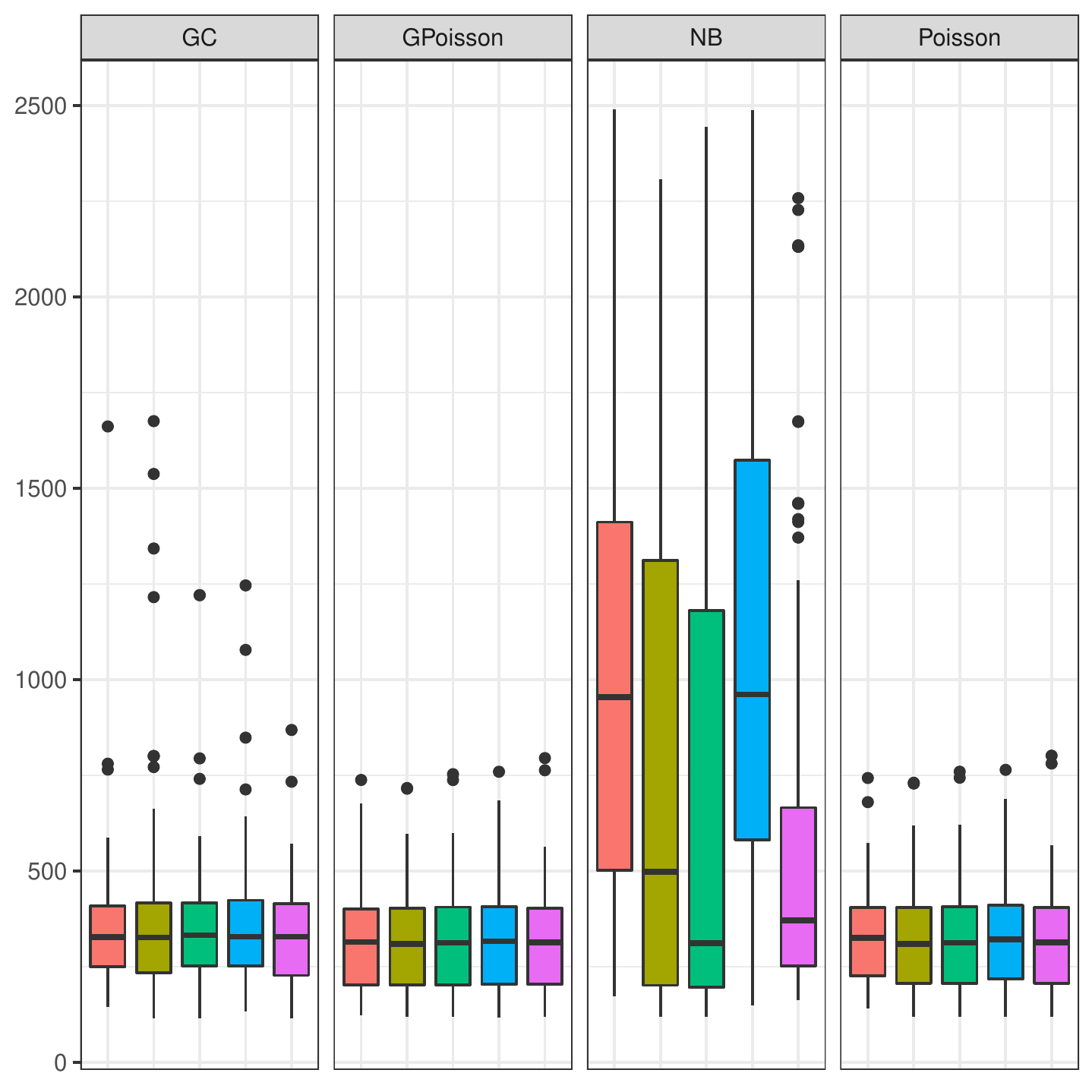} & 
\includegraphics[width=155pt,height=6.8pc]{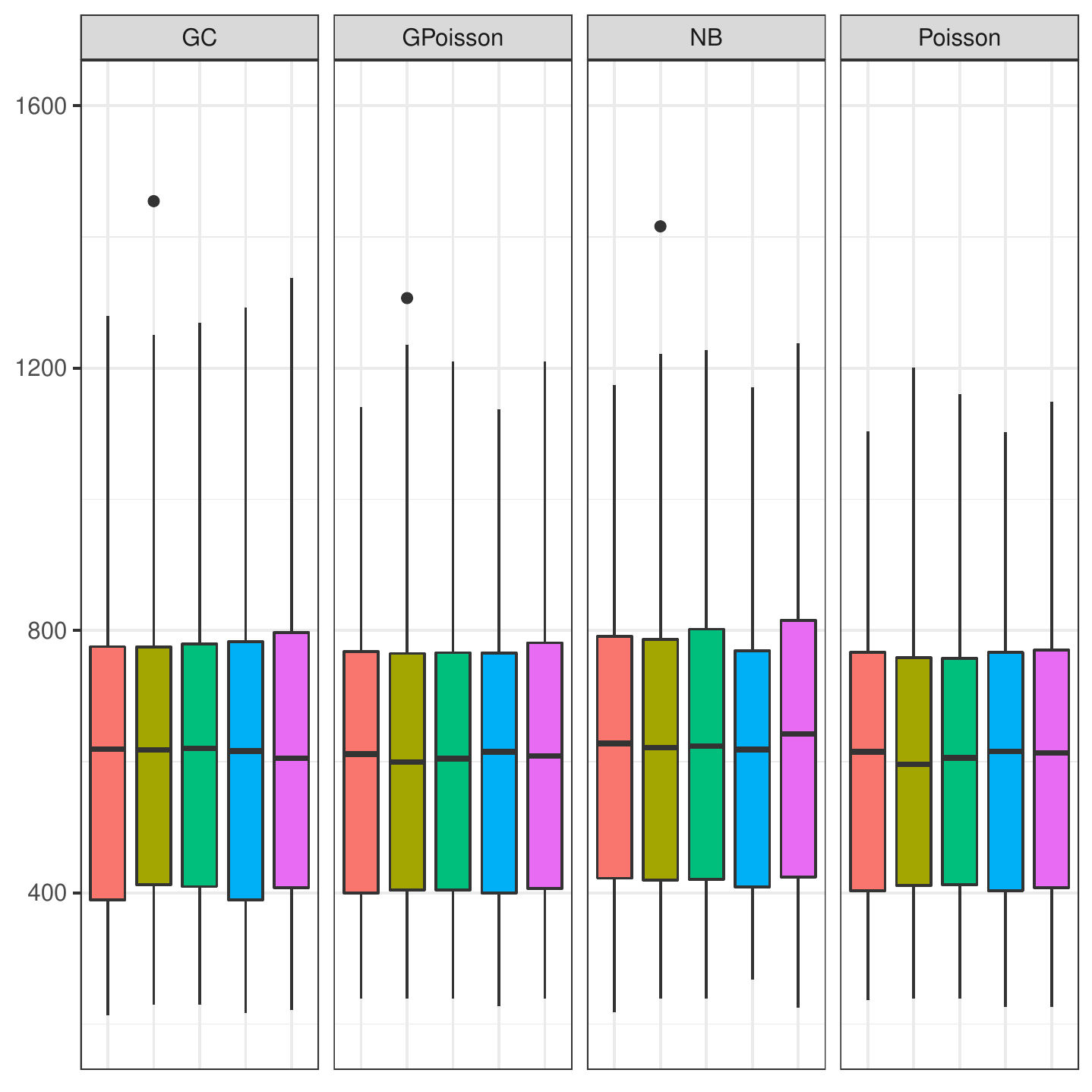}&
\includegraphics[width=155pt,height=6.8pc]{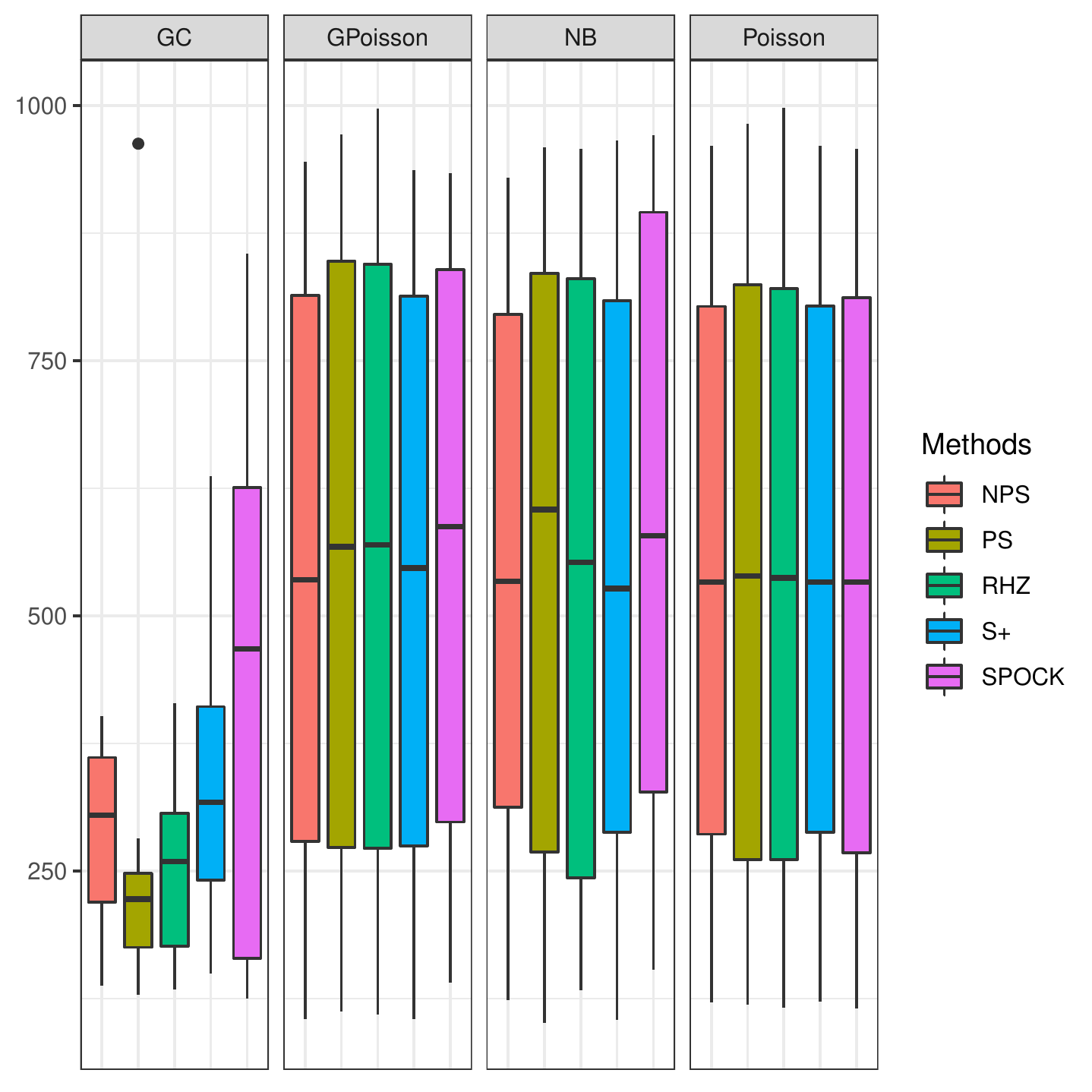}\\ 
\includegraphics[width=155pt,height=6.8pc]{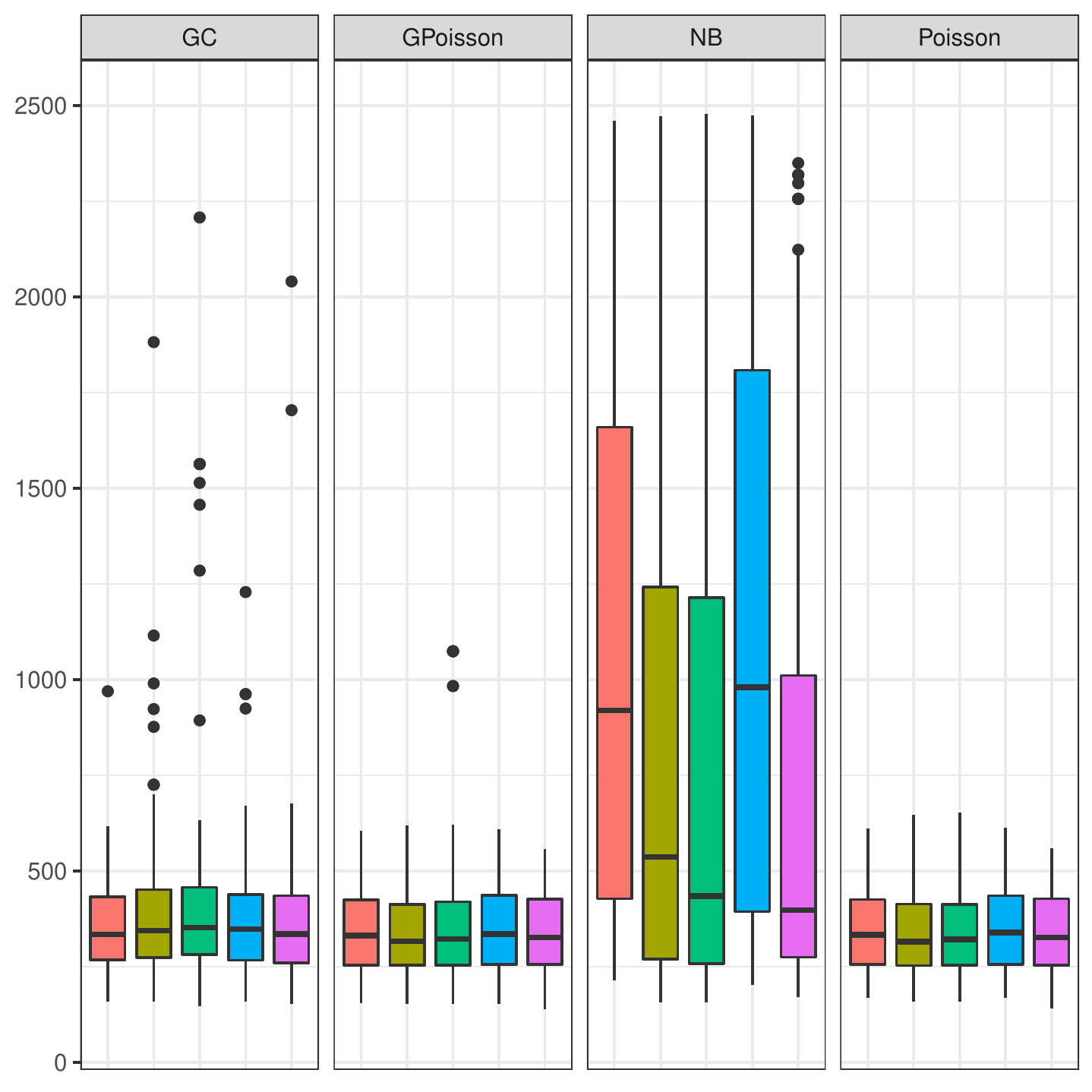} & 
\includegraphics[width=155pt,height=6.8pc]{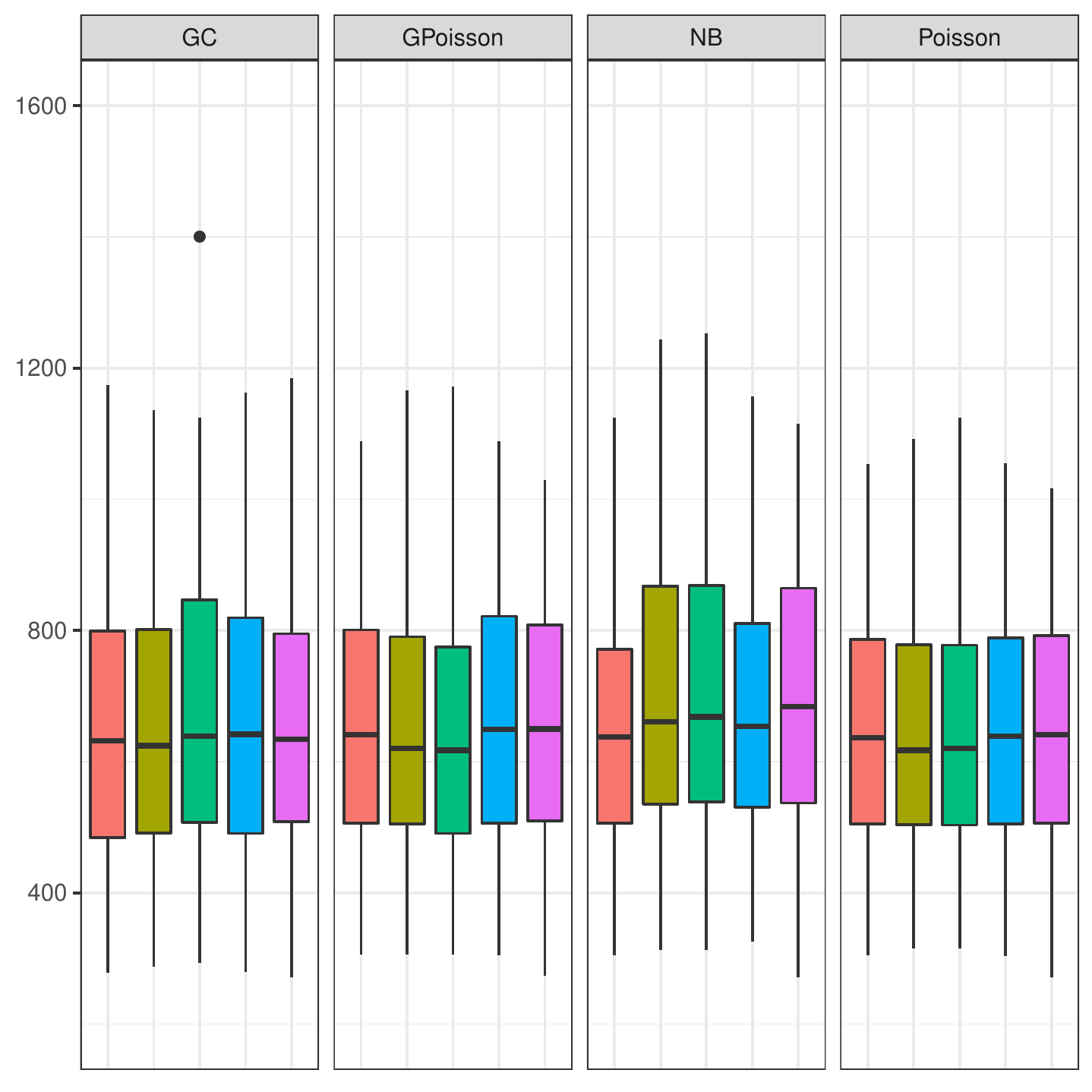}&
\includegraphics[width=155pt,height=6.8pc]{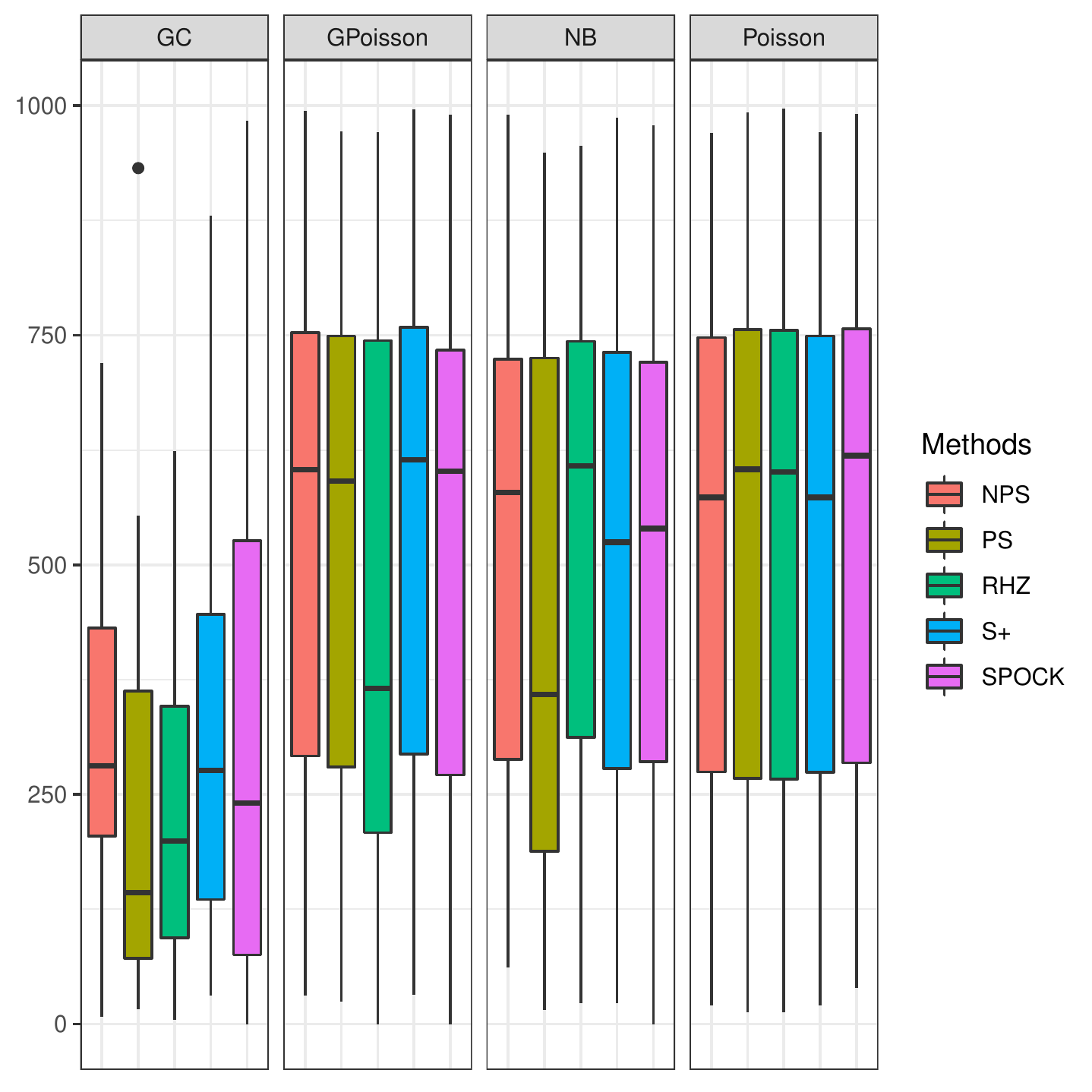}\\ 
\includegraphics[width=155pt,height=6.8pc]{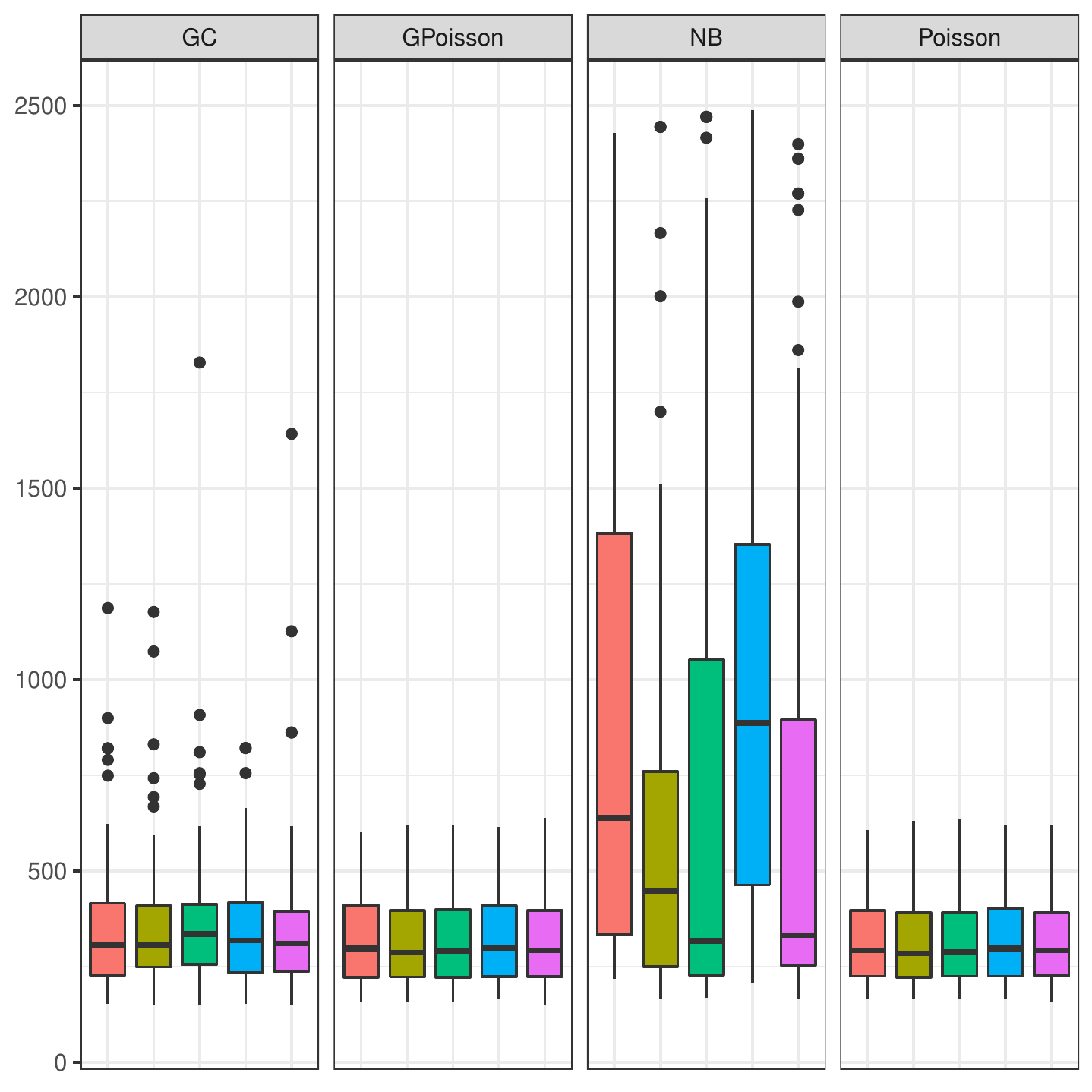} & 
\includegraphics[width=155pt,height=6.8pc]{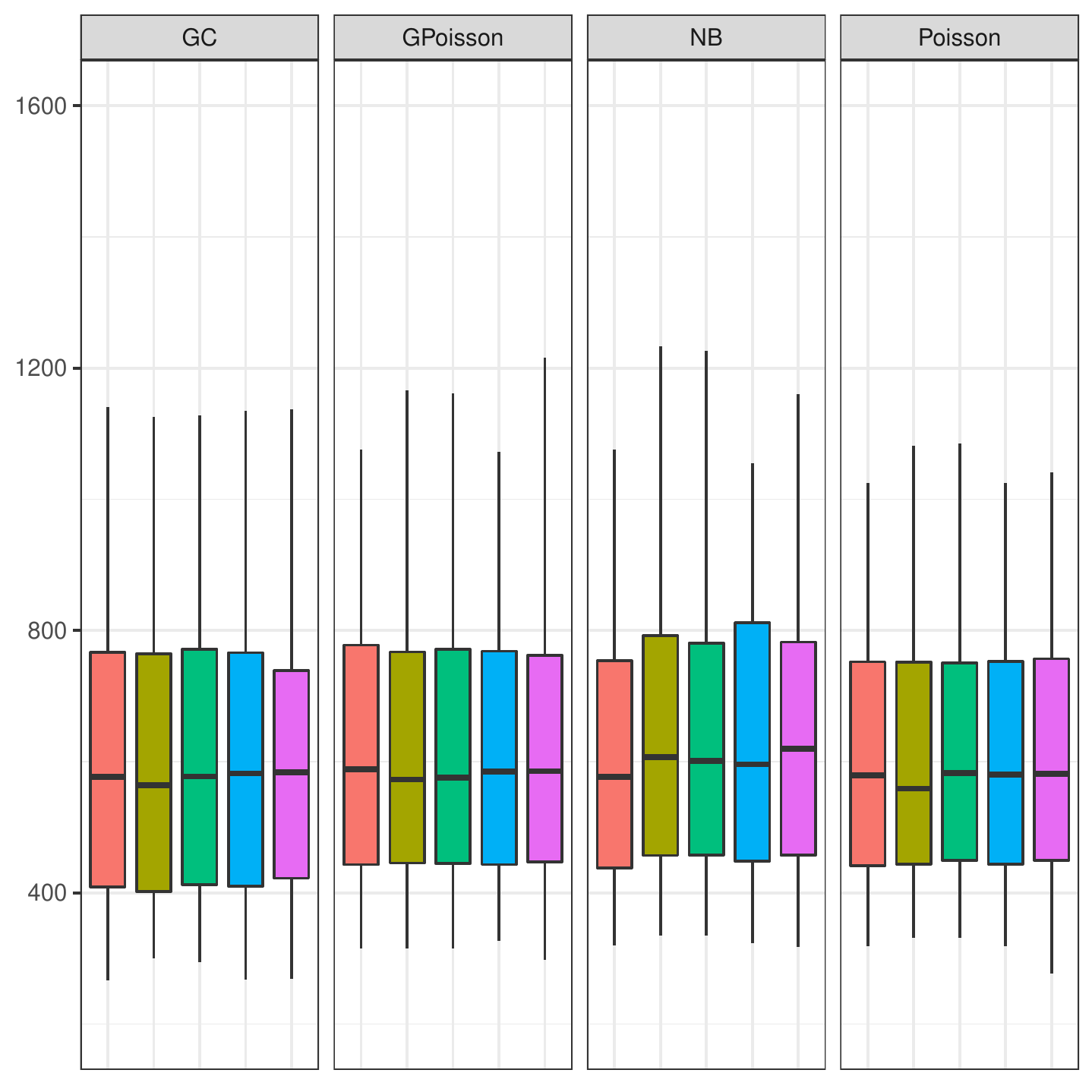}&
\includegraphics[width=155pt,height=6.8pc]{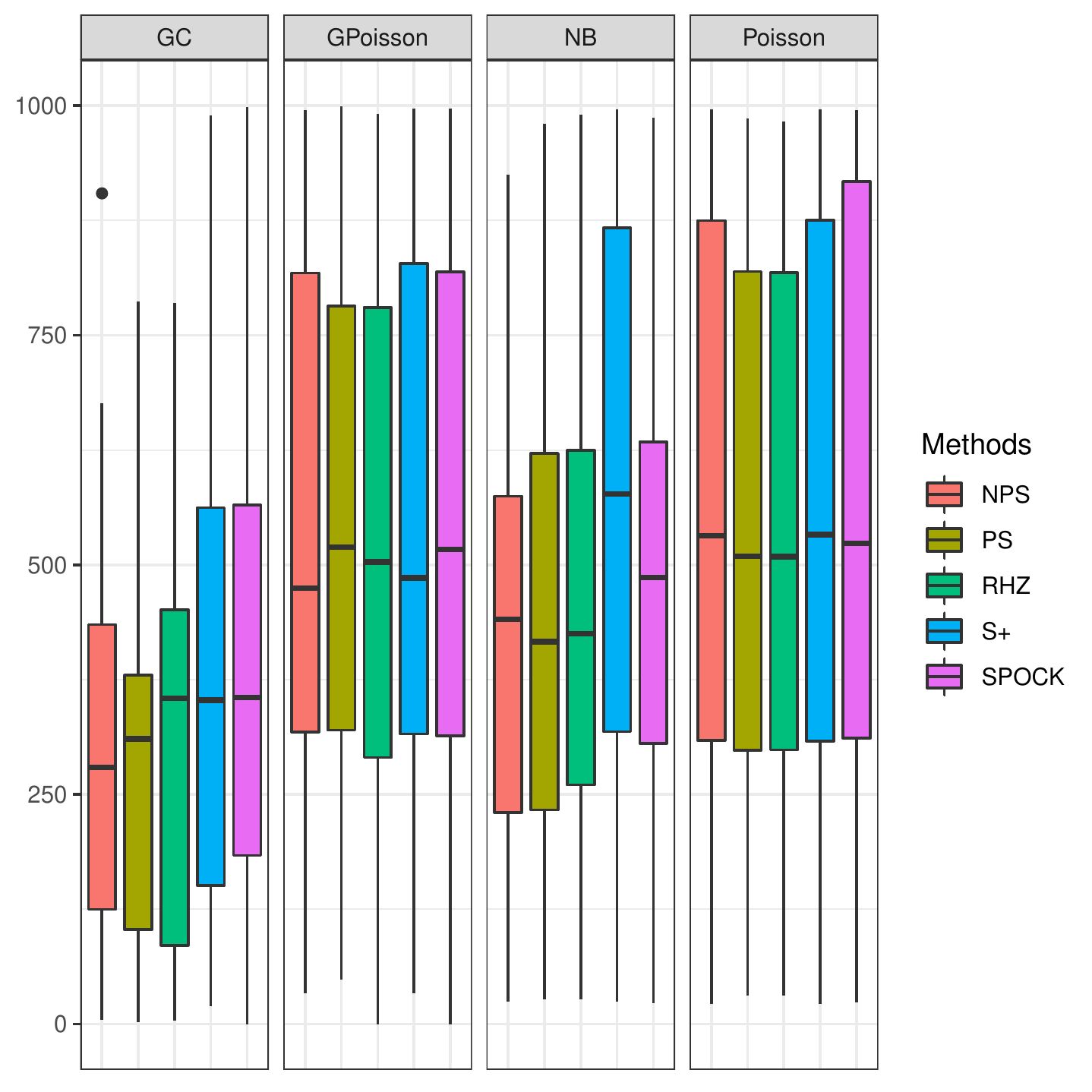} 
\end{tabular} 
\end{center} 
\caption{LS (left column), WAIC (middle column) and 
DIC (right column) for each 
model with $\tau_x=1$ for scenarios: over-dispersion (first row), equivalent-dispersion (second row), under-dispersion (third row; $\alpha=1.3$) and under-dispersion (forth row; $\alpha=2$).\label{fig13}} 
\end{figure} 
\begin{table}[h!]
\centering\caption{\label{Tab4}MSE  and coverage rate for a nominal rate of 95$\%$ for all scenarios for Poisson model.}
\begin{scriptsize}
\begin{tabular}{lllllllllllllll}
  \hline
\textbf{Family} &\textbf{$\alpha$}&\textbf{Model} &\multicolumn{6}{c}{\textbf{$\beta_1$}}&\multicolumn{6}{c}{\textbf{$\beta_2$}}\\
 \cmidrule[0.7pt](lr{0.125em}){4-9}\cmidrule[0.7pt](lr{0.125em}){10-15}
   &&&\multicolumn{6}{c}{$\tau_x$}&\multicolumn{6}{c}{$\tau_x$}\\
  & & &\multicolumn{2}{c}{11}&\multicolumn{2}{c}{4}&\multicolumn{2}{c}{1}&\multicolumn{2}{c}{11}&\multicolumn{2}{c}{4}&\multicolumn{2}{c}{1}\\
 \cmidrule[0.7pt](lr{0.125em}){4-5}\cmidrule[0.7pt](lr{0.125em}){6-7}  \cmidrule[0.7pt](lr{0.125em}){8-9}\cmidrule[0.7pt](lr{0.125em}){10-11}  \cmidrule[0.7pt](lr{0.125em}){12-13}\cmidrule[0.7pt](lr{0.125em}){14-15}
  &&& MSE & CR & MSE & CR & MSE& CR & MSE & CR & MSE & CR & MSE & CR \\ 
   \hline
  &&NPS & 0.20 & 0.96 & 0.03 & 0.96 & 0.02 & 0.96 & 1.27 & 0.66 & 1.27 & 0.66 & 1.27 & 0.66 \\ 
  &&S+ & 0.26 & 0.94 & 0.02 & 0.94 & 0.02 & 0.94 & 0.49 & 0.49 & 0.49 & 0.49 & 0.49 & 0.49 \\ 
  &0.5&PS & 0.10 & 0.96 & 0.03 & 0.96 & 0.03 & 0.96 & 1.32 & 0.68 & 1.32 & 0.68 & 1.32 & 0.68 \\ 
  &&SPOCK & 0.15 & 0.96 & 0.05 & 0.96 & 0.03 & 0.96 & 0.08 & 0.74 & 0.08 & 0.74 & 0.08 & 0.74 \\ 
  &&RHZ & 0.14 & 0.96 & 0.03 & 0.96 & 0.03 & 0.96 & 0.11 & 0.71 & 0.11 & 0.71 & 0.11 & 0.71 \\ 
  [3mm]
 && NPS & 0.28 & 0.95 & 0.02 & 0.95 & 0.02 & 0.95 & 1.41 & 0.80 & 1.41 & 0.80 & 1.41 & 0.80 \\ 
 && S+ & 0.30 & 0.93 & 0.02 & 0.93 & 0.02 & 0.93 & 0.32 & 0.66 & 0.32 & 0.66 & 0.32 & 0.66 \\ 
 &1& PS & 0.33 & 0.95 & 0.03 & 0.95 & 0.03 & 0.95 & 1.40 & 0.85 & 1.40 & 0.85 & 1.40 & 0.85 \\ 
 && SPOCK & 0.33 & 0.95 & 0.04 & 0.95 & 0.03 & 0.95 & 0.03 & 0.90 & 0.03 & 0.90 & 0.03 & 0.90 \\ 
\textbf{Poisson} && RHZ & 0.32 & 0.94 & 0.03 & 0.94 & 0.03 & 0.94 & 0.03 & 0.82 & 0.03 & 0.82 & 0.03 & 0.82 \\ 
 [3mm] 
  &&NPS & 0.28 & 0.98 & 0.06 & 0.98 & 0.07 & 0.98 & 1.40 & 0.90 & 1.40 & 0.90 & 1.40 & 0.90 \\ 
  &&S+ & 0.34 & 0.92 & 0.11 & 0.92 & 0.14 & 0.92 & 0.24 & 0.84 & 0.24 & 0.84 & 0.24 & 0.84 \\ 
 &1.3& PS & 0.26 & 0.92 & 0.06 & 0.92 & 0.04 & 0.92 & 1.40 & 0.92 & 1.40 & 0.92 & 1.40 & 0.92 \\ 
 && SPOCK & 0.20 & 0.95 & 0.09 & 0.95 & 0.06 & 0.95 & 0.11 & 0.93 & 0.11 & 0.93 & 0.11 & 0.93 \\ 
  &&RHZ & 0.21 & 0.94 & 0.04 & 0.94 & 0.03 & 0.94 & 0.04 & 0.90 & 0.04 & 0.90 & 0.04 & 0.90 \\ 
  [3mm]
  &&NPS & 0.34 & 0.94 & 0.03 & 0.94 & 0.02 & 0.94 & 1.44 & 0.83 & 1.44 & 0.83 & 1.44 & 0.83 \\ 
  &&S+ & 0.33 & 0.92 & 0.03 & 0.92 & 0.02 & 0.92 & 0.29 & 0.66 & 0.29 & 0.66 & 0.29 & 0.66 \\ 
  &2&PS & 0.35 & 0.96 & 0.03 & 0.96 & 0.02 & 0.96 & 1.31 & 0.85 & 1.31 & 0.85 & 1.31 & 0.85 \\ 
 && SPOCK & 0.35 & 0.96 & 0.03 & 0.96 & 0.02 & 0.96 & 0.03 & 0.91 & 0.03 & 0.91 & 0.03 & 0.91 \\ 
 && RHZ & 0.35 & 0.96 & 0.03 & 0.96 & 0.02 & 0.96 & 0.03 & 0.85 & 0.03 & 0.85 & 0.03 & 0.85 \\ 
  \hline
\end{tabular}
\end{scriptsize}
\end{table}
\begin{table}[h!]
\centering\caption{\label{Tab5}MSE  and coverage rate for a nominal rate of 95$\%$ for all scenarios for NB model.}
\begin{scriptsize}
\begin{tabular}{lllllllllllllll}
  \hline
\textbf{Family} &\textbf{$\alpha$}&\textbf{Model} &\multicolumn{6}{c}{\textbf{$\beta_1$}}&\multicolumn{6}{c}{\textbf{$\beta_2$}}\\
\cmidrule[0.7pt](lr{0.125em}){4-9}\cmidrule[0.7pt](lr{0.125em}){10-15}
   &&&\multicolumn{6}{c}{$\tau_x$}&\multicolumn{6}{c}{$\tau_x$}\\
  & & &\multicolumn{2}{c}{11}&\multicolumn{2}{c}{4}&\multicolumn{2}{c}{1}&\multicolumn{2}{c}{11}&\multicolumn{2}{c}{4}&\multicolumn{2}{c}{1}\\
  \cmidrule[0.7pt](lr{0.125em}){4-5}\cmidrule[0.7pt](lr{0.125em}){6-7}  \cmidrule[0.7pt](lr{0.125em}){8-9}\cmidrule[0.7pt](lr{0.125em}){10-11}  \cmidrule[0.7pt](lr{0.125em}){12-13}\cmidrule[0.7pt](lr{0.125em}){14-15}
 &&& MSE & CR & MSE & CR & MSE& CR & MSE & CR & MSE & CR & MSE & CR \\ 
   \hline
 && NPS & 0.25 & 0.94 & 0.02 & 0.94 & 0.02 & 0.94 & 1.22 & 0.94 & 1.22 & 0.94 & 1.22 & 0.94 \\ 
&&  S+ & 0.36 & 0.94 & 0.02 & 0.94 & 0.01 & 0.94 & 0.57 & 0.56 & 0.57 & 0.56 & 0.57 & 0.56 \\ 
 &0.5& PS & 0.26 & 0.92 & 0.02 & 0.92 & 0.03 & 0.92 & 1.26 & 0.92 & 1.26 & 0.92 & 1.26 & 0.92 \\ 
 && SPOCK & 0.26 & 0.92 & 0.03 & 0.92 & 0.02 & 0.92 & 0.06 & 0.93 & 0.06 & 0.93 & 0.06 & 0.93 \\ 
  &&RHZ & 0.26 & 0.92 & 0.03 & 0.92 & 0.02 & 0.92 & 0.06 & 0.90 & 0.06 & 0.90 & 0.06 & 0.90 \\ 
  [3mm]
 && NPS & 0.26 & 0.97 & 0.02 & 0.97 & 0.01 & 0.97 & 1.59 & 0.87 & 1.59 & 0.87 & 1.59 & 0.87 \\ 
 && S+ & 0.33 & 0.96 & 0.02 & 0.96 & 0.01 & 0.96 & 0.46 & 0.42 & 0.46 & 0.42 & 0.46 & 0.42 \\ 
 &1& PS & 0.33 & 0.96 & 0.02 & 0.96 & 0.02 & 0.96 & 1.58 & 0.85 & 1.58 & 0.85 & 1.58 & 0.85 \\ 
  &&SPOCK & 0.33 & 0.96 & 0.03 & 0.96 & 0.02 & 0.96 & 0.08 & 0.86 & 0.08 & 0.86 & 0.08 & 0.86 \\ 
 \textbf{NB} &&RHZ & 0.33 & 0.96 & 0.02 & 0.96 & 0.02 & 0.96 & 0.08 & 0.85 & 0.08 & 0.85 & 0.08 & 0.85 \\
  [3mm]
  &&NPS & 0.35 & 0.96 & 0.04 & 0.96 & 0.07 & 0.96 & 1.43 & 0.91 & 1.43 & 0.91 & 1.43 & 0.91 \\ 
 && S+ & 0.36 & 0.94 & 0.07 & 0.94 & 0.05 & 0.94 & 0.42 & 0.60 & 0.42 & 0.60 & 0.42 & 0.60 \\ 
 &1.3& PS & 0.36 & 0.94 & 0.08 & 0.94 & 0.06 & 0.94 & 1.51 & 0.88 & 1.51 & 0.88 & 1.51 & 0.88 \\ 
 && SPOCK & 0.27 & 0.93 & 0.05 & 0.93 & 0.01 & 0.93 & 0.17 & 0.85 & 0.17 & 0.85 & 0.17 & 0.85 \\ 
 && RHZ & 0.31 & 0.95 & 0.04 & 0.95 & 0.05 & 0.95 & 0.21 & 0.88 & 0.21 & 0.88 & 0.21 & 0.88 \\ 
  [3mm]
 && NPS & 0.31 & 0.97 & 0.02 & 0.97 & 0.01 & 0.97 & 1.57 & 0.90 & 1.57 & 0.90 & 1.57 & 0.90 \\ 
 && S+ & 0.33 & 0.96 & 0.02 & 0.96 & 0.01 & 0.96 & 0.44 & 0.49 & 0.44 & 0.49 & 0.44 & 0.49 \\ 
 &2& PS & 0.27 & 0.96 & 0.02 & 0.96 & 0.02 & 0.96 & 1.55 & 0.88 & 1.55 & 0.88 & 1.55 & 0.88 \\ 
 && SPOCK & 0.27 & 0.96 & 0.02 & 0.96 & 0.02 & 0.96 & 0.08 & 0.88 & 0.08 & 0.88 & 0.08 & 0.88 \\ 
  &&RHZ & 0.27 & 0.96 & 0.02 & 0.96 & 0.02 & 0.96 & 0.07 & 0.88 & 0.07 & 0.88 & 0.07 & 0.88 \\ 
  \hline
\end{tabular}
\end{scriptsize}
\end{table}
\begin{table}[h!]
\centering\caption{\label{Tab6}MSE  and coverage rate for a nominal rate of 95$\%$ for all scenarios for GP model.}
\begin{scriptsize}
\begin{tabular}{lllllllllllllll}
  \hline
\textbf{Family} &\textbf{$\alpha$}&\textbf{Model} &\multicolumn{6}{c}{\textbf{$\beta_1$}}&\multicolumn{6}{c}{\textbf{$\beta_2$}}\\
\cmidrule[0.7pt](lr{0.125em}){4-9}\cmidrule[0.7pt](lr{0.125em}){10-15}
   &&&\multicolumn{6}{c}{$\tau_x$}&\multicolumn{6}{c}{$\tau_x$}\\
  & & &\multicolumn{2}{c}{11}&\multicolumn{2}{c}{4}&\multicolumn{2}{c}{1}&\multicolumn{2}{c}{11}&\multicolumn{2}{c}{4}&\multicolumn{2}{c}{1}\\
   \cmidrule[0.7pt](lr{0.125em}){4-5}\cmidrule[0.7pt](lr{0.125em}){6-7}  \cmidrule[0.7pt](lr{0.125em}){8-9}\cmidrule[0.7pt](lr{0.125em}){10-11}  \cmidrule[0.7pt](lr{0.125em}){12-13}\cmidrule[0.7pt](lr{0.125em}){14-15}
 &&& MSE & CR & MSE & CR & MSE& CR & MSE & CR & MSE & CR & MSE & CR \\ 
   \hline
 && NPS & 0.18 & 0.94 & 0.02 & 0.94 & 0.01 & 0.94 & 1.24 & 0.68 & 1.24 & 0.68 & 1.24 & 0.68 \\ 
 && S+ & 0.22 & 0.94 & 0.02 & 0.94 & 0.01 & 0.94 & 0.33 & 0.40 & 0.33 & 0.40 & 0.33 & 0.40 \\ 
  &0.5&PS & 0.23 & 0.93 & 0.02 & 0.93 & 0.02 & 0.93 & 1.22 & 0.76 & 1.22 & 0.76 & 1.22 & 0.76 \\ 
&&  SPOCK & 0.21 & 0.94 & 0.04 & 0.94 & 0.01 & 0.94 & 0.05 & 0.80 & 0.05 & 0.80 & 0.05 & 0.80 \\ 
  &&RHZ & 0.22 & 0.94 & 0.02 & 0.94 & 0.02 & 0.94 & 0.02 & 0.76 & 0.02 & 0.76 & 0.02 & 0.76 \\ 
  [3mm]
  &&NPS & 0.21 & 0.96 & 0.02 & 0.96 & 0.01 & 0.96 & 1.40 & 0.57 & 1.40 & 0.57 & 1.40 & 0.57 \\ 
  &&S+ & 0.36 & 0.97 & 0.02 & 0.97 & 0.01 & 0.97 & 0.69 & 0.34 & 0.69 & 0.34 & 0.69 & 0.34 \\ 
&1& PS & 0.36 & 0.94 & 0.03 & 0.94 & 0.02 & 0.94 & 1.40 & 0.61 & 1.40 & 0.61 & 1.40 & 0.61 \\ 
  &&SPOCK & 0.36 & 0.95 & 0.05 & 0.95 & 0.01 & 0.95 & 0.33 & 0.66 & 0.33 & 0.66 & 0.33 & 0.66 \\ 
 \textbf{GP}&&RHZ & 0.36 & 0.92 & 0.03 & 0.92 & 0.02 & 0.92 & 0.31 & 0.56 & 0.31 & 0.56 & 0.31 & 0.56 \\ 
  [3mm]
 && NPS & 0.15 & 0.95 & 0.06 & 0.95 & 0.09 & 0.95 & 1.37 & 0.74 & 1.37 & 0.74 & 1.37 & 0.74 \\ 
  &&S+ & 0.27 & 0.94 & 0.17 & 0.94 & 0.15 & 0.94 & 0.52 & 0.61 & 0.52 & 0.61 & 0.52 & 0.61 \\ 
  &1.3&PS & 0.18 & 0.91 & 0.10 & 0.91 & 0.04 & 0.91 & 1.29 & 0.83 & 1.29 & 0.83 & 1.29 & 0.83 \\ 
  &&SPOCK & 0.10 & 0.90 & 0.10 & 0.90 & 0.04 & 0.90 & 0.25 & 0.82 & 0.25 & 0.82 & 0.25 & 0.82 \\ 
  &&RHZ & 0.07 & 0.91 & 0.16 & 0.91 & 0.04 & 0.91 & 0.18 & 0.83 & 0.18 & 0.83 & 0.18 & 0.83 \\ 
  [3mm]
 && NPS & 0.29 & 0.96 & 0.02 & 0.96 & 0.01 & 0.96 & 1.39 & 0.59 & 1.39 & 0.59 & 1.39 & 0.59 \\ 
  &&S+ & 0.32 & 0.97 & 0.02 & 0.97 & 0.01 & 0.97 & 0.67 & 0.34 & 0.67 & 0.34 & 0.67 & 0.34 \\ 
 &2& PS & 0.33 & 0.96 & 0.02 & 0.96 & 0.01 & 0.96 & 1.40 & 0.62 & 1.40 & 0.62 & 1.40 & 0.62 \\ 
  &&SPOCK & 0.33 & 0.96 & 0.04 & 0.96 & 0.01 & 0.96 & 0.29 & 0.68 & 0.29 & 0.68 & 0.29 & 0.68 \\ 
  &&RHZ & 0.33 & 0.95 & 0.02 & 0.95 & 0.01 & 0.95 & 0.27 & 0.60 & 0.27 & 0.60 & 0.27 & 0.60 \\ 
   \hline
\end{tabular}
\end{scriptsize}
\end{table}
\end{document}